\documentclass[aps,prd,twocolumn,10pt,amsmath,reprint,superscriptaddress,amssymb,amsfonts,nofootinbib,longbibliography]{revtex4-1}

\usepackage{graphicx}
\usepackage{hyperref}
\usepackage{amssymb}
\usepackage{amsmath}
\usepackage{longtable}
\usepackage[utf8]{inputenc} 
\usepackage{color}
\usepackage{supertabular}
\usepackage{dcolumn}
\usepackage{ wasysym }
\usepackage{subfigure}
\usepackage[utf8]{inputenc}
\usepackage{xr}
\usepackage{multirow}

\begin{document}
\date{\today}
\title{Characterizing the continuous gravitational-wave signal from boson clouds around Galactic isolated black holes}
\author{Sylvia J. Zhu}
\email{sylvia.zhu@desy.de}
\affiliation{Max-Planck-Institut f{\"u}r Gravitationsphysik, Callinstra{$\beta$}e 38, D-30167, Hannover, Germany}
\affiliation{Leibniz Universit{\"a}t Hannover, D-30167 Hannover, Germany}
\affiliation{DESY, D-15738 Zeuthen, Germany}
\author{Masha Baryakhtar}
\email{mbaryakhtar@nyu.edu}
\affiliation{Center for Cosmology and Particle Physics, Department of Physics, New York University, New York, NY 10003, USA}
\author{Maria Alessandra Papa}
\email{maria.alessandra.papa@aei.mpg.de}
\affiliation{Max-Planck-Institut f{\"u}r Gravitationsphysik, Callinstra{$\beta$}e 38, 30167, Hannover, Germany}
\affiliation{Leibniz Universit{\"a}t Hannover, D-30167 Hannover, Germany}
\affiliation{University of Wisconsin-Milwaukee, Milwaukee, Wisconsin 53201, USA}
\author{Daichi Tsuna}
\email{tsuna@resceu.s.u-tokyo.ac.jp}
\affiliation{Research Center for the Early Universe (RESCEU), the University of Tokyo, Hongo, Tokyo 113-0033, Japan}
\affiliation{Department of Physics, School of Science, the University of Tokyo, Hongo, Tokyo 113-0033, Japan}
\author{Norita Kawanaka}
\affiliation{Department of Astronomy, Graduate School of Science, Kyoto University, Kitashirakawa Oiwake-cho, Sakyo-ku Kyoto, 606-8502, Japan}
\affiliation{Hakubi Center, Yoshida Honmachi, Sakyo-ku, Kyoto 606-8501, Japan}
\author{Heinz-Bernd Eggenstein}
\affiliation{Max-Planck-Institut f{\"u}r Gravitationsphysik, Callinstra{$\beta$}e 38, D-30167, Hannover, Germany}
\affiliation{Leibniz Universit{\"a}t Hannover, D-30167 Hannover, Germany}

\begin{abstract}
Ultralight bosons can form large clouds around stellar-mass black holes via the superradiance instability. Through processes such as annihilation, these bosons can source continuous gravitational wave signals with frequencies within the range of LIGO and Virgo.  If boson annihilation occurs, then the Galactic black hole population will give rise to many gravitational signals; we refer to this as the ensemble signal. We characterize the ensemble signal as observed by the gravitational-wave detectors; this is important because the ensemble signal carries the primary signature that a continuous wave signal has a boson annihilation origin. We explore how a broad set of black hole population parameters affects the resulting spin-0 boson annihilation signal and consider its detectability by recent searches for continuous gravitational waves. A population of $10^8$ black holes with masses up to $30\mathrm{M}_\odot$ and a flat dimensionless initial spin distribution between zero and unity produces up to a thousand signals loud enough to be in principle detected by these searches. For a more moderately spinning population the number of signals drops by about an order of magnitude, still yielding up to a hundred detectable signals for some boson masses. A non-detection of annihilation signals at frequencies between 100 and 1200~Hz  disfavors the existence of scalar bosons with rest energies between $2\times10^{-13}$ and $2.5\times10^{-12}$~eV.  Finally we show that, depending on the black hole population parameters, care must be taken in assuming that the continuous wave upper limits from searches for isolated signals are still valid for signals that are part of a dense ensemble: Between 200 and 300 Hz, we urge caution when interpreting a null result for bosons between 4 and $6\times10^{-13}$~eV.
\end{abstract}


\maketitle

\section{Introduction}
\label{sec:introduction}
In the last few years, transient and rapidly evolving gravitational waves
have been observed from the mergers of stellar-mass compact objects \cite{LVC_GWTC1}.
Persistent and slowly evolving sources of gravitational waves have
been predicted as well and have yet to be detected.
These continuous gravitational waves are expected to be much weaker than
the transient events and have durations much longer than the typical observation time; searches for
continuous waves integrate over long periods of time to extract signal from background.

Canonical sources of continuous waves include neutron stars with mass deformations
or internal fluid oscillation modes (see \cite{Riles2017} for a recent review).
Scenarios involving more exotic emitters of continuous waves are also being explored and
can provide evidence for --- or disfavor the presence of --- new physics beyond
the Standard Model of particle physics \cite{Arvanitaki+2010, Arvanitaki+2011,Pierce:2018xmy, Guo:2019ker, Horowitz:2019pru}. A particularly well-motivated target is the axion, proposed to solve the strong-CP problem in particle physics
\cite{Peccei+1977, Weinberg1978, Wilczek1978}; axions and axion-like particles
are also promising dark matter candidates (see e.g. \cite{Essig:2013lka} for a
review). 

Bosons such as axions or axion-like particles can form ``clouds'' with
enormous occupation numbers around rotating black holes, and do so rapidly on astrophysical timescales when the black
hole size is similar to the boson's Compton wavelength \cite{Arvanitaki+2010, Arvanitaki+2011}.
 These boson--black hole systems can be thought of as gravitational ``atoms,'' and within this
scenario, boson annihilations and level transitions
source monochromatic gravitational-wave emission \cite{Arvanitaki+2010, Arvanitaki+2011}. Annihilations in particular 
produce signals that may be loud enough to be detected in standard continuous wave searches on data from the current generation of
ground-based interferometers \cite{Arvanitaki+2015, Arvanitaki+2017, Brito+2017,  Brito+2017a}. 
Bosons with masses of
$\sim\!10^{-13}$ to $4\times10^{-12}$~eV produce annihilation signals with frequencies between $\sim\!50$ and 2000~Hz, in the LIGO and Virgo band.

The boson cloud forms by extracting energy and angular momentum from the black hole, which loses a significant fraction of its natal spin~\cite{Arvanitaki+2010, Arvanitaki+2011}.  Measurements of the spins of several old, highly rotating black
holes in X-ray binaries \cite{Miller:2014aaa,McClintock:2013vwa} have been
used to disfavor the range of masses $6\times10^{-13}~\mathrm{eV}
\lesssim\mu_\mathrm{b}\lesssim 2\times10^{-11}$~eV \cite{Arvanitaki+2015, Cardoso+2018}. 
However, since the spin measurements come with a set of systematic uncertainties,
it is very important to undertake complementary searches. 
Direct continuous wave searches for the annihilation signal in LIGO and Virgo data are an
independent test in the boson mass range disfavored by rapidly rotating black holes, and may be able to extend the reach to lower masses, as well as discover a new particle.

In order to disfavor a range of boson masses or characterize a potential continuous wave signal, it is crucial to consider the {\it{ensemble}} signal from the population of sources as a whole. The properties of the annihilation signal depend strongly on the mass, spin, distance, and age
of each black hole, and so the properties of the ensemble signal depend
on the properties of the black hole population. Moreover, to leading order, the annihilation signal frequency is set by the boson
rest mass, and so the emission from all boson clouds falls within a small
frequency range.  This is in contrast to continuous waves from neutron stars, which are expected to span
a broad range of frequencies depending on the rotation rates of the individual stars.
The clustering of signals in a narrow frequency band could reduce the effectiveness of continuous wave search methods --- which are optimized for weak, isolated
signals --- in identifying the
annihilations, and may recommend the use of other methods entirely (e.g., \cite{Brito+2017a,Tsukada+2019}).

In this paper, we study the expected boson annihilation
signal from the population of isolated black holes in the Galaxy,  of which there are expected to be up to $\sim 10^8$.
In order to take all effects into account, we use simulated populations
of $10^8$ black holes and calculate the expected signal from all of the systems
for bosons with energies between $1\times10^{-13}$ and $4\times10^{-12}$~eV,
corresponding approximately to gravitational-wave frequencies between 50 and
2000 Hz. We investigate the detectability of the ensemble signal in current LIGO data
by broad continuous gravitational wave surveys  \cite{FalconO1, Dergachev:2019oyu, RomeBosonClouds, Abbott:2017pqa}, and its dependence on black hole population assumptions.

We review the theory of boson cloud formation and continuous wave emission in Section~\ref{sec:theory} and present the Galactic isolated
black hole population in Section~\ref{sec:blackHoles}. We explore the resultant
ensemble signal from the entire population in Section~\ref{sec:results},  study its detectability in Section~\ref{sec:detectability}, compare our results to current literature in Section~\ref{sec:discussion}, and summarize in Section~\ref{sec:conclusions}.

\section{Signal model}
\label{sec:theory}

``Gravitational atoms''---macroscopic, gravitationally-bound states of ultralight bosons around astrophysical black holes---form rapidly via the superradiance instability. The bosons subsequently annihilate, sourcing coherent, monochromatic, and long-lasting gravitational waves \cite{Arvanitaki+2010,Arvanitaki+2011, Arvanitaki+2015}.
The frequency of the signal at leading order is given by twice the boson's rest energy, 
  \begin{align}\label{eqn:freq0}
    f_\mathrm{GW}^0 = \frac{2\mu_\mathrm{b}}{h} \approx 48.3~\mathrm{Hz}\left( \frac{\mu_b}{10^{-13} ~ \textrm{eV}}\right), \end{align}
with $\mu_\mathrm{b}= m_\mathrm{b} c^2$, $m_\mathrm{b} $ the boson mass, $c$ the speed of light, and $h$ the  Planck constant.
In this work, we focus on signals from gravitationally-interacting scalar (spin-0) bosons around stellar-mass black holes; we use ``black hole'' to exclusively
refer to stellar-mass black holes throughout the text. The growth and annihilation timescales of vector (spin-1) \cite{Rosa+2012, Pani+2012, Endlich+2017, Baryakhtar+2017, East+2017} and spin-2 \cite{Brito+2020} bosons
are shorter and require a separate analysis.
We only consider annihilation signals from the fastest-growing bound state and limit our analysis to isolated black holes without external effects such as accretion  or binary companions. 

We summarize the necessary background below and provide further details in App.~\ref{sec:app}; see, e.g.,~\cite{Arvanitaki+2011, Arvanitaki+2015}
for more details and~\cite{Brito+lectureNotes} for a review.

\subsection{Cloud formation}
\label{sec:cloudFormation}

Black hole superradiance is a purely kinematic process \cite{Zeldovich1971} whereby a wave
scattering off a rapidly rotating black hole increases in amplitude by
extracting some of the black hole's angular momentum~\cite{Misner1972, Starobinskii1973}. 
The gravitational potential of the black hole enables massive particle bound states, and the amplitudes of bound states that satisfy the ``superradiance condition'' increase~\cite{Damour:1976kh,Ternov:1978gq,Zouros:1979iw,Detweiler1980}. For bosons, the growth is exponential and results in a ``cloud'' with a macroscopic number of particles all occupying the same state. The initial seed can be a vacuum fluctuation, so the process need not rely on an existing abundance of bosons in the black hole's vicinity.

\begin{figure}[t!]
  \includegraphics[width=0.95\columnwidth]{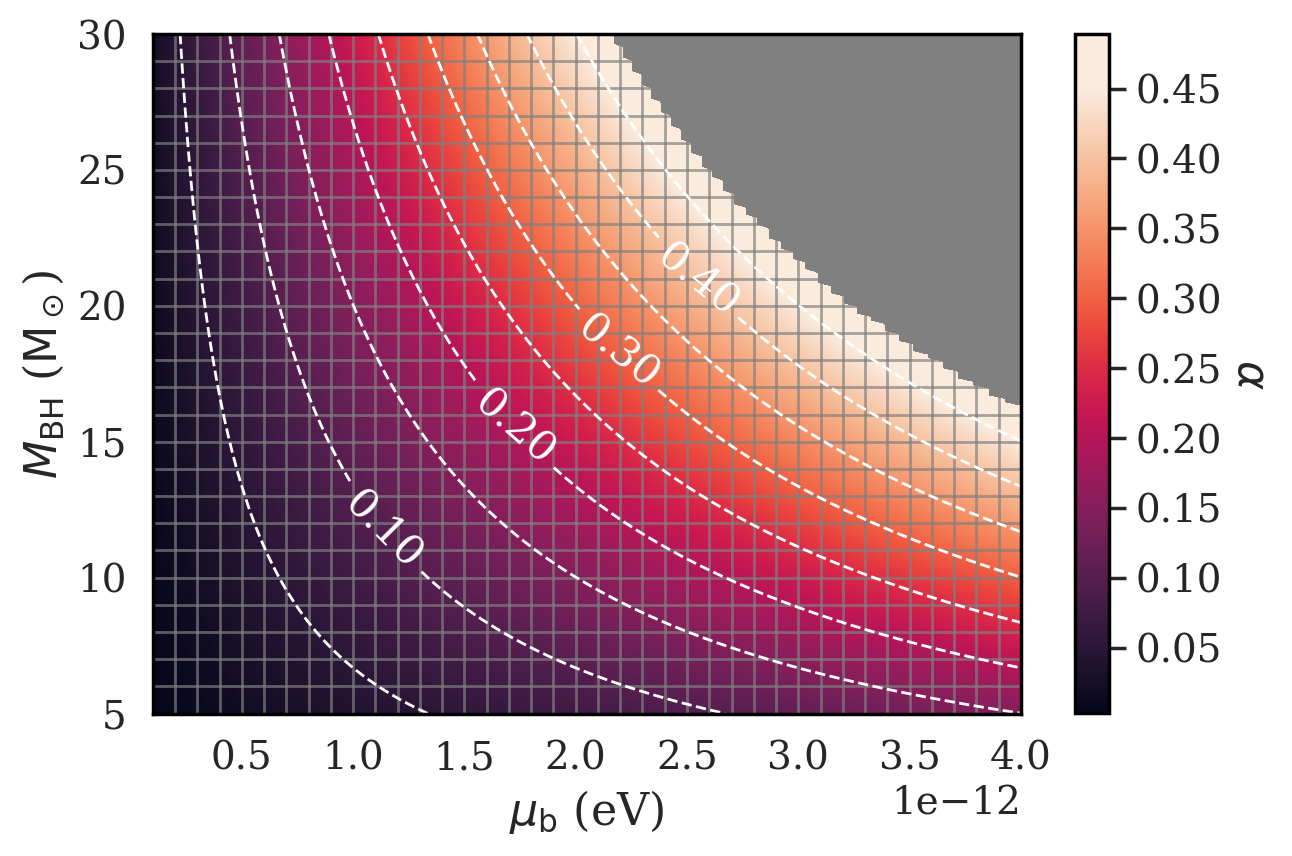}
  \caption{Gravitational ``fine structure constant'' $\alpha~\equiv~{G M_\mathrm{BH}\mu_\mathrm{b} }/{(\hbar c)}$. Given the superradiance condition, the $(n,\ell,m)=(0,1,1)$ bound state cannot form for $\alpha\geq0.5$, shown in gray. 
    }
 \label{fig:alphas}
\end{figure}

The bound states are approximated by hydrogenic wavefunctions, characterized by radial, orbital and azimuthal quantum numbers $(n, \ell, m)$ and the gravitational analog of the ``fine structure constant,'' $\alpha$, 
\begin{align}\label{eqn:alpha}
  \alpha & \equiv \frac{G M_\mathrm{BH}\mu_\mathrm{b} }{\hbar c^3}
  \approx 0.0075 \left(\frac{M_\mathrm{BH}}{10 \mathrm{M}_\odot}\right)
  \left(\frac{\mu_\mathrm{b}}{10^{-13}~\,\mathrm{eV}}\right),
\end{align}
with $G$  the gravitational constant, $\hbar=h/2\pi$, $\mathrm{M}_\odot$ the mass of the Sun, and $M_\mathrm{BH}$ the mass of the black hole  (Fig.~\ref{fig:alphas}).

\begin{figure}[t!]
  \includegraphics[width=0.95\columnwidth]{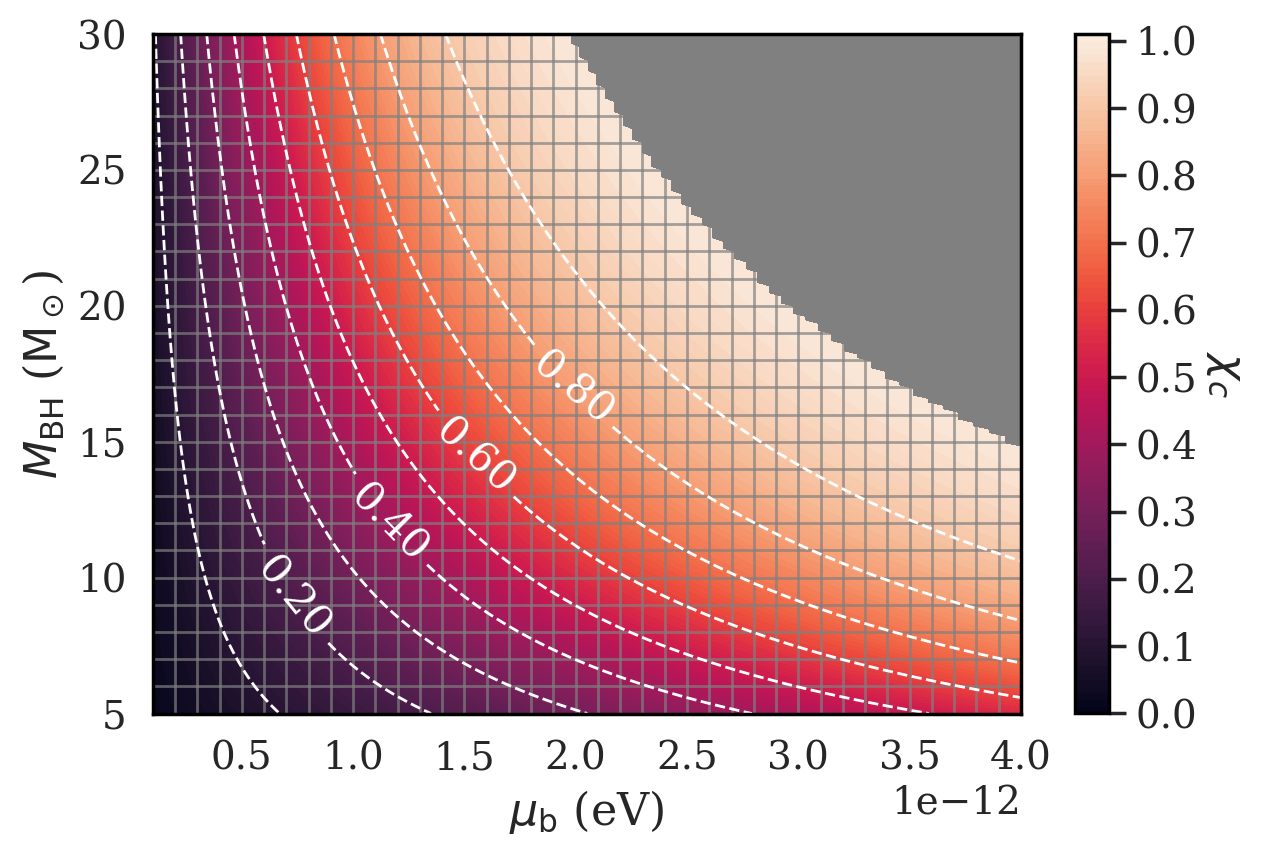}
  \caption{Critical spin $\chi_c$ for the $(n,\ell,m)=(0,1,1)$ bound state. The critical spin is a function of the gravitational fine structure  constant $\alpha$ and the bound state quantum numbers. A cloud of bosons with rest energy
    $\mu_\mathrm{b}$ will only form around a black hole of mass $M_\mathrm{BH}$
    and initial spin $\chi_i$ if $\chi_i > \chi_c$ (Eq.~\eqref{eqn:chi_c}).
    }
  \label{fig:chic}
\end{figure}

\begin{figure}[t!]
  \includegraphics[width=0.95\columnwidth]{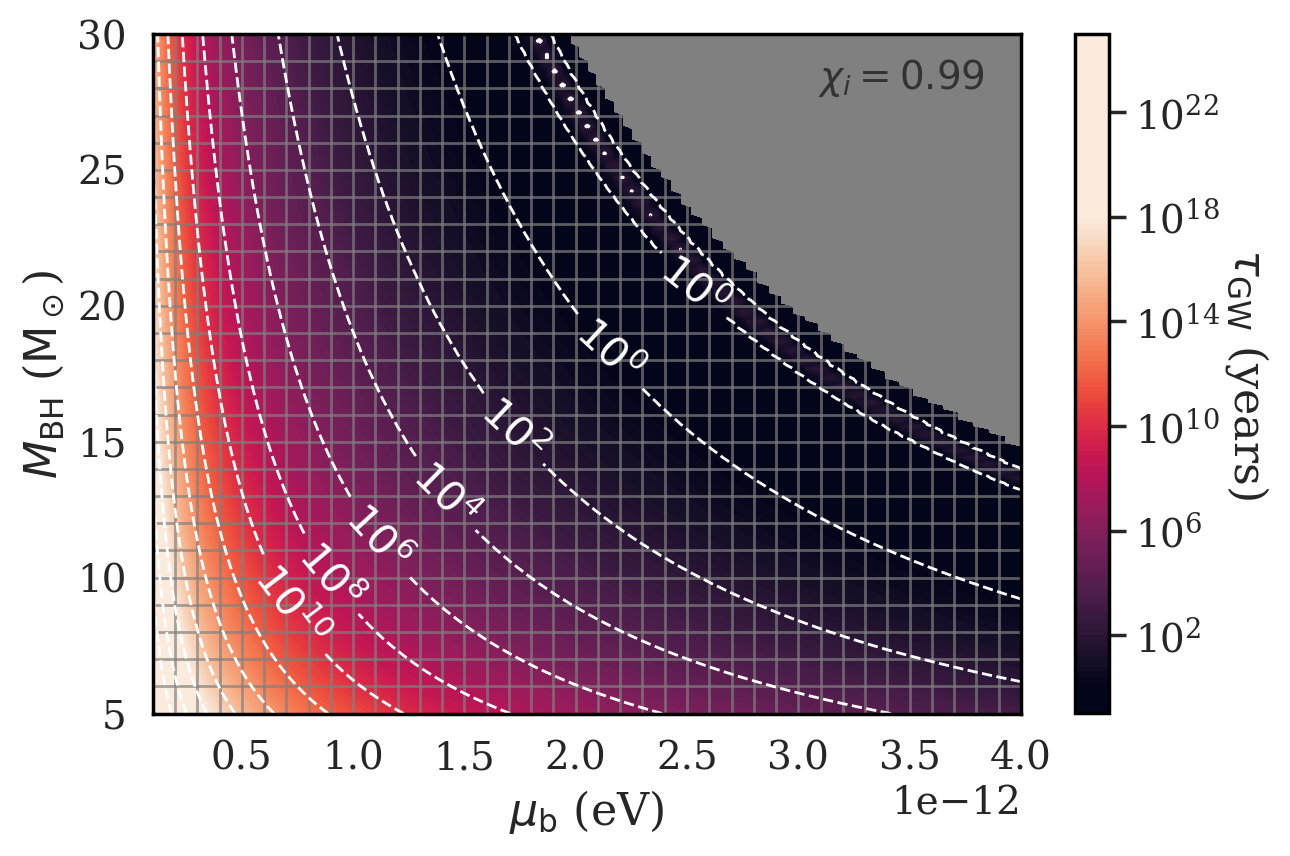}
  \caption{The gravitational-wave signal amplitude decay half-time $\tau_\mathrm{GW}$ (Eq.~\eqref{eqn:h0vsTime}) ranges from longer than the Hubble time for lighter bosons to less than 100 years for heavier bosons. The longer signals from lighter bosons are weaker compared to shorter signals from heavier bosons (Eq.~\eqref{eqn:peakh0}). Similarly to $h_0$, $\tau_\mathrm{gw}$ is a non-monotonic function of $\alpha$ above $\alpha \gtrsim 0.35$ (Sec.~\ref{app:gwsig}).}
  \label{fig:tauGWsGrid}
\end{figure}

Unless otherwise specified, we consider only the first superradiant level, with $(n, \ell, m) = (0,1,1).$\footnote{Some literature uses the principal number, $\bar{n} = n + \ell+1$ with $(\bar{n}, \ell, m) = (2,1,1)$ the fastest-growing state.}  The level will grow if the initial spin of the black hole,  $\chi_i$, satisfies the superradiance condition,
\begin{align}\label{eqn:alphac}\alpha \lesssim \frac{1}{2}\frac{\chi_i}{1+\sqrt{1-\chi_i^2}}.\end{align}
 The inequality is equivalent to the angular velocity  of the black hole exceeding the angular velocity of the cloud. For a given $\alpha$, this is true when the initial spin of the black hole $\chi_i$ is greater than the critical spin $\chi_c$ (Fig.~\ref{fig:chic}),
\begin{align}\label{eqn:chi_c}
  \chi_c  \approx \frac{4\alpha}{1+4\alpha^2}.
\end{align}
(See Eq.~\eqref{eqn:chi_c_app} for gravitational potential energy corrections and  dependence on bound state quantum numbers.) Thus, the first level will form for $\alpha\lesssim 0.5$, smaller for lower initial spins. However, only the range $0.03\lesssim\alpha\lesssim 0.2$ contributes to signals that are simultaneously loud enough (Fig.~\ref{fig:h0sGrid})  and sufficiently long-lasting (Fig.~\ref{fig:tauGWsGrid}) to be potentially detectable.
For $\alpha\in[0.03,0.2]$, $\chi_c\in[0.1,0.7]$.

The $(n, \ell, m) = (0,1,1)$ level has the largest annihilation power and the fastest
growth time, with $e$-folding time for the number of particles $\tau_{\mathrm{inst}}$ given by 
\begin{align}\label{eqn:growthTime}
  \tau_\mathrm{inst} \approx 14\,~\mathrm{days}
  \left(\frac{M_\mathrm{BH}}{10 \mathrm{M}_\odot}\right)
  \left(\frac{0.1}{\alpha}\right)^9
  \frac{1}{\chi_i},
\end{align}
at leading order in $\alpha$, see also Eq.~\eqref{eqn:time_app}. 
The instability timescale is very sensitive to the boson and black hole masses
and systems with lighter bosons and lighter black holes take
 longer to form.
 
The cloud extracts angular momentum from the black hole until $\chi\simeq\chi_c$, at which point the cloud ceases to grow. The black hole loses mass as well as spin in this process, so the values of $\alpha$ and $\chi_c$ decrease slightly as the cloud grows; we take this evolution into account as described in  App.~\ref{sec:app}. At this point, the 
masses of the black hole and of the cloud are approximately
\begin{align}\label{eqn:MBH_final}
 M_{\mathrm{BH},f} &\approx M_\mathrm{BH} \left( 1 - \alpha\left(\chi_i - \chi_c\right)\right),\\
   M_{\mathrm{cloud}} &\approx M_\mathrm{BH} \alpha \left(\chi_i - \chi_c\right),
\end{align}
where $M_\mathrm{BH}$ is the black hole mass at the onset of the cloud formation. 
For the systems of interest, $M_\mathrm{cloud}~\approx~0.1$--$5\%~M_\mathrm{BH}$, which corresponds to 
\begin{align}\label{eqn:Nc}
N\simeq\frac{M_\mathrm{cloud}}{\mu_b/c^2}\approx 10^{77}({M_\mathrm{BH}}/{10 \mathrm{M}_\odot})^2
\end{align}
particles in the cloud. The timescale to fully populate the level is then
$\ln(N)\tau_\mathrm{inst}\sim 180\,\tau_\mathrm{inst}$.

\subsection{Gravitational wave emission}
\label{sec:annihilations}

We study gravitational-wave signals emitted from the resultant bound state:
two bosons within the cloud can annihilate into a single graviton
in the presence of the black hole gravitational field. ``Transition'' signals can also occur if multiple states are simultaneously populated, but are less promising at current sensitivities \cite{Arvanitaki+2015}.
We assume the bosons only interact via gravity\footnote{For example, the QCD axion in this mass range has self-interaction scale $\gtrsim\!10^{18}~\mathrm{GeV}$ so this assumption is valid.}. The presence of self-interactions
can change the evolution and limit the size of the cloud
\cite{Gruzinov2016, Baryakhtar+2020}, or potentially cause the collapse of the cloud in a ``bosenova'' \cite{Arvanitaki+2011, Yoshino+2012, Yoshino+2015}. 
The signal from the bosons in a single macroscopic bound state is coherent, monochromatic, and evolves slowly, thus providing an ideal target for continuous wave searches. 
For lighter bosons and black holes, the signal properties are essentially unchanged over a Hubble time.

\begin{figure}[t!]
  \includegraphics[width=.95\columnwidth]{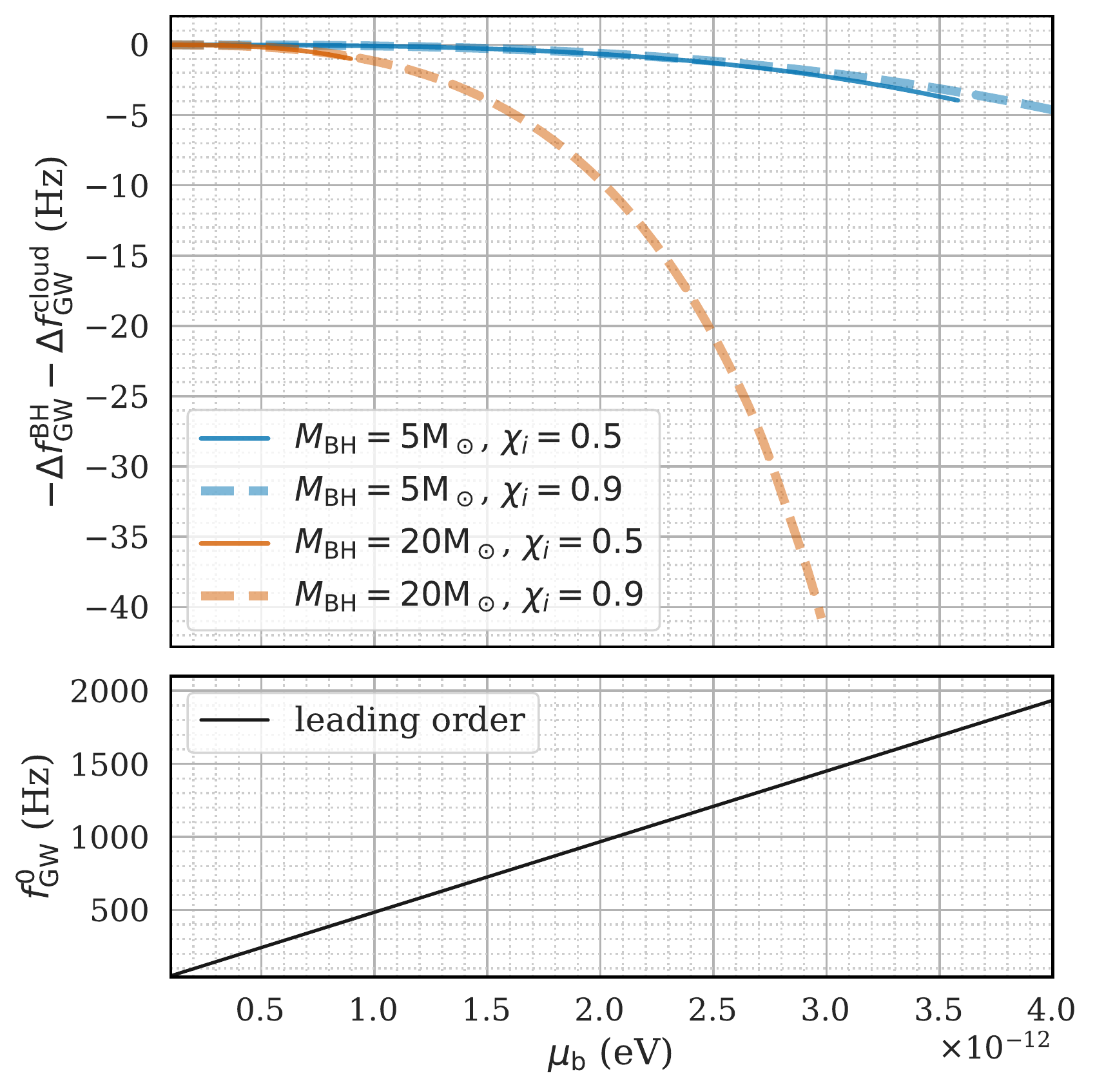}
  \caption{\textit{Bottom}: The leading-order gravitational-wave frequency $f_\mathrm{GW}^0$ is proportional to the boson rest energy $\mu_\mathrm{b}$ (Eq.~\eqref{eqn:freq}). \textit{Top}: Systems with heavier
    black holes (orange curves) produce GW signals with lower frequencies
    than do systems with lighter black holes (blue curves) due to the larger
    gravitational binding energy. Systems with larger spins (dashed curves)
    result in slightly higher frequencies than systems with smaller spins (solid curves)
    due to a positive spin-orbit energy. 
The two solid curves --- corresponding to 
    $\chi_i = 0.5$ --- stop at intermediate boson masses;
    for heavier bosons, $\chi_c > 0.5$, and the systems never form.     The contribution from the cloud self-energy is of order $10^{-3}$ or less of the black hole's gravitational potential energy.
}
  \label{fig:mbfgw}
\end{figure}

The frequency $f_\mathrm{GW}$ of the emitted gravitational wave is given by  
\begin{align}\label{eqn:freq}
  f_\mathrm{GW} =  f_\mathrm{GW}^0 - \Delta f_\mathrm{GW}^\mathrm{BH} - \Delta f_\mathrm{GW}^\mathrm{cloud},
  \end{align}
where  the corrections to the leading order frequency, Eq.~\eqref{eqn:freq0}, are due to the gravitational potential of the black hole \cite{Dolan2007, Baumann+2019}, see Fig.~\ref{fig:mbfgw}, and the cloud itself,
  \begin{align}\label{eqn:dfreq}
\Delta f_\mathrm{GW}^\mathrm{BH}&\approx  f_\mathrm{GW}^0 \left(\frac{\alpha^2}{8} + \frac{17\alpha^4}{128} - \frac{\chi_i \alpha^5}{12}\right),\\
\Delta f_\mathrm{GW}^\mathrm{cloud}&\approx  f_\mathrm{GW}^0 \left(0.2{\alpha^2}\frac{M_{\mathrm{cloud}}}{M_{\mathrm{BH}}}\right).\end{align}

The power emitted in gravitational waves for $\alpha \ll 1$ is given by \cite{Brito:2014wla},
\begin{align}\label{eqn:power}
  P_{\mathrm{GW}} \approx 0.025 \frac{c^5}{G}\alpha^{14} \frac{M_\mathrm{cloud}^2}{M_\mathrm{BH}^2}.
\end{align}
The corrections at larger $\alpha$ are significant, and the power in this regime has been computed numerically \cite{Brito+2017,Yoshino+2014}  (see Eq.~\eqref{eqn:powernum}). The gravitational waves are produced approximately in a background defined by the final black hole mass, and we use the final value of $\alpha$ to evaluate the power, strain, and timescale expressions.

\begin{figure}[t!]
  \includegraphics[width=0.95\columnwidth]{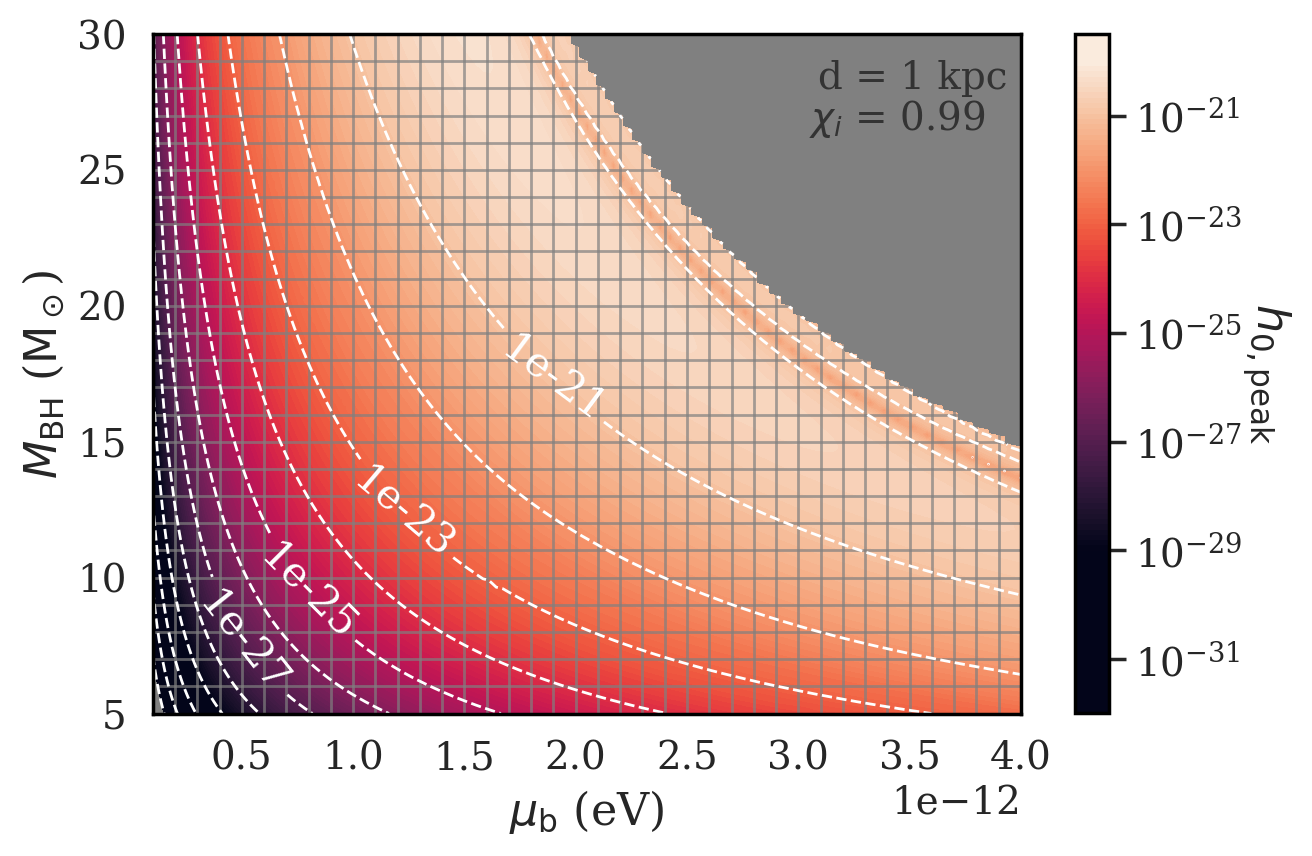}
  \caption{Maximum characteristic strain $h_{0,\mathrm{peak}}$ for a rapidly rotating black hole as a function of black hole mass $M_\mathrm{BH} $ and boson mass $\mu_\mathrm{b} $ at a distance of $d=1$~kpc. The peak strain scales as $1/d$. The value of $h_0$ increases with $\alpha$ until $\alpha \sim 0.35$; the quadrupolar gravitational wave power  and thus the observed strain changes non-monotonically at large $\alpha$ (Sec.~\ref{app:gwsig}).}
  \label{fig:h0sGrid}
\end{figure}

The characteristic strain of the annihilation signal is largest
when the cloud first reaches its maximum occupation number. At leading order,
\begin{align}\label{eqn:peakh0}
  h_{0,\mathrm{peak}} \approx 3\times10^{-24}
  \left(\frac{\alpha}{0.1}\right)^7
  \left(\frac{\chi_i-\chi_c}{0.5}\right)
  \left(\frac{M_\mathrm{BH}}{10 \mathrm{M}_\odot}\right)
  \left(\frac{1\,\mathrm{kpc}}{d}\right),
\end{align}
where $h_0$ is the maximal intrinsic gravitational wave amplitude (Eqs.~\eqref{eqn:peakh0_precise_app},\eqref{eqn:peakh0_def}).  The strain $h_{0,\mathrm{peak}}$ thus rapidly increases for larger values of $\mu_b,\,M_\mathrm{BH}$ (Fig.~\ref{fig:h0sGrid}).  For reference, recent all-sky continuous gravitational wave searches have reported $h_0$ upper limits of $\approx\!\{2\times 10^{-25}, 3 \times 10^{-25}, 6\times 10^{-25}, 1\times 10^{-24}\}$  at frequencies of $\{180,600,1000,2000\}$ Hz, respectively  \cite{O1AS20-100, LVC_O1AS, FalconO1, Dergachev:2019oyu}.

As the bosons annihilate and the cloud is depleted, the signal strength $h_0(t)$
decreases from its peak as \cite{Arvanitaki+2011}
\begin{align}\label{eqn:h0vsTime}
  h_0(t) = \frac{h_{0,\mathrm{peak}}}{1 + t/\tau_\mathrm{GW}},
\end{align}
where $\tau_\mathrm{GW} = M_{\mathrm{cloud}}^{\mathrm{peak}}c^2/  P_{\mathrm{GW}}^{\mathrm{peak}}$ is the signal ``half-time'',
\begin{align}\label{eqn:tauGW}
  \tau_\mathrm{GW} \approx 5\times10^{5}\,\mathrm{yr}
  \left(\frac{M_\mathrm{BH}}{10\mathrm{M}_\odot}\right)
  \left(\frac{0.1}{\alpha}\right)^{15}
  \left(\frac{0.5}{\chi_i-\chi_c}\right).
\end{align}

The self-binding energy of the cloud decreases with decreasing cloud mass (Eq.~\eqref{eqn:dfreq}), giving a small
\textit{positive} first frequency derivative (also see Eq.~\eqref{eqn:spinup2}),
\begin{align}\label{eqn:spinup}
  \dot{f}_\mathrm{gw}(t) \approx 0.2\alpha \frac{f_\mathrm{GW}}{\tau_\mathrm{GW}}
    \left(\frac{M_{\mathrm{cloud}}(t)}{M_{\mathrm{BH}}}\right)^2.
\end{align}
In the relevant part of parameter space, 
the frequency drift is $  \dot{f}_\mathrm{gw}\lesssim 10^{-12}$~Hz/s Fig.~\ref{fig:fdots_shrinkingCloud}),which is smaller than what current all-sky continuous wave searches can resolve
  in their initial stages
 \cite{O1AS20-100, LVC_O1AS}.
The frequency drift is discussed further in Sec.~\ref{sec:procedure}.

\subsection{Second fastest growing bound state}
\label{sec:secondEigenstate}

The second fastest growing bound state with
($n=0$, $\ell = m = 2$) has parametrically longer growth and annihilation timescales (App.~\ref{sec:app}).
It also corresponds to a lower critical spin and thus will reduce the black hole spin below
the first level's critical value $\chi_c$. The black hole then
begins absorbing the first level bosons \cite{Arvanitaki+2011,Bosch:2016vcp}, and the ``spin-up'' rate due to first level absorption balances the spin-down rate caused by
the growth of the second level. Ultimately, enough of the first level
is absorbed by the black hole that the annihilations effectively
cease, and continuous wave emission shuts off \cite{Arvanitaki+2011}. 
We take this effect into account by
setting $h_0(t) = 0$ when the second level fully populates, $\sim\ln(N) \,\tau_\mathrm{inst}^{022}$ after the black hole formation, where
\begin{align}\label{eqn:growthTime022}
  \tau_\mathrm{inst}^{022} \approx 2\times 10^{5}~\mathrm{yrs}
  \left(\frac{M_\mathrm{BH}}{10 \mathrm{M}_\odot}\right)
  \left(\frac{0.1}{\alpha}\right)^{13}
  \frac{1}{\chi_i(1+3\chi_i^2)}.
\end{align}
The cutoff time in the signal becomes important at $\alpha\gtrsim0.1$, reducing the number of signals.

The second level also produces continuous waves, and given the lower critical spin, clouds can form in systems with smaller 
$\chi_i$ or larger boson masses. However, the strain is significantly weaker than for the first level; as described in App.~\ref{sec:app}, $h_{0,\mathrm{peak}}^{011} / h_{0,\mathrm{peak}}^{022} \sim 90/\alpha^2$ \cite{Yoshino+2014}. 
In addition, the emission from the ($n=0$, $\ell = m = 2$) level is no longer dominantly quadrupolar \cite{Yoshino+2014}, which would require an ad-hoc continuous wave search. 
For these reasons, we do not consider the gravitational-wave
emission from the second fastest growing state in our study, although 
it would be interesting to do so in the future.

\section{The black hole population}
\label{sec:blackHoles}

For a given boson mass, the annihilation signal properties depend on
properties of the black hole: the distance $d$, velocity $v$, initial
$M_\mathrm{BH}$, and initial spin $\chi_i$ (Eqs.~\eqref{eqn:freq},~\eqref{eqn:peakh0}),
and age $\tau_\mathrm{BH}$. The total number of black holes in
the Galaxy is estimated to be around
$10^8$ (e.g., \cite{ShapiroTeukolsky1983,Caputo2017}); based on black hole
formation rates of $\sim 0.1-0.9$ per Milky-Way type galaxy per century, the total number
is typically estimated to vary between $10^7$ and $10^8$ \cite{Heger:2002by,Zhang:2007nw,Prantzos:2003ph,Diehl:2006cf,Olejak+2020},
with some estimates being as high as $10^9$ \cite{Agol+2002}.
The number of {\it{isolated}} black holes
is unknown, but is expected to be the same order of magnitude as the
total number of black holes (e.g., \cite{Agol+2002, Wiktorowicz+2019, Olejak+2020}) 
We therefore take $10^8$ as the benchmark number of isolated black holes in the Galaxy,
  consistent with the latest simulations of Galactic black holes \cite{Olejak+2020};
as the total number is uncertain, our results can be rescaled to draw conclusions on populations with a different number of black holes.
To manage computation time, we simulate the
positions and velocities of $10^6$ black holes and perform
a ``resampling'' to scale up to
$10^8$ isolated black holes in the Galaxy (Sec.~\ref{sec:positions}). 

\subsection{Simulating positions and velocities}
\label{sec:positions}

The spatial and velocity distribution of the current population of isolated Galactic black
holes is not known. However, since the black holes formed from massive stars at an earlier time, we approximate the black holes'
spatial distribution at birth with the current stellar distribution and
evolve their trajectories, including initial bulk velocities and natal kicks, though the Galactic gravitational potential using the procedure described in Tsuna et al.~\cite{Tsuna+2018}. The
Galactic gravitational potential is divided into the bulge, disc, and
halo components; the black holes are born in the bulge and disc according
to their respective birth rates in the regions, and their positions and
velocities are evolved through the Galactic gravitational potential. (See
\cite{Tsuna+2018} and references therein for more details.) Direct measurement of isolated Galactic black holes, such as by radio observations \cite{Tsuna+2018,Tsuna+2019}, would significantly reduce the uncertainties in their properties.

The disk is defined by a cylindrical coordinate system with radial coordinate $r$, height $z$,
and azimuth $\phi$, while the halo is defined by a spherical coordinate system with radial coordinate
$R$. The bulge is described by both cylindrical and spherical coordinate systems. All
coordinates are defined with the Galactic Center at the origin.
The black hole birth rate per unit volume in the bulge
(subscript ``b'') and disk (``d'') are given the following prescriptions
\cite{Licquia+2015, Sofue2013}:
\begin{align}
  \dot\rho_b(R) & \propto \exp(-R/R_b), \label{eqn:BHbirth_b} \\
  \dot\rho_d(r) & \propto \exp(-r/r_d) \label{eqn:BHbirth_d}
\end{align}
where $R_b = 120$~pc and $r_d = 2.15$~kpc; $\dot\rho_d$ is uniform along $z$ for $|z| < 75$~pc.
The proportionality factors in Eqs.~\eqref{eqn:BHbirth_b} and
\eqref{eqn:BHbirth_d} are determined by requiring that the total number of black holes
born be $10^6$, with 15\% born in the bulge and 85\%
in the disc \cite{Licquia+2015}. $\dot\rho_d$ is taken
to be constant over time, while $\dot\rho_b$ is nonzero and constant between 10 and 8 Gyrs ago \cite{Nataf2016}.

When a black hole is first born, it is given an initial velocity 
composed of a bulk velocity as well as an individual kick. The bulk velocity
in the disk is a piecewise function of distance from the
Galactic center, in the $\phi$ direction \cite{Irrgang+2013}:
\begin{align}\label{eqn:bulkVelocity_d}
  v_\phi\,(\mathrm{km\,s}^{-1}) = \begin{cases}
    265 - 1875(r-0.2)^2 & r < 0.2 \\
    225 + 15.625(r-1.8)^2 & 0.2 < r < 1.8 \\
    225 + 3.75(r-1.8) & 1.8 < r < 5.8 \\
    240  & r > 5.8
  \end{cases}
\end{align}
where $r$ is in kpc.  The bulk velocity in the bulge is taken from a
Maxwell-Boltzmann distribution with a mean of $130$~km/s~\cite{Kunder+2012}.

Asymmetries in the supernova explosion can impart a ``kick''
to the newly formed compact object. The natal kicks of Galactic neutron
stars are known to be as large as hundreds of km/s \cite{Fryer+1998}.
Since black holes are many times heavier than neutron stars,
relatively smaller natal kicks are expected, depending on the kick mechanism 
\cite{Nordhaus+2010, Janka2013}.
Recent studies of known
Galactic stellar-mass black holes find that their properties are
consistent with much smaller natal kicks than the ones attributed to
neutron stars \cite{Mandel2016, Atri+2019}
(although \cite{Repetto+2012} finds similar natal
  kick distributions for black holes and neutron stars).

The black holes'
initial kicks in our simulation follow a Maxwell-Boltzmann distribution with average 3D velocity of 50 km/s, the smallest average kick velocity used in \cite{Tsuna+2018}.
This means that black holes tend to stay in the Galactic component in which they were born, and few black holes migrate to high Galactic latitudes. We also consider the impact of larger kick velocities in App.~\ref{sec:appHigherNatalKicks}, finding
that natal kicks of 100 km/s give slightly reduced signal numbers due to black holes
with larger kicks having larger distance $d$ from the solar system, on
average;  $\langle1/d\rangle \sim (7.8 \,\mathrm{kpc} )^{-1}$ for the
50 km/s population  and $\langle1/d\rangle \sim (8.3 \,\mathrm{kpc}  )^{-1}$
for the 100 km/s population, corresponding to the signals from the latter population being about $10\%$
weaker on average. See App.~\ref{sec:appHigherNatalKicks} for further discussion.

After a black hole is born in the simulation,
its trajectory is determined by the Galactic gravitational potential.
The gravitational potential of the bulge, disc, and halo (subscript ``h'')
are defined as follows \cite{Irrgang+2013}:
\begin{align}
  \phi_b(r,z) & = -\frac{G M_b}{\sqrt{r^2 + \left(a_b + \sqrt{z^2 + b_b^2}\right)^2}},
    \label{eqn:gravPotential_b} \\
    \phi_d(r,z) & = -\frac{G M_d}{\sqrt{r^2 + \left(a_d + \sqrt{z^2 + b_d^2}\right)^2}},
    \label{eqn:gravPotential_d} \\
  \phi_h(R) & = -\frac{GM_h}{R_h} \ln
  \left(\frac{\sqrt{R^2 + R_h^2} + R_h}{R}\right),
    \label{eqn:gravPotential_h}
\end{align}
where $R \equiv \sqrt{r^2 + z^2}$, and 
the constants are,
\[ \begin{array}{lll}%
M_b = 4.07\times10^9 \mathrm{M}_\odot    &  a_b = 0\,\mathrm{kpc}    &  b_b = 0.184\,\mathrm{kpc}\\
M_d = 6.58\times10^{10}\mathrm{M}_\odot  &  a_d = 4.85\,\mathrm{kpc} &  b_d = 0.305\,\mathrm{kpc}\\
M_h = 1.62\times10^{12}\mathrm{M}_\odot  &  R_h = 200\,\mathrm{kpc}
\end{array}\]
as determined by fits to observed Galactic properties (e.g., rotation curves)
in \cite{Irrgang+2013}.

\begin{figure}[t!]
  \includegraphics[width=0.7\columnwidth]{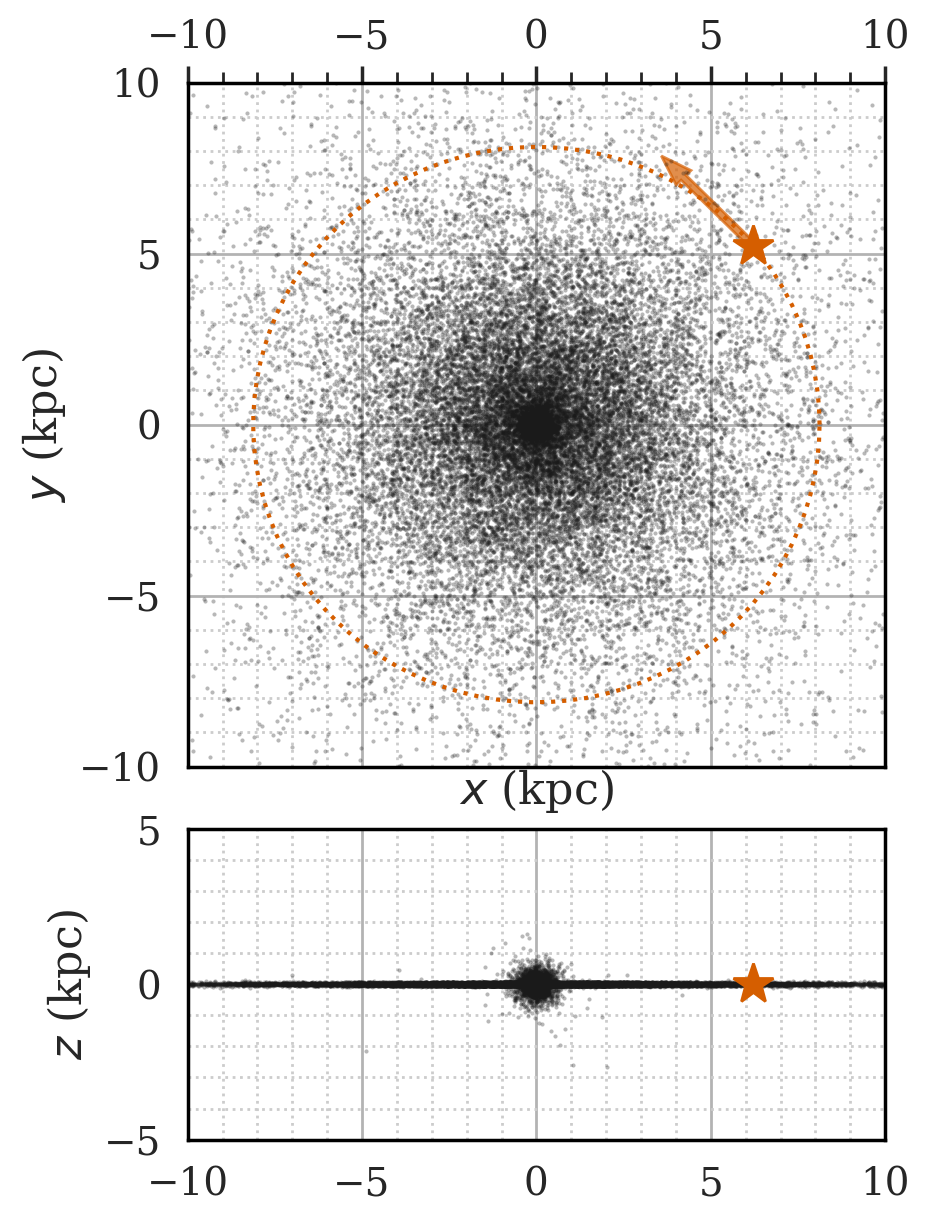}
  \caption{
Top-down (top panel) and edge-on (bottom panel) view of the Galaxy, showing a random sample of 30,000 black holes from our simulated population. The disc and bulge components are visually   distinct. Our simulations assume axial symmetry, and so the results do not change if the position of the Sun is chosen arbitrarily on the circle of distance
    $\sim\!8$~kpc from the Galactic center (orange circle). As explained in the text, we exploit this freedom to simulate $10^8$ black holes starting from a set of $10^6$. The orange star marks one sample position and the arrow shows its velocity in the disk, tangential to the circle.
    }
  \label{fig:BHlocations}
\end{figure}

\begin{figure}[t!]
  \includegraphics[width=0.9\columnwidth]{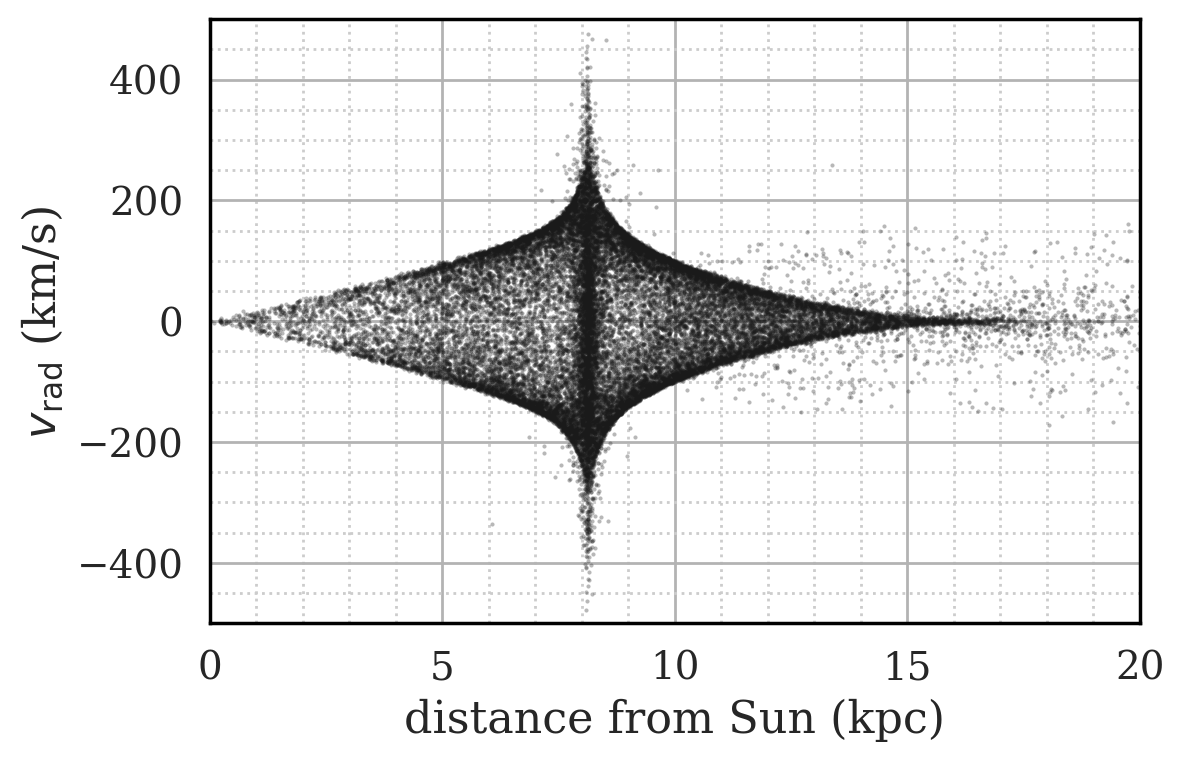}
  \caption{
Radial velocities of the black holes of Fig.~\ref{fig:BHlocations}, as a
    function of their distance from the Sun. The distribution is symmetric around zero km/s
    and is widest at 8~kpc, corresponding to the distance of the Galactic Center, as the
    bulk velocity in the bulge is uniformly distributed in direction. In contrast,
    since the bulk velocity in the disk is in the $\phi$ direction within the Galactic plane,
    the radial velocity of black holes in the disk (i.e., not near 8~kpc) has smaller magnitude.
    The scatter in
    velocities at large distances from the Sun is due to the black holes that have
  wandered into the halo and no longer follow the bulk motion of the disk.}
  \label{fig:BHvelocities}
\end{figure}

\begin{figure}[h]
  \includegraphics[width=0.9\columnwidth]{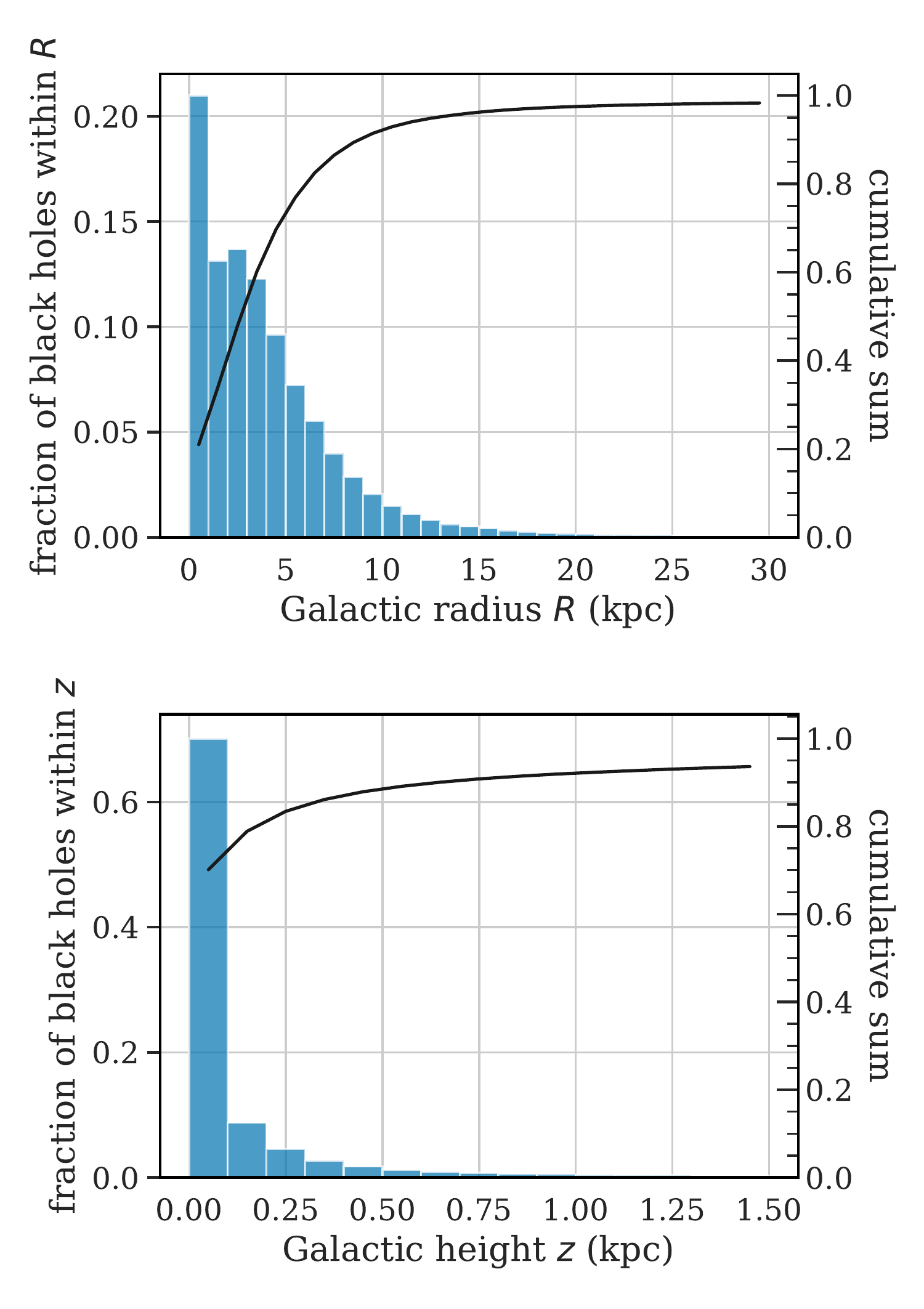}
  \caption{Most black holes in our simulation are located close to the
Galactic center and at low Galactic latitudes. These regions tend to be
    metal-rich, which can limit the maximum black hole mass
  $M_\mathrm{BH,max}$ that can form from stars.}
  \label{fig:BHspatialDistributions}
\end{figure}

With this procedure, we obtain the final positions and velocities of $10^6$ black holes
with respect to the Galactic Center, but require 
a full population of $10^8$ black holes with positions and velocities
as measured from the Solar System Barycenter (SSB). 
Since our simulations assume axial symmetry (i.e., we do
not include any structure like the spiral arms), we are free to choose
the position of the SSB to be anywhere 8.12~kpc away from the center in the disk, with
a velocity prescribed by Eq.~\eqref{eqn:bulkVelocity_d}. 
We therefore randomly select 100 points from the circle of
radius 8.12~kpc, as illustrated in Fig.~\ref{fig:BHlocations} and
 build up a sample of $10^8$ black holes at minimal
computational cost increase. For the resulting black holes, we compute the distances $d$ and
radial and tangential velocities $v_\mathrm{rad}$ and $v_\mathrm{tan}$
as measured from the SSB.

We require that the individual black holes be independent; to check that the distribution is not oversampled, 
we estimate that on average, the distance between one SSB position assignment and the next is 500 pc. In comparison, the average separation between black holes is $\mathcal{O}$(10~pc), assuming the $10^8$ black holes are uniformly
distributed in the Galactic plane.
The average separation between black holes is therefore over an order of
magnitude smaller than the average separation between SSB position assignments,
and so the local black hole populations that are ``seen'' at given
SSB position assignments are largely independent of one another.

While this procedure effectively reproduces the spatial and velocity
distribution of $10^8$ black holes using only $10^6$ simulated objects, it does
{\it not} accurately reproduce the age distribution of young black holes. Given the assumption of a constant formation rate in the last 8 Gyrs, a population of  $10^6$ black holes underestimates the number with age $10^4$ years or less compared to that which would be produced by a full simulation of $10^8$ objects. This effect is relevant for heavier bosons, which have short radiation timescales compared to the typical age of a black hole, and is discussed further in
Sec.~\ref{sec:detectableBHs}.

\subsection{Choice of mass distribution}
\label{sec:masses}

For the mass distribution, we use the Salpeter function, 
$dN/dM \propto M^{-2.35}$ \cite{Salpeter}, 
an empirically determined function that has been shown to apply to
Galactic stars, especially those more massive than the Sun \cite{Bastian+2010}.
 We choose this distribution for simplicity, as the
true mass distribution for isolated black holes is unknown. In applying the Salpeter function to black holes we implicitly ignore
effects such as mass loss due to stellar winds, which has a larger effect
on heavier stars. To better approximate the true mass distribution, we introduce  minimum
and maximum black hole masses for the population based on both observational and theoretical arguments.
We use
the same mass distribution for both the bulge and the disk
\cite{Wegg+2017}, and assume it is unchanged over the lifetime
of the Galaxy \cite{Bastian+2010}.

Two populations of stellar mass black holes have been detected.
All known Galactic black holes reside in binaries, and most are observable
in X-rays due to accreting material from their stellar companions;
these black holes tend to have masses $\lesssim\!20\mathrm{M}_\odot$ \cite{Miller:2014aaa,McClintock:2013vwa,Reynolds:2013qqa}
(with the possible exception of LB-1, which has
  been claimed to have a mass of $70\mathrm{M}_\odot$ \cite{LB1_massive},
  although this is controversial \cite{LB1_notmassive}\cite{LB1_notmassive_origauthors}). 
  In contrast,
the mass distribution of black holes in merging binaries detected by
LIGO and Virgo is consistent with a mass cutoff of approximately
$M_\mathrm{BH} \sim 40 \mathrm{M}_\odot$ and a power-law index of
around 2 \cite{Fishbach:2017zga,LVC_BBHProperties,Roulet:2018jbe};
however, it is difficult to extrapolate from the
properties of the gravitationally detected black holes, as these are found
in other Galaxies with unknown star formation histories and metallicities.
Indeed, studies suggest that low metallicity
environments are required to produce these more massive black holes
\cite{LVC_GW150914_astro, Belczynski+2016}. In contrast, most of the black
holes in our simulation are at low Galactic latitudes
and close to the Galactic Center (Fig.~\ref{fig:BHspatialDistributions}),
where stars are metal-rich  \cite{metallicity_disk, metallicity_bulge}, 
although the metallicity could have been different in the past
\cite{metallicity_evolution}.

\begin{table}[t]
\begin{tabular}{|l|c|c|}
\hline
&  $M_\mathrm{BH}$ & $\chi_{i}$ \\ \hline
standard population & $[5\mathrm{M}_\odot, 20\mathrm{M}_\odot]$ & $[0, 1.0]$ \\
heavy population & $[5\mathrm{M}_\odot, 30\mathrm{M}_\odot]$ & $[0, 1.0]$ \\
moderate spins & $[5\mathrm{M}_\odot, 20\mathrm{M}_\odot]$ & $[0, 0.5]$ \\
heavy with moderate spins & $[5\mathrm{M}_\odot, 30\mathrm{M}_\odot]$ & $[0, 0.5]$ \\
pessimistic spins & $[5\mathrm{M}_\odot, 20\mathrm{M}_\odot]$ & $[0, 0.3]$ \\
 \hline
\end{tabular}
\caption{We consider five black hole populations. The black hole mass
  $M_\mathrm{BH}$ is drawn from an $M^{-2.35}$ distribution with
  the minimum and maximum masses given by the second column (Sec.~\ref{sec:masses}).
  The initial black hole spin $\chi_i$ is drawn from a uniform distribution
  with limits given by the third column (Sec.~\ref{sec:otherAssumptions}).
}
\label{tab:BHPopulations}
\end{table}

With these considerations in mind, we  take
$[5\mathrm{M}_\odot, 20\mathrm{M}_\odot]$ as the standard
distribution of black hole masses (Table~\ref{tab:BHPopulations}).
This range includes the known Galactic black holes. A maximum mass of $20\mathrm{M}_\odot$ is
also consistent with simulations, which suggest that the heaviest
black hole that can form at Galactic metallicities is around
$20\mathrm{M}_\odot$ \cite{Belczynski+2016}.

Our results are not sensitive to the choice of minimum black hole mass for all but the heaviest boson masses; the loudest signals come from the heavier black holes, and bosons heavy enough to produce large strains around light black holes source few observable signals. See App.~\ref{app:lightBHPopulation} for more discussion on the effect of decreasing the minimum black hole mass.

The choice of maximum black hole mass more directly impacts our simulated
signals, as both the signal strength and duration are highly sensitive to
the black hole mass. Since the loudest signals come from systems with the heaviest black holes,
our choice of $20\mathrm{M}_\odot$ is conservative. To evaluate the dependence of our conclusions on the maximum mass, we also consider heavy distributions with a maximum mass of $30\mathrm{M}_\odot$ (Table~\ref{tab:BHPopulations}).

\subsection{Spin magnitude and orientation}
\label{sec:otherAssumptions}

The strain observed at the detector will in general be smaller
than the characteristic strain, $h_0$, by a factor that depends on the
geometry of the detector and the source. Relevant parameters are the
inclination angle $\iota$ (or, more directly, $\cos \iota$)
and the polarization angle $\theta$, defined by the orientation of the spin axis
of the black hole. We assume that the black hole spin axis is primarily determined by the
angular momentum axis of the progenitor star. Since there is no known favored
spin direction of stars in the Galaxy, we choose $\cos\iota$ and $\theta$ to be
uniformly distributed in $[-1,1]$ and $[-\pi/4,\pi/4]$, respectively. We
take $\phi_0$ (the phase of the gravitational wave at a chosen reference time)
to be uniformly distributed in $[0, 2\pi]$.

The distribution of black hole spin magnitudes at birth is not well understood; in particular, there is no observational data for the spins of isolated Galactic black holes.   Future analysis on the detectability of the boson ensemble signal in binaries would be interesting, as many high-spin channels are thought to occur in binary systems. Some 1D stellar evolution models indicate that a majority of black holes tend to be born with minimal spins (e.g.~\cite{Fuller+2019}); this is also compatible with the measurements of spins of pulsars extrapolated to their natal spins \cite{faucher2006birth,Miller:2014aaa}. Models also predict a potentially small fraction (order $10^{-3}$ -- $10^{-2}$) of black holes to be born with very high spins ($\chi>0.9$) through the chemically homogenous channel, thought to be associated with gamma-ray bursts \cite{Yoon:2005tv,Woosley:2005gy}. 

On the other hand, there are measurements of black hole spins in binary systems. Black holes in binaries observed to merge in LIGO have tended to low spins, although the error on measurements of initial spins is quite large; the inferred spin magnitudes of 90\% of black holes are found to be below $0.6^{+0.24}_{-0.28}$  or $0.8^{+0.15}_{-0.24}$ in the aligned or isotropic spin scenarios, respectively \cite{LVC_BBHProperties}, and typical spin magnitudes have a preferred low central value of $\sim 0.2 \pm 0.2$ \cite{Roulet:2018jbe,Wysocki:2018mpo}. Black holes in X-ray binary systems have a range of spin values, from low to $\chi>0.9$, with order half with spin above 0.5~\cite{Miller:2014aaa,McClintock:2013vwa,Reynolds:2013qqa}. While accretion can contribute to the spin-up of a black hole, the stellar companions in most of these systems are not long lived or not massive enough to substantially increase the black hole spin from its natal spin \cite{mendez2016limits,Miller:2014aaa,McClintock:2013vwa}. 

Given these uncertainties and range of observations, we take a distribution uniform in the range $[0,1]$ as a standard
 and $\chi_{i,\mathrm{max}} =0.5,\,0.3$ as moderate and pessimistic spin distributions, respectively (Table~\ref{tab:BHPopulations}). We also discuss our conclusions in the case in which a subpopulation of black holes has moderate or high spin.

\subsection{Frequency derivatives}
\label{sec:fdots}

Any astrophysical signal has a small apparent $\dot{f}$ due
to the radial velocity relative to the observer changing over time and thus a time-dependent Doppler shift,
  $\dot{f}_\mathrm{app} \simeq \frac{\Delta v_\mathrm{rad}}{c \Delta t}f$,
  where $v_\mathrm{rad}$ the radial velocity of the source, and $d$ the distance to the source.
For distances that are large compared to the
distance the black hole travels over the observational period, as is the case here, 
the change in the radial velocity can be related to the velocity tangential to the line of sight $v_\mathrm{tan}$ and distance to the source $d$ to give 
$\dot{f}_\mathrm{app} \approx v_\mathrm{tan}^2 f / cd$. 
 At the highest
frequencies, $\dot{f}_\mathrm{app} < 10^{-19}$~Hz/s for the black hole
population.

As discussed in Section~\ref{sec:annihilations}, the decreasing mass of the boson
cloud also causes a small spin-\emph{up}. For a given $M_\mathrm{BH}$, the
magnitude of $\dot{f}$ is largest for large values of $\mu_\mathrm{b}$.
For the detectable systems in our search,  the frequency derivatives fall in the range $10^{-14}$~Hz/s $ \lesssim\dot{f}< 6\times10^{-13}$~Hz/s for both the standard and heavy black hole populations
(Fig.~\ref{fig:fdots_shrinkingCloud}).

These $\dot{f}$ values are smaller than
the minimum $|\dot{f}|$ sensitivity of a typical semi-coherent all-sky search. In general, an all-sky search for the boson annihilation signal does not need to include
 frequency derivatives. On the other hand,  follow-up searches of interesting signal candidates may
need to take $\dot{f}$ into account and could detect a small positive frequency drift,  pointing toward the signal origin being a boson cloud.

\subsection{Systems outside the Milky Way}
\label{sec:otherSystems}
We have focused on the signal from Galactic black holes, as they produce the loudest signals.
We can also ask whether the black holes in the closest neighboring Galaxy,
Andromeda, will produce a detectable signal. The distance to Andromeda is around
770~kpc \cite{AndromedaDistance}. The signal strains would
drop by almost three orders of magnitude compared to what is plotted in
Fig.~\ref{fig:h0sGrid}, so that systems in the top right corner of the
plot (roughly speaking, $\mu_\mathrm{b} > 1.4\times10^{-12}$~eV and
$M_\mathrm{BH} > 15\mathrm{M}_\odot$ would produce detectable signals. These systems
correspond to $\alpha \gtrsim 0.17$ (Fig.~\ref{fig:alphas}) which
have timescales of $\tau_\mathrm{GW} < 10^4$~years (Fig.~\ref{fig:tauGWsGrid}).
It is therefore unlikely for annihilation signals from Andromeda to be
detectable by the current generation of detectors.

Closer to home, the Large (LMC)
and Small Magellanic Clouds (SMC) are potential sources of 
signals. The contribution from either of these two dwarf galaxies would be an additional
over-density of signals in a particular part of the sky and within a small range of Doppler shifts, since the black
holes  are traveling at the respective galaxy's bulk velocity
(68~km/s for the LMC and 5 km/s for the SMC \cite{LGVelocities}). The LMC and SMC are an order of magnitude further than the Galactic Center \cite{LMCDistance,SMCDistance}, so the peak strains are  an order of magnitude
below the peak of the Milky Way distribution. Since the expected signal is small and the star formation histories of the LMC and SMC are very uncertain, 
we do not include them in our simulations.

\section{The ensemble signal}
\label{sec:results}

\begin{figure*}[t!]
  \includegraphics[width=0.8\textwidth]{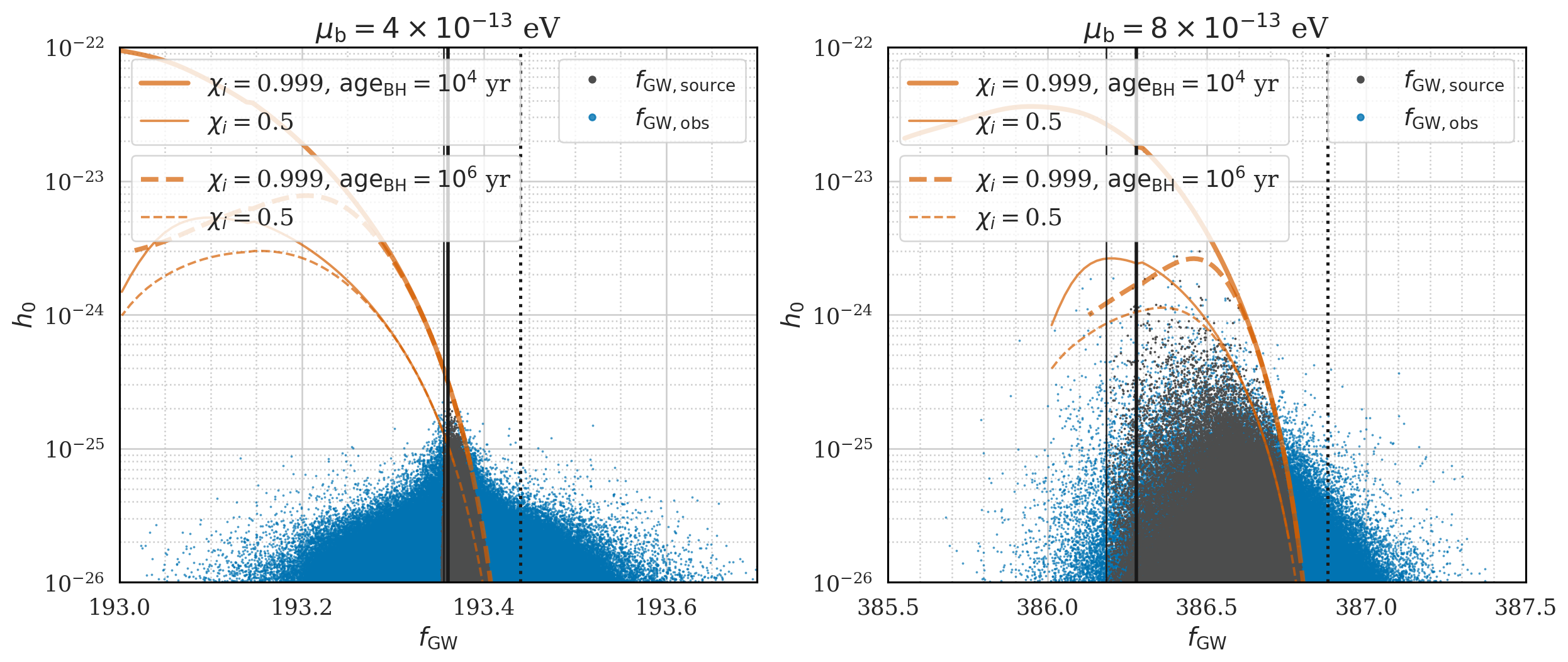}
  \caption{The distribution of signal strains as a function of source frequency (gray dots) and observed frequency with Doppler shift (blue dots). The strains are selected for signals with $d \geq 1$~kpc to facilitate comparison to the strain as a function of source frequency for a system at a distance $d=1$~kpc: the upper envelope for two sets of black hole spin and age parameters is shown in orange. The black vertical lines show the boundaries of the source frequency distribution: the vertical dotted line shows the rest mass frequency $f_\mathrm{GW}^0$, and the thick and thin vertical lines correspond to $f_\mathrm{GW,source}$ for ($20\mathrm{M}_\odot$, $\chi_i=0.999$) and ($20\mathrm{M}_\odot, \chi_i = 0.5$), respectively, yielding successively more negative potential energy corrections. (In the left panel, the two solid vertical lines lie close to each other.)}
  \label{fig:ensembleExplanation}
\end{figure*}

\begin{figure*}[t!]
  \includegraphics[width=0.95\textwidth]{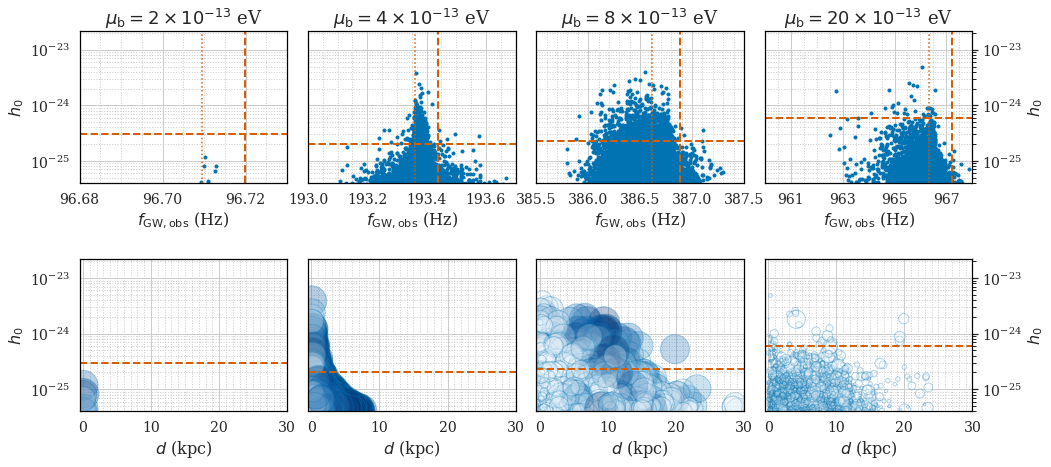}
  \caption{The ensemble signals for the ``standard'' black hole population (Table~\ref{tab:BHPopulations}), and boson mass (left to right) $\mu_\mathrm{b}~=~2\times10^{-13},~4\times10^{-13},
    ~8\times10^{-13}$, and $20\times10^{-13}$~eV.
    \textbf{Top row:} Each point represents the annihilation signal from an individual system, illustrating
    the range of strains and frequencies of the signals. The dashed horizontal red line corresponds to 
    the approximate upper limit at that frequency reported by recent all-sky continuous wave searches
    \cite{LVC_O1AS, LVC_O2AS, FalconO1,Dergachev:2019oyu}. 
    The thick dashed vertical red line marks the values of $f_\mathrm{GW}^0$, and
    the dotted red line shows $f_\mathrm{GW}$ calculated using the mode of the $M_\mathrm{BH}$ distribution
    (Fig.~\ref{fig:ensembleBHs_MassSpin}).
    \textbf{Second row:} Distribution of strains as a function of distance
    and black hole mass. Smaller, whiter circles correspond
    to lighter black holes (minimum $5 M_\odot$) and larger bluer circles
    correspond to heavier black holes (maximum $20 M_\odot$); the differences
    in circle sizes are exaggerated for visual effect and do not scale
    linearly with $M_\mathrm{BH}$. The horizontal dashed red line again represents current search upper limits.
  } 
\label{fig:ensembleExamples}
\end{figure*}
In this Section we consider the ensemble of signals produced by
annihilations of bosons with masses in the range $1\times 10^{-13}$ to $4\times 10^{-12}$ eV,
in clouds around Galactic isolated black holes. 
  To facilitate reproducibility and further studies, we provide the full set of simulated
  signal populations in an online repository~\cite{onlinerepository}.

We focus here on the standard black hole population (Table~\ref{tab:BHPopulations}).

\begin{figure*}[t!]
  \includegraphics[width=0.8\textwidth]{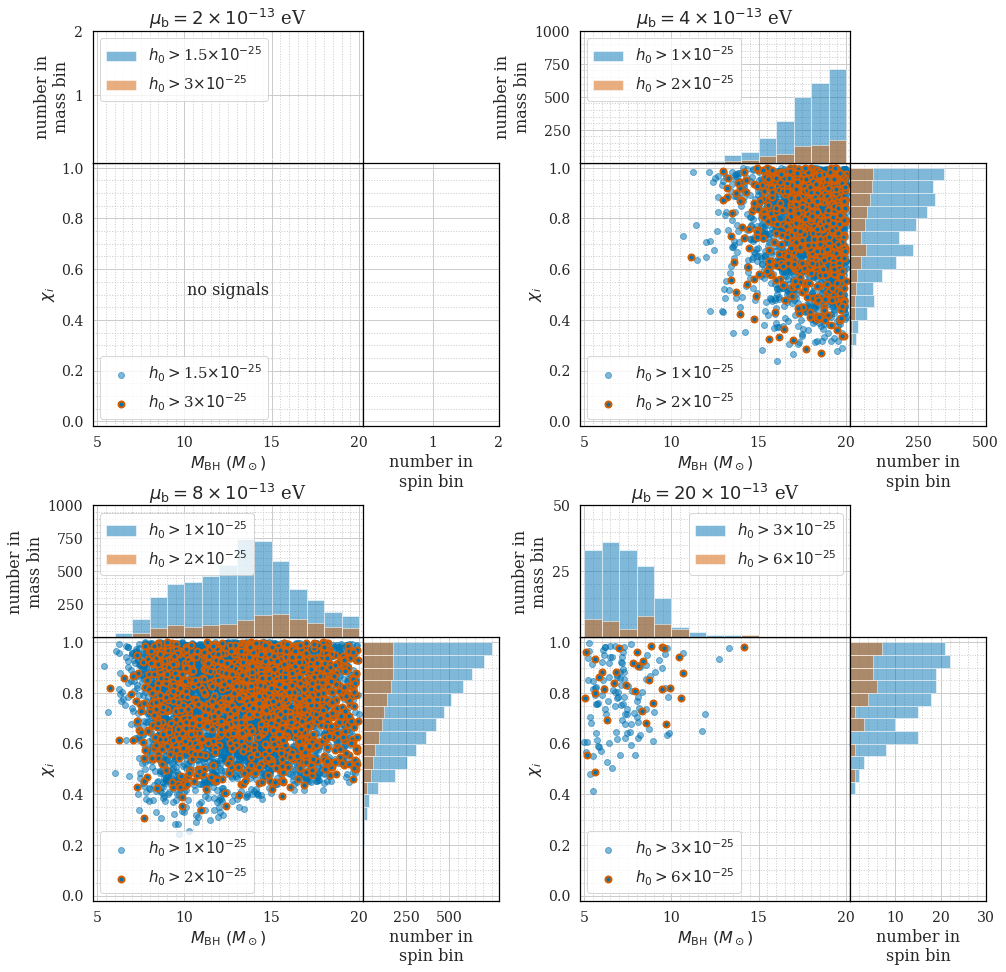}
  \caption{Each of the four panels shows the mass $M_\mathrm{BH}$ and initial spin $\chi_i$
    of the black holes that produce the ensemble signals for the four boson masses from
    Fig.~\ref{fig:ensembleExamples}, assuming the standard black hole population
    (Table~\ref{tab:BHPopulations}). In each panel, the scatterplot shows $M_\mathrm{BH}$
    and $\chi_i$ for the systems that are detectable with the current search sensitivities
    (blue circles with orange outlines) and a factor of 2 improvement in sensitivity
    (blue circles). The histograms show the distributions of $M_\mathrm{BH}$ (top) and
    $\chi_i$ (right) for these sets of systems. The black holes are assigned spins uniformly
    distributed in $[0,1]$ and only systems satisfying Eq.~\eqref{eqn:chi_c} will
    form clouds. The upper-left set of plots ($\mu_\mathrm{b} = 2\times10^{-13}$~eV)
      contains no signals at these values of $h_0$, suggesting that a
      larger increase in sensitivity is necessary
        to detect the annihilation signals produced by the lightest bosons.}
  \label{fig:ensembleBHs_MassSpin}
\end{figure*}

\begin{figure*}[t!]
  \includegraphics[width=0.8\textwidth]{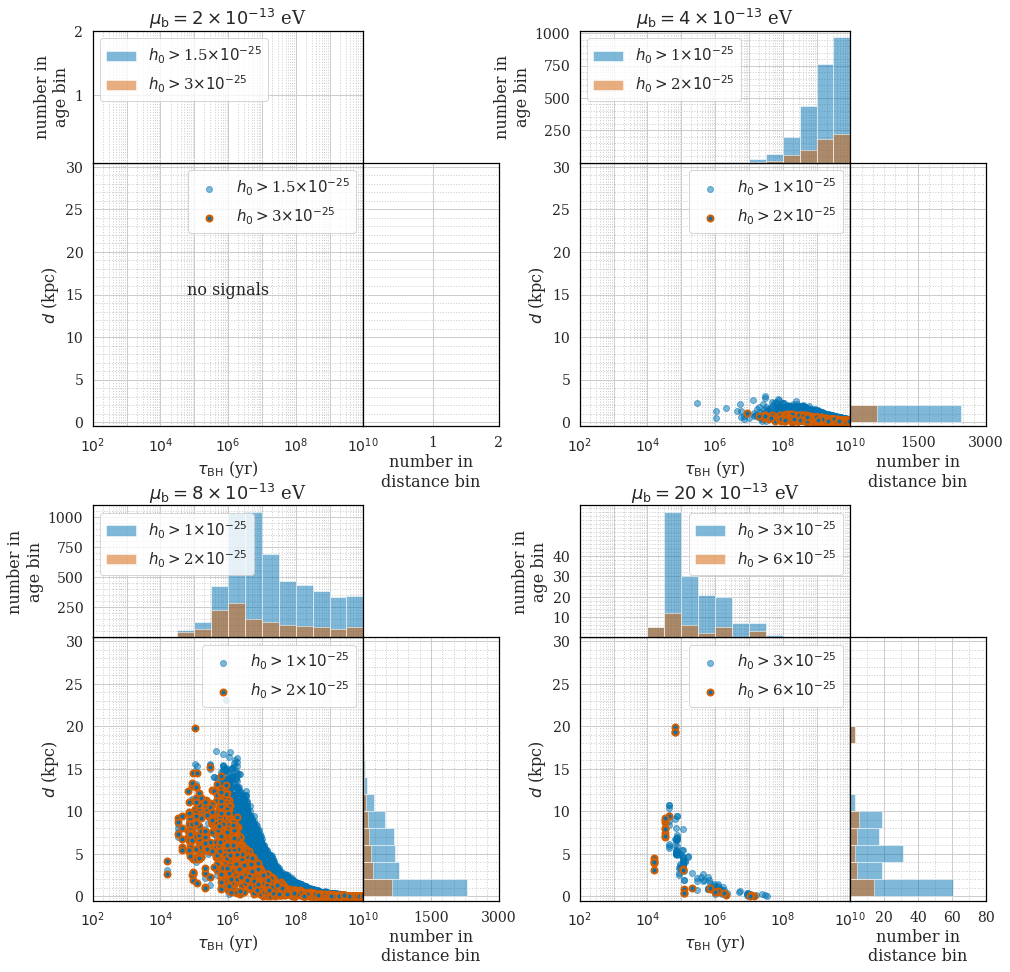}
  \caption{Each of the four panels shows the age $\tau_\mathrm{BH}$ and
    distance $d$ of the black holes that
    produce the ensemble signals for the four boson masses in Fig.~\ref{fig:ensembleExamples},
    assuming the standard black hole population (Table~\ref{tab:BHPopulations}).
    In each panel, the scatterplot shows $\tau_\mathrm{BH}$ and $d$ for the systems that
    are detectable with the current search sensitivities (blue circles with orange outlines) and
    a factor of 2 improvement in sensitivity (blue circles). 
    The histograms show the distributions of $\tau_\mathrm{BH}$ (top) and $d$ (right)
    for these sets of systems. As in Fig.~\ref{fig:ensembleBHs_MassSpin}, the upper-left set 
    of plots ($\mu_\mathrm{b} = 2\times10^{-13}$~eV)
      contains no signals at these values of $h_0$.}
  \label{fig:ensembleBHs_AgeDistance}
\end{figure*}

\subsection{Signal simulation procedure}
\label{sec:procedure}

Each of the $10^{8}$ black holes is tagged with a unique position, velocity, and age.
For a given run (i.e., choice of $\mu_\mathrm{b}$), we
randomly assign  a mass and
spin to each black hole as described in Sections~\ref{sec:masses} and \ref{sec:otherAssumptions} and summarized in Table~\ref{tab:BHPopulations}.
For each boson-black hole system,
we compare the initial spin $\chi_i$ to the critical spin $\chi_c$ (Eq.~\eqref{eqn:chi_c}) to determine
whether the cloud forms. If $\chi_i > \chi_c$, we calculate the instability
timescale $\tau_\mathrm{inst}$ (Eq.~\eqref{eqn:growthTime}) which defines the
e-folding time for the cloud to grow, and take $\ln(N)\tau_\mathrm{inst}
\approx 180 \tau_\mathrm{inst}$ as the timescale for
the cloud to reach its maximum size (discussed in Sec.~\ref{sec:cloudFormation}),
keeping systems for which this time is less than the black hole age. 

To check whether the second level has fully formed and therefore caused the
continuous wave emission from the first level to cease, we also calculate $\tau_\mathrm{inst}^{022}$ for the
($n$ = 0, $\ell$ = $m$ = 2) level; for systems with black holes 
older than $\ln(N_{022})\tau_\mathrm{inst}^{022}$,
we set $h_0 = 0$ since the first level is no longer radiating.

For systems that pass these checks, we calculate $h_{0,\mathrm{peak}}$
and $\tau_\mathrm{GW}$ (Eqs.~\eqref{eqn:peakh0_precise_app},~\eqref{eqn:tauGW_precise_app}),
and use the time evolution of  $h_0$ (Eq.~\eqref{eqn:h0vsTime}) to determine its current value.
The relevant time $t$ is the time since cloud formation minus
  the time $d/c$ required for the GW emission to travel the distance $d$ to Earth; if
 the emission since cloud formation has not yet reached Earth, we set $h_0 = 0$.
We compute the frequency
$f_\mathrm{GW}$ at cloud formation, Eq.~\eqref{eqn:freq}, and apply the frequency drift (Eq.~\eqref{eqn:spinup}) over
the lifetime of the cloud, which is the black hole age minus $\ln(N)\tau_\mathrm{inst}$.
Finally, we determine the apparent signal  frequency by applying a Doppler shift
\begin{align}\label{eqn:DopplerShift}
  f_\mathrm{GW, obs} &= \left(1 - \frac{v_\mathrm{rad}}{c}\right) f_\mathrm{GW} \approx
  \left(1 \pm 10^{-4}\right) f_\mathrm{GW},
\end{align}
where $f_\mathrm{GW, obs}$ is the observed gravitational-wave frequency
due to the source's radial velocity as measured at the SSB. The maximum value of $\left|v_\mathrm{rad}/c\right|$ for our
black hole population is 0.0025, and 90\% of black holes have values of
$\left|v_\mathrm{rad}/c\right| < 6 \times 10^{-4}$.

In Table~\ref{tab:exampleSystems}, we list a set of example systems
  representative of the parameter space on which this study focuses.

\begin{table*}[t]
\begin{tabular}{|c|c||c|c|c|c|c|c|c|}
\hline
$\mu_\mathrm{b}$ (eV) &  $M_\mathrm{BH}$ ($\mathrm{M}_\odot$)
  & $\alpha_i$ & $\chi_{c,i}$ & $\tau_\mathrm{inst}$ (yr) & $\tau_\mathrm{inst}^{022}$ (yr) & $h_{0,\mathrm{peak}}$ at 1 kpc & $\tau_\mathrm{GW}$ (yr) & $f_\mathrm{GW}$ (Hz) \\ \hline
\multirow{4}{*}{$2\times10^{-13}$}
& 5 & 0.00749 & 0.030 & $5.3\times10^{10}$ & $5.9\times10^{23}$ & $6.4\times10^{-32}$ & $3.2\times10^{21}$ & $96.72$ \\
& 10 & 0.0150 & 0.060 & $2.2\times10^{8}$ & $7.9\times10^{19}$ &  $1.5\times10^{-29}$ & $2.3\times10^{17}$ & $96.72$\\
& 20 & 0.0300 & 0.12 & $1.0\times10^{6}$ & $1.1\times10^{16}$ & $3.2\times10^{-27}$ & $1.9\times10^{13}$ & $96.71$\\
& 30 & 0.0449 & 0.18 & $4.5\times10^{4}$ & $6.8\times10^{13}$ & $6.8\times10^{-26}$ & $8.7\times10^{10}$ & $96.70$ \\
\hline
\multirow{4}{*}{$5\times10^{-13}$}
& 5 & 0.0187 & 0.075 & $1.5\times10^7$ & $1.8\times10^{18}$ & $3.4\times10^{-29}$ & $4.3\times10^{15}$ & 241.8\\
& 10 & 0.0374 & 0.15 & $7.2\times10^4$ & $2.7\times10^{14}$ & $6.9\times10^{-27}$ & $3.9\times10^{11}$ & 241.8\\
& 20 & 0.0749 & 0.29 & $4.1\times10^2$ & $4.6\times10^{10}$ & $1.2\times10^{-24}$ & $4.7\times10^{7}$ & 241.7\\
& 30 & 0.112 & 0.43 & $2.4\times10^1$ & $3.1\times10^{8}$ & $2.0\times10^{-23}$ & $2.9\times10^5$ & 241.5\\
\hline
\multirow{4}{*}{$8\times10^{-13}$}
& 5 & 0.0300 & 0.12 & $2.5\times10^{5}$ & $2.8\times10^{15}$ & $7.9\times10^{-28}$ & $4.7\times10^{12}$ & 386.8\\
& 10 & 0.0599 & 0.24 & $1.3\times10^3$ & $4.6\times10^{11}$ & $1.4\times10^{-25}$ & $5.1\times10^{8} $ & 386.7\\
& 20 & 0.120 & 0.45 & $9.6\times10^0$ & $8.8\times10^7$ & $1.8\times10^{-23}$ & $9.4\times10^4 $ & 386.3\\
& 30 & 0.180 & 0.63 & $7.9\times10^{-1}$ & $6.1\times10^5$ & $2.7\times10^{-22}$ & $7.1\times10^2 $ & 385.5\\
\hline
\multirow{4}{*}{$1.5\times10^{-12}$}
& 5 & 0.056 & 0.22 & $1.1\times10^3$ & $5.4\times10^{11}$ & $4.7\times10^{-26}$ & $6.3\times10^8$ & 725.1\\
& 10 & 0.112 & 0.43 & $7.8\times10^0$ & $1.0\times10^8$ & $6.7\times10^{-24}$ & $9.6\times10^4$ & 724.4\\
& 20 & 0.225 & 0.74 & $1.4\times10^{-1}$ & $2.0\times10^4$ & $5.7\times10^{-22}$ & $2.7\times10^1$ & 721.2\\
& 30\rlap{$^*$} & 0.34 & 0.92 & n/a & n/a & n/a & n/a & n/a \\
\hline
\multirow{4}{*}{$2.0\times10^{-12}$}
& 5 & 0.0749 & 0.29 & $10.\times10^1$ & $1.1\times10^{10}$ & $2.9\times10^{-25}$ & $1.2\times10^{7}$ & 966.6\\
& 10 & 0.150 & 0.55 & $9.1\times10^{-1}$ & $2.3\times10^6$ & $2.9\times10^{-23}$ & $3.3\times10^{3}$ & 964.7\\
& 20 & 0.299 & 0.88 & $7.1\times10^{-2}$ & $3.4\times10^2$ & $7.9\times10^{-22}$ & $2.5\times10^0$ & 956.0\\
& 30\rlap{$^*$} & 0.449 & 0.99 & n/a & n/a & n/a & n/a & n/a\\
\hline
\multirow{4}{*}{$3.5\times10^{-12}$}
& 5 & 0.131 & 0.49 & $1.2\times10^{0}$ & $6.6\times10^{6}$ & $7.6\times10^{-24}$ & $7.4\times10^{3}$ & 1689\\
& 10 & 0.262 & 0.82 & $3.4\times10^{-2}$ & $1.2\times10^{3}$ & $4.9\times10^{-22}$ & $2.5\times10^{0}$ & 1679\\
& 20\rlap{$^\dagger$} & 0.52 & n/a & n/a & n/a & n/a & n/a & n/a\\
& 30\rlap{$^\dagger$} & 0.79 & n/a & n/a & n/a & n/a & n/a & n/a\\
\hline
\end{tabular}
\caption{Representative values of relevant parameters and timescales for a selection of boson and black hole masses of interest. 
  All black holes in the table possess initial
  spins of $\chi_i = 0.9$. The dominant effect of $\chi_i$ is the determination of whether the cloud forms; the approximate formulations for the cloud and gravitational-wave parameters have linear dependencies
  on $\chi_i$. The listed values of $\alpha$ and $\chi_c$ correspond to the initial values at the time of black hole formation. $\tau_\mathrm{GW}$ is the decay rate of the gravitational-wave signal strength from its peak $h_{0,\mathrm{peak}}$
    while $\tau_\mathrm{inst}^{022}$ is a strict cutoff on any gravitational-wave emission ($h_0 = 0$ at $t = \tau_\mathrm{inst}^{022}$).
    Black hole masses marked with $^*$ indicate that $\chi_i < \chi_{c,i}$ for these systems, and those marked
with $^\dagger$ indicate that $\alpha > 0.49$; the cloud does not form in either scenario.}
\label{tab:exampleSystems}
\end{table*}

\subsection{Ensemble signal properties}
\label{sec:morphology}

{\bf Ensemble signal shape:} The source-frame frequency of the GW signal, $f_\mathrm{GW,source}$, has a small
dependence on $M_\mathrm{BH}$ through $\alpha$, resulting in different
`potential energy' corrections for different black hole masses.
For a fixed boson mass, increasing the black hole mass produces a signal with a lower frequency (Eq.~\eqref{eqn:dfreq}) and a higher peak strain (Eq.~\eqref{eqn:peakh0}). Thus, in a given ensemble,
the signals with lower frequencies have higher strains at cloud formation. The signal half-time $\tau_{\mathrm{GW}}$ decreases rapidly with increasing black hole mass (Eq.~\eqref{eqn:tauGW}).  If  $\tau_{\mathrm{GW}}$ is shorter than the black hole age $\tau_\mathrm{BH}$, the signal strain today is suppressed by $\tau_\mathrm{GW}/\tau_\mathrm{BH}$; increasing the black hole mass above the value that gives $\tau_\mathrm{GW}\sim\tau_\mathrm{BH}$ increases the peak strain but decreases the strain at the current time. The current strain as a function of frequency for a fixed black hole age and distance as seen in the source frame is shown in the orange curves of Fig.~\ref{fig:ensembleExplanation}. 

The maximum frequency in an ensemble in the source frame is $f_\mathrm{GW}^0$, while the minimum value depends on multiple factors: there are no signals below a certain source frequency if 
  a) there are no heavier black holes due to the imposed maximum black hole mass cutoff; b) heavier black holes result in prohibitively large critical spins  $\chi_c$ (Eq.~\eqref{eqn:chi_c})  hence systems with heavier black holes never form; 
or c) heavier black holes have short superradiance times for the $\ell=m=2$ level, $\tau^{022}_{\mathrm{inst}}$ (Eq.~\eqref{eqn:time_app}), cutting off the signal before today.  The ensemble-signal shape as seen in the source frame is shown in the gray dots in Fig.~\ref{fig:ensembleExplanation}.

The line-of-sight BH velocity  produces an additional Doppler shift for each source, proportional to $v_\mathrm{rad}/c$, which ``smears'' the ensemble signal's distribution in frequency (blue dots in Fig.~\ref{fig:ensembleExplanation}). Nearby black holes produce stronger signals than farther ones, while farther black holes tend to have larger line-of-sight velocities (Fig.~\ref{fig:BHvelocities}). This means that weaker (more distant) signals tend to have larger relative Doppler shifts.
 The result is that, at a fixed boson mass, the ensemble of signal frequencies observed at the detector ($f_\mathrm{GW,obs}$) is shaped roughly like a triangle, with a wider base and a central `peak'.
 In Fig.~\ref{fig:ensembleExamples} (top row) we show the ensemble signals from the standard population of $10^8$ black holes at a few illustrative boson masses.

The spread in $f_\mathrm{GW,obs}$ is greater for heavier bosons for several reasons. First, heavier bosons produce GW emission with larger values of $f_\mathrm{GW}^0$, and as both the Doppler shift (Eq.~\eqref{eqn:DopplerShift}) and potential energy shift (Eq.~\eqref{eqn:dfreq}) terms are proportional to $f_\mathrm{GW}^0$, higher frequencies result in a larger \textit{absolute} frequency spread, $\Delta f_\mathrm{GW}$. In addition, increasing the boson mass from $2\times 10^{-13}$~eV to $8\times 10^{-13}$~eV increases the \textit{relative} frequency spread, $\Delta f_\mathrm{GW}/ f_\mathrm{GW}^0$ for two reasons: Within this mass range, 1) heavier bosons form systems with a larger range of black hole masses (Fig.~\ref{fig:ensembleBHs_MassSpin}), leading to a larger range
of potential energy corrections (Eq.~\eqref{eqn:dfreq}), and  2) heavier bosons produce detectable signals at larger distances (Fig.~\ref{fig:ensembleExamples}, bottom row), resulting in larger Doppler corrections.  Increasing the boson mass further starts to reduce these effects; the relative frequency spread decreases while the absolute frequency spread continues to increase.
In all cases, the signal frequencies lie in the range $f\in f_\mathrm{GW}^0\times\{1-0.01, 1+0.0025\}$,
where the upper end of the range is given by the maximum Doppler shift in the ensemble and the lower end of the range  is given by the largest negative Doppler shift plus the largest observed potential energy correction with $\alpha \sim 0.25$.

{\bf  Maximum signal strength:} The power emitted by the  boson cloud in gravitational waves is set by  the total energy of the cloud, and the timescale $\tau_{GW}$ over which the annihilations take place,  $P\sim M_{\mathrm{cloud}}/\tau_{\mathrm{GW}}$, providing an estimate of the maximum strain in the detector today:
\begin{align}
h_{0} &\lesssim\left(\frac{5 G M_{\mathrm{cloud}}/\tau_{\mathrm{GW}}}{2\pi^2 c\,f_\mathrm{GW}^2r^2}\right)^{1/2} \nonumber\\
&\lesssim 10^{-23} \left(\frac{M_{\mathrm{cloud}}}{0.05M_{\mathrm{BH}}}\right)^{1/2}\left(\frac{M_{\mathrm{BH}}}{20\mathrm{M}_\odot}\right)^{1/2}\left(\frac{4\times 10^{-13}~\mathrm{eV}}{\mu_\mathrm{b}}\right)\nonumber\\
&\quad\times \left(\frac{10^5~\mathrm{years}}{\tau_{\mathrm{GW}}}\right)^{1/2}\left(\frac{2 \mathrm{kpc}}{r}\right).
\end{align}
Here we have chosen near-maximal parameters for the cloud and black hole mass, as well as the lightest axion mass that produces observable signals; note that $M_{\mathrm{cloud}}$ and $\tau_{\mathrm{GW}}$ are not independent parameters and cannot reach the optimal values for most ($\mu_{\mathrm{b}},~ M_{\mathrm{BH}}$) combinations. In addition, given the finite birth rate of black holes and their spatial distribution, the nearby black holes tend to be older, while the small number of young black holes are on average farther away (Fig.~\ref{fig:ensembleBHs_AgeDistance}).
We also see that black holes younger than $10^5$ years contribute to the potentially detectable signal for heavier bosons, but the strain is reduced due to the higher signal frequency and farther typical black hole distance. Thus it is very unlikely to find strains larger than $h_{0} \sim10^{-23}$.
The typical distribution from black holes at $\gtrsim\!1 $~kpc  as well as the analytical strain amplitudes for different black hole age and spin parameter choices are shown in Fig.~\ref{fig:ensembleExplanation}.

\subsection{Properties of black holes in potentially detectable systems}
\label{sec:detectableBHs}

The detectability of a signal depends on the exact search involved. As a guideline, we take the approximate search sensitivities of recent all-sky searches in Advanced LIGO data \cite{FalconO1, Dergachev:2019oyu, RomeBosonClouds} at the relevant frequencies.

{{\bf Black hole distances:}} By construction, all the signals in the ensemble arise from black holes within the Milky Way (Fig.~\ref{fig:ensembleBHs_AgeDistance}). 
The number of black holes increases rapidly as a function of distance, and in general, many distant black holes produce signals below the typical search sensitivity (Fig.~\ref{fig:ensembleExamples}). 

If the maximum black hole mass $M_\mathrm{BH,max} = 20\mathrm{M}_\odot$, then for the lighter bosons, the signals are very weak and are only observable from short distances (e.g., less than 2 kpc for  $\mu_\mathrm{b} = 4\times10^{-13}$~eV). Increasing the boson mass increases the signal power, and at $\mu_\mathrm{b} = 8\times10^{-13}$~eV, signals at 10 kpc are observable, although
most signals are still produced by systems within 2~kpc. If we allow for $M_\mathrm{BH,max} = 30\mathrm{M}_\odot$, systems as far as 15~kpc are detectable for $\mu_\mathrm{b} \geq 8\times10^{-13}$~eV (Fig.~\ref{fig:ensembleBHs_ageDistance_heavy}).

{{\bf Black hole masses:}} We consider black holes with masses between $5\mathrm{M}_\odot$ and $20\mathrm{M}_\odot$ ($30\mathrm{M}_\odot$) in our standard (heavy) population. For lighter bosons, the signals are weak and the distribution of black holes is therefore peaked toward the heavier black holes (Fig.~\ref{fig:ensembleBHs_MassSpin}). The signal strain increases with increasing black hole mass, $h_0\propto M_\mathrm{BH}^8$ (Eq.~\eqref{eqn:peakh0}),  and the total number of signals is thus sensitive to the upper cutoff of the mass distribution. This is especially evident for $\mu_\mathrm{b} = 2\times10^{-13}$~eV; while the standard population produces no signals with $h_0>1.5\times10^{-25}$ (Fig.~\ref{fig:ensembleBHs_MassSpin}), the heavy black hole population produces tens of such signals (Fig.~\ref{fig:ensembleBHs_massSpin_heavy}).

In contrast, for heavier bosons, there are fewer signals from heavier black holes because of prohibitively high critical spins $\chi_c$ or short instability ($\tau_\mathrm{inst}^{022}$) and gravitational wave ($\tau_\mathrm{GW}$) timescales, as discussed in Section~\ref{sec:morphology}. 
Because of these effects, for $\mu_\mathrm{b} \gtrsim 8\times10^{-13}$~eV, the fraction of signals with black holes masses of $20\mathrm{M}_\odot$ or above goes to zero (Fig.~\ref{fig:ensembleBHs_MassSpin}).

The effect of the minimum black hole mass is relatively weak: we find that  $5-6\mathrm{M}_\odot$ black holes do not contribute any signals above current search sensitivities for  boson masses up to $8\times10^{-13}$~eV.  Black holes as light as $5\mathrm{M}_\odot$ can form systems with $\alpha >0.075$ for $\mu_\mathrm{b} \gtrsim 2\times10^{-12}$~eV, and comprise $\sim\!\!20\%$ of the population of detectable signals at $\mu_\mathrm{b} \sim 2\times10^{-12}$~eV. Thus, decreasing  $M_\mathrm{BH,min}$ would give a slightly larger number of signals  for boson masses $\gtrsim\!2\times10^{-12}$~eV. However, the instrument sensitivity is lower at the high frequencies corresponding to these heavier bosons, and few clouds are radiating today (Fig.~\ref{fig:countingClouds}). 

\begin{figure}[t!]
  \includegraphics[width=\columnwidth]{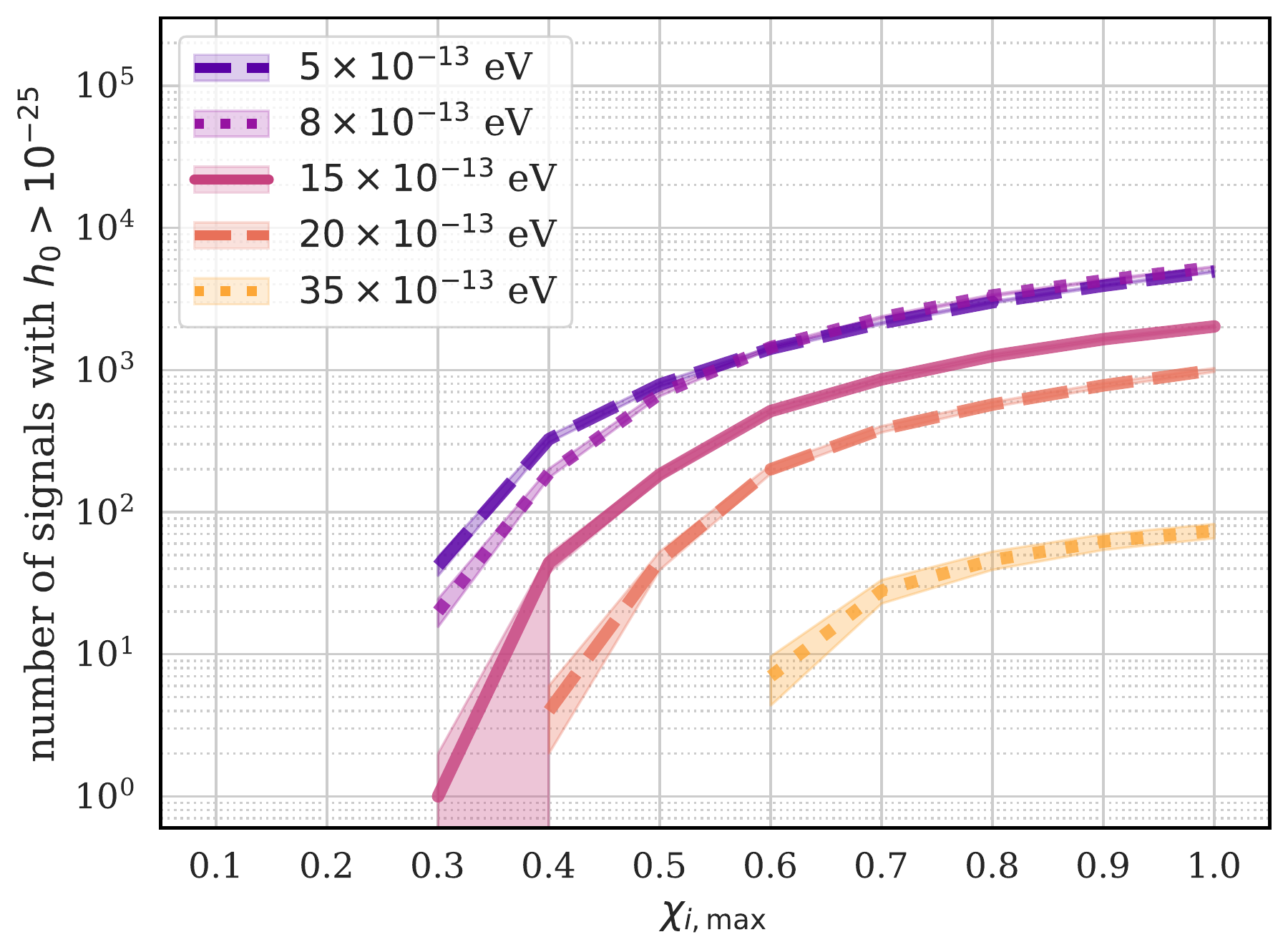}
  \caption{The number of signals above $h_0>10^{-25}$ depends on the maximum of the
    initial spin distribution, $\chi_{i,\mathrm{max}}$. For a cloud to form, the initial
    spin must be larger than the critical spin (Eq.~\eqref{eqn:chi_c}). Since $\chi_c$
    increases with $\alpha$, this also results in fewer signals above a given strain
    for heavier bosons.  The contours show Gaussian statistical uncertainties. There are no signals with amplitude $h_0>10^{-25}$ for $\mu_b=2 \times 10^{-13}$~eV for any value of $\chi_{i,\mathrm{max}}$.} 
  \label{fig:numSignalsVsMaxSpin}
\end{figure}

{{\bf Black hole spins:}}  We find that only systems with initial spins of $\chi_i > 0.25$ produce strains above current search sensitivities (Figs.~\ref{fig:ensembleBHs_MassSpin},\ref{fig:numSignalsVsMaxSpin}). Systems with $\chi < 0.25$ require $\alpha \lesssim 0.06$ for the cloud to form, producing signals that are too weak to observe.  For $\mu_\mathrm{b} \geq 8\times10^{-13}$~eV, there is a trend toward higher black hole spins as the black hole mass increases (Fig.~\ref{fig:ensembleBHs_MassSpin}) due to the critical spin increasing for larger $\alpha$.

We show the number of signals with $h_0>10^{-25}$ as a function of
$\chi_\mathrm{i, max}$ for six example boson masses in Fig.~\ref{fig:numSignalsVsMaxSpin}.
The lightest boson mass produces few to no
signals with $h_0 > 10^{-25}$; for the heavier five boson masses, the number of signals with $h_0 > 10^{-25}$ drops by at least an order of magnitude in reducing the maximum spin from $\chi_{i,\mathrm{max}} = 1.0$ to
$\chi_{i,\mathrm{max}} = 0.5$ and two orders of magnitude when $\chi_{i,\mathrm{max}}$
    decreases to $0.3$. For $\mu_\mathrm{b} = 2\times10^{-12}$~eV and $3.5\times10^{-12}$~eV, there are no signals at or below $\chi_{i,\mathrm{max}} = 0.3$ and $0.5$, respectively, since these critical spins would require black holes with mass below $5\mathrm{M}_\odot$ to form clouds.
    
{{\bf Black hole ages:}} Given the approximately constant-in-time black hole formation rate, young
black holes are proportionally less common than old ones. For the lighter bosons, e.g. $\mu_\mathrm{b} = 4\times10^{-13}$~eV, detectable signals arise from heavier black holes (Fig.~\ref{fig:ensembleBHs_MassSpin}) which are relatively rare; since black holes that are both heavy \emph{and} young are doubly uncommon, the
black holes that produce detectable signals for this boson mass are predominantly $10^8$--$10^{10}$ years old, Fig.~\ref{fig:ensembleBHs_AgeDistance}. Bosons with
$\mu_\mathrm{b} = 8\times10^{-13}$~eV produce detectable signals around black holes with masses across the entire range, and have intermediate signal times, so these black holes span the entire range of ages. Finally, for the heaviest bosons, e.g. $\mu_\mathrm{b} = 2\times10^{-12}$~eV, the shortness of the gravitational-wave timescale (Eq.~\eqref{eqn:tauGW})
and the instability timescale of the second level (Eq.~\eqref{eqn:growthTime022}) means that very old systems have stopped radiating, so that only young black holes produce detectable signals.

As discussed in Sec.~\ref{sec:positions}, our method of producing $10^8$ black holes results in fewer young black holes than the population would statistically contain. Specifically, considering our black hole formation rates, on average we expect $\mathcal{O}(10)$ black holes of  $\tau_\mathrm{BH}\sim10^3$~years, and $\mathcal{O}(1)$ black holes of $\tau_\mathrm{BH}\sim10^2$ years. On the other hand our example population overestimates the number of $10^4$ year black holes, as the initial population included an upward fluctuation of the number of these black holes. Thus our population includes 200 black holes that are younger than $10^4$ years old, compared to an average expectation of 62. This disproportionately affects heavier bosons; at $\mu_\mathrm{b} = 2\times10^{-12}$~eV, of order half of the signals that  are detectable with current search sensitivities are produced by $\sim\!10^4$ year old black holes.  Therefore, our sampling procedure introduces an order-1 uncertainty in the number of signals produced by bosons with $\mu_\mathrm{b} \geq 2\times10^{-12}$~eV.

\subsection{Number and density of signals as a function of GW amplitude}
\label{sec:numberanddensity}

\begin{figure}[t!]
  \includegraphics[width=\columnwidth]{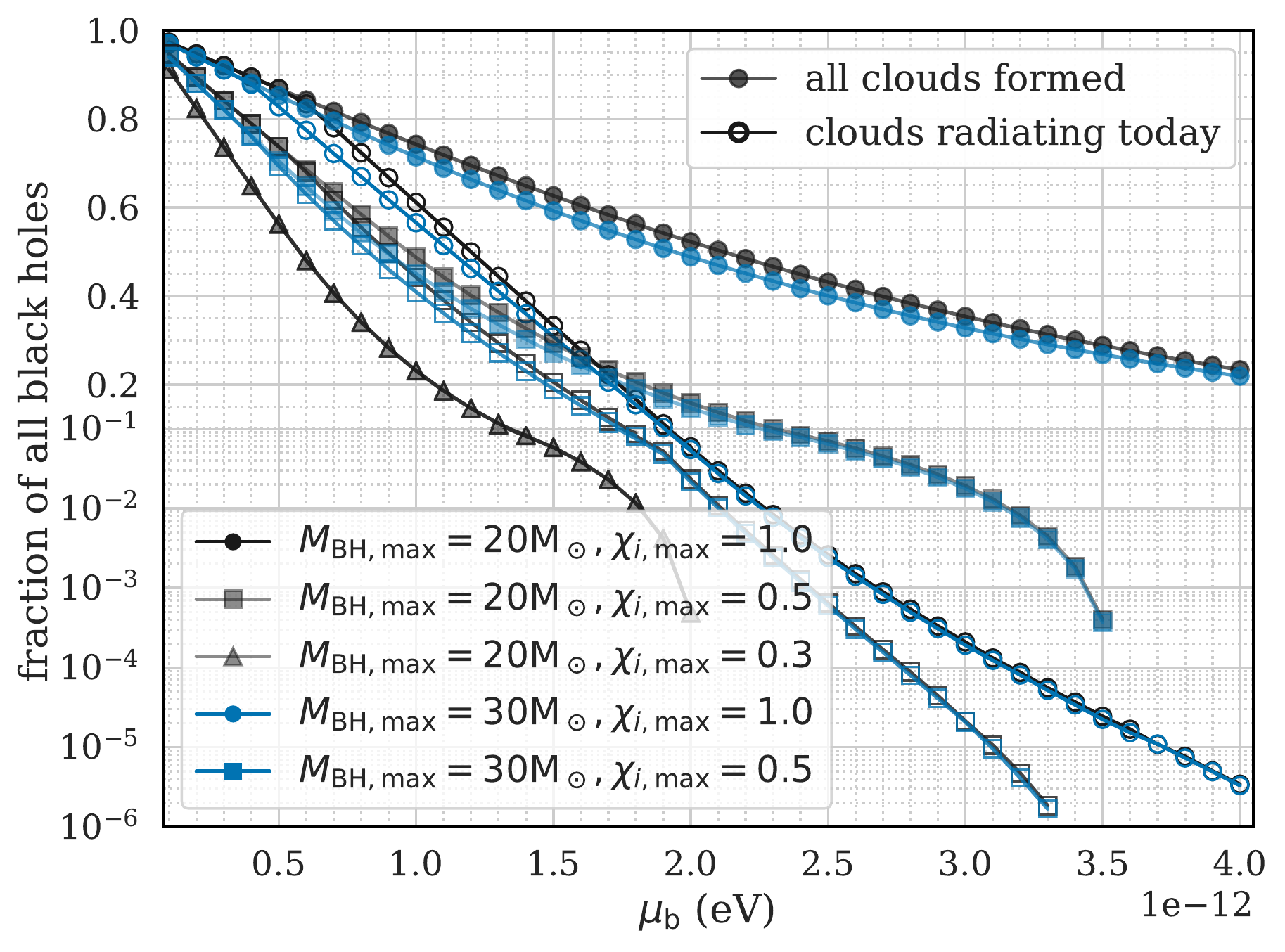}
  \caption{For a given boson mass, a fraction of black holes is born with spins high enough to support
    cloud formation (filled markers). Gravitational-wave emission stops when the second level is fully populated, so a subset of the initial clouds are still emitting today (unfilled markers). We consider the five black hole populations  of Table~\ref{tab:BHPopulations}:  standard (black circles),    heavy (blue circles), moderate spins (gray squares), heavy with moderate spins
    (blue squares), and pessimistic spins (gray triangles) populations. The percentage
    of systems still emitting today decreases rapidly with increasing boson mass; at
    $\mu_\mathrm{b} = 4\times10^{-12}$~eV, only up to one in every million black holes in the Galaxy is in
   a currently emitting system.}
  \label{fig:countingClouds}
\end{figure}

\begin{figure}[t!]
  \includegraphics[width=\columnwidth]{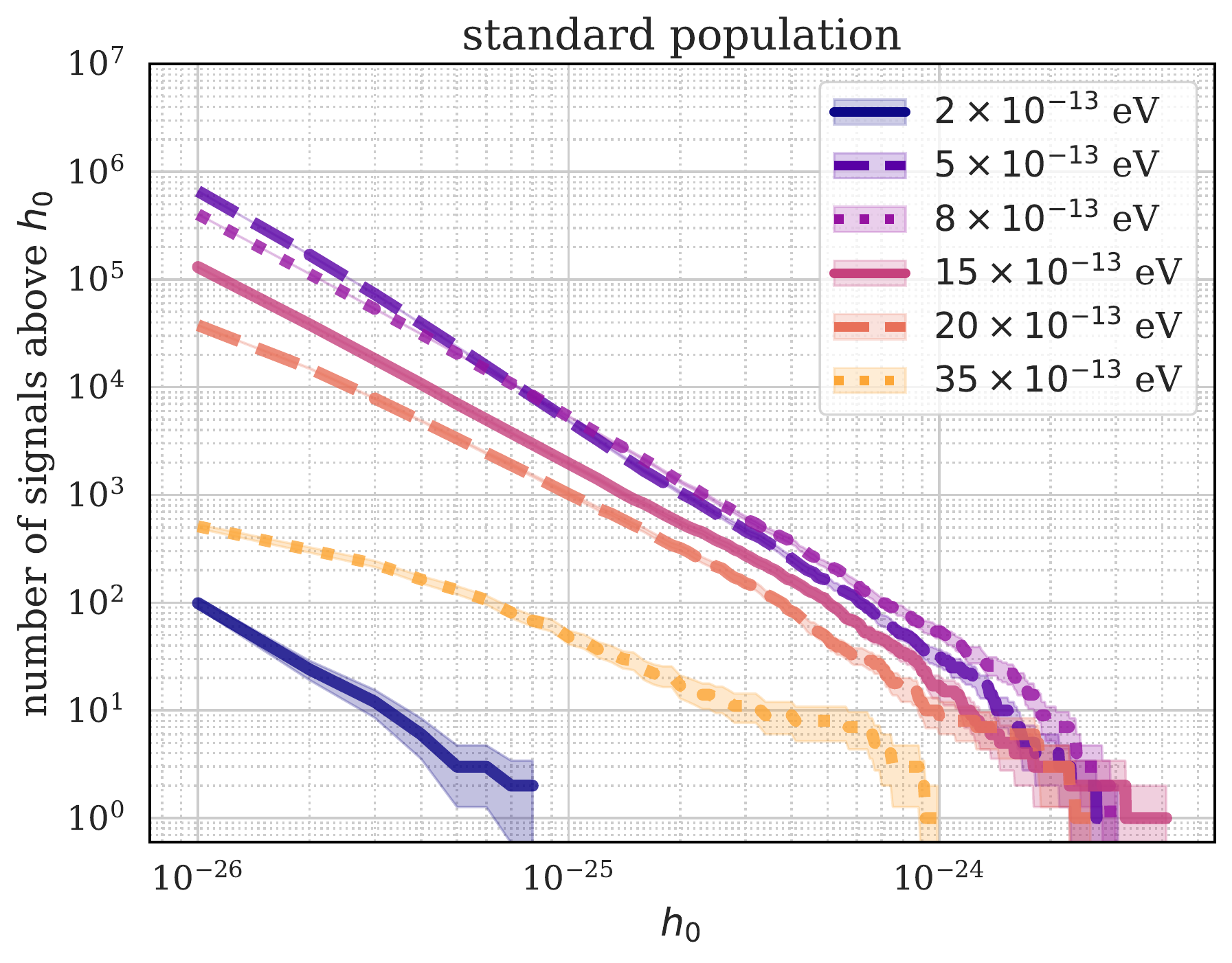}
  \caption{
   The number of signals with intrinsic
    amplitude above a given $h_0$ value. The signal number is generally largest at $\mu_\mathrm{b} = 8\times10^{-13}$~eV
    ($f_\mathrm{GW} \approx 400$~Hz) for the standard black hole population
    (Table~\ref{tab:BHPopulations}). At
    higher boson masses, the number of signals decreases because both the signal
    half-time (Eq.~\ref{eqn:tauGW}) and the signal cutoff time due to the formation of the
    second level (Eq.~\eqref{eqn:growthTime022}) decrease with increasing frequency.
    The contours show Gaussian statistical uncertainties, an underestimate at small
    signal number.
  }
  \label{fig:numSignals}
\end{figure}

\begin{figure}[t!]
  \includegraphics[width=\columnwidth]{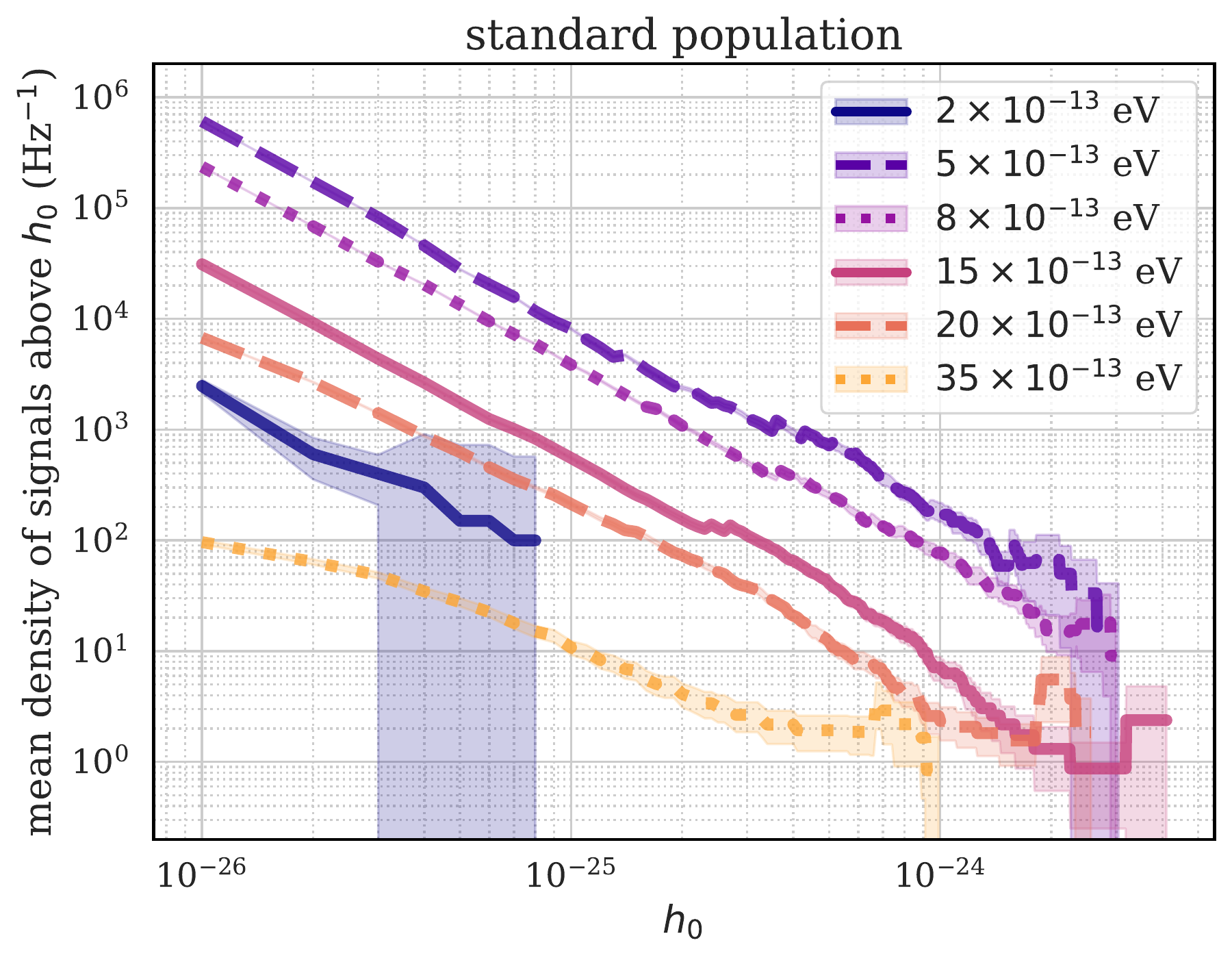}
  \caption{The mean density of signals above a given $h_0$.
    The mean density is largest for
    $\mu_\mathrm{b} = 5\times10^{-13}$~eV ($f_\mathrm{GW} \approx 250$~Hz)
    for the standard black hole population (Table~\ref{tab:BHPopulations}).
    For heavier bosons, the overall density \emph{decreases}; the number of
    signals above a given $h_0$ drops due to the combination of effects
    described in Fig.~\ref{fig:numSignals} while the frequency range spanned
    by the ensemble signal continues to increase. The contours show the
    uncertainty $\sqrt{N}/\Delta f$, where $N$ is the number of signals
    and $\Delta f$ is the frequency range spanned by the $N$ signals. Note that this
    does not take into account the additional uncertainty from
    $\Delta f$, which is itself correlated with $N$.}
  \label{fig:densities}
\end{figure}

\begin{figure}[t!]
  \includegraphics[width=\columnwidth]{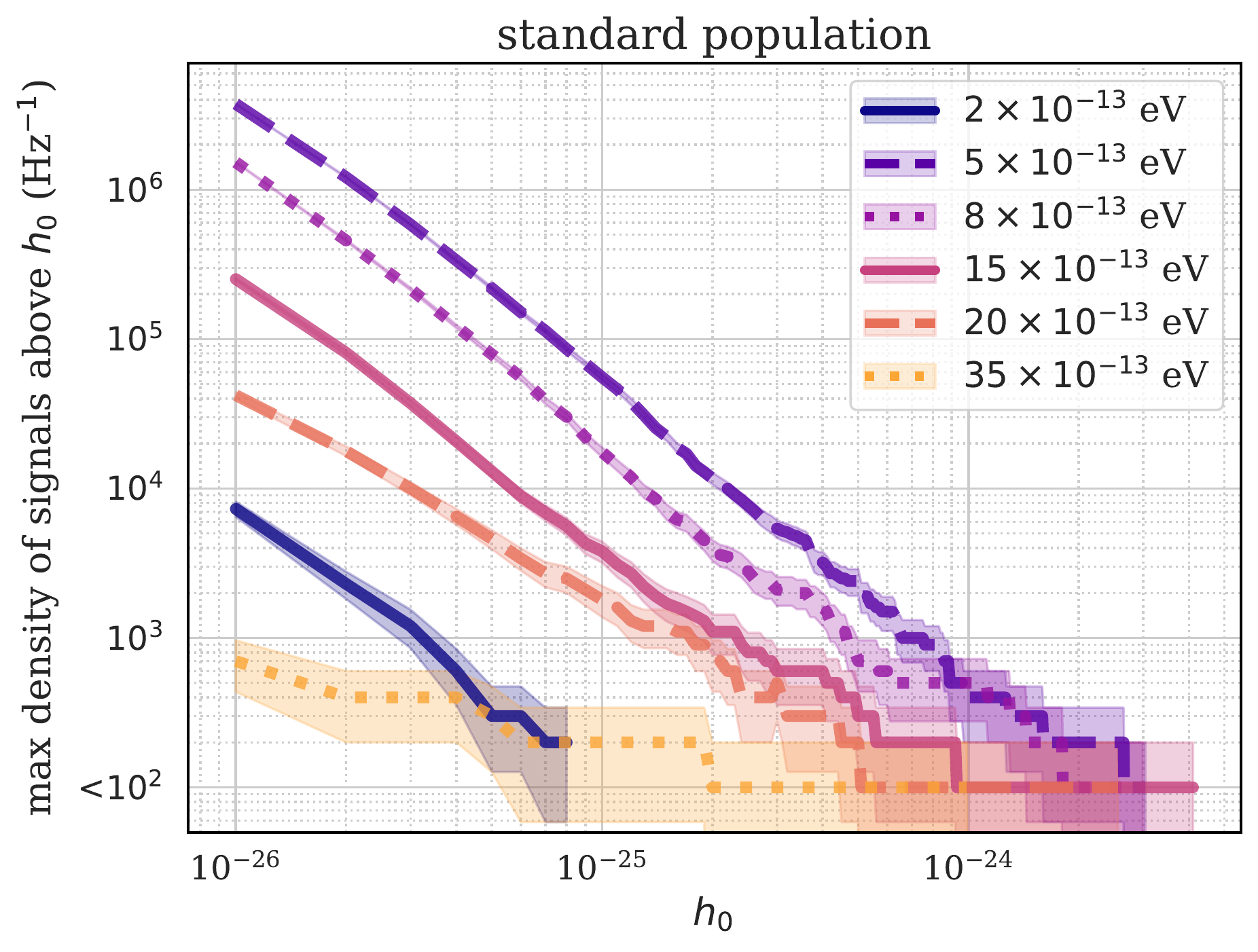}
  \caption{The maximum density of signals above a given $h_0$ is up to an order of magnitude larger than the mean
    density (Fig.~\ref{fig:densities}). Frequency ranges with sufficiently high signal
    densities can cause complications for standard searches for continuous wave signals (Sec.~\ref{sec:detectability}).
    The contours show the uncertainty $\sqrt{N}/\Delta f$, where $N$ is the
    number of signals and $\Delta f = 0.01$~Hz is the size of the frequency bin
    over which the maximum density is calculated.
  }
  \label{fig:densities_max}
\end{figure}

Figure~\ref{fig:numSignals} shows the total number of signals with
characteristic strains above given values of $h_0 \in [10^{-26},10^{-23}]$ for six reference
boson masses. For most values of $h_0$, the number of signals above a given $h_0$ is highest
for $\mu_\mathrm{b} = 8\times10^{-13}$~eV ($f_\mathrm{GW} \approx 400$~Hz).

For the `standard' black hole population ($M_\mathrm{BH,max} = 20\mathrm{M}_\odot$),  annihilation signals from the lightest bosons are unlikely to be detectable by current continuous wave searches: for $\mu_\mathrm{b} = 2\times10^{-13}$~eV (corresponding to
$f_\mathrm{GW} \sim 100$~Hz) few or no systems produce GW emission with intrinsic amplitude above $10^{-25}$. The loudest signal in
our example has a strain a factor of a few below the current sensitivity thresholds  (Fig.~\ref{fig:ensembleExamples}).

At fixed black hole mass, increasing the boson mass increases the strain as $\mu_\mathrm{b}^9$. Thus, increasing the boson mass by a factor of 2 to $\mu_\mathrm{b} = 4\times10^{-13}$~eV ($f_\mathrm{GW} \approx 200$~Hz),
greatly increases the number of signals that would be detectable by current searches ($h_0 >2\times10^{-25}$ near this frequency).

Although the peak strain increases with boson mass, fewer signals are expected at heavier boson masses: Fig.~\ref{fig:countingClouds} illustrates that both the percentage of systems with fully formed clouds and the percentage of systems that are still radiating continuous waves today decrease with boson mass. The further effect of the signal half-time is not included in Fig.~\ref{fig:countingClouds} as it depends on the search strain sensitivity and the black hole's age.
Fig.~\ref{fig:numSignals} shows that for bosons with $\mu_\mathrm{b} > 8\times10^{-13}$~eV, the number of signals
above a given $h_0$ begins to decrease with increasing boson mass.

Figures~\ref{fig:densities} and \ref{fig:densities_max} show the mean and maximum
density of signals in frequency space above a characteristic
strain $h_0$, for the same six reference bosons as in Figure~\ref{fig:numSignals}. Both the mean and maximum signal density are largest for $\mu_\mathrm{b} = 5\times10^{-13}$~eV. 
This is in contrast with the fact that the \emph{number}
of signals is typically larger for $8\times10^{-13}$~eV (Fig.~\ref{fig:numSignals}). Signals from heavier bosons are spread over a larger frequency range  as described in Section~\ref{sec:morphology}, thereby decreasing the signal density.

Finally, Figs.~\ref{fig:numSignals},~\ref{fig:densities}, and~\ref{fig:densities_max} demonstrate how signal number and density increase with decreasing strain. The number of signals  and the maximum signal density for a future search sensitivity an order of magnitude below the current continuous wave searches can be expected to increase by up to two orders of magnitude. Thus we highlight that as the gravitational wave observatories improve further continuous wave searches for bosons can be increasingly promising and less reliant on assumptions on BH populations. On the other hand, the issues of signal detectability as detailed in the following section will only become more important to consider in the design and interpretation of continuous wave wave searches for boson ensemble signals.

\section{Signal detectability}
\label{sec:detectability}

We now  consider the signal detectability in the context of the results of current continuous wave searches with LIGO data, using several black hole mass and spin distributions (Table~\ref{tab:BHPopulations}).

\subsection{Number of detectable signals}
\label{sec:howManyDetectable}

\begin{figure*}[t!]
  \includegraphics[width=0.9\textwidth]{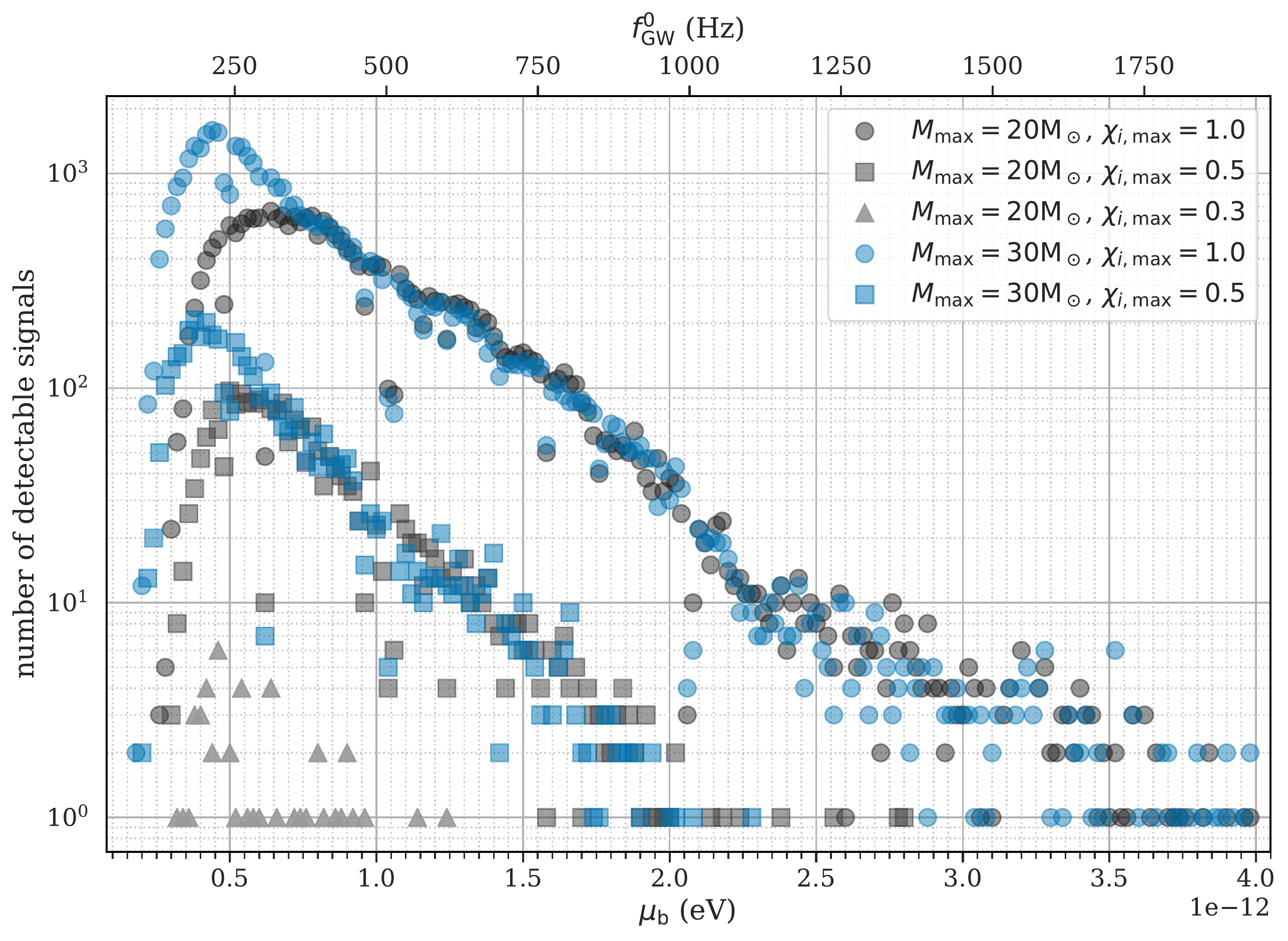}
  \caption{Number of signals with amplitudes above the detectability level of recent all-sky searches for
    continuous gravitational waves \cite{FalconO1,Dergachev:2019oyu, RomeBosonClouds}, for bosons in the mass range $1\times10^{-13}$~eV
    to $4\times10^{-12}$~eV. The Falcon search \cite{FalconO1,Dergachev:2019oyu} provides 95\% confidence {\it strict} upper limits on continuous wave emission
    with near-zero frequency derivative between 20 and 600 Hz. These upper limits are valid for any sky position and polarization of the source, even the ones with the most unfavorable coupling to the detectors. In contrast the upper limits from the Frequency Hough
    search \cite{RomeBosonClouds} hold for 95\% of the {\it entire} population.
    For this reason, even at the same sensitivity, the Frequency Hough upper limits are lower than the Falcon upper limits.
    We use the Falcon upper limits for bosons with $\mu_\mathrm{b} <1.2\times10^{-12}$~eV
    and the Frequency Hough upper limits for bosons with $\mu_\mathrm{b} \geq 1.2\times10^{-12}$~eV.
    We consider the standard (black circles) and heavy (blue circles) black hole
    populations, as well as the moderate (gray squares), heavy moderate (blue squares) and pessimistic (gray triangles)
    spin populations. The fluctuations are due to 
    stationary lines and other artifacts in the detectors, which decrease the continuous wave search 
    sensitivity in nearby frequency bins.
    }
  \label{fig:detectableSignals}
\end{figure*}

For bosons in the mass range $\mu_\mathrm{b} \in [1\times10^{-13}, 4\times10^{-12}$]~eV,
in increments of $2\times10^{-14}$~eV,
  we calculate the number of detectable
signals for the five black hole populations (Table~\ref{tab:BHPopulations}) 
at the corresponding frequencies  $f_\mathrm{GW,obs} \in [50, 2000]$~Hz. For a signal to be
detectable, we require that its intrinsic strain $h_0$ is greater than the 95\%
upper limit value in the $\{f_\mathrm{GW,obs}\}$ frequency bin.
Finally, we multiply the total number of detectable signals for each boson mass by 95\%.

We use the
95\% upper limits for the low-$\dot{f}$ continuous wave search in the 20--600 Hz range \cite{FalconO1,Dergachev:2019oyu}
and the upper limits for the continuous wave search in the 10--2048 Hz range \cite{RomeBosonClouds}. For ease of reading, we refer to the searches  \cite{FalconO1,Dergachev:2019oyu} and \cite{RomeBosonClouds} by their nicknames ``Falcon'' and ``Freq Hough,'' respectively. 
 For bosons with $\mu_\mathrm{b} < 1.2\times10^{-12}$~eV,
we use the near-zero spin-down upper limits from the Falcon searches and for $\mu_\mathrm{b} \geq 1.2\times10^{-12}$~eV, we use the
upper limits from the very broad spin-down Freq Hough search.


The number of detectable signals for the five black hole populations is 
shown in Fig.~\ref{fig:detectableSignals}. All the curves peak at intermediate values of $\mu_\mathrm{b}$. As discussed in Sections~\ref{sec:morphology} and \ref{sec:numberanddensity}, the signal strength grows with boson mass but the number of emitting systems decreases with increasing  boson mass. The two competing effects combined with the sensitivity curve of the LIGO detectors (the noise is smallest at $\sim\!170$~Hz) determine the boson mass $\mu_\mathrm{b}$ with the largest number of detectable signals.

For a given spin distribution, a heavier black hole population
($M_\mathrm{BH,max} = 30\mathrm{M}_\odot$) produces more signals for light bosons
($\mu_\mathrm{b} < 8\times10^{-13}$~eV) than does a lighter black hole population 
($M_\mathrm{BH,max} = 20\mathrm{M}_\odot$). For the heavy black hole population,
the largest number of detectable signals occurs for $\mu_\mathrm{b} = 4\times10^{-13}$~eV;
for the lighter black hole population, the largest
number corresponds to $\mu_\mathrm{b} = 6\times10^{-13}$~eV.

In the case that $M_\mathrm{BH,max} < 20\mathrm{M}_\odot$, fewer signals would be produced  at the lightest boson masses.
At a fixed value of $\alpha$, the strain is  linearly
proportional to black hole mass, $h_\mathrm{0,peak} \propto M_\mathrm{BH}$, so even
decreasing the maximum black hole mass by a factor of 1.5 (already in tension with known
Galactic black holes \cite{Ozel+2010}) would  shift the peak of the detectable signals to bosons a factor of $1.5$ heavier.

For $\mu_\mathrm{b} > 8\times10^{-13}$~eV, the two black hole populations produce
the same number of detectable signals. For these heavier bosons, systems with
heavier black holes form less frequently and have significantly shorter emission lifetimes  and do not contribute to the event rates.

For a given black hole mass distribution, the number of detectable signals is most sensitive to the 
black hole spin distribution, through the superradiance condition of Eq.~\eqref{eqn:chi_c}.
We consider three different uniform spin distributions: a standard distribution with $\chi_i \in [0, 1]$, a moderate distribution with $\chi_i \in [0, 0.5]$, 
and a pessimistic distribution with $\chi_{i} \in [0, 0.3]$.

The standard and moderate spin distributions produce detectable signals for boson masses
$2\times10^{-13}~\mathrm{eV}\lesssim\mu_\mathrm{b}\lesssim 2.5\times10^{-12}$~eV, although
the number of detectable signals decreases by an order of magnitude from $\chi_{i,\mathrm{max}} = 1$
to 0.5, as fewer black holes have initial spins above the critical spin.  For the pessimistic spin distribution, only a handful of signals are detectable, and only
for boson masses $4\times10^{-13}~\mathrm{eV}\lesssim\mu_\mathrm{b}\lesssim1.3\times10^{-12}$~eV. The small signal number, in conjunction with there being no signals with $h_0>10^{-25}$ for  $\chi_{i,\mathrm{max}} <0.3$ (Fig.~\ref{fig:numSignalsVsMaxSpin}, suggests that if 
all black holes are born with spins smaller than 0.3, no boson mass will produce a significant number of signals detectable by current continuous wave searches.

\subsection{Density of signals}
\label{sec:density}

Since searches are designed under specific assumptions about signal properties, their performance may be degraded when the signals do not meet those assumptions. 
An important property of the target signal population is the number of detectable signals in a parameter space search cell.
If the total signal is an incoherent superposition of many signals that cannot be resolved individually, then the total signal is 
a stochastic process that can best be detected with cross-correlation methods \cite{Allen:1997ad}.
On the other hand, if the signals are sparse enough that they
can be individually resolved, the matched filtering and semi-coherent techniques that are commonly used to search for
continuous wave signals should be applied \cite{Walsh:2016hyc,Walsh:2019nmr,Piccinni:2018akm,FalconO1}. In between these two extremes is a regime
with some loud and resolvable signals on a background of weaker and unresolved ones, that may act like confusion noise.

\begin{figure}[t!]
\includegraphics[width=\columnwidth]{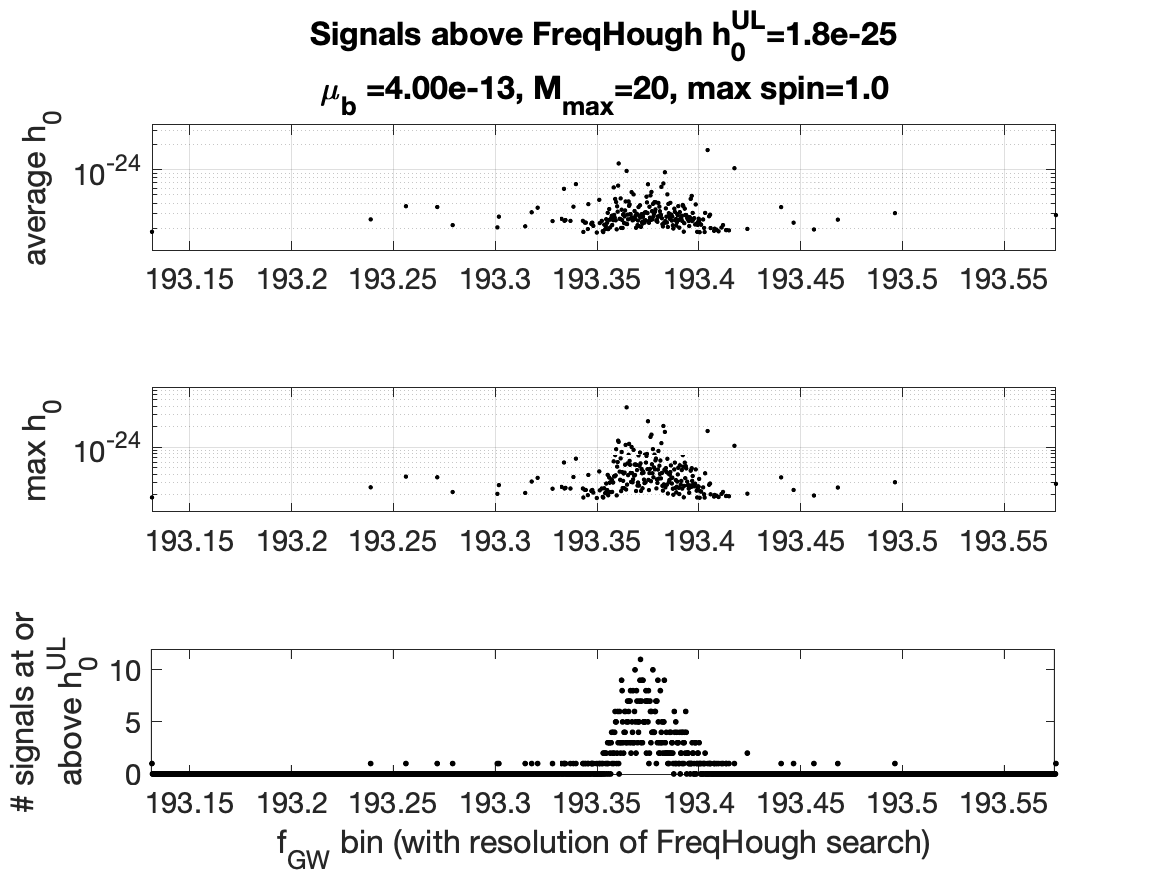}
\includegraphics[width=\columnwidth]{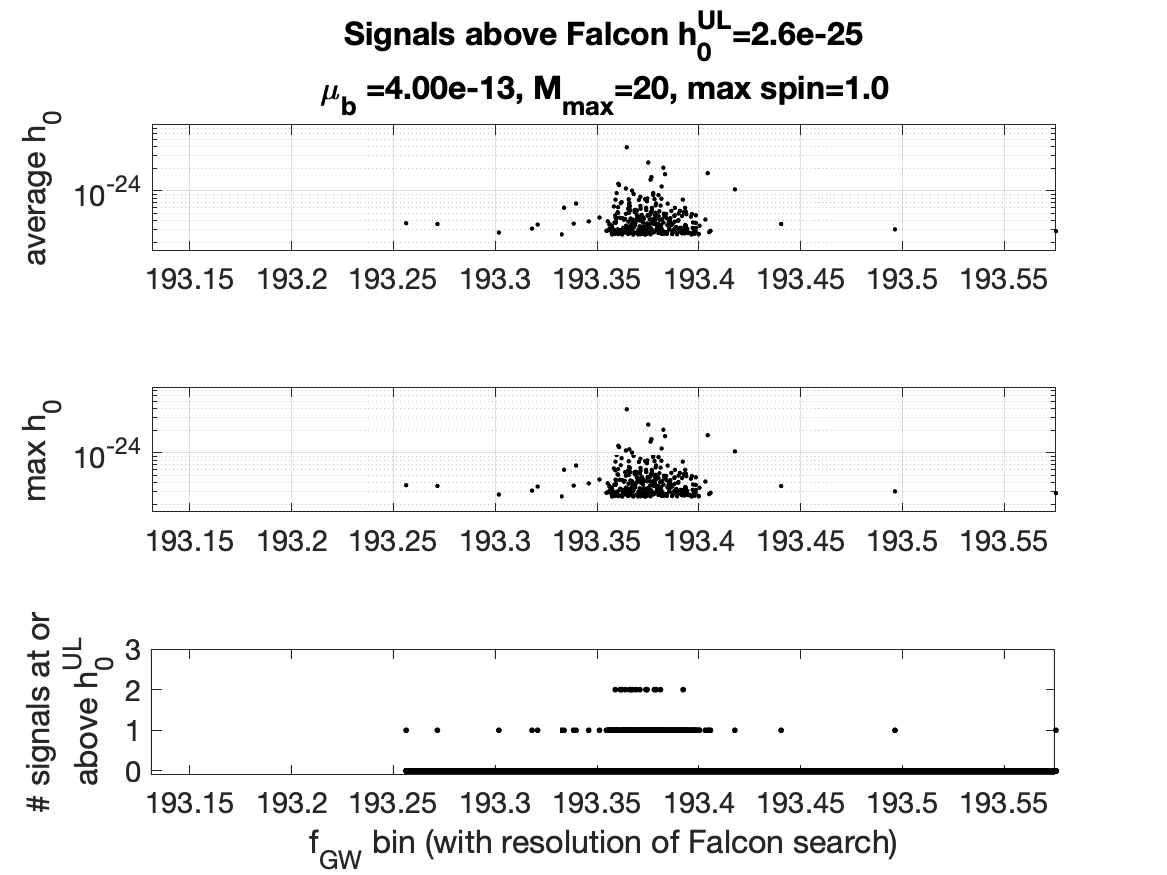}
  \caption{The x-axes show the signal frequency binned with the resolution of the two searches: $2.4 \times 10^{-4}$ Hz for the FreqHough search and $1.7 \times 10^{-5}$ Hz for the Falcon search. From the bottom panel going up, the y axis shows the number of detectable signals in each bin, their maximum intrinsic amplitude and their average amplitude. The main reason why there are fewer signals per frequency bin in the Falcon searches \cite{FalconO1,Dergachev:2019oyu} than in the FreqHough search \cite{RomeBosonClouds} is that the frequency bin of FreqHough is significantly larger (by a factor of 14) than the frequency bin of Falcon. The maximum initial spin of the black hole population is taken to be 1.}
  \label{fig:detectableSignalStatsPerBin}
\end{figure}

\begin{figure}[t!]
\includegraphics[width=\columnwidth]{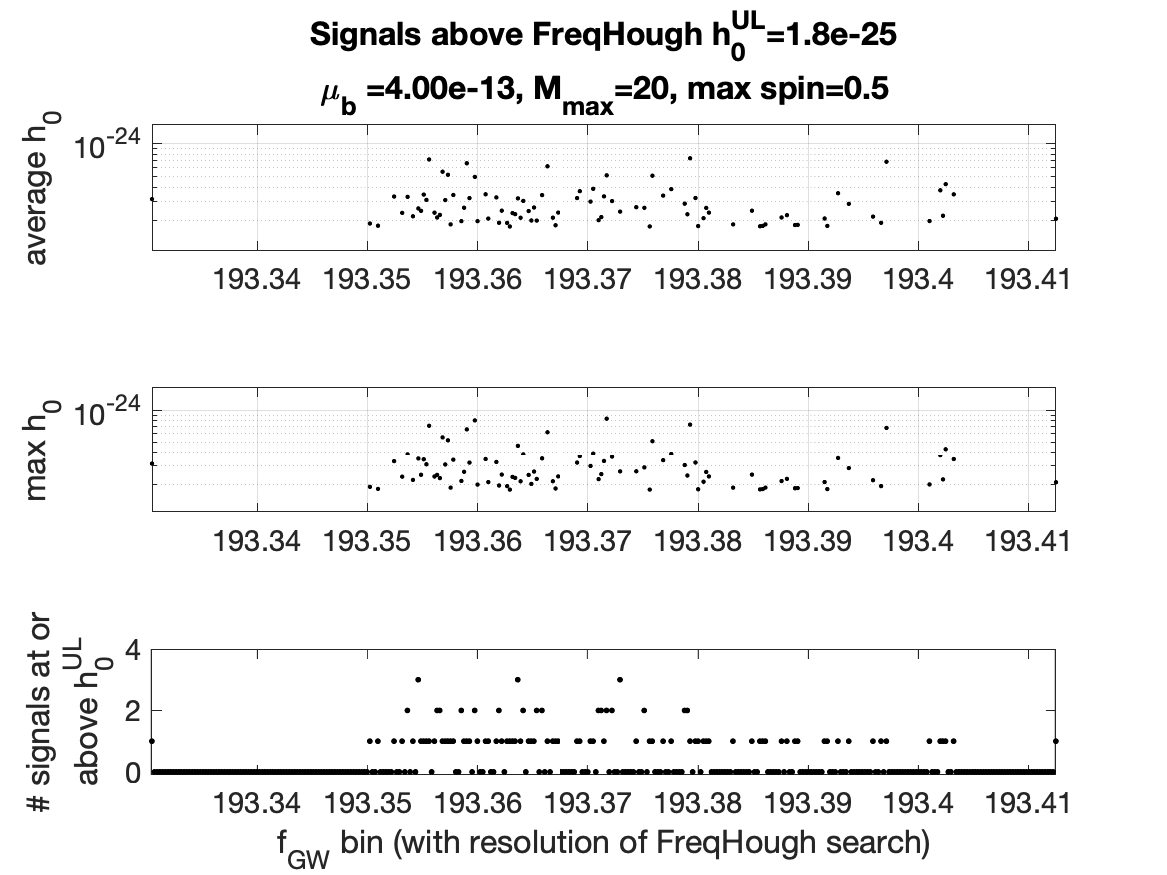}
\includegraphics[width=\columnwidth]{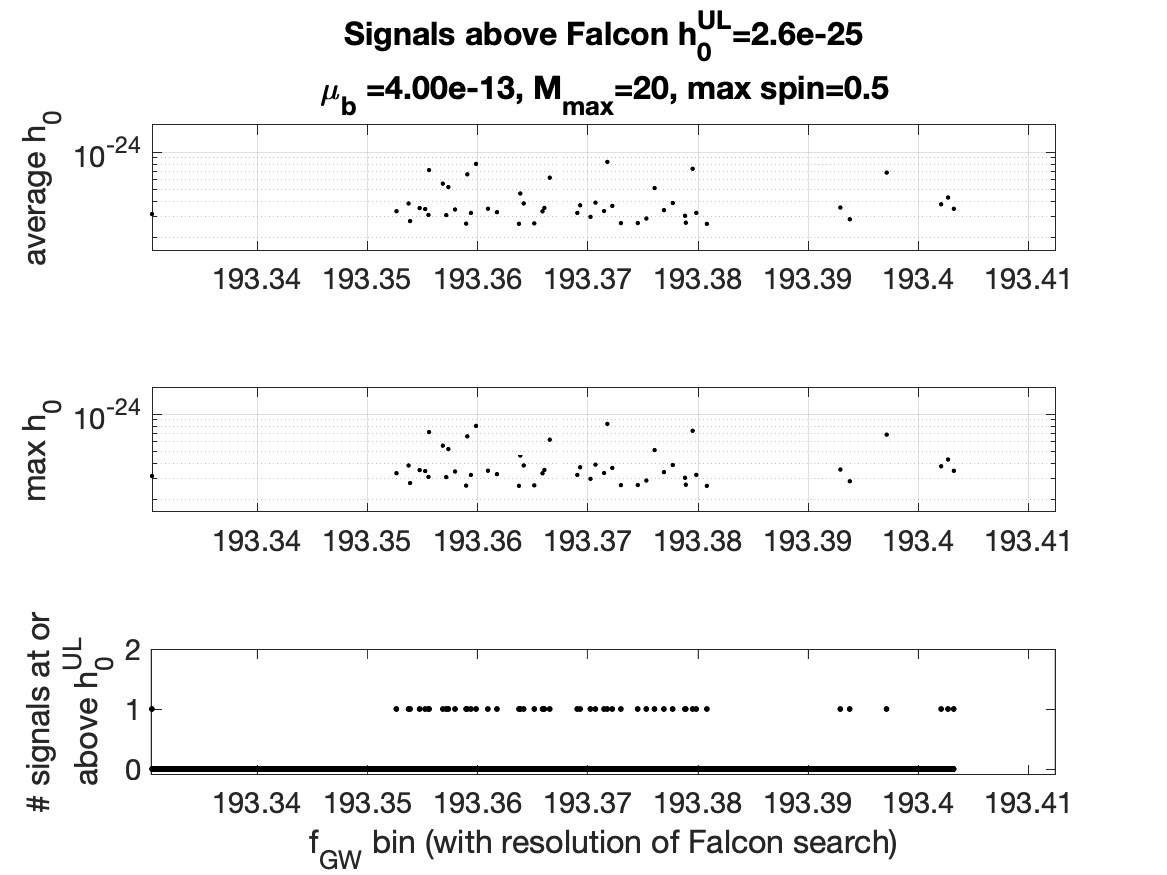}
  \caption{Same plots as those of Fig. \ref{fig:detectableSignalStatsPerBin} but having assumed that the maximum initial spin of the black hole population is 0.5.}
  \label{fig:detectableSignalStatsPerBin0p5}
\end{figure}

We now want to understand what situation applies here. To address this question we consider the FreqHough \cite{RomeBosonClouds} and Falcon searches \cite{FalconO1,Dergachev:2019oyu}. 

We bin the signal frequency with the resolution of the first stage of the respective continuous wave search. We compute the ensemble signal for a standard black hole population and count the number of signals in each frequency bin with amplitude equal to or greater than the amplitude upper limit from the given continuous wave search, as a function of the signal frequency. 
For every bin we also determine the mean and maximum signal amplitude over the ensemble signals with frequency in that bin and amplitude greater than the search upper limit amplitude. Figures~\ref{fig:detectableSignalStatsPerBin} and \ref{fig:detectableSignalStatsPerBin0p5} show these quantities for $\mu_b = 4\times 10^{-13}$ eV which is close to the boson mass yielding the highest number of  detectable signals per bin. The maximum number of signals in a single FreqHough bin is $\sim$ 10, whereas for Falcon it is 2. This difference is mostly due to the fact that the FreqHough frequency bin is 14 times larger than the Falcon bin. We show the equivalent figures for a range of boson masses in App.~\ref{appendix:signalsAboveUL}. As the number of signals scales linearly with the number of black holes, a smaller number of black holes could still yield multiple signals in a single bin for some combination of boson mass and maximum black hole mass. For instance a population of $10^7$ black holes produces multiple signals in a single FreqHough bin for boson masses of $3\times10^{-13}$~eV assuming $M_\mathrm{BH,max} = 30 \mathrm{M}_\odot$ \ref{fig:aboveULmub3}.

Although there are many FreqHough signal-frequency bins that contain multiple signals, these signals are in principle still resolvable by the search because they come from different sky locations. However, due to large-scale parameter correlations of the detection statistics, sufficiently strong signals produce high detection statistic values even at template parameters far from the signal parameters. This is evident for instance in \cite{Dergachev:2019pgs}; in spite of this being a search aimed at the Galactic Center, it still detects all LIGO hardware-injected fake signals at sky locations far from the Galactic Center. This means that when the ensemble signal comprises many loud signals in nearby frequency bins, even if each signal comes from a different sky location, it will contribute to the detection statistics at the templates of other signals, possibly giving rise to contamination/confusion noise. So the high concentration of detectable signals in many neighboring frequency bins calls for caution whilst interpreting results from continuous wave searches for rare signals as boson-annihilation signals.

A large signal density has consequences because the complex all-sky continuous wave search pipelines \cite{Abbott:2017pqa,FalconO1,Dergachev:2019oyu,RomeBosonClouds,Ming:2019xse} comprise several follow-up stages of increasing coherence lengths, with vetoes and thresholds designed to reject noise fluctuations and coherent disturbances while preserving signal candidates. The procedures are carefully calibrated with extensive Monte Carlos on fake {\it isolated} signals added to real noise. It is unlikely that settings that have been tuned to isolated signals would be equally effective for the ensemble signals explored in this study, especially in regions of high signal density.
We expect that the detection efficiency of searches for which the density and number of detectable signals {\it per bin} is larger will be impacted more severely; so, for example, the impact will be greater for the Freq. Hough search \cite{RomeBosonClouds} than the Falcon search \cite{FalconO1,Dergachev:2019oyu}.

We note that in Fig.~\ref{fig:detectableSignalStatsPerBin}, the frequencies where the signals are the strongest are also the frequencies with the highest density of signals. This is not the case in general. As discussed in Section~\ref{sec:morphology}, for a given boson mass, systems with heavier black holes produce signals with lower frequencies and larger strains; at the same time, heavier black holes are rarer, and have larger minimum spins required for cloud formation. This causes the typical ``peaked'' shape (Fig.~\ref{fig:ensembleExplanation}) of the ensemble signal, so that the loudest signals tend to be located at or below the frequencies with the densest clustering of
signals, as can be seen in App.~\ref{appendix:signalsAboveUL}.

A high density of loud signals in a broad enough frequency range could also have a more subtle negative impact on the ability of standard continuous wave searches to detect signals from the ensemble.  The reason is that the average noise level data is ``normalized out" of the detection statistic, but if this is not done carefully, there is a danger of reducing the signal. Most continuous wave searches (and all of the ones that we refer to in this paper) whiten the data
at an early point in the analysis by dividing the Fast Fourier Transform of the data taken over some short time-baseline (SFT)
with the amplitude spectral density estimated from the data itself.
The noise level is estimated through an average of the noise power over tens of SFT-bins. Under the assumption that the data is noise-dominated, with at most a single detectable signal concentrated in an SFT-bin or two,  the ``average" over tens of bins is insensitive to the signal. For instance, FreqHough \cite{RomeBosonClouds} searches 
  use an autoregressive average on the power spectral density over $\lesssim$ 0.02 Hz \cite{Astone_2005, Astone:2019} and the
  Einstein@Home searches \cite{Abbott:2017pqa,Ming:2019xse} use running medians over
  $\lesssim\!0.06$ Hz frequency bins.

\begin{figure}
\includegraphics[width=\columnwidth]{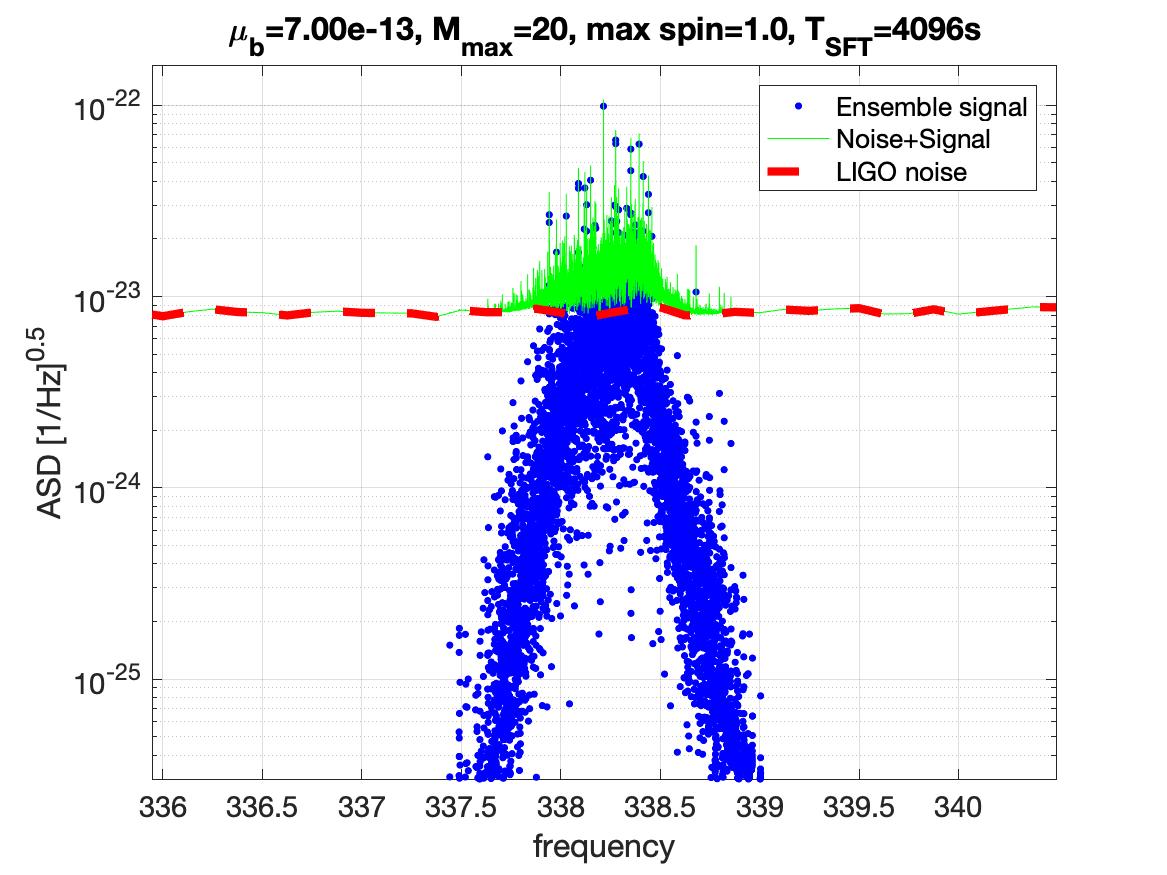}
\caption{The amplitude spectral density of the LIGO O2 data alone (dashed red line) and of an ensemble signal (top green points) assuming $\mu_b=7\times 10^{-13}$ eV and a galactic black hole population with maximum mass 20 M$_\odot$ and maximum initial spin of 1. The time-baseline assumed is 4096s as used by the Freq. Hough search \cite{RomeBosonClouds}.}
\label{fig:EnsembleSignalASD}
\end{figure}

If the ensemble signal is a collection of many signals loud enough to increase the apparent noise floor in a broad enough band in the input data (the SFTs), then the noise estimate will be sensitive to this increase, the search normalization procedure will down-weight the data containing the signals
and this might decrease the detection statistic associated with some of the ensemble-signals. In this case the upper limits derived for isolated signals should be used with caution when re-interpreted to apply to ensemble signals. 
  In Fig. \ref{fig:EnsembleSignalASD} we plot the amplitude spectral density (ASD) expected for the ensemble signal corresponding to $\mu_\mathrm{b}=7\times 10^{-13}$ eV. There clearly is additional power due to the ensemble signal extending over $\gtrsim\!1$ Hz, this excess would also appear in the noise estimate and for the reason explained above could change the detection efficiency curve as a function of signal amplitude, with respect to the isolated signal case. More plots like that of Fig.~\ref{fig:EnsembleSignalASD} are given in App. \ref{app:ASDPlots}, for different combinations of boson mass, maximum black hole mass and maximum initial black hole spin. For $3 ~\leq \mu_b \leq 7 ~ \times 10^{-13}$~eV, an increased ASD on the SFT timescale is visible, except for ensembles with initial maximum black hole spin 0.3 or less {\it{and}} the maximum black hole mass of $20\mathrm{M}_\odot$.


\section{Existing literature and results}
\label{sec:discussion}

Several works have made projections of the detectability of the boson
annihilation signal at Advanced LIGO and Virgo 
\cite{Arvanitaki+2015, Arvanitaki+2017, Brito+2017}.
These studies predict tens to thousands of signals detectable with  Advanced LIGO/Virgo continuous wave searches, with a range of assumptions about black hole mass and spin distributions in combination with often only idealized search concepts. 
Other works have searched the gravitational wave data for boson annihilation signals leveraging existing search pipelines and/or results from other astrophysical searches.

 All-sky O2 continuous wave survey non-detection results of ~\cite{RomeBosonClouds} are used  to exclude bosons with $1.1\times10^{-13}$~eV $< \mu_\mathrm{b} < 4\times10^{-13}$~eV
assuming all black holes are formed with spin $0.998$, or $1.2\times10^{-13}$~eV
$< \mu_\mathrm{b} < 1.8\times10^{-13}$~eV assuming all black holes are formed with spin 
$0.6$,  and a range of 3--100 $M_\odot$ black hole masses in both cases.
We find that the number of signals below $\sim\!3\times10^{-13}$~eV falls off
very rapidly for both $M_\mathrm{max,BH} = 30\mathrm{M}_\odot$ and $M_\mathrm{max,BH} = 20\mathrm{M}_\odot$
(Fig.~\ref{fig:detectableSignals}), indicating that the excluded masses of
\cite{RomeBosonClouds} rely on a speculative population of heavy black holes. Our analysis of the ensemble signal shows that any bound derived from a single spin or distance assumption can result in misleading conclusions. Furthermore, the ensemble signal from bosons with mass between 3 and $4\times 10^{-13}$ eV would likely be detected with a lower efficiency than that of an isolated signal and hence non-detections from a standard continuous wave search cannot automatically be translated in exclusions for boson signals at the same frequency at the same $h_0$ level.

A cross-correlation search technique is considered in \cite{Tsukada+2019}: the boson annihilation ensemble signal from outside the Milky Way  is treated as a stochastic background \cite{Brito+2017a}. Based on  null results, \cite{Tsukada+2019} exclude the range 2.0--3.8 $\times10^{-13}$ eV under  an ``optimistic''  black hole spin distribution (defined as unity maximum dimensionless spin and letting the minimum vary) but not under a ``pessimistic" one (zero minimum spin and letting the maximum vary). The range of black hole masses considered is 3--$100\mathrm{M}_\odot$. The amplitude spectral density estimates based on our ensemble signal suggest that their null result could exclude Galactic signals from some boson masses even for maximum initial black hole spins of 0.5, as long as the black hole masses go at least up to $30M_\odot$.

A way to bypass the uncertainties in the system's parameters is to consider a {\it known} black hole, for example the remnant of a merger of two compact objects \cite{Arvanitaki+2017}. For the LIGO black holes the formation time is precisely known, and the spin and mass are measured with good precision but the remnant is too far to yield a signal detectable by the current interferometers \cite{Arvanitaki+2017,Isi+2019}. Turning to galactic sources, \cite{Sun+2019} have searched for  continuous wave emission from Cygnus X-1, a well known system: a black hole of $\sim\!15 \mathrm{M}_\odot$, 5 million years old at a distance of 1.86 kpc. The absence of a detection disfavors the existence of bosons in the range 5.8--$8.6\times10^{-13}$ eV. This result however assumes a near-extremal initial black hole spin of 0.99 and implies a low current spin, $\chi \in [0.25, 0.36]$, which is in tension with continuum and reflection measurements of the spin, $\chi \gtrsim 0.95$ \cite{CygnusX1_spin,Gou:2013dna}.

\section{Conclusions}
\label{sec:conclusions} 

In this work, we simulate the ensemble annihilation signal from a population of
isolated Galactic black holes and predict the signal detectability
using recent all-sky continuous wave searches. We study the dependence of the ensemble signal  on black
hole population properties.  We show how population assumptions, in particular on black hole mass and spin distributions, can strongly
affect the detectability of the ensemble signal. We propose that the results of future boson annihilation signal searches 
are interpreted using the ensemble-signal paradigm that we present here. We make the ensemble signal population parameters used in this work publicly available to facilitate future studies \cite{onlinerepository}.

Our analysis is the first study to characterize
the unique shape of the ensemble signal --- constructed from a range of distributions
of the underlying parameters --- and suggests a clear way to distinguish between
these signals and the isolated signals produced by rotating neutron stars. 

We choose a population of $10^8$ black holes as a benchmark. For a population with the same mass, age and initial spin distribution but a different number of black holes, simple scaling arguments can be applied to our results in order to predict the results for that population. A study of the impact of different black hole population parameters on the boson mass constraints is very interesting but outside of the scope of this paper. On the other hand the ensemble signal approach that we propose here is exactly meant to enable this type of study.

We have considered the signals from five different black hole populations.
We use two mass distributions: the `standard' black hole mass distribution (maximum black hole mass $M_\mathrm{BH,max}=20\mathrm{M}_\odot$) covers the range of masses for known Galactic black holes \cite{Ozel+2010} and is therefore conservative, while the heavy black hole population ($M_\mathrm{BH,max}=30\mathrm{M}_\odot$) allows us to test how the properties of the ensemble signal depend on the heaviest black holes.
We also consider three different initial spin distributions: the `optimistic'
  spin distribution of $\chi_i \in [0,1]$, the `moderate' spin distribution of $\chi_i \in [0,0.5]$, and
the `pessimistic' spin distribution of $\chi_i \in [0,0.3]$.
For all populations, we assume a total of $10^8$ Galactic black holes.
The predicted number typically varies between $10^7$ and $10^8$ and is highly uncertain, as there are no direct observations;  the number of signals scales proportionally to black hole number. 

 Assuming there are $10^8$ isolated black holes
in the Galaxy with $M_\mathrm{BH} \in [5\mathrm{M}_\odot, 30\mathrm{M}_\odot]$
and initial spin $\chi_i \in [0,1]$, bosons with masses
  $\sim\!2.5$--$17\times10^{-13}$ eV
produce 100 or more signals with amplitude above the upper limits in the O1/O2  continuous wave searches, and masses $\sim\!4.5\times 10^{-13}$ eV produce over 1000 signals above those upper limits 
\cite{LVC_O1AS, LVC_O2AS, FalconO1,Dergachev:2019oyu}.  For a lighter population of black holes ($M_\mathrm{BH} \in [5\mathrm{M}_\odot, 20\mathrm{M}_\odot]$), the boson mass range yielding signal amplitudes above existing upper limits is not significantly different than the range for the heavier population  --- the lowest boson mass yielding 100 or more signals increases to $\sim\!3.5\times10^{-13}$~eV --- and the highest number of signals is a factor of $\sim\!2$ lower. 

If we take the maximum black hole spin at birth to be $0.5$ rather than $1.0$, keeping
$M_\mathrm{BH,max} = 20\mathrm{M}_\odot$, we find that the number of detectable signals
drops by an order of magnitude. This suggests that black holes with $\chi_i > 0.5$ produce
90\% of the ``detectable" signals, so  even if only 10\% of black holes are born with $\chi_i > 0.5$, we predict  tens to hundreds of ``detectable" signals
 for $3\times10^{-13}$~eV $\lesssim \mu_\mathrm{b} \lesssim 17\times10^{-13}$~eV.
The number of ``detectable" signals drops by at least another order of
magnitude if the maximum spin at birth is $0.3$, and black holes with
$\chi_i$ between 0.3 and 0.5 produce $\sim 10$\% of the detectable signals.
If instead {\it{all}} black holes are born with spins  $\lesssim 0.3$, we
expect ten or fewer signals across the entire frequency range. Given other uncertainties
such as the overall number of Galactic black holes, in this regime even a single detectable signal is not assured.

We caution that for boson masses in the range $\sim 3$--$10\times 10^{-13}$ eV, the sensitivity of standard continuous wave search methods 
that have been carefully tuned for the regime of quiet and 
very sparse signals, should be characterized again on a dense ensemble of loud signals. The interplay between 
the potential down-weighting of the signal from the normalization of the data and the overlap in parameter space of the detection statistic from 
different signals in the ensemble and how these elements factor in a multi-stage follow-up process, needs to be investigated. 
For boson masses between $4$--$6 \times 10^{-13}$ eV, even for moderately rotating black hole populations, the GW ensemble signal is very prominent, 
which is a dramatically different regime than the one assumed by standard continuous wave searches.

While the detection of a single particularly loud signal from a nearby black hole could serendipitously occur even with a highly ``mismatched" search, ultimately the identification of a boson annihilation signature --- as opposed to a continuous wave signal from a compact rotating object --- lies in the identification of the ensemble signal. Searches for gravitational-wave signals from boson clouds around black holes
are only starting to be explored. This work lays the foundations for this type of detectability study.
\\

\acknowledgments
The authors thank Richard Brito for fruitful discussions, and for providing
the numerical results and uncertainties of the annihilation
power. We thank  Bruce Allen, Pia Astone, Banafsheh Beheshtipour, Alessandra Buonanno,  Horng Sheng Chia, Vladimir Dergachev, Badri Krishnan, Andrew MacFadyen, and Cole Miller for useful discussions. MB is supported by the James Arthur Postdoctoral Fellowship.
DT is supported by the Advanced Leading Graduate Course for Photon Science
at the University of Tokyo, and by JSPS KAKENHI grant No JP19J21578.
NK is supported by the Hakubi project at Kyoto University.
\\ \\

\bibliographystyle{apsrev4-1}
\bibliography{BosonCloudsSignal}

\begin{thebibliography}{114}%
\makeatletter
\providecommand \@ifxundefined [1]{%
 \@ifx{#1\undefined}
}%
\providecommand \@ifnum [1]{%
 \ifnum #1\expandafter \@firstoftwo
 \else \expandafter \@secondoftwo
 \fi
}%
\providecommand \@ifx [1]{%
 \ifx #1\expandafter \@firstoftwo
 \else \expandafter \@secondoftwo
 \fi
}%
\providecommand \natexlab [1]{#1}%
\providecommand \enquote  [1]{``#1''}%
\providecommand \bibnamefont  [1]{#1}%
\providecommand \bibfnamefont [1]{#1}%
\providecommand \citenamefont [1]{#1}%
\providecommand \href@noop [0]{\@secondoftwo}%
\providecommand \href [0]{\begingroup \@sanitize@url \@href}%
\providecommand \@href[1]{\@@startlink{#1}\@@href}%
\providecommand \@@href[1]{\endgroup#1\@@endlink}%
\providecommand \@sanitize@url [0]{\catcode `\\12\catcode `\$12\catcode
  `\&12\catcode `\#12\catcode `\^12\catcode `\_12\catcode `\%12\relax}%
\providecommand \@@startlink[1]{}%
\providecommand \@@endlink[0]{}%
\providecommand \url  [0]{\begingroup\@sanitize@url \@url }%
\providecommand \@url [1]{\endgroup\@href {#1}{\urlprefix }}%
\providecommand \urlprefix  [0]{URL }%
\providecommand \Eprint [0]{\href }%
\providecommand \doibase [0]{http://dx.doi.org/}%
\providecommand \selectlanguage [0]{\@gobble}%
\providecommand \bibinfo  [0]{\@secondoftwo}%
\providecommand \bibfield  [0]{\@secondoftwo}%
\providecommand \translation [1]{[#1]}%
\providecommand \BibitemOpen [0]{}%
\providecommand \bibitemStop [0]{}%
\providecommand \bibitemNoStop [0]{.\EOS\space}%
\providecommand \EOS [0]{\spacefactor3000\relax}%
\providecommand \BibitemShut  [1]{\csname bibitem#1\endcsname}%
\let\auto@bib@innerbib\@empty
\bibitem [{\citenamefont {Abbott}\ \emph {et~al.}(2018)\citenamefont {Abbott}
  \emph {et~al.}}]{LVC_GWTC1}%
  \BibitemOpen
  \bibfield  {author} {\bibinfo {author} {\bibfnamefont {B.~P.}\ \bibnamefont
  {Abbott}} \emph {et~al.},\ }\href@noop {} {\enquote {\bibinfo {title}
  {{GWTC-1: A Gravitational-Wave Transient Catalog of Compact Binary Mergers
  Observed by LIGO and Virgo during the First and Second Observing Runs}},}\ }
  (\bibinfo {year} {2018}),\ \bibinfo {note} {{submitted}},\ \Eprint
  {http://arxiv.org/abs/1811.12907} {arXiv:1811.12907 [astro-ph.HE]}
  \BibitemShut {NoStop}%
\bibitem [{\citenamefont {Riles}(2017)}]{Riles2017}%
  \BibitemOpen
  \bibfield  {author} {\bibinfo {author} {\bibfnamefont {K.}~\bibnamefont
  {Riles}},\ }\href@noop {} {\bibfield  {journal} {\bibinfo  {journal} {{Modern
  Physics Letters A}}\ }\textbf {\bibinfo {volume} {32}} (\bibinfo {year}
  {2017})}\BibitemShut {NoStop}%
\bibitem [{\citenamefont {Arvanitaki}\ \emph {et~al.}(2010)\citenamefont
  {Arvanitaki}, \citenamefont {Dimopoulos}, \citenamefont {Dubovsky},
  \citenamefont {Kaloper},\ and\ \citenamefont
  {March-Russell}}]{Arvanitaki+2010}%
  \BibitemOpen
  \bibfield  {author} {\bibinfo {author} {\bibfnamefont {A.}~\bibnamefont
  {Arvanitaki}}, \bibinfo {author} {\bibfnamefont {S.}~\bibnamefont
  {Dimopoulos}}, \bibinfo {author} {\bibfnamefont {S.}~\bibnamefont
  {Dubovsky}}, \bibinfo {author} {\bibfnamefont {N.}~\bibnamefont {Kaloper}}, \
  and\ \bibinfo {author} {\bibfnamefont {J.}~\bibnamefont {March-Russell}},\
  }\href {\doibase 10.1103/PhysRevD.81.123530} {\bibfield  {journal} {\bibinfo
  {journal} {Phys. Rev.}\ }\textbf {\bibinfo {volume} {D81}},\ \bibinfo {pages}
  {123530} (\bibinfo {year} {2010})},\ \Eprint {http://arxiv.org/abs/0905.4720}
  {arXiv:0905.4720 [hep-th]} \BibitemShut {NoStop}%
\bibitem [{\citenamefont {Arvanitaki}\ and\ \citenamefont
  {Dubovsky}(2011)}]{Arvanitaki+2011}%
  \BibitemOpen
  \bibfield  {author} {\bibinfo {author} {\bibfnamefont {A.}~\bibnamefont
  {Arvanitaki}}\ and\ \bibinfo {author} {\bibfnamefont {S.}~\bibnamefont
  {Dubovsky}},\ }\href {\doibase 10.1103/PhysRevD.83.044026} {\bibfield
  {journal} {\bibinfo  {journal} {Phys. Rev.}\ }\textbf {\bibinfo {volume}
  {D83}},\ \bibinfo {pages} {044026} (\bibinfo {year} {2011})},\ \Eprint
  {http://arxiv.org/abs/1004.3558} {arXiv:1004.3558 [hep-th]} \BibitemShut
  {NoStop}%
\bibitem [{\citenamefont {Pierce}\ \emph {et~al.}(2018)\citenamefont {Pierce},
  \citenamefont {Riles},\ and\ \citenamefont {Zhao}}]{Pierce:2018xmy}%
  \BibitemOpen
  \bibfield  {author} {\bibinfo {author} {\bibfnamefont {A.}~\bibnamefont
  {Pierce}}, \bibinfo {author} {\bibfnamefont {K.}~\bibnamefont {Riles}}, \
  and\ \bibinfo {author} {\bibfnamefont {Y.}~\bibnamefont {Zhao}},\ }\href
  {\doibase 10.1103/PhysRevLett.121.061102} {\bibfield  {journal} {\bibinfo
  {journal} {Phys. Rev. Lett.}\ }\textbf {\bibinfo {volume} {121}},\ \bibinfo
  {pages} {061102} (\bibinfo {year} {2018})},\ \Eprint
  {http://arxiv.org/abs/1801.10161} {arXiv:1801.10161 [hep-ph]} \BibitemShut
  {NoStop}%
\bibitem [{\citenamefont {Guo}\ \emph {et~al.}(2019)\citenamefont {Guo},
  \citenamefont {Riles}, \citenamefont {Yang},\ and\ \citenamefont
  {Zhao}}]{Guo:2019ker}%
  \BibitemOpen
  \bibfield  {author} {\bibinfo {author} {\bibfnamefont {H.-K.}\ \bibnamefont
  {Guo}}, \bibinfo {author} {\bibfnamefont {K.}~\bibnamefont {Riles}}, \bibinfo
  {author} {\bibfnamefont {F.-W.}\ \bibnamefont {Yang}}, \ and\ \bibinfo
  {author} {\bibfnamefont {Y.}~\bibnamefont {Zhao}},\ }\href@noop {} {\
  (\bibinfo {year} {2019})},\ \Eprint {http://arxiv.org/abs/1905.04316}
  {arXiv:1905.04316 [hep-ph]} \BibitemShut {NoStop}%
\bibitem [{\citenamefont {Horowitz}\ \emph {et~al.}(2020)\citenamefont
  {Horowitz}, \citenamefont {Papa},\ and\ \citenamefont
  {Reddy}}]{Horowitz:2019pru}%
  \BibitemOpen
  \bibfield  {author} {\bibinfo {author} {\bibfnamefont {C.~J.}\ \bibnamefont
  {Horowitz}}, \bibinfo {author} {\bibfnamefont {M.~A.}\ \bibnamefont {Papa}},
  \ and\ \bibinfo {author} {\bibfnamefont {S.}~\bibnamefont {Reddy}},\ }\href
  {\doibase 10.1016/j.physletb.2019.135072} {\bibfield  {journal} {\bibinfo
  {journal} {Phys. Lett.}\ }\textbf {\bibinfo {volume} {B800}},\ \bibinfo
  {pages} {135072} (\bibinfo {year} {2020})},\ \Eprint
  {http://arxiv.org/abs/1902.08273} {arXiv:1902.08273 [gr-qc]} \BibitemShut
  {NoStop}%
\bibitem [{\citenamefont {Peccei}\ and\ \citenamefont
  {Quinn}(1977)}]{Peccei+1977}%
  \BibitemOpen
  \bibfield  {author} {\bibinfo {author} {\bibfnamefont {R.~D.}\ \bibnamefont
  {Peccei}}\ and\ \bibinfo {author} {\bibfnamefont {H.~R.}\ \bibnamefont
  {Quinn}},\ }\href@noop {} {\bibfield  {journal} {\bibinfo  {journal}
  {Physical Review Letters}\ }\textbf {\bibinfo {volume} {38}},\ \bibinfo
  {pages} {1440} (\bibinfo {year} {1977})}\BibitemShut {NoStop}%
\bibitem [{\citenamefont {Weinberg}(1978)}]{Weinberg1978}%
  \BibitemOpen
  \bibfield  {author} {\bibinfo {author} {\bibfnamefont {S.}~\bibnamefont
  {Weinberg}},\ }\href@noop {} {\bibfield  {journal} {\bibinfo  {journal}
  {{Physical Review Letters}}\ }\textbf {\bibinfo {volume} {40}},\ \bibinfo
  {pages} {{223}} (\bibinfo {year} {1978})}\BibitemShut {NoStop}%
\bibitem [{\citenamefont {Wilczek}(1978)}]{Wilczek1978}%
  \BibitemOpen
  \bibfield  {author} {\bibinfo {author} {\bibfnamefont {F.}~\bibnamefont
  {Wilczek}},\ }\href@noop {} {\bibfield  {journal} {\bibinfo  {journal}
  {{Physical Review Letters}}\ }\textbf {\bibinfo {volume} {40}},\ \bibinfo
  {pages} {279} (\bibinfo {year} {1978})}\BibitemShut {NoStop}%
\bibitem [{\citenamefont {Essig}\ \emph {et~al.}(2013)\citenamefont {Essig}
  \emph {et~al.}}]{Essig:2013lka}%
  \BibitemOpen
  \bibfield  {author} {\bibinfo {author} {\bibfnamefont {R.}~\bibnamefont
  {Essig}} \emph {et~al.}\ }(\bibinfo {year} {2013})\ \Eprint
  {http://arxiv.org/abs/1311.0029} {arXiv:1311.0029 [hep-ph]} \BibitemShut
  {NoStop}%
\bibitem [{\citenamefont {Arvanitaki}\ \emph {et~al.}(2015)\citenamefont
  {Arvanitaki}, \citenamefont {Baryakhtar},\ and\ \citenamefont
  {Huang}}]{Arvanitaki+2015}%
  \BibitemOpen
  \bibfield  {author} {\bibinfo {author} {\bibfnamefont {A.}~\bibnamefont
  {Arvanitaki}}, \bibinfo {author} {\bibfnamefont {M.}~\bibnamefont
  {Baryakhtar}}, \ and\ \bibinfo {author} {\bibfnamefont {X.}~\bibnamefont
  {Huang}},\ }\href {\doibase 10.1103/PhysRevD.91.084011} {\bibfield  {journal}
  {\bibinfo  {journal} {Phys. Rev.}\ }\textbf {\bibinfo {volume} {D91}},\
  \bibinfo {pages} {084011} (\bibinfo {year} {2015})},\ \Eprint
  {http://arxiv.org/abs/1411.2263} {arXiv:1411.2263 [hep-ph]} \BibitemShut
  {NoStop}%
\bibitem [{\citenamefont {Arvanitaki}\ \emph {et~al.}(2017)\citenamefont
  {Arvanitaki}, \citenamefont {Baryakhtar}, \citenamefont {Dimopoulos},
  \citenamefont {Dubovsky},\ and\ \citenamefont {Lasenby}}]{Arvanitaki+2017}%
  \BibitemOpen
  \bibfield  {author} {\bibinfo {author} {\bibfnamefont {A.}~\bibnamefont
  {Arvanitaki}}, \bibinfo {author} {\bibfnamefont {M.}~\bibnamefont
  {Baryakhtar}}, \bibinfo {author} {\bibfnamefont {S.}~\bibnamefont
  {Dimopoulos}}, \bibinfo {author} {\bibfnamefont {S.}~\bibnamefont
  {Dubovsky}}, \ and\ \bibinfo {author} {\bibfnamefont {R.}~\bibnamefont
  {Lasenby}},\ }\href {\doibase 10.1103/PhysRevD.95.043001} {\bibfield
  {journal} {\bibinfo  {journal} {Phys. Rev.}\ }\textbf {\bibinfo {volume}
  {D95}},\ \bibinfo {pages} {043001} (\bibinfo {year} {2017})},\ \Eprint
  {http://arxiv.org/abs/1604.03958} {arXiv:1604.03958 [hep-ph]} \BibitemShut
  {NoStop}%
\bibitem [{\citenamefont {Brito}\ \emph
  {et~al.}(2017{\natexlab{a}})\citenamefont {Brito}, \citenamefont {Ghosh},
  \citenamefont {Barausse}, \citenamefont {Berti}, \citenamefont {Cardoso},
  \citenamefont {Dvorkin}, \citenamefont {Klein},\ and\ \citenamefont
  {Pani}}]{Brito+2017}%
  \BibitemOpen
  \bibfield  {author} {\bibinfo {author} {\bibfnamefont {R.}~\bibnamefont
  {Brito}}, \bibinfo {author} {\bibfnamefont {S.}~\bibnamefont {Ghosh}},
  \bibinfo {author} {\bibfnamefont {E.}~\bibnamefont {Barausse}}, \bibinfo
  {author} {\bibfnamefont {E.}~\bibnamefont {Berti}}, \bibinfo {author}
  {\bibfnamefont {V.}~\bibnamefont {Cardoso}}, \bibinfo {author} {\bibfnamefont
  {I.}~\bibnamefont {Dvorkin}}, \bibinfo {author} {\bibfnamefont
  {A.}~\bibnamefont {Klein}}, \ and\ \bibinfo {author} {\bibfnamefont
  {P.}~\bibnamefont {Pani}},\ }\href@noop {} {\bibfield  {journal} {\bibinfo
  {journal} {Physical Review D}\ }\textbf {\bibinfo {volume} {96}},\ \bibinfo
  {pages} {064050} (\bibinfo {year} {2017}{\natexlab{a}})}\BibitemShut
  {NoStop}%
\bibitem [{\citenamefont {Brito}\ \emph
  {et~al.}(2017{\natexlab{b}})\citenamefont {Brito}, \citenamefont {Ghosh},
  \citenamefont {Barausse}, \citenamefont {Berti}, \citenamefont {Cardoso},
  \citenamefont {Dvorkin}, \citenamefont {Klein},\ and\ \citenamefont
  {Pani}}]{Brito+2017a}%
  \BibitemOpen
  \bibfield  {author} {\bibinfo {author} {\bibfnamefont {R.}~\bibnamefont
  {Brito}}, \bibinfo {author} {\bibfnamefont {S.}~\bibnamefont {Ghosh}},
  \bibinfo {author} {\bibfnamefont {E.}~\bibnamefont {Barausse}}, \bibinfo
  {author} {\bibfnamefont {E.}~\bibnamefont {Berti}}, \bibinfo {author}
  {\bibfnamefont {V.}~\bibnamefont {Cardoso}}, \bibinfo {author} {\bibfnamefont
  {I.}~\bibnamefont {Dvorkin}}, \bibinfo {author} {\bibfnamefont
  {A.}~\bibnamefont {Klein}}, \ and\ \bibinfo {author} {\bibfnamefont
  {P.}~\bibnamefont {Pani}},\ }\href {\doibase 10.1103/PhysRevLett.119.131101}
  {\bibfield  {journal} {\bibinfo  {journal} {Phys. Rev. Lett.}\ }\textbf
  {\bibinfo {volume} {119}},\ \bibinfo {pages} {131101} (\bibinfo {year}
  {2017}{\natexlab{b}})},\ \Eprint {http://arxiv.org/abs/1706.05097}
  {arXiv:1706.05097 [gr-qc]} \BibitemShut {NoStop}%
\bibitem [{\citenamefont {Miller}\ and\ \citenamefont
  {Miller}(2014)}]{Miller:2014aaa}%
  \BibitemOpen
  \bibfield  {author} {\bibinfo {author} {\bibfnamefont {M.~C.}\ \bibnamefont
  {Miller}}\ and\ \bibinfo {author} {\bibfnamefont {J.~M.}\ \bibnamefont
  {Miller}},\ }\href {\doibase 10.1016/j.physrep.2014.09.003} {\bibfield
  {journal} {\bibinfo  {journal} {Phys. Rept.}\ }\textbf {\bibinfo {volume}
  {548}},\ \bibinfo {pages} {1} (\bibinfo {year} {2014})}\BibitemShut {NoStop}%
\bibitem [{\citenamefont {McClintock}\ \emph {et~al.}(2014)\citenamefont
  {McClintock}, \citenamefont {Narayan},\ and\ \citenamefont
  {Steiner}}]{McClintock:2013vwa}%
  \BibitemOpen
  \bibfield  {author} {\bibinfo {author} {\bibfnamefont {J.~E.}\ \bibnamefont
  {McClintock}}, \bibinfo {author} {\bibfnamefont {R.}~\bibnamefont {Narayan}},
  \ and\ \bibinfo {author} {\bibfnamefont {J.~F.}\ \bibnamefont {Steiner}},\
  }\href {\doibase 10.1007/s11214-013-0003-9} {\bibfield  {journal} {\bibinfo
  {journal} {Space Sci. Rev.}\ }\textbf {\bibinfo {volume} {183}},\ \bibinfo
  {pages} {295} (\bibinfo {year} {2014})}\BibitemShut {NoStop}%
\bibitem [{\citenamefont {Cardoso}\ \emph {et~al.}(2018)\citenamefont
  {Cardoso}, \citenamefont {Dias}, \citenamefont {Hartnett}, \citenamefont
  {Middleton}, \citenamefont {Pani},\ and\ \citenamefont
  {Santos}}]{Cardoso+2018}%
  \BibitemOpen
  \bibfield  {author} {\bibinfo {author} {\bibfnamefont {V.}~\bibnamefont
  {Cardoso}}, \bibinfo {author} {\bibfnamefont {O.~J.~C.}\ \bibnamefont
  {Dias}}, \bibinfo {author} {\bibfnamefont {G.~S.}\ \bibnamefont {Hartnett}},
  \bibinfo {author} {\bibfnamefont {M.}~\bibnamefont {Middleton}}, \bibinfo
  {author} {\bibfnamefont {P.}~\bibnamefont {Pani}}, \ and\ \bibinfo {author}
  {\bibfnamefont {J.~E.}\ \bibnamefont {Santos}},\ }\href@noop {} {\bibfield
  {journal} {\bibinfo  {journal} {Journal of Cosmology and Astroparticle
  Physics}\ ,\ \bibinfo {pages} {043}} (\bibinfo {year} {2018})}\BibitemShut
  {NoStop}%
\bibitem [{\citenamefont {Tsukada}\ \emph {et~al.}(2019)\citenamefont
  {Tsukada}, \citenamefont {Callister}, \citenamefont {Matas},\ and\
  \citenamefont {Meyers}}]{Tsukada+2019}%
  \BibitemOpen
  \bibfield  {author} {\bibinfo {author} {\bibfnamefont {L.}~\bibnamefont
  {Tsukada}}, \bibinfo {author} {\bibfnamefont {T.}~\bibnamefont {Callister}},
  \bibinfo {author} {\bibfnamefont {A.}~\bibnamefont {Matas}}, \ and\ \bibinfo
  {author} {\bibfnamefont {P.}~\bibnamefont {Meyers}},\ }\href {\doibase
  10.1103/PhysRevD.99.103015} {\bibfield  {journal} {\bibinfo  {journal} {Phys.
  Rev.}\ }\textbf {\bibinfo {volume} {D99}},\ \bibinfo {pages} {103015}
  (\bibinfo {year} {2019})},\ \Eprint {http://arxiv.org/abs/1812.09622}
  {arXiv:1812.09622 [astro-ph.HE]} \BibitemShut {NoStop}%
\bibitem [{\citenamefont {Dergachev}\ and\ \citenamefont
  {Papa}(2019)}]{FalconO1}%
  \BibitemOpen
  \bibfield  {author} {\bibinfo {author} {\bibfnamefont {V.}~\bibnamefont
  {Dergachev}}\ and\ \bibinfo {author} {\bibfnamefont {M.~A.}\ \bibnamefont
  {Papa}},\ }\href@noop {} {\bibfield  {journal} {\bibinfo  {journal}
  {{Physical Review Letters}}\ }\textbf {\bibinfo {volume} {123}},\ \bibinfo
  {pages} {101101} (\bibinfo {year} {2019})}\BibitemShut {NoStop}%
\bibitem [{\citenamefont {Dergachev}\ and\ \citenamefont
  {Papa}(2020{\natexlab{a}})}]{Dergachev:2019oyu}%
  \BibitemOpen
  \bibfield  {author} {\bibinfo {author} {\bibfnamefont {V.}~\bibnamefont
  {Dergachev}}\ and\ \bibinfo {author} {\bibfnamefont {M.~A.}\ \bibnamefont
  {Papa}},\ }\href {\doibase 10.1103/PhysRevD.101.022001} {\bibfield  {journal}
  {\bibinfo  {journal} {Phys. Rev.}\ }\textbf {\bibinfo {volume} {D101}},\
  \bibinfo {pages} {022001} (\bibinfo {year} {2020}{\natexlab{a}})},\ \Eprint
  {http://arxiv.org/abs/1909.09619} {arXiv:1909.09619 [gr-qc]} \BibitemShut
  {NoStop}%
\bibitem [{\citenamefont {Palomba}\ \emph {et~al.}(2019)\citenamefont {Palomba}
  \emph {et~al.}}]{RomeBosonClouds}%
  \BibitemOpen
  \bibfield  {author} {\bibinfo {author} {\bibfnamefont {C.}~\bibnamefont
  {Palomba}} \emph {et~al.},\ }\href {\doibase 10.1103/PhysRevLett.123.171101}
  {\bibfield  {journal} {\bibinfo  {journal} {Phys. Rev. Lett.}\ }\textbf
  {\bibinfo {volume} {123}},\ \bibinfo {pages} {171101} (\bibinfo {year}
  {2019})},\ \Eprint {http://arxiv.org/abs/1909.08854} {arXiv:1909.08854
  [astro-ph.HE]} \BibitemShut {NoStop}%
\bibitem [{\citenamefont {Abbott}\ \emph
  {et~al.}(2017{\natexlab{a}})\citenamefont {Abbott} \emph
  {et~al.}}]{Abbott:2017pqa}%
  \BibitemOpen
  \bibfield  {author} {\bibinfo {author} {\bibfnamefont {B.~P.}\ \bibnamefont
  {Abbott}} \emph {et~al.} (\bibinfo {collaboration} {LIGO Scientific,
  Virgo}),\ }\href {\doibase 10.1103/PhysRevD.96.122004} {\bibfield  {journal}
  {\bibinfo  {journal} {Phys. Rev.}\ }\textbf {\bibinfo {volume} {D96}},\
  \bibinfo {pages} {122004} (\bibinfo {year} {2017}{\natexlab{a}})},\ \Eprint
  {http://arxiv.org/abs/1707.02669} {arXiv:1707.02669 [gr-qc]} \BibitemShut
  {NoStop}%
\bibitem [{\citenamefont {Rosa}\ and\ \citenamefont {Dolan}(2012)}]{Rosa+2012}%
  \BibitemOpen
  \bibfield  {author} {\bibinfo {author} {\bibfnamefont {J.~G.}\ \bibnamefont
  {Rosa}}\ and\ \bibinfo {author} {\bibfnamefont {S.~R.}\ \bibnamefont
  {Dolan}},\ }\href@noop {} {\bibfield  {journal} {\bibinfo  {journal}
  {{Physical Review D}}\ }\textbf {\bibinfo {volume} {85}},\ \bibinfo {pages}
  {044043} (\bibinfo {year} {2012})}\BibitemShut {NoStop}%
\bibitem [{\citenamefont {Pani}\ \emph {et~al.}(2012)\citenamefont {Pani},
  \citenamefont {Cardoso}, \citenamefont {Gualtieri}, \citenamefont {Berti},\
  and\ \citenamefont {Ishibashi}}]{Pani+2012}%
  \BibitemOpen
  \bibfield  {author} {\bibinfo {author} {\bibfnamefont {P.}~\bibnamefont
  {Pani}}, \bibinfo {author} {\bibfnamefont {V.}~\bibnamefont {Cardoso}},
  \bibinfo {author} {\bibfnamefont {L.}~\bibnamefont {Gualtieri}}, \bibinfo
  {author} {\bibfnamefont {E.}~\bibnamefont {Berti}}, \ and\ \bibinfo {author}
  {\bibfnamefont {A.}~\bibnamefont {Ishibashi}},\ }\href@noop {} {\bibfield
  {journal} {\bibinfo  {journal} {{Physical Review D}}\ }\textbf {\bibinfo
  {volume} {86}},\ \bibinfo {pages} {104017} (\bibinfo {year}
  {2012})}\BibitemShut {NoStop}%
\bibitem [{\citenamefont {Endlich}\ and\ \citenamefont
  {Penco}(2017)}]{Endlich+2017}%
  \BibitemOpen
  \bibfield  {author} {\bibinfo {author} {\bibfnamefont {S.}~\bibnamefont
  {Endlich}}\ and\ \bibinfo {author} {\bibfnamefont {R.}~\bibnamefont
  {Penco}},\ }\href@noop {} {\bibfield  {journal} {\bibinfo  {journal} {Journal
  of High Energy Physics}\ }\textbf {\bibinfo {volume} {2017}} (\bibinfo {year}
  {2017})}\BibitemShut {NoStop}%
\bibitem [{\citenamefont {Baryakhtar}\ \emph {et~al.}(2017)\citenamefont
  {Baryakhtar}, \citenamefont {Lasenby},\ and\ \citenamefont
  {Teo}}]{Baryakhtar+2017}%
  \BibitemOpen
  \bibfield  {author} {\bibinfo {author} {\bibfnamefont {M.}~\bibnamefont
  {Baryakhtar}}, \bibinfo {author} {\bibfnamefont {R.}~\bibnamefont {Lasenby}},
  \ and\ \bibinfo {author} {\bibfnamefont {M.}~\bibnamefont {Teo}},\ }\href
  {\doibase 10.1103/PhysRevD.96.035019} {\bibfield  {journal} {\bibinfo
  {journal} {Phys. Rev.}\ }\textbf {\bibinfo {volume} {D96}},\ \bibinfo {pages}
  {035019} (\bibinfo {year} {2017})},\ \Eprint
  {http://arxiv.org/abs/1704.05081} {arXiv:1704.05081 [hep-ph]} \BibitemShut
  {NoStop}%
\bibitem [{\citenamefont {East}(2017)}]{East+2017}%
  \BibitemOpen
  \bibfield  {author} {\bibinfo {author} {\bibfnamefont {W.~E.}\ \bibnamefont
  {East}},\ }\href {\doibase 10.1103/PhysRevD.96.024004} {\bibfield  {journal}
  {\bibinfo  {journal} {Phys. Rev.}\ }\textbf {\bibinfo {volume} {D96}},\
  \bibinfo {pages} {024004} (\bibinfo {year} {2017})}\BibitemShut {NoStop}%
\bibitem [{\citenamefont {Brito}\ \emph {et~al.}(2020)\citenamefont {Brito},
  \citenamefont {Grillo},\ and\ \citenamefont {Pani}}]{Brito+2020}%
  \BibitemOpen
  \bibfield  {author} {\bibinfo {author} {\bibfnamefont {R.}~\bibnamefont
  {Brito}}, \bibinfo {author} {\bibfnamefont {S.}~\bibnamefont {Grillo}}, \
  and\ \bibinfo {author} {\bibfnamefont {P.}~\bibnamefont {Pani}},\ }\href@noop
  {} {\  (\bibinfo {year} {2020})},\ \Eprint {http://arxiv.org/abs/2002.04055}
  {arXiv:2002.04055 [gr-qc]} \BibitemShut {NoStop}%
\bibitem [{\citenamefont {Brito}\ \emph
  {et~al.}(2015{\natexlab{a}})\citenamefont {Brito}, \citenamefont {Cardoso},\
  and\ \citenamefont {Pani}}]{Brito+lectureNotes}%
  \BibitemOpen
  \bibfield  {author} {\bibinfo {author} {\bibfnamefont {R.}~\bibnamefont
  {Brito}}, \bibinfo {author} {\bibfnamefont {V.}~\bibnamefont {Cardoso}}, \
  and\ \bibinfo {author} {\bibfnamefont {P.}~\bibnamefont {Pani}},\ }\href
  {\doibase 10.1007/978-3-319-19000-6} {\emph {\bibinfo {title}
  {{Superradiance}: {Energy Extraction, Black-Hole Bombs and Implications for
  Astrophysics and Particle Physics}}}},\ Vol.\ \bibinfo {volume} {906}\
  (\bibinfo  {publisher} {Springer},\ \bibinfo {year} {2015})\ \Eprint
  {http://arxiv.org/abs/1501.06570} {arXiv:1501.06570 [gr-qc]} \BibitemShut
  {NoStop}%
\bibitem [{\citenamefont {Zeldovich}(1971)}]{Zeldovich1971}%
  \BibitemOpen
  \bibfield  {author} {\bibinfo {author} {\bibfnamefont {Y.~B.}\ \bibnamefont
  {Zeldovich}},\ }\href@noop {} {\bibfield  {journal} {\bibinfo  {journal}
  {Journal of Experimental and Theoretical Physics Letters}\ }\textbf {\bibinfo
  {volume} {14}},\ \bibinfo {pages} {180} (\bibinfo {year} {1971})}\BibitemShut
  {NoStop}%
\bibitem [{\citenamefont {Misner}(1972)}]{Misner1972}%
  \BibitemOpen
  \bibfield  {author} {\bibinfo {author} {\bibfnamefont {C.~W.}\ \bibnamefont
  {Misner}},\ }\href@noop {} {\bibfield  {journal} {\bibinfo  {journal}
  {Physical Review Letters}\ }\textbf {\bibinfo {volume} {28}},\ \bibinfo
  {pages} {994} (\bibinfo {year} {1972})}\BibitemShut {NoStop}%
\bibitem [{\citenamefont {Starobinskii}(1973)}]{Starobinskii1973}%
  \BibitemOpen
  \bibfield  {author} {\bibinfo {author} {\bibfnamefont {A.~A.}\ \bibnamefont
  {Starobinskii}},\ }\href@noop {} {\bibfield  {journal} {\bibinfo  {journal}
  {Soviet Phys JETP}\ }\textbf {\bibinfo {volume} {37}},\ \bibinfo {pages} {28}
  (\bibinfo {year} {1973})}\BibitemShut {NoStop}%
\bibitem [{\citenamefont {Damour}\ \emph {et~al.}(1976)\citenamefont {Damour},
  \citenamefont {Deruelle},\ and\ \citenamefont {Ruffini}}]{Damour:1976kh}%
  \BibitemOpen
  \bibfield  {author} {\bibinfo {author} {\bibfnamefont {T.}~\bibnamefont
  {Damour}}, \bibinfo {author} {\bibfnamefont {N.}~\bibnamefont {Deruelle}}, \
  and\ \bibinfo {author} {\bibfnamefont {R.}~\bibnamefont {Ruffini}},\ }\href
  {\doibase 10.1007/BF02725534} {\bibfield  {journal} {\bibinfo  {journal}
  {Lett. Nuovo Cim.}\ }\textbf {\bibinfo {volume} {15}},\ \bibinfo {pages}
  {257} (\bibinfo {year} {1976})}\BibitemShut {NoStop}%
\bibitem [{\citenamefont {Ternov}\ \emph {et~al.}(1978)\citenamefont {Ternov},
  \citenamefont {Khalilov}, \citenamefont {Chizhov},\ and\ \citenamefont
  {Gaina}}]{Ternov:1978gq}%
  \BibitemOpen
  \bibfield  {author} {\bibinfo {author} {\bibfnamefont {I.~M.}\ \bibnamefont
  {Ternov}}, \bibinfo {author} {\bibfnamefont {V.~R.}\ \bibnamefont
  {Khalilov}}, \bibinfo {author} {\bibfnamefont {G.~A.}\ \bibnamefont
  {Chizhov}}, \ and\ \bibinfo {author} {\bibfnamefont {A.~B.}\ \bibnamefont
  {Gaina}},\ }\href {\doibase 10.1007/BF00894575} {\bibfield  {journal}
  {\bibinfo  {journal} {Sov. Phys. J.}\ }\textbf {\bibinfo {volume} {21}},\
  \bibinfo {pages} {1200} (\bibinfo {year} {1978})},\ \bibinfo {note} {[Izv.
  Vuz. Fiz.21N9,109(1978)]}\BibitemShut {NoStop}%
\bibitem [{\citenamefont {Zouros}\ and\ \citenamefont
  {Eardley}(1979)}]{Zouros:1979iw}%
  \BibitemOpen
  \bibfield  {author} {\bibinfo {author} {\bibfnamefont {T.~J.~M.}\
  \bibnamefont {Zouros}}\ and\ \bibinfo {author} {\bibfnamefont {D.~M.}\
  \bibnamefont {Eardley}},\ }\href {\doibase 10.1016/0003-4916(79)90237-9}
  {\bibfield  {journal} {\bibinfo  {journal} {Annals Phys.}\ }\textbf {\bibinfo
  {volume} {118}},\ \bibinfo {pages} {139} (\bibinfo {year}
  {1979})}\BibitemShut {NoStop}%
\bibitem [{\citenamefont {Detweiler}(1980)}]{Detweiler1980}%
  \BibitemOpen
  \bibfield  {author} {\bibinfo {author} {\bibfnamefont {S.~L.}\ \bibnamefont
  {Detweiler}},\ }\href@noop {} {\bibfield  {journal} {\bibinfo  {journal}
  {Physical Review D}\ }\textbf {\bibinfo {volume} {22}},\ \bibinfo {pages}
  {2323} (\bibinfo {year} {1980})}\BibitemShut {NoStop}%
\bibitem [{\citenamefont {Gruzinov}(2016)}]{Gruzinov2016}%
  \BibitemOpen
  \bibfield  {author} {\bibinfo {author} {\bibfnamefont {A.}~\bibnamefont
  {Gruzinov}},\ }\href@noop {} {\enquote {\bibinfo {title} {{Black Hole
  Spindown by Light Bosons}},}\ } (\bibinfo {year} {2016}),\ \Eprint
  {http://arxiv.org/abs/1604.06422} {arXiv:1604.06422 [astro-ph.HE]}
  \BibitemShut {NoStop}%
\bibitem [{\citenamefont {Baryakhtar}\ \emph {et~al.}()\citenamefont
  {Baryakhtar}, \citenamefont {Galanis}, \citenamefont {Lasenby},\ and\
  \citenamefont {Simon}}]{Baryakhtar+2020}%
  \BibitemOpen
  \bibfield  {author} {\bibinfo {author} {\bibfnamefont {M.}~\bibnamefont
  {Baryakhtar}}, \bibinfo {author} {\bibfnamefont {M.}~\bibnamefont {Galanis}},
  \bibinfo {author} {\bibfnamefont {R.}~\bibnamefont {Lasenby}}, \ and\
  \bibinfo {author} {\bibfnamefont {O.}~\bibnamefont {Simon}},\ }\href@noop {}
  {\enquote {\bibinfo {title} {{Black hole superradiance of self-interacting
  scalar fields}},}\ }\bibinfo {note} {In preparation}\BibitemShut {NoStop}%
\bibitem [{\citenamefont {Yoshino}\ and\ \citenamefont
  {Kodama}(2012)}]{Yoshino+2012}%
  \BibitemOpen
  \bibfield  {author} {\bibinfo {author} {\bibfnamefont {H.}~\bibnamefont
  {Yoshino}}\ and\ \bibinfo {author} {\bibfnamefont {H.}~\bibnamefont
  {Kodama}},\ }\href@noop {} {\bibfield  {journal} {\bibinfo  {journal}
  {Progress of Theoretical and Experimental Physics}\ }\textbf {\bibinfo
  {volume} {128}},\ \bibinfo {pages} {153} (\bibinfo {year}
  {2012})}\BibitemShut {NoStop}%
\bibitem [{\citenamefont {Yoshino}\ and\ \citenamefont
  {Kodama}(2015)}]{Yoshino+2015}%
  \BibitemOpen
  \bibfield  {author} {\bibinfo {author} {\bibfnamefont {H.}~\bibnamefont
  {Yoshino}}\ and\ \bibinfo {author} {\bibfnamefont {H.}~\bibnamefont
  {Kodama}},\ }\href@noop {} {\bibfield  {journal} {\bibinfo  {journal}
  {Classical and Quantum Gravity}\ }\textbf {\bibinfo {volume} {32}},\ \bibinfo
  {pages} {214001} (\bibinfo {year} {2015})}\BibitemShut {NoStop}%
\bibitem [{\citenamefont {Dolan}(2007)}]{Dolan2007}%
  \BibitemOpen
  \bibfield  {author} {\bibinfo {author} {\bibfnamefont {S.}~\bibnamefont
  {Dolan}},\ }\href@noop {} {\bibfield  {journal} {\bibinfo  {journal}
  {{Physical Review D}}\ }\textbf {\bibinfo {volume} {76}},\ \bibinfo {pages}
  {084001} (\bibinfo {year} {2007})}\BibitemShut {NoStop}%
\bibitem [{\citenamefont {Baumann}\ \emph {et~al.}(2019)\citenamefont
  {Baumann}, \citenamefont {Chia},\ and\ \citenamefont {Porto}}]{Baumann+2019}%
  \BibitemOpen
  \bibfield  {author} {\bibinfo {author} {\bibfnamefont {D.}~\bibnamefont
  {Baumann}}, \bibinfo {author} {\bibfnamefont {H.~S.}\ \bibnamefont {Chia}}, \
  and\ \bibinfo {author} {\bibfnamefont {R.~A.}\ \bibnamefont {Porto}},\
  }\href@noop {} {\bibfield  {journal} {\bibinfo  {journal} {Physical Review
  D}\ }\textbf {\bibinfo {volume} {99}},\ \bibinfo {pages} {044001} (\bibinfo
  {year} {2019})}\BibitemShut {NoStop}%
\bibitem [{\citenamefont {Brito}\ \emph
  {et~al.}(2015{\natexlab{b}})\citenamefont {Brito}, \citenamefont {Cardoso},\
  and\ \citenamefont {Pani}}]{Brito:2014wla}%
  \BibitemOpen
  \bibfield  {author} {\bibinfo {author} {\bibfnamefont {R.}~\bibnamefont
  {Brito}}, \bibinfo {author} {\bibfnamefont {V.}~\bibnamefont {Cardoso}}, \
  and\ \bibinfo {author} {\bibfnamefont {P.}~\bibnamefont {Pani}},\ }\href
  {\doibase 10.1088/0264-9381/32/13/134001} {\bibfield  {journal} {\bibinfo
  {journal} {Class. Quant. Grav.}\ }\textbf {\bibinfo {volume} {32}},\ \bibinfo
  {pages} {134001} (\bibinfo {year} {2015}{\natexlab{b}})},\ \Eprint
  {http://arxiv.org/abs/1411.0686} {arXiv:1411.0686 [gr-qc]} \BibitemShut
  {NoStop}%
\bibitem [{\citenamefont {Yoshino}\ and\ \citenamefont
  {Kodama}(2014)}]{Yoshino+2014}%
  \BibitemOpen
  \bibfield  {author} {\bibinfo {author} {\bibfnamefont {H.}~\bibnamefont
  {Yoshino}}\ and\ \bibinfo {author} {\bibfnamefont {H.}~\bibnamefont
  {Kodama}},\ }\href {\doibase 10.1093/ptep/ptu029} {\bibfield  {journal}
  {\bibinfo  {journal} {PTEP}\ }\textbf {\bibinfo {volume} {2014}},\ \bibinfo
  {pages} {043E02} (\bibinfo {year} {2014})},\ \Eprint
  {http://arxiv.org/abs/1312.2326} {arXiv:1312.2326 [gr-qc]} \BibitemShut
  {NoStop}%
\bibitem [{\citenamefont {LIGO}\ and\ \citenamefont
  {Collaborations}(2017)}]{O1AS20-100}%
  \BibitemOpen
  \bibfield  {author} {\bibinfo {author} {\bibnamefont {LIGO}}\ and\ \bibinfo
  {author} {\bibfnamefont {V.}~\bibnamefont {Collaborations}},\ }\href@noop {}
  {\bibfield  {journal} {\bibinfo  {journal} {Physics Review D}\ }\textbf
  {\bibinfo {volume} {96}},\ \bibinfo {pages} {122004} (\bibinfo {year}
  {2017})}\BibitemShut {NoStop}%
\bibitem [{\citenamefont {Abbott}\ \emph
  {et~al.}(2017{\natexlab{b}})\citenamefont {Abbott} \emph
  {et~al.}}]{LVC_O1AS}%
  \BibitemOpen
  \bibfield  {author} {\bibinfo {author} {\bibfnamefont {B.~P.}\ \bibnamefont
  {Abbott}} \emph {et~al.},\ }\href@noop {} {\bibfield  {journal} {\bibinfo
  {journal} {{Physical Review D}}\ }\textbf {\bibinfo {volume} {96}},\ \bibinfo
  {pages} {062002} (\bibinfo {year} {2017}{\natexlab{b}})}\BibitemShut
  {NoStop}%
\bibitem [{\citenamefont {Bosch}\ \emph {et~al.}(2016)\citenamefont {Bosch},
  \citenamefont {Green},\ and\ \citenamefont {Lehner}}]{Bosch:2016vcp}%
  \BibitemOpen
  \bibfield  {author} {\bibinfo {author} {\bibfnamefont {P.}~\bibnamefont
  {Bosch}}, \bibinfo {author} {\bibfnamefont {S.~R.}\ \bibnamefont {Green}}, \
  and\ \bibinfo {author} {\bibfnamefont {L.}~\bibnamefont {Lehner}},\ }\href
  {\doibase 10.1103/PhysRevLett.116.141102} {\bibfield  {journal} {\bibinfo
  {journal} {Phys. Rev. Lett.}\ }\textbf {\bibinfo {volume} {116}},\ \bibinfo
  {pages} {141102} (\bibinfo {year} {2016})},\ \Eprint
  {http://arxiv.org/abs/1601.01384} {arXiv:1601.01384 [gr-qc]} \BibitemShut
  {NoStop}%
\bibitem [{\citenamefont {Shapiro}\ and\ \citenamefont
  {Teukolsky}(1983)}]{ShapiroTeukolsky1983}%
  \BibitemOpen
  \bibfield  {author} {\bibinfo {author} {\bibfnamefont {S.~L.}\ \bibnamefont
  {Shapiro}}\ and\ \bibinfo {author} {\bibfnamefont {S.~A.}\ \bibnamefont
  {Teukolsky}},\ }\href@noop {} {\emph {\bibinfo {title} {{Black holes, white
  dwarfs, and neutron stars: the physics of compact objects}}}}\ (\bibinfo
  {publisher} {Wiley},\ \bibinfo {year} {1983})\BibitemShut {NoStop}%
\bibitem [{\citenamefont {Caputo}\ \emph {et~al.}(2017)\citenamefont {Caputo},
  \citenamefont {de~Vries}, \citenamefont {Patruno},\ and\ \citenamefont
  {Portegies~Zwart}}]{Caputo2017}%
  \BibitemOpen
  \bibfield  {author} {\bibinfo {author} {\bibfnamefont {D.~P.}\ \bibnamefont
  {Caputo}}, \bibinfo {author} {\bibfnamefont {N.}~\bibnamefont {de~Vries}},
  \bibinfo {author} {\bibfnamefont {A.}~\bibnamefont {Patruno}}, \ and\
  \bibinfo {author} {\bibfnamefont {S.}~\bibnamefont {Portegies~Zwart}},\
  }\href {\doibase 10.1093/mnras/stw3336} {\bibfield  {journal} {\bibinfo
  {journal} {Monthly Notices of the Royal Astronomical Society}\ }\textbf
  {\bibinfo {volume} {468}},\ \bibinfo {pages} {4000} (\bibinfo {year}
  {2017})},\ \Eprint
  {http://arxiv.org/abs/http://oup.prod.sis.lan/mnras/article-pdf/468/4/4000/13953577/stw3336.pdf}
  {http://oup.prod.sis.lan/mnras/article-pdf/468/4/4000/13953577/stw3336.pdf}
  \BibitemShut {NoStop}%
\bibitem [{\citenamefont {Heger}\ \emph {et~al.}(2003)\citenamefont {Heger},
  \citenamefont {Fryer}, \citenamefont {Woosley}, \citenamefont {Langer},\ and\
  \citenamefont {Hartmann}}]{Heger:2002by}%
  \BibitemOpen
  \bibfield  {author} {\bibinfo {author} {\bibfnamefont {A.}~\bibnamefont
  {Heger}}, \bibinfo {author} {\bibfnamefont {C.~L.}\ \bibnamefont {Fryer}},
  \bibinfo {author} {\bibfnamefont {S.~E.}\ \bibnamefont {Woosley}}, \bibinfo
  {author} {\bibfnamefont {N.}~\bibnamefont {Langer}}, \ and\ \bibinfo {author}
  {\bibfnamefont {D.~H.}\ \bibnamefont {Hartmann}},\ }\href {\doibase
  10.1086/375341} {\bibfield  {journal} {\bibinfo  {journal} {Astrophys. J.}\
  }\textbf {\bibinfo {volume} {591}},\ \bibinfo {pages} {288} (\bibinfo {year}
  {2003})},\ \Eprint {http://arxiv.org/abs/astro-ph/0212469}
  {arXiv:astro-ph/0212469 [astro-ph]} \BibitemShut {NoStop}%
\bibitem [{\citenamefont {Zhang}\ \emph {et~al.}(2008)\citenamefont {Zhang},
  \citenamefont {Woosley},\ and\ \citenamefont {Heger}}]{Zhang:2007nw}%
  \BibitemOpen
  \bibfield  {author} {\bibinfo {author} {\bibfnamefont {W.-Q.}\ \bibnamefont
  {Zhang}}, \bibinfo {author} {\bibfnamefont {S.~E.}\ \bibnamefont {Woosley}},
  \ and\ \bibinfo {author} {\bibfnamefont {A.}~\bibnamefont {Heger}},\ }\href
  {\doibase 10.1086/526404} {\bibfield  {journal} {\bibinfo  {journal}
  {Astrophys. J.}\ }\textbf {\bibinfo {volume} {679}},\ \bibinfo {pages} {639}
  (\bibinfo {year} {2008})},\ \Eprint {http://arxiv.org/abs/astro-ph/0701083}
  {arXiv:astro-ph/0701083 [astro-ph]} \BibitemShut {NoStop}%
\bibitem [{\citenamefont {Prantzos}\ and\ \citenamefont
  {Boissier}(2003)}]{Prantzos:2003ph}%
  \BibitemOpen
  \bibfield  {author} {\bibinfo {author} {\bibfnamefont {N.}~\bibnamefont
  {Prantzos}}\ and\ \bibinfo {author} {\bibfnamefont {S.}~\bibnamefont
  {Boissier}},\ }\href {\doibase 10.1051/0004-6361:20030717} {\bibfield
  {journal} {\bibinfo  {journal} {Astron. Astrophys.}\ }\textbf {\bibinfo
  {volume} {406}},\ \bibinfo {pages} {259} (\bibinfo {year} {2003})},\ \Eprint
  {http://arxiv.org/abs/astro-ph/0305376} {arXiv:astro-ph/0305376 [astro-ph]}
  \BibitemShut {NoStop}%
\bibitem [{\citenamefont {Diehl}\ \emph {et~al.}(2006)\citenamefont {Diehl}
  \emph {et~al.}}]{Diehl:2006cf}%
  \BibitemOpen
  \bibfield  {author} {\bibinfo {author} {\bibfnamefont {R.}~\bibnamefont
  {Diehl}} \emph {et~al.},\ }\href {\doibase 10.1038/nature04364} {\bibfield
  {journal} {\bibinfo  {journal} {Nature}\ }\textbf {\bibinfo {volume} {439}},\
  \bibinfo {pages} {45} (\bibinfo {year} {2006})},\ \Eprint
  {http://arxiv.org/abs/astro-ph/0601015} {arXiv:astro-ph/0601015 [astro-ph]}
  \BibitemShut {NoStop}%
\bibitem [{\citenamefont {Olejak}\ \emph {et~al.}(2020)\citenamefont {Olejak},
  \citenamefont {Belczynski}, \citenamefont {Bulik},\ and\ \citenamefont
  {Sobolewska}}]{Olejak+2020}%
  \BibitemOpen
  \bibfield  {author} {\bibinfo {author} {\bibfnamefont {A.}~\bibnamefont
  {Olejak}}, \bibinfo {author} {\bibfnamefont {K.}~\bibnamefont {Belczynski}},
  \bibinfo {author} {\bibfnamefont {T.}~\bibnamefont {Bulik}}, \ and\ \bibinfo
  {author} {\bibfnamefont {M.}~\bibnamefont {Sobolewska}},\ }\href@noop {}
  {\bibfield  {journal} {\bibinfo  {journal} {Astronomy \& Astrophysics}\ }
  (\bibinfo {year} {2020})},\ \Eprint {http://arxiv.org/abs/1908.08775}
  {arXiv:1908.08775 [astro-ph.SR]} \BibitemShut {NoStop}%
\bibitem [{\citenamefont {Agol}\ and\ \citenamefont
  {Kamionkowski}(2002)}]{Agol+2002}%
  \BibitemOpen
  \bibfield  {author} {\bibinfo {author} {\bibfnamefont {E.}~\bibnamefont
  {Agol}}\ and\ \bibinfo {author} {\bibfnamefont {M.}~\bibnamefont
  {Kamionkowski}},\ }\href@noop {} {\bibfield  {journal} {\bibinfo  {journal}
  {{MNRAS}}\ }\textbf {\bibinfo {volume} {334}},\ \bibinfo {pages} {553}
  (\bibinfo {year} {2002})}\BibitemShut {NoStop}%
\bibitem [{\citenamefont {Wiktorowicz}\ \emph {et~al.}(2019)\citenamefont
  {Wiktorowicz} \emph {et~al.}}]{Wiktorowicz+2019}%
  \BibitemOpen
  \bibfield  {author} {\bibinfo {author} {\bibfnamefont {G.}~\bibnamefont
  {Wiktorowicz}} \emph {et~al.},\ }\href@noop {} {\enquote {\bibinfo {title}
  {{Populations of stellar mass Black holes from binary systems}},}\ }
  (\bibinfo {year} {2019}),\ \Eprint {http://arxiv.org/abs/1907.11431}
  {arXiv:1907.11431 [astro-ph.HE]} \BibitemShut {NoStop}%
\bibitem [{\citenamefont {Tsuna}\ \emph {et~al.}(2018)\citenamefont {Tsuna},
  \citenamefont {Kawanaka},\ and\ \citenamefont {Totani}}]{Tsuna+2018}%
  \BibitemOpen
  \bibfield  {author} {\bibinfo {author} {\bibfnamefont {D.}~\bibnamefont
  {Tsuna}}, \bibinfo {author} {\bibfnamefont {N.}~\bibnamefont {Kawanaka}}, \
  and\ \bibinfo {author} {\bibfnamefont {T.}~\bibnamefont {Totani}},\
  }\href@noop {} {\bibfield  {journal} {\bibinfo  {journal} {{MNRAS}}\ }\textbf
  {\bibinfo {volume} {477}},\ \bibinfo {pages} {791} (\bibinfo {year}
  {2018})}\BibitemShut {NoStop}%
\bibitem [{\citenamefont {Tsuna}\ and\ \citenamefont
  {Kawanaka}(2019)}]{Tsuna+2019}%
  \BibitemOpen
  \bibfield  {author} {\bibinfo {author} {\bibfnamefont {D.}~\bibnamefont
  {Tsuna}}\ and\ \bibinfo {author} {\bibfnamefont {N.}~\bibnamefont
  {Kawanaka}},\ }\href@noop {} {\bibfield  {journal} {\bibinfo  {journal}
  {{Monthly Notices of the Royal Astronomical Society}}\ }\textbf {\bibinfo
  {volume} {488}},\ \bibinfo {pages} {2099} (\bibinfo {year}
  {2019})}\BibitemShut {NoStop}%
\bibitem [{\citenamefont {Licquia}\ \emph {et~al.}(2015)\citenamefont {Licquia}
  \emph {et~al.}}]{Licquia+2015}%
  \BibitemOpen
  \bibfield  {author} {\bibinfo {author} {\bibfnamefont {T.~C.}\ \bibnamefont
  {Licquia}} \emph {et~al.},\ }\href@noop {} {\bibfield  {journal} {\bibinfo
  {journal} {{The Astrophysical Journal}}\ }\textbf {\bibinfo {volume} {806}},\
  \bibinfo {pages} {96} (\bibinfo {year} {2015})}\BibitemShut {NoStop}%
\bibitem [{\citenamefont {Sofue}(2013)}]{Sofue2013}%
  \BibitemOpen
  \bibfield  {author} {\bibinfo {author} {\bibfnamefont {Y.}~\bibnamefont
  {Sofue}},\ }\href@noop {} {\bibfield  {journal} {\bibinfo  {journal}
  {Publications of the Astronomical Society of Japan}\ }\textbf {\bibinfo
  {volume} {65}},\ \bibinfo {pages} {118} (\bibinfo {year} {2013})}\BibitemShut
  {NoStop}%
\bibitem [{\citenamefont {Nataf}(2016)}]{Nataf2016}%
  \BibitemOpen
  \bibfield  {author} {\bibinfo {author} {\bibfnamefont {D.~M.}\ \bibnamefont
  {Nataf}},\ }\href@noop {} {\bibfield  {journal} {\bibinfo  {journal}
  {Publications of the Astronomical Society of Australia}\ }\textbf {\bibinfo
  {volume} {33}} (\bibinfo {year} {2016})}\BibitemShut {NoStop}%
\bibitem [{\citenamefont {Irrgang}(2013)}]{Irrgang+2013}%
  \BibitemOpen
  \bibfield  {author} {\bibinfo {author} {\bibfnamefont {A.}~\bibnamefont
  {Irrgang}},\ }\href@noop {} {\bibfield  {journal} {\bibinfo  {journal}
  {Astronomy \& Astrophysics}\ }\textbf {\bibinfo {volume} {549}} (\bibinfo
  {year} {2013})}\BibitemShut {NoStop}%
\bibitem [{\citenamefont {Kunder}\ \emph {et~al.}(2012)\citenamefont {Kunder}
  \emph {et~al.}}]{Kunder+2012}%
  \BibitemOpen
  \bibfield  {author} {\bibinfo {author} {\bibfnamefont {A.}~\bibnamefont
  {Kunder}} \emph {et~al.},\ }\href@noop {} {\bibfield  {journal} {\bibinfo
  {journal} {The Astronomical Journal}\ }\textbf {\bibinfo {volume} {143}}
  (\bibinfo {year} {2012})}\BibitemShut {NoStop}%
\bibitem [{\citenamefont {Fryer}\ \emph {et~al.}(1998)\citenamefont {Fryer},
  \citenamefont {Burrows},\ and\ \citenamefont {Benz}}]{Fryer+1998}%
  \BibitemOpen
  \bibfield  {author} {\bibinfo {author} {\bibfnamefont {C.}~\bibnamefont
  {Fryer}}, \bibinfo {author} {\bibfnamefont {A.}~\bibnamefont {Burrows}}, \
  and\ \bibinfo {author} {\bibfnamefont {W.}~\bibnamefont {Benz}},\ }\href@noop
  {} {\bibfield  {journal} {\bibinfo  {journal} {{The Astrophysical Journal}}\
  }\textbf {\bibinfo {volume} {496}},\ \bibinfo {pages} {333} (\bibinfo {year}
  {1998})}\BibitemShut {NoStop}%
\bibitem [{\citenamefont {Nordhaus}\ \emph {et~al.}(2010)\citenamefont
  {Nordhaus}, \citenamefont {Brandt}, \citenamefont {Burrows}, \citenamefont
  {Livne},\ and\ \citenamefont {Ott}}]{Nordhaus+2010}%
  \BibitemOpen
  \bibfield  {author} {\bibinfo {author} {\bibfnamefont {J.}~\bibnamefont
  {Nordhaus}}, \bibinfo {author} {\bibfnamefont {T.~D.}\ \bibnamefont
  {Brandt}}, \bibinfo {author} {\bibfnamefont {A.}~\bibnamefont {Burrows}},
  \bibinfo {author} {\bibfnamefont {E.}~\bibnamefont {Livne}}, \ and\ \bibinfo
  {author} {\bibfnamefont {C.~D.}\ \bibnamefont {Ott}},\ }\href@noop {}
  {\bibfield  {journal} {\bibinfo  {journal} {{Physical Review D}}\ }\textbf
  {\bibinfo {volume} {82}},\ \bibinfo {pages} {103016} (\bibinfo {year}
  {2010})}\BibitemShut {NoStop}%
\bibitem [{\citenamefont {Janka}(2013)}]{Janka2013}%
  \BibitemOpen
  \bibfield  {author} {\bibinfo {author} {\bibfnamefont {H.~T.}\ \bibnamefont
  {Janka}},\ }\href@noop {} {\bibfield  {journal} {\bibinfo  {journal}
  {{MNRAS}}\ }\textbf {\bibinfo {volume} {434}},\ \bibinfo {pages} {1355}
  (\bibinfo {year} {2013})}\BibitemShut {NoStop}%
\bibitem [{\citenamefont {Mandel}(2016)}]{Mandel2016}%
  \BibitemOpen
  \bibfield  {author} {\bibinfo {author} {\bibfnamefont {I.}~\bibnamefont
  {Mandel}},\ }\href@noop {} {\bibfield  {journal} {\bibinfo  {journal}
  {MNRAS}\ }\textbf {\bibinfo {volume} {456}},\ \bibinfo {pages} {578}
  (\bibinfo {year} {2016})}\BibitemShut {NoStop}%
\bibitem [{\citenamefont {Atri}\ \emph {et~al.}(2019)\citenamefont {Atri} \emph
  {et~al.}}]{Atri+2019}%
  \BibitemOpen
  \bibfield  {author} {\bibinfo {author} {\bibfnamefont {P.}~\bibnamefont
  {Atri}} \emph {et~al.},\ }\href@noop {} {\bibfield  {journal} {\bibinfo
  {journal} {{MNRAS}}\ }\textbf {\bibinfo {volume} {489}},\ \bibinfo {pages}
  {3116} (\bibinfo {year} {2019})}\BibitemShut {NoStop}%
\bibitem [{\citenamefont {Repetto}\ \emph {et~al.}(2012)\citenamefont {Repetto}
  \emph {et~al.}}]{Repetto+2012}%
  \BibitemOpen
  \bibfield  {author} {\bibinfo {author} {\bibfnamefont {S.}~\bibnamefont
  {Repetto}} \emph {et~al.},\ }\href@noop {} {\bibfield  {journal} {\bibinfo
  {journal} {MNRAS}\ }\textbf {\bibinfo {volume} {425}},\ \bibinfo {pages}
  {2799} (\bibinfo {year} {2012})}\BibitemShut {NoStop}%
\bibitem [{\citenamefont {Salpeter}(1955)}]{Salpeter}%
  \BibitemOpen
  \bibfield  {author} {\bibinfo {author} {\bibfnamefont {E.~E.}\ \bibnamefont
  {Salpeter}},\ }\href@noop {} {\bibfield  {journal} {\bibinfo  {journal}
  {{Astrophysical Journal}}\ }\textbf {\bibinfo {volume} {121}},\ \bibinfo
  {pages} {161} (\bibinfo {year} {1955})}\BibitemShut {NoStop}%
\bibitem [{\citenamefont {Bastian}\ \emph {et~al.}(2010)\citenamefont
  {Bastian}, \citenamefont {Covey},\ and\ \citenamefont
  {Meyer}}]{Bastian+2010}%
  \BibitemOpen
  \bibfield  {author} {\bibinfo {author} {\bibfnamefont {N.}~\bibnamefont
  {Bastian}}, \bibinfo {author} {\bibfnamefont {K.~R.}\ \bibnamefont {Covey}},
  \ and\ \bibinfo {author} {\bibfnamefont {M.~R.}\ \bibnamefont {Meyer}},\
  }\href@noop {} {\bibfield  {journal} {\bibinfo  {journal} {{Annual Review of
  Astronomy and Astrophysics}}\ }\textbf {\bibinfo {volume} {48}},\ \bibinfo
  {pages} {339} (\bibinfo {year} {2010})}\BibitemShut {NoStop}%
\bibitem [{\citenamefont {Wegg}\ \emph {et~al.}(2017)\citenamefont {Wegg} \emph
  {et~al.}}]{Wegg+2017}%
  \BibitemOpen
  \bibfield  {author} {\bibinfo {author} {\bibfnamefont {C.}~\bibnamefont
  {Wegg}} \emph {et~al.},\ }\href@noop {} {\bibfield  {journal} {\bibinfo
  {journal} {{The Astrophysical Journal Letters}}\ }\textbf {\bibinfo {volume}
  {843}},\ \bibinfo {pages} {5} (\bibinfo {year} {2017})}\BibitemShut {NoStop}%
\bibitem [{\citenamefont {Reynolds}(2014)}]{Reynolds:2013qqa}%
  \BibitemOpen
  \bibfield  {author} {\bibinfo {author} {\bibfnamefont {C.~S.}\ \bibnamefont
  {Reynolds}},\ }\href {\doibase 10.1007/s11214-013-0006-6} {\bibfield
  {journal} {\bibinfo  {journal} {Space Sci. Rev.}\ }\textbf {\bibinfo {volume}
  {183}},\ \bibinfo {pages} {277} (\bibinfo {year} {2014})},\ \Eprint
  {http://arxiv.org/abs/1302.3260} {arXiv:1302.3260 [astro-ph.HE]} \BibitemShut
  {NoStop}%
\bibitem [{\citenamefont {Liu}\ \emph {et~al.}(2019)\citenamefont {Liu} \emph
  {et~al.}}]{LB1_massive}%
  \BibitemOpen
  \bibfield  {author} {\bibinfo {author} {\bibfnamefont {J.}~\bibnamefont
  {Liu}} \emph {et~al.},\ }\href@noop {} {\bibfield  {journal} {\bibinfo
  {journal} {Nature}\ }\textbf {\bibinfo {volume} {575}},\ \bibinfo {pages}
  {618} (\bibinfo {year} {2019})}\BibitemShut {NoStop}%
\bibitem [{\citenamefont {El-Badry}\ and\ \citenamefont
  {Quataert}(2020)}]{LB1_notmassive}%
  \BibitemOpen
  \bibfield  {author} {\bibinfo {author} {\bibfnamefont {K.}~\bibnamefont
  {El-Badry}}\ and\ \bibinfo {author} {\bibfnamefont {E.}~\bibnamefont
  {Quataert}},\ }\href@noop {} {\bibfield  {journal} {\bibinfo  {journal}
  {Monthly Notices of the Royal Astronomical Society: Letters}\ }\textbf
  {\bibinfo {volume} {493}},\ \bibinfo {pages} {L22} (\bibinfo {year}
  {2020})}\BibitemShut {NoStop}%
\bibitem [{\citenamefont {Liu}\ \emph {et~al.}(2020)\citenamefont {Liu} \emph
  {et~al.}}]{LB1_notmassive_origauthors}%
  \BibitemOpen
  \bibfield  {author} {\bibinfo {author} {\bibfnamefont {J.}~\bibnamefont
  {Liu}} \emph {et~al.},\ }\href@noop {} {\enquote {\bibinfo {title}
  {{Phase-dependent study of near-infrared disk emission lines in LB-1}},}\ }
  (\bibinfo {year} {2020}),\ \Eprint {http://arxiv.org/abs/2005.12595}
  {arXiv:2005.12595 [astro-ph.SR]} \BibitemShut {NoStop}%
\bibitem [{\citenamefont {Fishbach}\ and\ \citenamefont
  {Holz}(2017)}]{Fishbach:2017zga}%
  \BibitemOpen
  \bibfield  {author} {\bibinfo {author} {\bibfnamefont {M.}~\bibnamefont
  {Fishbach}}\ and\ \bibinfo {author} {\bibfnamefont {D.~E.}\ \bibnamefont
  {Holz}},\ }\href {\doibase 10.3847/2041-8213/aa9bf6} {\bibfield  {journal}
  {\bibinfo  {journal} {Astrophys. J.}\ }\textbf {\bibinfo {volume} {851}},\
  \bibinfo {pages} {L25} (\bibinfo {year} {2017})},\ \Eprint
  {http://arxiv.org/abs/1709.08584} {arXiv:1709.08584 [astro-ph.HE]}
  \BibitemShut {NoStop}%
\bibitem [{\citenamefont {Abbott}\ \emph
  {et~al.}(2019{\natexlab{a}})\citenamefont {Abbott} \emph
  {et~al.}}]{LVC_BBHProperties}%
  \BibitemOpen
  \bibfield  {author} {\bibinfo {author} {\bibfnamefont {B.~P.}\ \bibnamefont
  {Abbott}} \emph {et~al.},\ }\href@noop {} {\bibfield  {journal} {\bibinfo
  {journal} {The Astrophysical Journal Letters}\ }\textbf {\bibinfo {volume}
  {882}},\ \bibinfo {pages} {24} (\bibinfo {year}
  {2019}{\natexlab{a}})}\BibitemShut {NoStop}%
\bibitem [{\citenamefont {Roulet}\ and\ \citenamefont
  {Zaldarriaga}(2019)}]{Roulet:2018jbe}%
  \BibitemOpen
  \bibfield  {author} {\bibinfo {author} {\bibfnamefont {J.}~\bibnamefont
  {Roulet}}\ and\ \bibinfo {author} {\bibfnamefont {M.}~\bibnamefont
  {Zaldarriaga}},\ }\href {\doibase 10.1093/mnras/stz226} {\bibfield  {journal}
  {\bibinfo  {journal} {Mon. Not. Roy. Astron. Soc.}\ }\textbf {\bibinfo
  {volume} {484}},\ \bibinfo {pages} {4216} (\bibinfo {year}
  {2019})}\BibitemShut {NoStop}%
\bibitem [{\citenamefont {Abbott}\ \emph {et~al.}(2016)\citenamefont {Abbott}
  \emph {et~al.}}]{LVC_GW150914_astro}%
  \BibitemOpen
  \bibfield  {author} {\bibinfo {author} {\bibfnamefont {B.~P.}\ \bibnamefont
  {Abbott}} \emph {et~al.},\ }\href@noop {} {\bibfield  {journal} {\bibinfo
  {journal} {Astrophysical Journal Letters}\ }\textbf {\bibinfo {volume} {818}}
  (\bibinfo {year} {2016})}\BibitemShut {NoStop}%
\bibitem [{\citenamefont {Belczynski}\ \emph {et~al.}(2016)\citenamefont
  {Belczynski}, \citenamefont {Holz}, \citenamefont {Bulik},\ and\
  \citenamefont {O'Shaughnessy}}]{Belczynski+2016}%
  \BibitemOpen
  \bibfield  {author} {\bibinfo {author} {\bibfnamefont {C.}~\bibnamefont
  {Belczynski}}, \bibinfo {author} {\bibfnamefont {D.~E.}\ \bibnamefont
  {Holz}}, \bibinfo {author} {\bibfnamefont {T.}~\bibnamefont {Bulik}}, \ and\
  \bibinfo {author} {\bibfnamefont {R.}~\bibnamefont {O'Shaughnessy}},\
  }\href@noop {} {\bibfield  {journal} {\bibinfo  {journal} {{Nature}}\
  }\textbf {\bibinfo {volume} {534}},\ \bibinfo {pages} {512} (\bibinfo {year}
  {2016})}\BibitemShut {NoStop}%
\bibitem [{\citenamefont {Toyouchi}\ and\ \citenamefont
  {Chiba}(2018)}]{metallicity_disk}%
  \BibitemOpen
  \bibfield  {author} {\bibinfo {author} {\bibfnamefont {D.}~\bibnamefont
  {Toyouchi}}\ and\ \bibinfo {author} {\bibfnamefont {M.}~\bibnamefont
  {Chiba}},\ }\href@noop {} {\bibfield  {journal} {\bibinfo  {journal} {The
  Astrophysical Journal}\ }\textbf {\bibinfo {volume} {855}},\ \bibinfo {pages}
  {104} (\bibinfo {year} {2018})}\BibitemShut {NoStop}%
\bibitem [{\citenamefont {Ness}\ and\ \citenamefont
  {Freeman}(2016)}]{metallicity_bulge}%
  \BibitemOpen
  \bibfield  {author} {\bibinfo {author} {\bibfnamefont {M.}~\bibnamefont
  {Ness}}\ and\ \bibinfo {author} {\bibfnamefont {K.}~\bibnamefont {Freeman}},\
  }\href@noop {} {\bibfield  {journal} {\bibinfo  {journal} {Publications of
  the Astronomical Society of Australia}\ }\textbf {\bibinfo {volume} {33}}
  (\bibinfo {year} {2016})}\BibitemShut {NoStop}%
\bibitem [{\citenamefont {Mackereth}\ \emph {et~al.}(2017)\citenamefont
  {Mackereth} \emph {et~al.}}]{metallicity_evolution}%
  \BibitemOpen
  \bibfield  {author} {\bibinfo {author} {\bibfnamefont {J.}~\bibnamefont
  {Mackereth}} \emph {et~al.},\ }\href@noop {} {\bibfield  {journal} {\bibinfo
  {journal} {Monthly Notices of the Royal Astronomical Society}\ }\textbf
  {\bibinfo {volume} {471}},\ \bibinfo {pages} {3057} (\bibinfo {year}
  {2017})}\BibitemShut {NoStop}%
\bibitem [{\citenamefont {Fuller}\ and\ \citenamefont
  {Ma}(2019)}]{Fuller+2019}%
  \BibitemOpen
  \bibfield  {author} {\bibinfo {author} {\bibfnamefont {J.}~\bibnamefont
  {Fuller}}\ and\ \bibinfo {author} {\bibfnamefont {L.}~\bibnamefont {Ma}},\
  }\href@noop {} {\bibfield  {journal} {\bibinfo  {journal} {{Astrophysical
  Journal Letters}}\ }\textbf {\bibinfo {volume} {881}},\ \bibinfo {pages} {1}
  (\bibinfo {year} {2019})}\BibitemShut {NoStop}%
\bibitem [{\citenamefont {Faucher-Giguere}\ and\ \citenamefont
  {Kaspi}(2006)}]{faucher2006birth}%
  \BibitemOpen
  \bibfield  {author} {\bibinfo {author} {\bibfnamefont {C.-A.}\ \bibnamefont
  {Faucher-Giguere}}\ and\ \bibinfo {author} {\bibfnamefont {V.~M.}\
  \bibnamefont {Kaspi}},\ }\href@noop {} {\bibfield  {journal} {\bibinfo
  {journal} {The Astrophysical Journal}\ }\textbf {\bibinfo {volume} {643}},\
  \bibinfo {pages} {332} (\bibinfo {year} {2006})}\BibitemShut {NoStop}%
\bibitem [{\citenamefont {Yoon}\ and\ \citenamefont
  {Langer}(2005)}]{Yoon:2005tv}%
  \BibitemOpen
  \bibfield  {author} {\bibinfo {author} {\bibfnamefont {S.-C.}\ \bibnamefont
  {Yoon}}\ and\ \bibinfo {author} {\bibfnamefont {N.}~\bibnamefont {Langer}},\
  }\href {\doibase 10.1051/0004-6361:20054030} {\bibfield  {journal} {\bibinfo
  {journal} {Astron. Astrophys.}\ }\textbf {\bibinfo {volume} {443}},\ \bibinfo
  {pages} {643} (\bibinfo {year} {2005})},\ \Eprint
  {http://arxiv.org/abs/astro-ph/0508242} {arXiv:astro-ph/0508242 [astro-ph]}
  \BibitemShut {NoStop}%
\bibitem [{\citenamefont {Woosley}\ and\ \citenamefont
  {Heger}(2006)}]{Woosley:2005gy}%
  \BibitemOpen
  \bibfield  {author} {\bibinfo {author} {\bibfnamefont {S.}~\bibnamefont
  {Woosley}}\ and\ \bibinfo {author} {\bibfnamefont {A.}~\bibnamefont
  {Heger}},\ }\href {\doibase 10.1086/498500} {\bibfield  {journal} {\bibinfo
  {journal} {Astrophys. J.}\ }\textbf {\bibinfo {volume} {637}},\ \bibinfo
  {pages} {914} (\bibinfo {year} {2006})},\ \Eprint
  {http://arxiv.org/abs/astro-ph/0508175} {arXiv:astro-ph/0508175 [astro-ph]}
  \BibitemShut {NoStop}%
\bibitem [{\citenamefont {Wysocki}\ \emph {et~al.}(2019)\citenamefont
  {Wysocki}, \citenamefont {Lange},\ and\ \citenamefont
  {O'Shaughnessy}}]{Wysocki:2018mpo}%
  \BibitemOpen
  \bibfield  {author} {\bibinfo {author} {\bibfnamefont {D.}~\bibnamefont
  {Wysocki}}, \bibinfo {author} {\bibfnamefont {J.}~\bibnamefont {Lange}}, \
  and\ \bibinfo {author} {\bibfnamefont {R.}~\bibnamefont {O'Shaughnessy}},\
  }\href {\doibase 10.1103/PhysRevD.100.043012} {\bibfield  {journal} {\bibinfo
   {journal} {Phys. Rev.}\ }\textbf {\bibinfo {volume} {D100}},\ \bibinfo
  {pages} {043012} (\bibinfo {year} {2019})},\ \Eprint
  {http://arxiv.org/abs/1805.06442} {arXiv:1805.06442 [gr-qc]} \BibitemShut
  {NoStop}%
\bibitem [{\citenamefont {M{\'e}ndez}\ and\ \citenamefont
  {Cantiello}(2016)}]{mendez2016limits}%
  \BibitemOpen
  \bibfield  {author} {\bibinfo {author} {\bibfnamefont {E.~M.}\ \bibnamefont
  {M{\'e}ndez}}\ and\ \bibinfo {author} {\bibfnamefont {M.}~\bibnamefont
  {Cantiello}},\ }\href@noop {} {\bibfield  {journal} {\bibinfo  {journal} {New
  Astronomy}\ }\textbf {\bibinfo {volume} {44}},\ \bibinfo {pages} {58}
  (\bibinfo {year} {2016})}\BibitemShut {NoStop}%
\bibitem [{\citenamefont {Ribas}\ \emph {et~al.}(2005)\citenamefont {Ribas}
  \emph {et~al.}}]{AndromedaDistance}%
  \BibitemOpen
  \bibfield  {author} {\bibinfo {author} {\bibfnamefont {I.}~\bibnamefont
  {Ribas}} \emph {et~al.},\ }\href@noop {} {\bibfield  {journal} {\bibinfo
  {journal} {{The Astrophysical Journal Letters}}\ }\textbf {\bibinfo {volume}
  {635}},\ \bibinfo {pages} {L37} (\bibinfo {year} {2005})}\BibitemShut
  {NoStop}%
\bibitem [{\citenamefont {McConnachie}(2012)}]{LGVelocities}%
  \BibitemOpen
  \bibfield  {author} {\bibinfo {author} {\bibfnamefont {A.~W.}\ \bibnamefont
  {McConnachie}},\ }\href@noop {} {\bibfield  {journal} {\bibinfo  {journal}
  {{The Astronomical Journal}}\ }\textbf {\bibinfo {volume} {144}},\ \bibinfo
  {pages} {4} (\bibinfo {year} {2012})}\BibitemShut {NoStop}%
\bibitem [{\citenamefont {Pietrzy\'{n}ski}\ \emph {et~al.}(2019)\citenamefont
  {Pietrzy\'{n}ski} \emph {et~al.}}]{LMCDistance}%
  \BibitemOpen
  \bibfield  {author} {\bibinfo {author} {\bibfnamefont {G.}~\bibnamefont
  {Pietrzy\'{n}ski}} \emph {et~al.},\ }\href@noop {} {\bibfield  {journal}
  {\bibinfo  {journal} {{Nature}}\ }\textbf {\bibinfo {volume} {567}},\
  \bibinfo {pages} {200} (\bibinfo {year} {2019})}\BibitemShut {NoStop}%
\bibitem [{\citenamefont {Graczyk}\ \emph {et~al.}(2014)\citenamefont {Graczyk}
  \emph {et~al.}}]{SMCDistance}%
  \BibitemOpen
  \bibfield  {author} {\bibinfo {author} {\bibfnamefont {D.}~\bibnamefont
  {Graczyk}} \emph {et~al.},\ }\href@noop {} {\bibfield  {journal} {\bibinfo
  {journal} {{The Astrophysical Journal}}\ }\textbf {\bibinfo {volume} {780}},\
  \bibinfo {pages} {59} (\bibinfo {year} {2014})}\BibitemShut {NoStop}%
\bibitem [{\citenamefont {Abbott}\ \emph
  {et~al.}(2019{\natexlab{b}})\citenamefont {Abbott} \emph
  {et~al.}}]{LVC_O2AS}%
  \BibitemOpen
  \bibfield  {author} {\bibinfo {author} {\bibfnamefont {B.~P.}\ \bibnamefont
  {Abbott}} \emph {et~al.},\ }\href@noop {} {\enquote {\bibinfo {title}
  {{All-sky search for continuous gravitational waves from isolated neutron
  stars using Advanced LIGO O2 data}},}\ } (\bibinfo {year}
  {2019}{\natexlab{b}}),\ \bibinfo {note} {{Accepted by PRD}},\ \Eprint
  {http://arxiv.org/abs/1903.01901} {arXiv:1903.01901 [{astro-ph.HE}]}
  \BibitemShut {NoStop}%
\bibitem [{onl()}]{onlinerepository}%
  \BibitemOpen
  \href@noop {} {}\bibinfo {howpublished}
  {\url{https://www.aei.mpg.de/continuouswaves/arxiv200303359}}\BibitemShut
  {NoStop}%
\bibitem [{\citenamefont {\:{O}zel}\ \emph {et~al.}(2010)\citenamefont
  {\:{O}zel}, \citenamefont {Psaltis}, \citenamefont {Narayan},\ and\
  \citenamefont {McClintock}}]{Ozel+2010}%
  \BibitemOpen
  \bibfield  {author} {\bibinfo {author} {\bibfnamefont {F.}~\bibnamefont
  {\:{O}zel}}, \bibinfo {author} {\bibfnamefont {D.}~\bibnamefont {Psaltis}},
  \bibinfo {author} {\bibfnamefont {R.}~\bibnamefont {Narayan}}, \ and\
  \bibinfo {author} {\bibfnamefont {J.~E.}\ \bibnamefont {McClintock}},\
  }\href@noop {} {\bibfield  {journal} {\bibinfo  {journal} {{The Astrophysical
  Journal}}\ }\textbf {\bibinfo {volume} {725}},\ \bibinfo {pages} {1918}
  (\bibinfo {year} {2010})}\BibitemShut {NoStop}%
\bibitem [{\citenamefont {Allen}\ and\ \citenamefont
  {Romano}(1999)}]{Allen:1997ad}%
  \BibitemOpen
  \bibfield  {author} {\bibinfo {author} {\bibfnamefont {B.}~\bibnamefont
  {Allen}}\ and\ \bibinfo {author} {\bibfnamefont {J.~D.}\ \bibnamefont
  {Romano}},\ }\href {\doibase 10.1103/PhysRevD.59.102001} {\bibfield
  {journal} {\bibinfo  {journal} {Phys. Rev.}\ }\textbf {\bibinfo {volume}
  {D59}},\ \bibinfo {pages} {102001} (\bibinfo {year} {1999})},\ \Eprint
  {http://arxiv.org/abs/gr-qc/9710117} {arXiv:gr-qc/9710117 [gr-qc]}
  \BibitemShut {NoStop}%
\bibitem [{\citenamefont {Walsh}\ \emph {et~al.}(2016)\citenamefont {Walsh}
  \emph {et~al.}}]{Walsh:2016hyc}%
  \BibitemOpen
  \bibfield  {author} {\bibinfo {author} {\bibfnamefont {S.}~\bibnamefont
  {Walsh}} \emph {et~al.},\ }\href {\doibase 10.1103/PhysRevD.94.124010}
  {\bibfield  {journal} {\bibinfo  {journal} {Phys. Rev.}\ }\textbf {\bibinfo
  {volume} {D94}},\ \bibinfo {pages} {124010} (\bibinfo {year} {2016})},\
  \Eprint {http://arxiv.org/abs/1606.00660} {arXiv:1606.00660 [gr-qc]}
  \BibitemShut {NoStop}%
\bibitem [{\citenamefont {Walsh}\ \emph {et~al.}(2019)\citenamefont {Walsh},
  \citenamefont {Wette}, \citenamefont {Papa},\ and\ \citenamefont
  {Prix}}]{Walsh:2019nmr}%
  \BibitemOpen
  \bibfield  {author} {\bibinfo {author} {\bibfnamefont {S.}~\bibnamefont
  {Walsh}}, \bibinfo {author} {\bibfnamefont {K.}~\bibnamefont {Wette}},
  \bibinfo {author} {\bibfnamefont {M.~A.}\ \bibnamefont {Papa}}, \ and\
  \bibinfo {author} {\bibfnamefont {R.}~\bibnamefont {Prix}},\ }\href {\doibase
  10.1103/PhysRevD.99.082004} {\bibfield  {journal} {\bibinfo  {journal} {Phys.
  Rev.}\ }\textbf {\bibinfo {volume} {D99}},\ \bibinfo {pages} {082004}
  (\bibinfo {year} {2019})},\ \Eprint {http://arxiv.org/abs/1901.08998}
  {arXiv:1901.08998 [astro-ph.IM]} \BibitemShut {NoStop}%
\bibitem [{\citenamefont {Piccinni}\ \emph {et~al.}(2019)\citenamefont
  {Piccinni}, \citenamefont {Astone}, \citenamefont {D'Antonio}, \citenamefont
  {Frasca}, \citenamefont {Intini}, \citenamefont {Leaci}, \citenamefont
  {Mastrogiovanni}, \citenamefont {Miller}, \citenamefont {Palomba},\ and\
  \citenamefont {Singhal}}]{Piccinni:2018akm}%
  \BibitemOpen
  \bibfield  {author} {\bibinfo {author} {\bibfnamefont {O.~J.}\ \bibnamefont
  {Piccinni}}, \bibinfo {author} {\bibfnamefont {P.}~\bibnamefont {Astone}},
  \bibinfo {author} {\bibfnamefont {S.}~\bibnamefont {D'Antonio}}, \bibinfo
  {author} {\bibfnamefont {S.}~\bibnamefont {Frasca}}, \bibinfo {author}
  {\bibfnamefont {G.}~\bibnamefont {Intini}}, \bibinfo {author} {\bibfnamefont
  {P.}~\bibnamefont {Leaci}}, \bibinfo {author} {\bibfnamefont
  {S.}~\bibnamefont {Mastrogiovanni}}, \bibinfo {author} {\bibfnamefont
  {A.}~\bibnamefont {Miller}}, \bibinfo {author} {\bibfnamefont
  {C.}~\bibnamefont {Palomba}}, \ and\ \bibinfo {author} {\bibfnamefont
  {A.}~\bibnamefont {Singhal}},\ }\href {\doibase 10.1088/1361-6382/aaefb5}
  {\bibfield  {journal} {\bibinfo  {journal} {Class. Quant. Grav.}\ }\textbf
  {\bibinfo {volume} {36}},\ \bibinfo {pages} {015008} (\bibinfo {year}
  {2019})},\ \Eprint {http://arxiv.org/abs/1811.04730} {arXiv:1811.04730
  [gr-qc]} \BibitemShut {NoStop}%
\bibitem [{\citenamefont {Dergachev}\ \emph {et~al.}(2019)\citenamefont
  {Dergachev}, \citenamefont {Papa}, \citenamefont {Steltner},\ and\
  \citenamefont {Eggenstein}}]{Dergachev:2019pgs}%
  \BibitemOpen
  \bibfield  {author} {\bibinfo {author} {\bibfnamefont {V.}~\bibnamefont
  {Dergachev}}, \bibinfo {author} {\bibfnamefont {M.~A.}\ \bibnamefont {Papa}},
  \bibinfo {author} {\bibfnamefont {B.}~\bibnamefont {Steltner}}, \ and\
  \bibinfo {author} {\bibfnamefont {H.-B.}\ \bibnamefont {Eggenstein}},\ }\href
  {\doibase 10.1103/PhysRevD.99.084048} {\bibfield  {journal} {\bibinfo
  {journal} {Phys. Rev.}\ }\textbf {\bibinfo {volume} {D99}},\ \bibinfo {pages}
  {084048} (\bibinfo {year} {2019})},\ \Eprint
  {http://arxiv.org/abs/1903.02389} {arXiv:1903.02389 [gr-qc]} \BibitemShut
  {NoStop}%
\bibitem [{\citenamefont {Ming}\ \emph {et~al.}(2019)\citenamefont {Ming} \emph
  {et~al.}}]{Ming:2019xse}%
  \BibitemOpen
  \bibfield  {author} {\bibinfo {author} {\bibfnamefont {J.}~\bibnamefont
  {Ming}} \emph {et~al.},\ }\href {\doibase 10.1103/PhysRevD.100.024063}
  {\bibfield  {journal} {\bibinfo  {journal} {Phys. Rev.}\ }\textbf {\bibinfo
  {volume} {D100}},\ \bibinfo {pages} {024063} (\bibinfo {year} {2019})},\
  \Eprint {http://arxiv.org/abs/1903.09119} {arXiv:1903.09119 [gr-qc]}
  \BibitemShut {NoStop}%
\bibitem [{\citenamefont {Astone}\ \emph {et~al.}(2005)\citenamefont {Astone},
  \citenamefont {Frasca},\ and\ \citenamefont {Palomba}}]{Astone_2005}%
  \BibitemOpen
  \bibfield  {author} {\bibinfo {author} {\bibfnamefont {P.}~\bibnamefont
  {Astone}}, \bibinfo {author} {\bibfnamefont {S.}~\bibnamefont {Frasca}}, \
  and\ \bibinfo {author} {\bibfnamefont {C.}~\bibnamefont {Palomba}},\ }\href
  {\doibase 10.1088/0264-9381/22/18/s34} {\bibfield  {journal} {\bibinfo
  {journal} {Classical and Quantum Gravity}\ }\textbf {\bibinfo {volume}
  {22}},\ \bibinfo {pages} {S1197} (\bibinfo {year} {2005})}\BibitemShut
  {NoStop}%
\bibitem [{Ast()}]{Astone:2019}%
  \BibitemOpen
  \href@noop {} {}\bibinfo {note} {Private communication.}\BibitemShut {Stop}%
\bibitem [{\citenamefont {Isi}\ \emph {et~al.}(2019)\citenamefont {Isi},
  \citenamefont {Sun}, \citenamefont {Brito},\ and\ \citenamefont
  {Melatos}}]{Isi+2019}%
  \BibitemOpen
  \bibfield  {author} {\bibinfo {author} {\bibfnamefont {M.}~\bibnamefont
  {Isi}}, \bibinfo {author} {\bibfnamefont {L.}~\bibnamefont {Sun}}, \bibinfo
  {author} {\bibfnamefont {R.}~\bibnamefont {Brito}}, \ and\ \bibinfo {author}
  {\bibfnamefont {A.}~\bibnamefont {Melatos}},\ }\href {\doibase
  10.1103/PhysRevD.99.084042} {\bibfield  {journal} {\bibinfo  {journal} {Phys.
  Rev.}\ }\textbf {\bibinfo {volume} {D99}},\ \bibinfo {pages} {084042}
  (\bibinfo {year} {2019})},\ \Eprint {http://arxiv.org/abs/1810.03812}
  {arXiv:1810.03812 [gr-qc]} \BibitemShut {NoStop}%
\bibitem [{\citenamefont {Sun}\ \emph {et~al.}(2019)\citenamefont {Sun},
  \citenamefont {Brito},\ and\ \citenamefont {Isi}}]{Sun+2019}%
  \BibitemOpen
  \bibfield  {author} {\bibinfo {author} {\bibfnamefont {L.}~\bibnamefont
  {Sun}}, \bibinfo {author} {\bibfnamefont {R.}~\bibnamefont {Brito}}, \ and\
  \bibinfo {author} {\bibfnamefont {M.}~\bibnamefont {Isi}},\ }\href@noop {}
  {\enquote {\bibinfo {title} {{Search for ultralight bosons in Cygnus X-1 with
  Advanced LIGO}},}\ } (\bibinfo {year} {2019}),\ \Eprint
  {http://arxiv.org/abs/1909.11267} {arXiv:1909.11267 [gr-qc]} \BibitemShut
  {NoStop}%
\bibitem [{\citenamefont {Gou}\ \emph {et~al.}(2011)\citenamefont {Gou} \emph
  {et~al.}}]{CygnusX1_spin}%
  \BibitemOpen
  \bibfield  {author} {\bibinfo {author} {\bibfnamefont {L.}~\bibnamefont
  {Gou}} \emph {et~al.},\ }\href@noop {} {\bibfield  {journal} {\bibinfo
  {journal} {{The Astrophysical Journal}}\ }\textbf {\bibinfo {volume} {742}}
  (\bibinfo {year} {2011})}\BibitemShut {NoStop}%
\bibitem [{\citenamefont {Gou}\ \emph {et~al.}(2014)\citenamefont {Gou} \emph
  {et~al.}}]{Gou:2013dna}%
  \BibitemOpen
  \bibfield  {author} {\bibinfo {author} {\bibfnamefont {L.}~\bibnamefont
  {Gou}} \emph {et~al.},\ }\href {\doibase 10.1088/0004-637X/790/1/29}
  {\bibfield  {journal} {\bibinfo  {journal} {Astrophys. J.}\ }\textbf
  {\bibinfo {volume} {790}},\ \bibinfo {pages} {29} (\bibinfo {year}
  {2014})}\BibitemShut {NoStop}%
\bibitem [{\citenamefont {Siemonsen}\ and\ \citenamefont
  {East}(2019)}]{Siemonsen:2019}%
  \BibitemOpen
  \bibfield  {author} {\bibinfo {author} {\bibfnamefont {N.}~\bibnamefont
  {Siemonsen}}\ and\ \bibinfo {author} {\bibfnamefont {W.~E.}\ \bibnamefont
  {East}},\ }\href@noop {} {\  (\bibinfo {year} {2019})},\ \Eprint
  {http://arxiv.org/abs/1910.09476} {arXiv:1910.09476 [gr-qc]} \BibitemShut
  {NoStop}%
\bibitem [{\citenamefont {Berti}\ \emph {et~al.}(2006)\citenamefont {Berti},
  \citenamefont {Cardoso},\ and\ \citenamefont {Casals}}]{Berti:2005gp}%
  \BibitemOpen
  \bibfield  {author} {\bibinfo {author} {\bibfnamefont {E.}~\bibnamefont
  {Berti}}, \bibinfo {author} {\bibfnamefont {V.}~\bibnamefont {Cardoso}}, \
  and\ \bibinfo {author} {\bibfnamefont {M.}~\bibnamefont {Casals}},\ }\href
  {\doibase 10.1103/PhysRevD.73.109902, 10.1103/PhysRevD.73.024013} {\bibfield
  {journal} {\bibinfo  {journal} {Phys. Rev.}\ }\textbf {\bibinfo {volume}
  {D73}},\ \bibinfo {pages} {024013} (\bibinfo {year} {2006})},\ \bibinfo
  {note} {[Erratum: Phys. Rev.D73,109902(2006)]},\ \Eprint
  {http://arxiv.org/abs/gr-qc/0511111} {arXiv:gr-qc/0511111 [gr-qc]}
  \BibitemShut {NoStop}%
\bibitem [{\citenamefont {Farr}\ \emph {et~al.}(2011)\citenamefont {Farr} \emph
  {et~al.}}]{Farr+2011}%
  \BibitemOpen
  \bibfield  {author} {\bibinfo {author} {\bibfnamefont {W.~M.}\ \bibnamefont
  {Farr}} \emph {et~al.},\ }\href@noop {} {\bibfield  {journal} {\bibinfo
  {journal} {{The Astrophysical Journal}}\ }\textbf {\bibinfo {volume} {741}}
  (\bibinfo {year} {2011})}\BibitemShut {NoStop}%
\bibitem [{\citenamefont {Dergachev}\ and\ \citenamefont
  {Papa}(2020{\natexlab{b}})}]{Dergachev:2020fli}%
  \BibitemOpen
  \bibfield  {author} {\bibinfo {author} {\bibfnamefont {V.}~\bibnamefont
  {Dergachev}}\ and\ \bibinfo {author} {\bibfnamefont {M.~A.}\ \bibnamefont
  {Papa}},\ }\href@noop {} {\  (\bibinfo {year} {2020}{\natexlab{b}})},\
  \Eprint {http://arxiv.org/abs/2004.08334} {arXiv:2004.08334 [gr-qc]}
  \BibitemShut {NoStop}%
\end{thebibliography}%

\appendix
\renewcommand\thefigure{A\arabic{figure}}
\section{General Formulae for Scalar Superradiance}
\label{sec:app}

In this Appendix we provide the more general forms for the equations that are used in the text. All analytic formulae are valid for $\alpha/\ell \ll 1$. We find that the potentially detectable signals in this work arise from systems with $0.03 \lesssim \alpha \lesssim 0.2$. We also comment on the assumptions and uncertainties that enter in the signal calculations.

\subsection{Signal Frequency}
\label{app:freq}

As the signal comes from the annihilation of two bosons, 
the frequency of the emitted gravitational wave $f_\mathrm{GW}$ is given by twice the boson energy $\omega_R$:
\begin{align}\label{eqn:freq_app}
  f_\mathrm{GW}^{n\ell m} & = \frac{\omega_\mathrm{GW}}{h} = \frac{ 2\omega_R}{h}.
  \end{align}
  
The energy of the boson is given by
\begin{align}\label{eqn:omega_r_app}
  \omega_R^{n\ell m} & = \mu_\mathrm{b}-\Delta\omega_{R,\mathrm{BH}}^{n\ell m} -\Delta\omega_{R,\mathrm{cloud}}^{n\ell m}, 
  \end{align}
where $ \Delta\omega_{R,\mathrm{BH}}^{n\ell m}  $ is the gravitational potential energy and $\Delta\omega_{R,\mathrm{cloud}}^{n\ell m} $ the gravitational self-energy of the cloud. The potential energy is given by \cite{Baumann+2019},
 \begin{align}\label{eqn:omega_r_bh_app}
&  \Delta\omega_{R,\mathrm{BH}}^{n\ell m}  \simeq \mu_\mathrm{b}
 \left(\frac{\alpha^2}{2(\ell + n + 1)^2}\right. \\
     &+\frac{\alpha^4}{8(\ell + n + 1)^4} - \frac{2\ell - 3(\ell + n + 1) + 1}{(\ell + n + 1)^4(\ell + 1/2)}\alpha^4\nonumber \\
     &- \left.\frac{2\chi_i m\alpha^5}{(\ell + n + 1)^3
     \ell(\ell + 1/2)(\ell + 1)}\right),\nonumber\end{align}
 and the uncertainty on the analytic calculation compared to the numerical result was shown to be $\lesssim 1\%$ for $\alpha \lesssim 0.3$, becoming an excellent approximation for $\alpha \lesssim 0.2$~\cite{Baumann+2019}.
 
The self-energy of the cloud depends on the mass of the cloud in the non-relativistic limit as~\cite{Isi+2019,Baryakhtar+2020},
  \begin{align}
  \label{eqn:omega_r_cloud_app}
\Delta\omega_{R,\mathrm{cloud}}^{n\ell m}  \approx \mu_\mathrm{b} \left(0.2\alpha^2\frac{M_{\mathrm{cloud}}}{M_{\mathrm{BH}}}\right).
 \end{align}
 The analogous calculation for a vector in the state $n=0, j=1, \ell=0, m=1$ varies by up to $50\%$ for $\alpha <0.3$ depending on whether the relativistic corrections of the wavefunction are included \cite{Siemonsen:2019}. We expect the uncertainty on our calculation to be smaller, as the vector states (with $\ell=0$) are localized closer to the black hole than the scalar states (with $\ell=1$), and therefore are more affected by relativistic effects.

  For $0.01 \lesssim \alpha \lesssim 0.3$, we find that the self energy contribution is much smaller than the black hole gravitational potential contribution, $ 10^{-6}\lesssim \Delta\omega_{R,\mathrm{cloud}}^{n\ell m}  / \Delta\omega_{R,\mathrm{BH}}^{n\ell m} \lesssim10^{-3}$. This is less than the error on the analytic estimate of \eqref{eqn:omega_r_bh_app};  nevertheless, as this quantity is time-dependent, it determines the frequency drift of the signal, see App.~\ref{app:drift}.

\subsection{Superradiance rate}
\label{app:srrate}
The instability rate of the boson is (\cite{Detweiler1980}, \cite{Cardoso+2018})
\begin{align}\label{eqn:omega_sr_app}
  \omega_{sr}^{n\ell m} & \approx 2\mu_bC_{n\ell m}(\alpha, \chi) \alpha^{4\ell + 4} \nonumber\\
  &\times (1 + \sqrt{1 - \chi^2}) \left(\frac{m \chi}{2 (1 + \sqrt{1 - \chi^2})} - \alpha_\omega\right),
\end{align}
where $\chi$ the dimensionless black hole spin and $C_{n\ell m}(\alpha, \chi)\ll1$ is a numerical coefficient which is a function of the bound state quantum numbers as well as $(\alpha, \chi)$,
\begin{align}\label{eqn:cnlm_sr_app}
&C_{n\ell m}(\alpha, \chi) = \frac{2^{4 \ell+2} (2 \ell+n+1)!}{n! (\ell+n+1)^{2 \ell+4}}\left(\frac{\ell!}{(2\ell)! (2 \ell+1)!}\right)^2\times \nonumber \\
&\prod _{j=1}^\ell\bigg(4 \left(\sqrt{1-\chi ^2}+1\right)^2 \bigg(\frac{m \chi }{2 \left(\sqrt{1-\chi ^2}+1\right)}-\alpha_\omega\bigg)^2+\nonumber\\
&+j^2 \left(1-\chi ^2\right)\bigg)
\end{align}
We define $\alpha_\omega \equiv \alpha \omega_R/\mu$, i.e., the $\alpha$ corrected for the potential energy contribution to the boson energy. The fastest-growing level for a light scalar, the $n=0,\ell=m=1$ level,
has $  \omega_{sr}^{011}  \simeq \frac{1}{24} \chi \alpha^8 \mu_b$.

The rate for  $\ell=1$ is larger than that for
$\ell=2$ by a factor of $\sim C_{011}/C_{022}\,\alpha^{-4}\sim 10^{3}\alpha^{-4}$.

The instability $e$-folding timescale is then given by
\begin{align}\label{eqn:time_app}
  \tau_{inst}^{n\ell m} = \frac{\hbar}{ \omega_{sr}^{n\ell m} },
\end{align}
with $\ln(N)=\ln(M_\mathrm{BH}c^2/\mu_b)\sim 180$ instability timescales required to fully saturate the cloud growth. The analytical form is accurate to within a factor of $2$ for $\alpha \lesssim 0.25$ and within $50\%$ for  $\alpha \lesssim 0.2$ \footnote{We thank Horng Sheng Chia for providing the numerical rate comparison.}.

\subsection{Final cloud mass}
Superradiance will start if the superradiance condition is satisfied, i.e. the initial spin of the black hole is above the critical spin,
\begin{align}\label{eqn:chi_c_app}
  \chi_c^{nlm} = \frac{4 M_\mathrm{BH} \omega_R m}{m^2 + 4M_\mathrm{BH}^2 \omega_R^2}
  \approx \frac{4\alpha m}{m^2+4\alpha^2}.
\end{align}
As the black hole loses mass and angular momentum, the superradiance condition and thus the final spin are affected. The equations governing the evolution are
\begin{align}\label{eqn:Mcevol}
   \dot{N}(t) &=  \omega_{sr}^{n\ell m}(t) N(t),\\
      \dot{M}(t) &= - \omega_R^{n\ell m}(t)  \dot{N}(t) \\
         \dot{J}(t) &= -    \dot{N}(t) 
\end{align}
where $N(t)$ is the number of particles in the cloud, $ M(t)$ is the black hole mass and $J(t) = G M(t)^2 \chi(t)$ is the angular momentum of the black hole. The superradiance rate and the energy of the boson depend on time through their implicit dependence on $\alpha(t) = G \mu M(t)$	and  $ \chi(t) = J(t)/(G M(t)^2)$.

The process saturates when
\begin{align}\label{eqn:chi_f_app}
  \chi_{f} = \frac{4 M_\mathrm{f} \omega_{R,f}}{1 + 4M_\mathrm{f}^2 \omega_{R,f}^2}.
\end{align}
By conservation of angular momentum, the number of particles in the cloud is given by the angular momentum lost by the black hole, $N_f = J_i - J_f$, with each particle carrying one unit of angular momentum (for $\ell=m=1$). The mass of the cloud is given by the number of particles times their energies,
\begin{align}\label{eqn:Mcapp}
   M_{\mathrm{cloud}} &= \omega_R N_f = \omega_R(\chi_i G M_i^2 - \chi_{f} G M_f^2).
   \end{align}
   
   In the limit $\alpha (\chi_i-\chi_c) \ll 1$ this reduces to
\begin{align}\label{eqn:Mcapp2}
   M_{\mathrm{cloud}} &= \alpha (\chi_i-\chi_c) M_i,
   \end{align}
   which is accurate to $1\%$ for $\alpha <0.03$ and underestimates the cloud mass by $\sim\!50\%$ for $\alpha\sim0.2, \chi\sim1$. We use the numerical evolution of $\alpha$ and $\chi$ to establish the final cloud mass.
 
\subsection{Gravitational-Wave Signal}
\label{app:gwsig}
The power emitted in gravitational waves for $\alpha \ll 1$ has been computed analytically in the Schwarzschild background and gives \cite{Brito:2014wla}
\begin{align}\label{eqn:powerapp}
  P_{\mathrm{GW}} \approx 0.025 \frac{c^5}{G}\alpha^{14} \frac{M_\mathrm{cloud}^2}{M_\mathrm{BH}^2}.
\end{align}
At larger $\alpha>0.1$, we use the power calculated numerically in \cite{Brito+2017}, 
\begin{align}\label{eqn:powernum}
  P(\alpha) & = \frac{1}{2\pi}\frac{c^5}{G} \frac{M_\mathrm{cloud}^2}{M_\mathrm{BH}^2}
    \left(\frac{\hbar c^3}{{\omega_\mathrm{GW}^{n\ell m }}G M_\mathrm{BH}}\right)^2 \mathcal{A}_{\tilde{ \ell} \tilde{m}}^{2}(\alpha, \chi_i)
\end{align}
where $\tilde{\ell},\tilde{m}$ are spherical harmonics modes of the emitted gravitational radiation,  $\mathcal{A}_{\tilde{ \ell} \tilde{ m}}$ is a dimensionless function that contains information about the fraction of energy deposited  in the $\tilde{\ell} = \tilde{m}=2$ mode. The uncertainty in $\mathcal{A}$ ranges between $\sim\!5\%$ at intermediate $\alpha$ up to $\sim15\%$ at $\alpha<0.1$ and/or large spin\footnote{We thank Richard Brito for discussions on the details of the calculation and uncertainties.}. We use a polynomial interpolation between the low-$\alpha$ (eq.~\eqref{eqn:power}) and large-$\alpha$ (eq.~\eqref{eqn:powernum}) regime.

For $\alpha \gtrsim 0.3$, the power emitted in modes with $\tilde{\ell}, \tilde{m}>2$ becomes dominant \cite{Yoshino+2014}, and defines the depletion rate of the cloud. Since we have only considered the $\tilde{\ell} = \tilde{m}=2$ mode in our power calculations, the gravitational-wave emission timescales $\tau_\mathrm{gw}$ calculated here are an overestimate. However, this does not change our conclusions with respect to the signal detectability, as only systems with $\alpha \lesssim 0.2$ are detectable using current searches.

\begin{figure}[t]
  \includegraphics[width=\columnwidth]{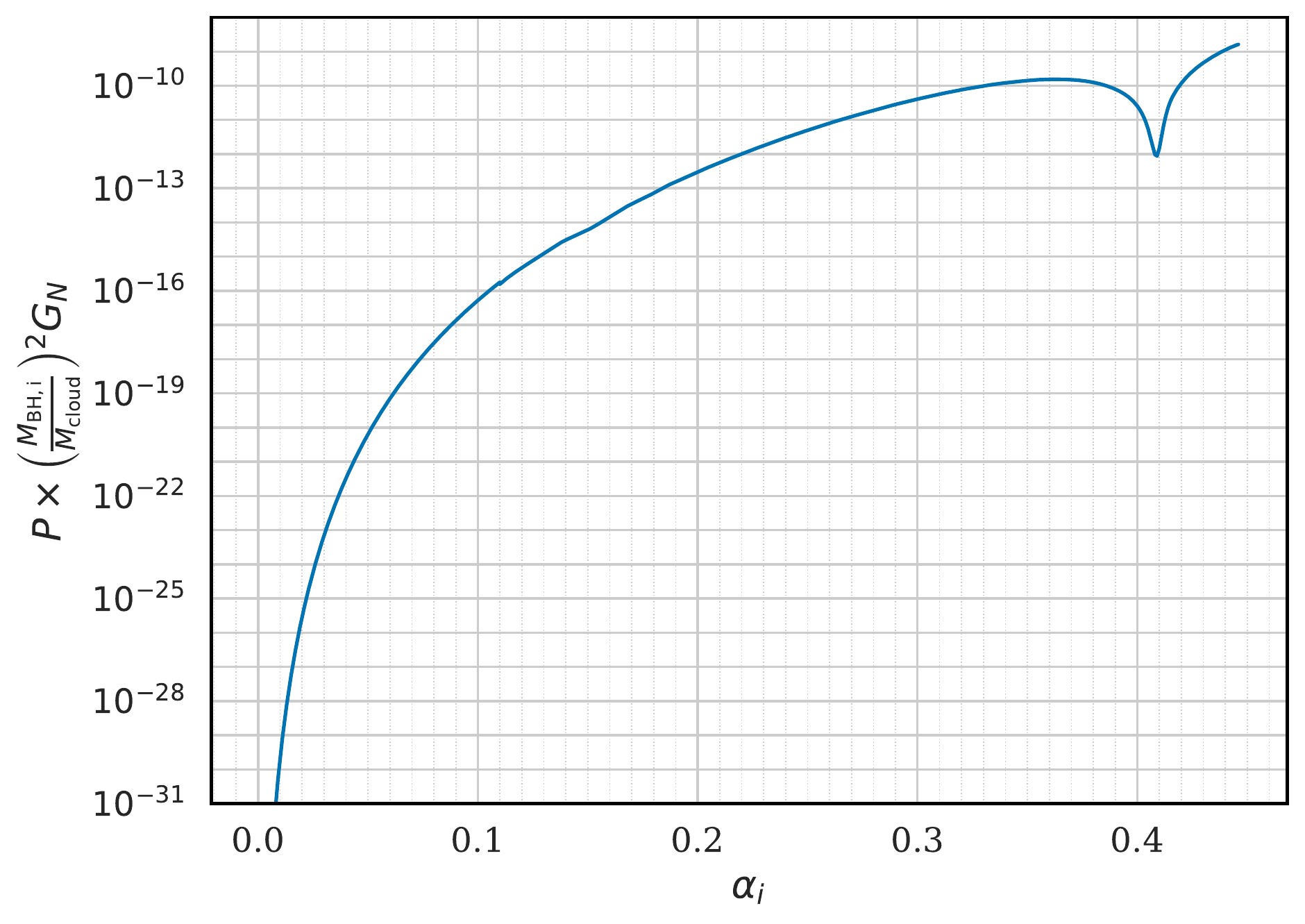}
  \caption{The power emitted in the $\tilde{\ell} = \tilde{m}=2$  spherical harmonics mode of the gravitational radiation, normalized by $\left(M_\mathrm{BH,i}/M_\mathrm{cloud}\right)G_N$ to be a dimensionless quantity. The initial black hole spin assumed here is $\chi_i = 0.99$; the curves for other values of $\chi_i$ are very similar and primarily differ in their maximum value of $\alpha_i$.
 The power for $\alpha_i > 0.1$ is calculated numerically (Eq.~\eqref{eqn:powernum}), and is suppressed for some high spins, resulting in a ``dip'' near $\alpha \sim 0.4$. This directly affects the shape of $h_0$ and $\tau_\mathrm{gw}$.}
\end{figure}

The maximal peak strain is related to the power emitted by the axion cloud as, 
\begin{align}\label{eqn:peakh0_precise_app}
  h_{0,\mathrm{peak}}(\alpha)  & =\left(\frac{10 G \hbar^2 \,P(\alpha)}{c^3\,\omega_\mathrm{GW}^2r^2}\right)^{1/2},\end{align}
  where the two gravitational wave polarizations are given by
\begin{align}\label{eqn:peakh0_def}
  h_{+}(t)  & =\frac{1}{2}h_0(1+\cos^2\iota)\cos \Phi(t)\nonumber\\
    h_{\times}(t)  & =h_0(\cos\iota)\sin \Phi(t).
  \end{align}
 
As the power is emitted in the vicinity of a Kerr black hole, the angular dependence is not exactly as defined in Eq.~\eqref{eqn:peakh0_def} but is instead specified by spin-weighted spheroidal harmonics, which depend on $\alpha$ and the black hole spin $\chi$ \cite{Brito+2017,Berti:2005gp}. However, for $\alpha, \chi_f$ corresponding to $\alpha \lesssim 0.3\, (0.2)$, the standard quadrupolar emission is an excellent approximation to within $5\%\, (3\%)$ or better in $h_{\times}$  and to  within $10\%\, (5\%)$ or better in $h_+$. Given that the exact angular power calculation is computationally intensive and the continuous wave pipelines are optimized for strain angular dependence according to Eq.~\eqref{eqn:peakh0_def}  we neglect the extra effect of spin in our analysis. We also focus only of the $\tilde{\ell} = \tilde{m}=2$ gravitational wave mode, which dominates the total power for $\chi\sim\chi_c$ and $\alpha \lesssim 0.35$ \cite{Yoshino+2014}.

We consider only the gravitational wave emission from the $n=0,\ell=m=1$ cloud; this bound state produces the largest strain: using the values of \cite{Yoshino+2014},
the strain of the first level is larger than that of the second by ${h_0}^{011}/{h_0}^{022} \sim 90\alpha^{-2}$. 

The time evolution of the signal as the cloud depletes through GW radiation is also related to the power emitted and is given by
\begin{align}\label{eqn:tauGW_precise_app}
  \tau_\mathrm{GW}(\alpha) & =   \frac{M_\mathrm{cloud}c^2}{P(\alpha)}.
\end{align}

The calculations of the power have been performed in the point particle approximation, i.e. the back-reaction of the  cloud on the metric is neglected. During the process of superradiance, the black hole spins down and loses mass to the cloud; thus the emission is taking place approximately in a background defined by the final black hole mass and we use the final, smaller, value of $\alpha$ to evaluate the strain and timescale expressions. The presence of the cloud may be viewed as an additional contribution to the mass in the Kerr metric, in which case the initial value of $\alpha$ could be used as an approximation  \cite{Siemonsen:2019}. Our expression is more conservative (as the final $\alpha$ is smaller and thus the power is reduced) and is correct in the limit when the cloud has a mass much smaller than that of the black hole, as is the case for many of our old, long-lasting signals. The difference between using the initial and final $\alpha$ gives approximately a $50\%$ change in the total power, which is a conservative estimate of the overall uncertainty in the rate.

\begin{figure}[t]
  \includegraphics[width=\columnwidth]{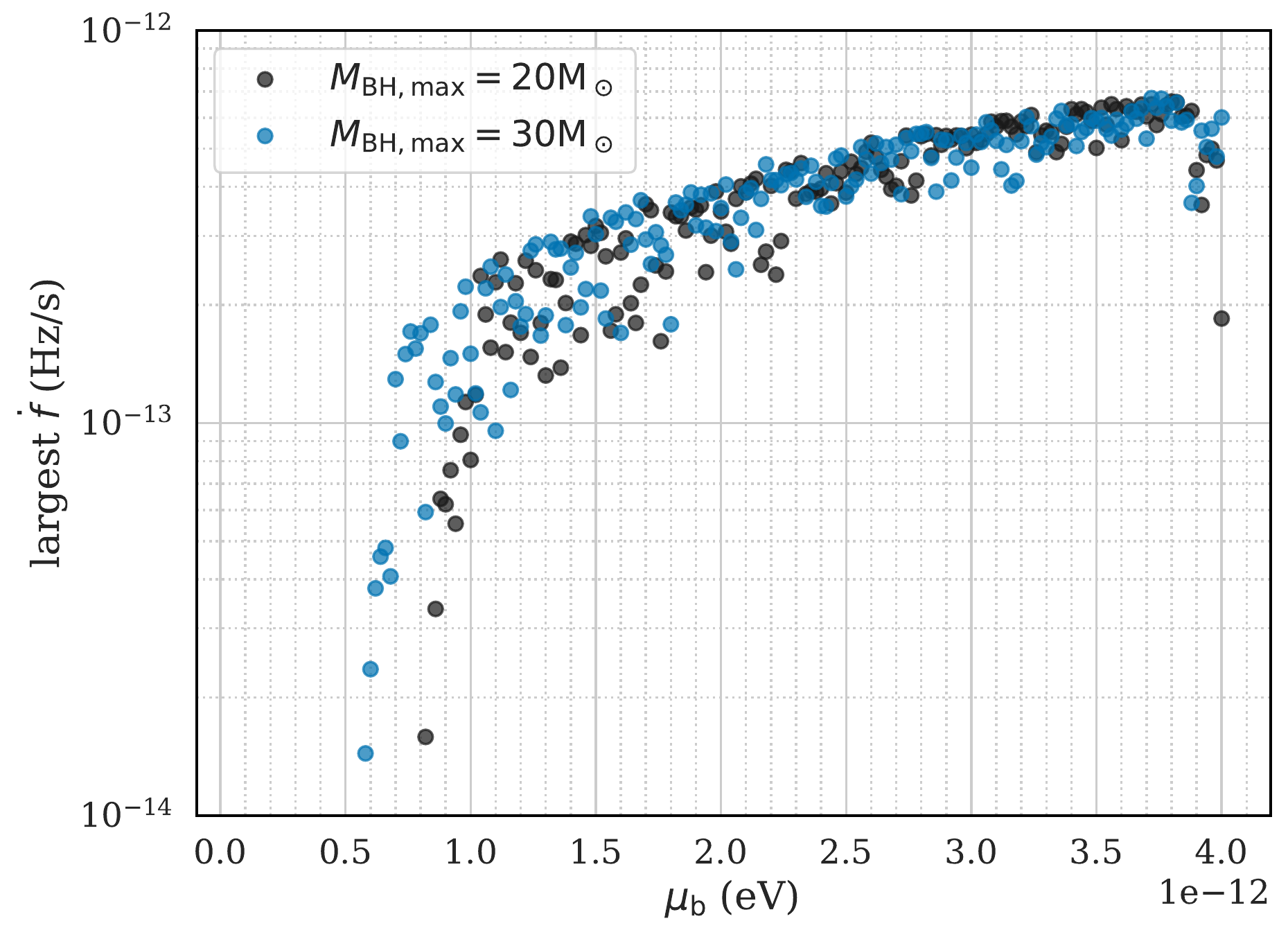}
  \caption{The spin-up $\dot{f}$ due to the decreasing binding energy of the
    cloud depends on $\mu_\mathrm{b}$ and
    $M_\mathrm{BH}$ (Eq.~\eqref{eqn:spinup}). 
    For both the standard and heavy black hole populations, the largest value
    of $\dot{f}$ is $6\times10^{-13}$ Hz/s.
  }
  \label{fig:fdots_shrinkingCloud}
\end{figure}

\subsection{Frequency drift}
\label{app:drift}
As the cloud annihilates to gravitational waves, its gravitational potential energy decreases, leading to a positive frequency drift,
\begin{align}\label{eqn:spinup2}
  \dot{f}_\mathrm{GW}(t) &\approx 5 \times10^{-15}{\mathrm{Hz}}/{{\mathrm{s}}}\\
  &\times \left( \frac{\alpha}{0.1}\right)^{19}\left(\frac{10\mathrm{M}_{\odot}}{M}\right)^2
    \left(\frac{\chi_i-\chi_c}{0.5}\right)^2\left(\frac{M_{\mathrm{cloud}}(t)}{M_{\mathrm{cloud}}^{\mathrm{max}}}\right)^2\nonumber.
\end{align}
There is a larger, \textit{negative} frequency drift at early times as the cloud is growing, but we neglect this in our analysis as the strain is small at these times.

In Fig.~\ref{fig:fdots_shrinkingCloud} we check that the $\dot{f}$ caused
by the cloud's decreasing mass is still smaller than can be resolved by
the first stage of current all-sky continuous wave searches. This is true for both
the standard ($M_\mathrm{BH,max} = 20\mathrm{M}_\odot$) and heavy
($M_\mathrm{BH,max} = 30\mathrm{M}_\odot$) populations. In both cases,
the maximum $\dot{f}$ from the changing cloud mass is $6\times10^{-13}$ Hz/s.

\section{Higher natal kicks}
\label{sec:appHigherNatalKicks}

Throughout this work we focus on a black hole population with a natal kicks given by a Maxwell-Boltzmann distribution with average 3D velocity of 50 km/s. Here we examine the ensemble signals that would be produced by black holes assuming they
are born with average natal kicks of 100 km/s.

The black holes in the 100 km/s population move faster on average, but the
apparent $\dot{f}$ due to the proper motion is still many orders of magnitude smaller than
that resolvable by continuous wave searches and is not a concern.

\begin{figure}[t]
  \includegraphics[width=.9\columnwidth]{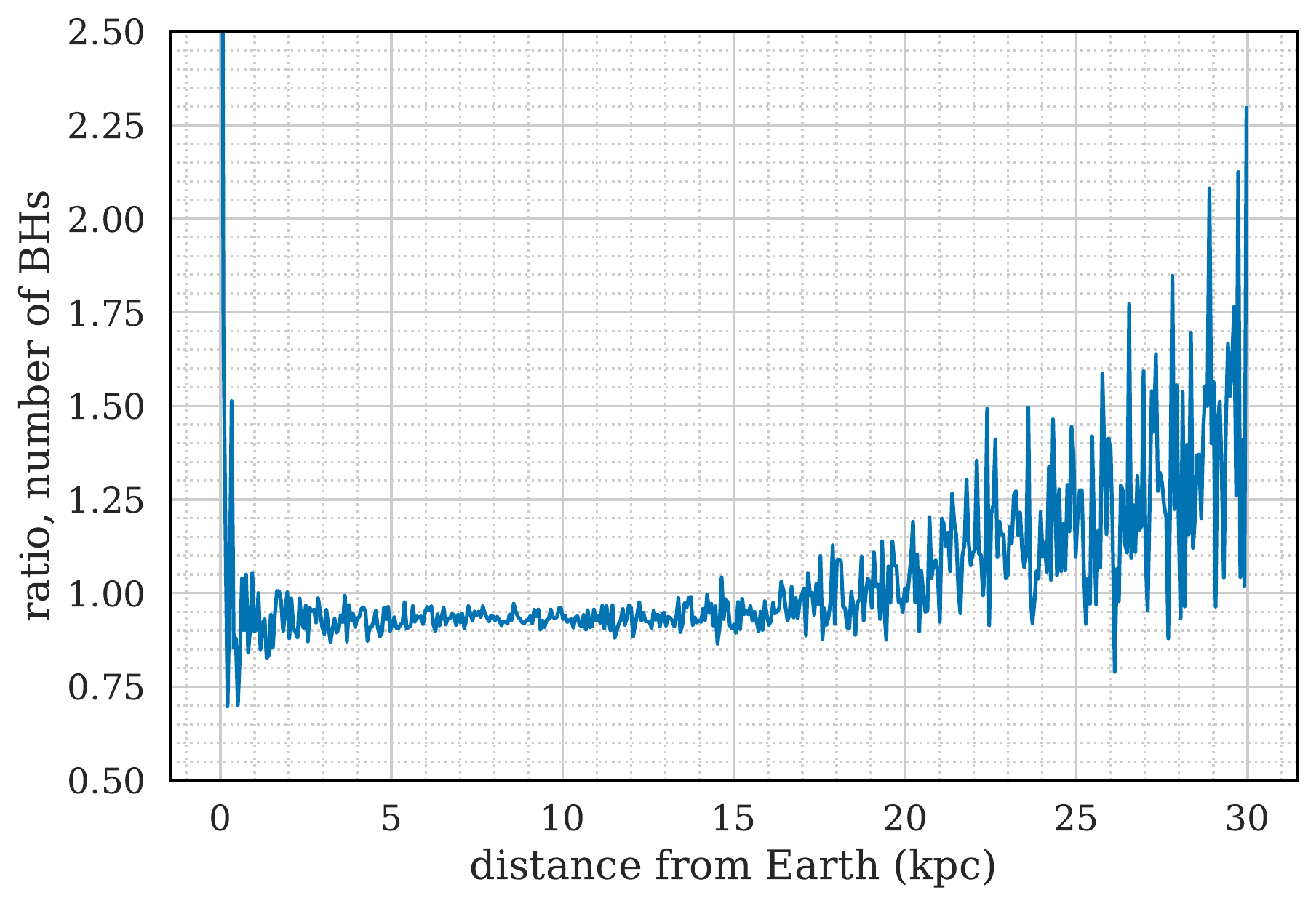}
  \caption{If black holes are born with faster natal kicks (100 km/s rather than
    50 km/s), more of them have sufficiently high speeds to escape the gravitational pull
    of the Galactic bulge and disk. This causes a slight deficit of black holes
    at small Galactic radii, which also means there are fewer BHs close to
    Earth, decreasing on average the number of loud signals.}
  \label{fig:BHdistanceRatios_100v50}
\end{figure}

\begin{figure}[t]
  \includegraphics[width=.9\columnwidth]{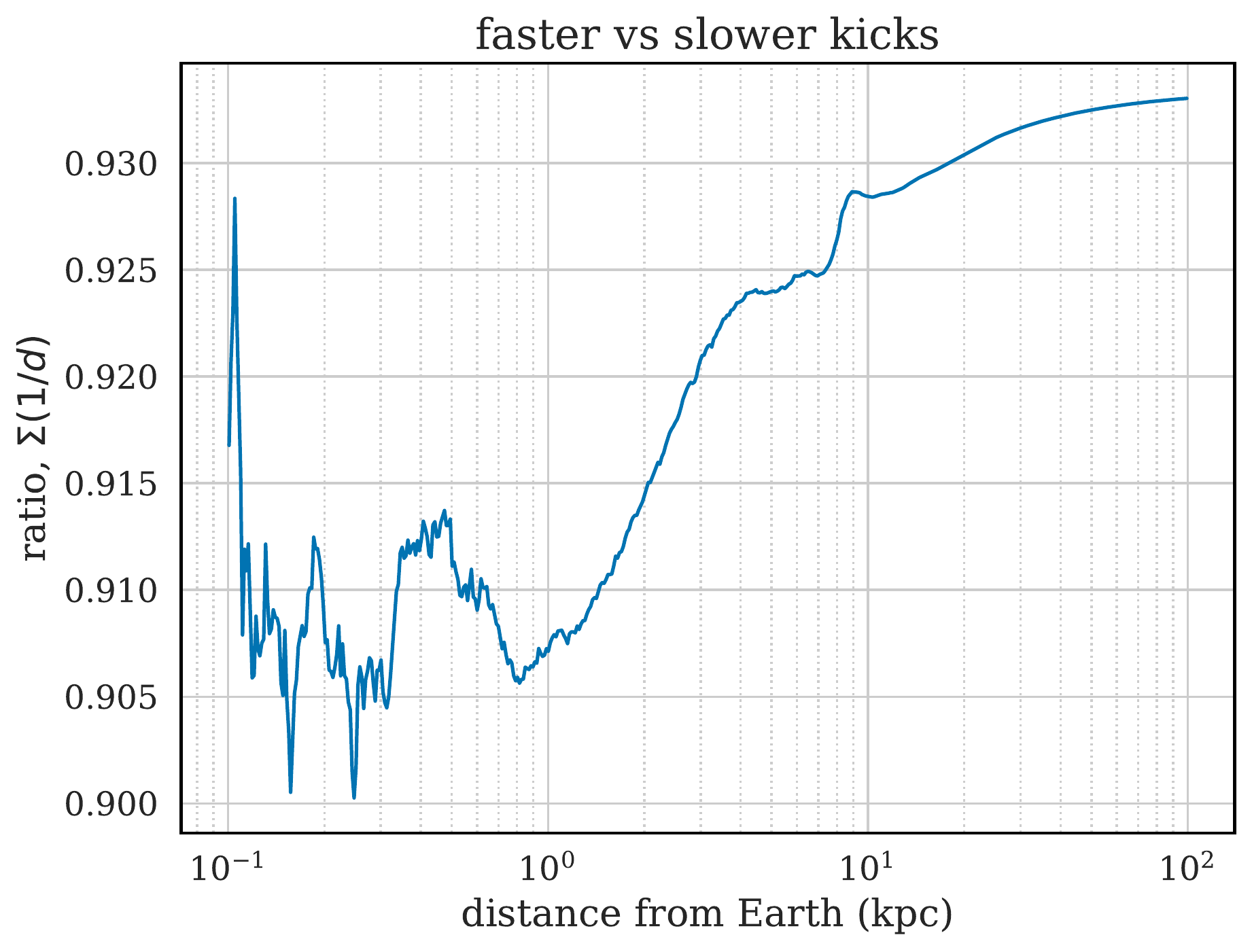}
  \caption{The ratio of the sum of $1/d$ for all the black holes within a distance
    $d$ provides a  comparison of the number of signals between the
    two populations. Here, we plot the ratio for the population with faster versus
    slower kicks. Overall, this ratio is smaller than unity and decreasing toward shorter distances; the ratio is affected by small number statistics at distances
    much less than $1$~kpc. The lower signal number is consistent with the fact that the population of black holes with faster kicks in general produces fewer signals above a given $h_0$. 
    (Fig.~\ref{fig:sumOneOverDRatios_100v50}).}
  \label{fig:sumOneOverDRatios_100v50}
\end{figure}

\begin{figure}[t]
  \includegraphics[width=.9\columnwidth]{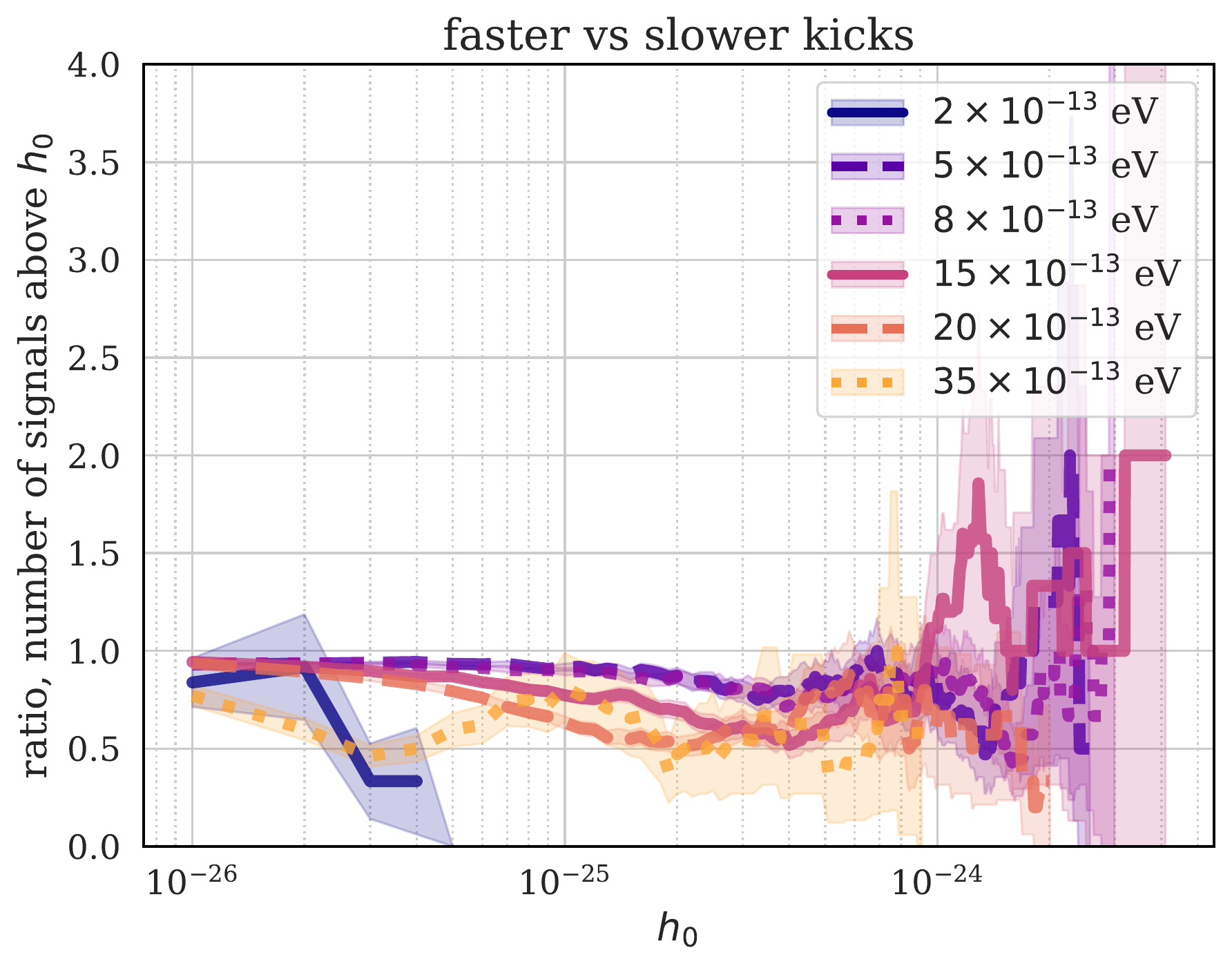}
  \caption{The loudest signals in an ensemble are in general produced by the closest
    black holes. A black hole population with slightly faster natal kicks produces
    a slight deficit of boson clouds close to Earth (Fig.~\ref{fig:BHdistanceRatios_100v50}),
    resulting in fewer signals at larger values of $h_0$. This effect is greater for
    heavier bosons, as the louder signals are preferentially produced by closer
    black holes.  }
  \label{fig:numSignalsRatios_100v50}
\end{figure}

Black holes born with faster natal kicks are more likely to have sufficiently
  high speeds to escape the Galactic bulge and disk and travel further on average. Thus, they are more likely to be
  found at larger distances from both the Galactic Center and the Earth (Fig.~\ref{fig:BHdistanceRatios_100v50}).
Fig.~\ref{fig:sumOneOverDRatios_100v50} shows the
  ratio of the sum of $1/d$ for all the black holes within a distance $d$; this quantity is smaller
  than one and decreases with decreasing distance from Earth until very small distances,
  where it is dominated by small number fluctuations. The ratio of the sum of $1/d$ approximates
  Fig.~\ref{fig:numSignalsRatios_100v50}, the ratio of the number of signals above a given value
  of $h_0$, which is in general less than one and decreases with increasing $h_0$ until the
  loudest signals ($h_0 > 10^{-24}$), a regime that is dominated by the few closest black holes.

The smaller number of signals above a given $h_0$ for the population with
  larger natal kicks also produces smaller 
   mean (Fig.~\ref{fig:ratio_densities_heavyVsStandard}).

\begin{figure}[t]
  \includegraphics[width=0.9\columnwidth]{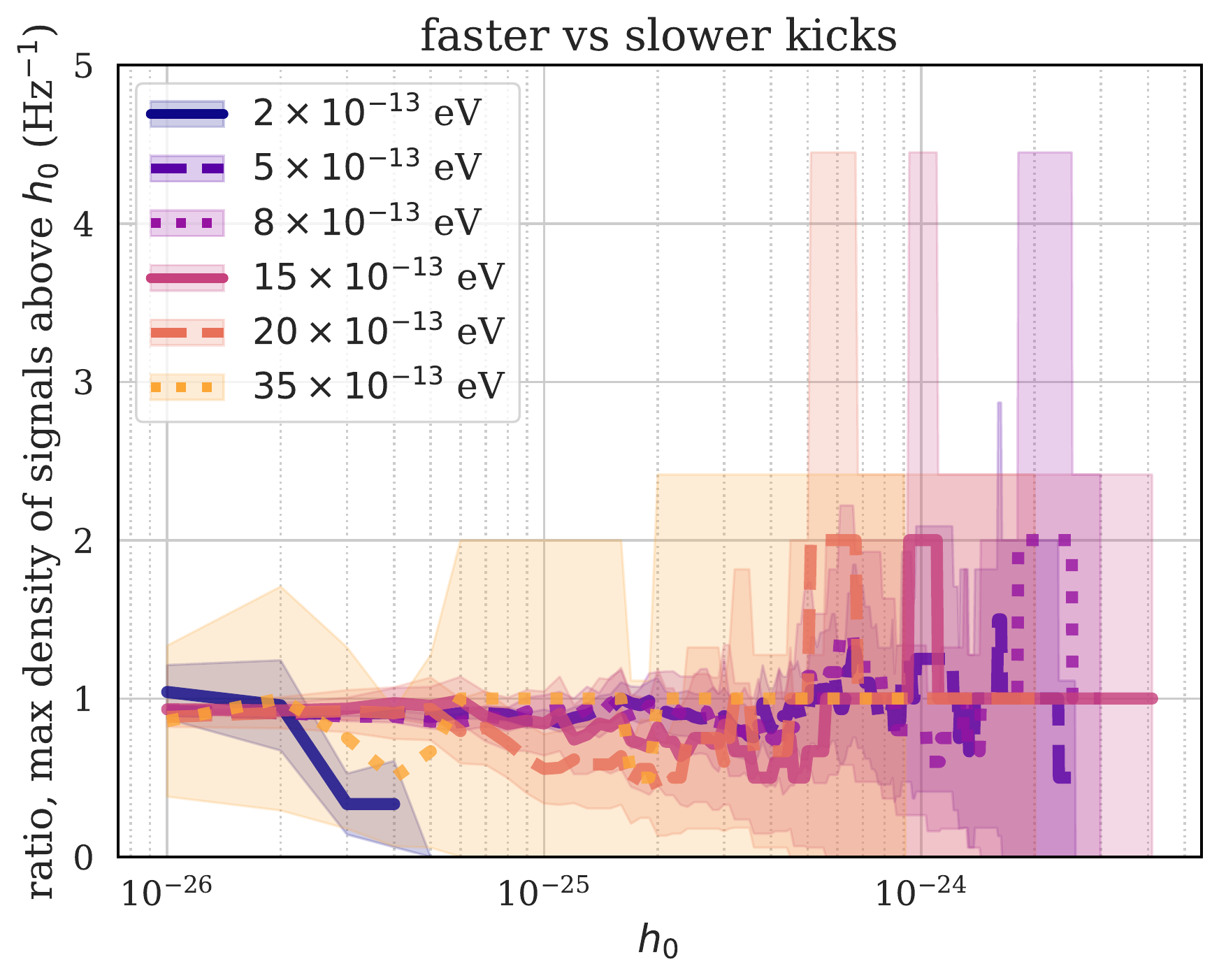}
  \caption{The ratio of maximum signal density above a given $h_0$ between
    the faster and slower kicks populations is consistent with unity.}
  \label{fig:densities_max_100v50}
\end{figure}

\begin{figure}[t]
  \includegraphics[width=0.9\columnwidth]{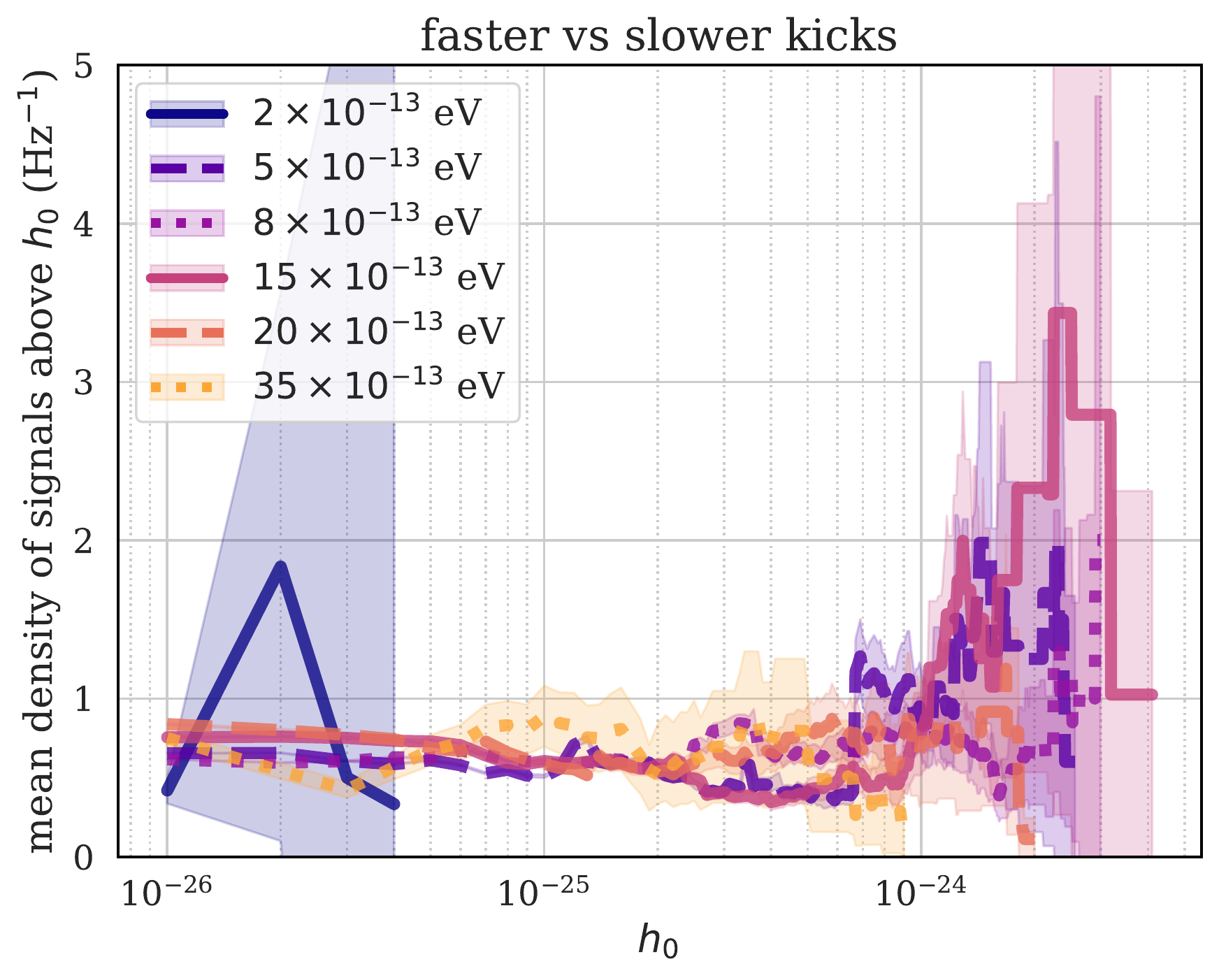}
  \caption{The ratio of mean signal density above a given $h_0$ between
    the faster and slower kicks populations reflects the number of signals
    above a given $h_0$ (Fig.~\ref{fig:numSignalsRatios_100v50}). In general,
    the black holes with faster natal kicks produce signals
    with larger Doppler shifts, thereby increasing the frequency range spanned
  by the ensemble and decreasing the mean signal density further.}
  \label{fig:densities_mean_100v50}
\end{figure}

\section{Light black hole population}
\label{app:lightBHPopulation}

  We compare the ensemble signals from the standard black hole population
($M_\mathrm{BH} \in [5,20]\mathrm{M}_\odot$) with the ensemble signals from a
lighter black hole population ($M_\mathrm{BH} \in [3,20]\mathrm{M}_\odot$). We maintain the
distribution shape (Salpeter function) as well as the total number of black holes ($10^8$);
reducing the minimum black hole mass therefore reduces the number of black holes of all other
masses.

\begin{figure}[t!]
  \includegraphics[width=0.9\columnwidth]{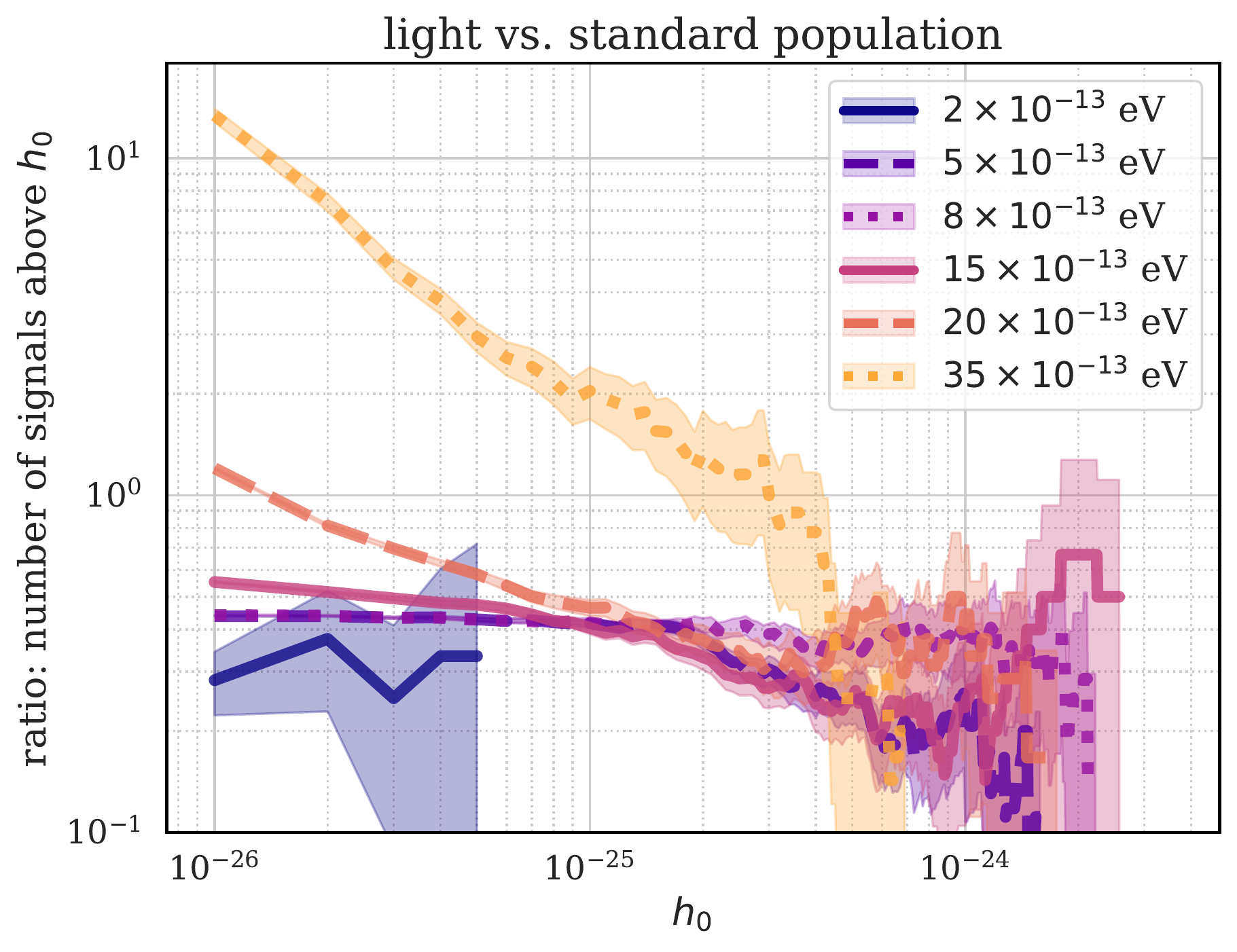}
  \caption{
      Increasing the number of light black holes ($M_\mathrm{BH} < 5\mathrm{M}_\odot$)
      while maintaining the same number of black holes effectively reduces the number of
      black holes at all other masses. Therefore, the number of
      signals above a given $h_0$ decreases by a factor of two for most of the boson
      masses. The exception is the heaviest boson ($\mu_\mathrm{b} = 3.5\times10^{-12}$~eV),
      for which the number of signals is \emph{larger} with the light black hole population
      for signals with $h_0 < 3\times10^{-25}$.
  }
  \label{fig:ratio_numSignals_lightVsStandard}
\end{figure}

For $\mu_\mathrm{b} = 3.5\times10^{-12}$~eV, the addition of the lighter black holes
results in a  larger number of signals with $h_0 \lesssim 3\times10^{-25}$; a similar
but smaller effect is seen for $\mu_\mathrm{b} = 2\times10^{-12}$~eV. This is due to the fact that at these heavy boson masses, lighter black holes can more easily satisfy the critical spin condition while still producing relatively large signals. However, the current
continuous wave search upper limits are around $10^{-24}$ near the signal frequencies produced
by these heavy bosons, so the addition of the lighter black holes does not affect our
estimates of the number of signals detectable by current continuous wave searches. 

The overall effect is that the total number of signals decreases by about a factor of two
for  all but the heaviest boson masses (Fig.~\ref{fig:ratio_numSignals_lightVsStandard}).   In reality, the black hole mass distribution should turn over at the lightest masses, i.e. the majority of black holes is still expected to have mass above $5\mathrm{M}_\odot$  \cite{Ozel+2010, Farr+2011}. Therefore, Fig.~\ref{fig:ratio_numSignals_lightVsStandard}  can be considered a lower bound on the ratio of the number of signals for the case that black holes lighter than $5\mathrm{M}_\odot$ exist in the Galaxy. 

\section{Heavy black hole population}
\label{app:heavyBHPopulation}

We examine the ensemble signal for a heavy population of black holes,
with $M_\mathrm{BH,max} = 30\mathrm{M}_\odot$, and compare the resultant
ensemble signals with the signals from the standard population.  

\begin{figure}[htbp!]
  \includegraphics[width=0.95\columnwidth]{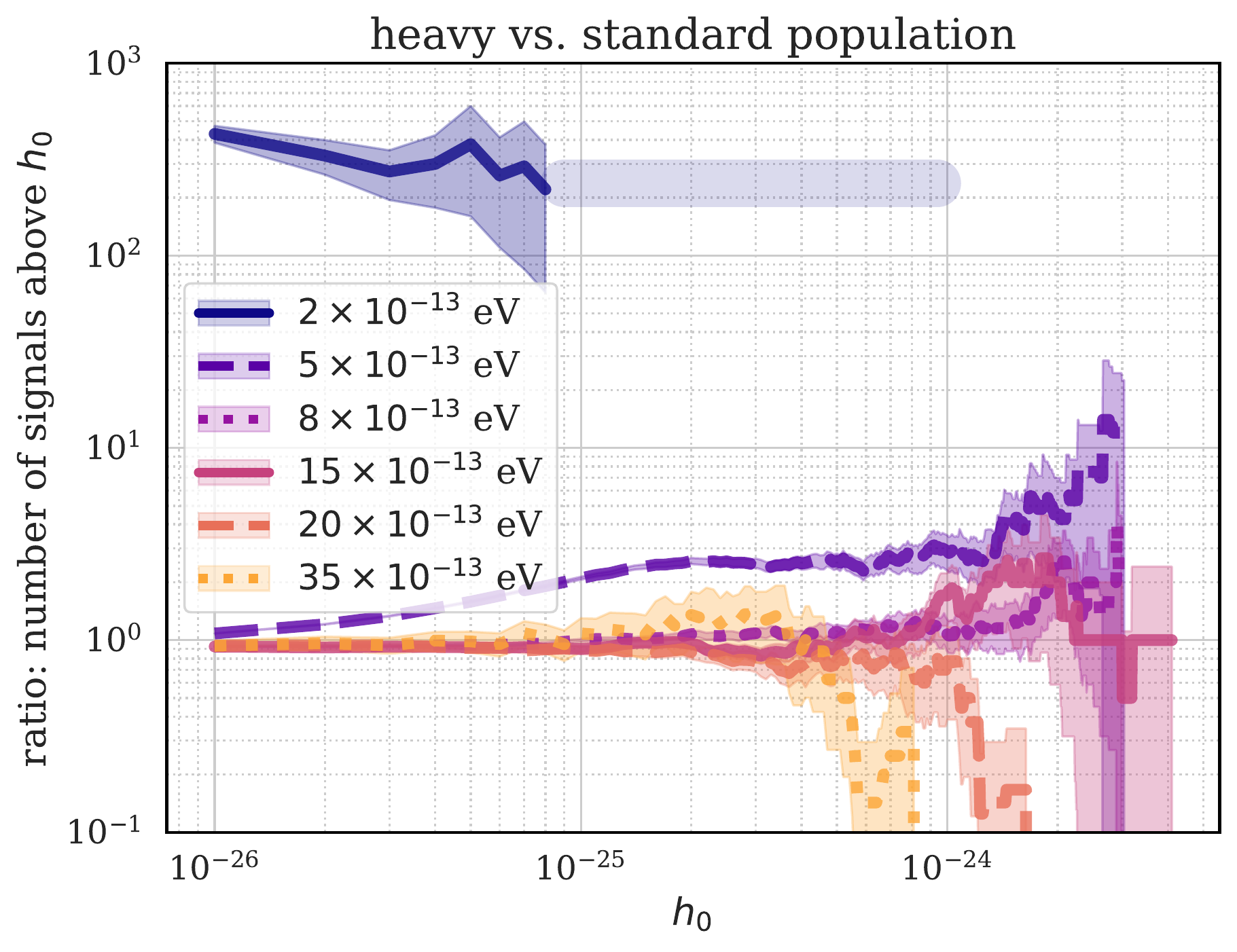}
  \caption{The ratio of the number of signals above a given $h_0$ for the
    heavy ($M_\mathrm{BH,max} = 30\mathrm{M}_\odot$) versus standard
    ($M_\mathrm{BH,max} = 20\mathrm{M}_\odot$) populations. The number of signals for the lighter bosons ($\mu_\mathrm{b} = 2\times10^{-13}$ eV and $5\times10^{-13}$ eV) increases in going from
    the standard to the heavy black hole population, while the
    number of signals for the heavier bosons stays approximately the same (Section~\ref{sec:howManyDetectable}).
    For $\mu_\mathrm{b} =2\times10^{-13}$~eV, the signals from the heavy
    black hole population extend to strains of $h_0 \approx 10^{-24}$ (indicated
    by the light, thick bar) while the signals from the standard black
    hole population only have $h_0 \leq 10^{-25}$.}
  \label{fig:ratio_numSignals_heavyVsStandard}
\end{figure}

\begin{figure}[htbp!]
  \includegraphics[width=0.95\columnwidth]{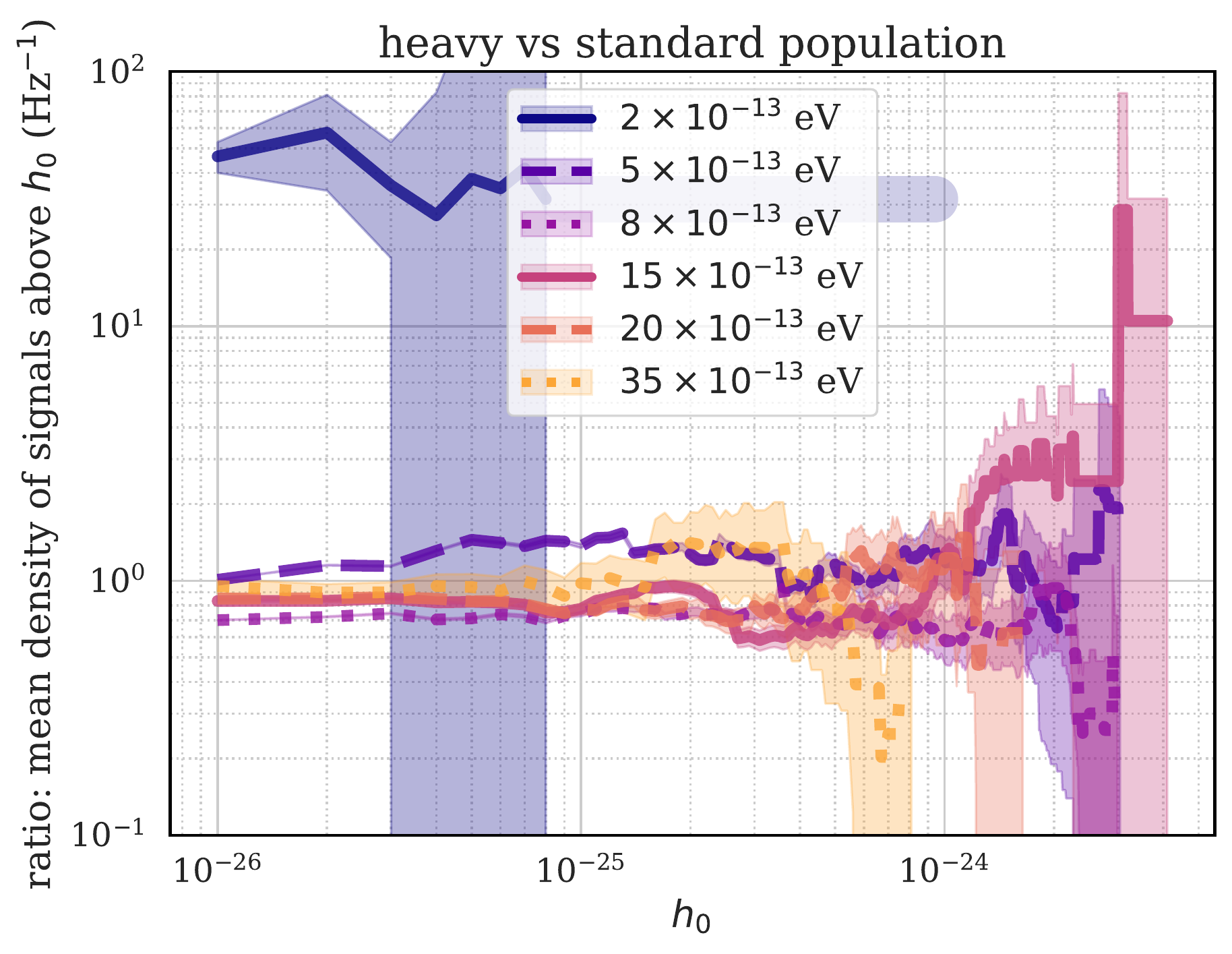}
  \caption{The ratio of the mean signal density above a given $h_0$ for
    the heavy versus standard black hole populations is approximately unity
    for the five heavier boson masses. For the lightest mass ($\mu_\mathrm{b}
    = 2\times10^{-13}$~eV), the mean density is over ten times larger for
  the heavier black hole population.}
  \label{fig:ratio_densities_heavyVsStandard}
\end{figure}

\begin{figure}[htbp!]
  \includegraphics[width=0.95\columnwidth]{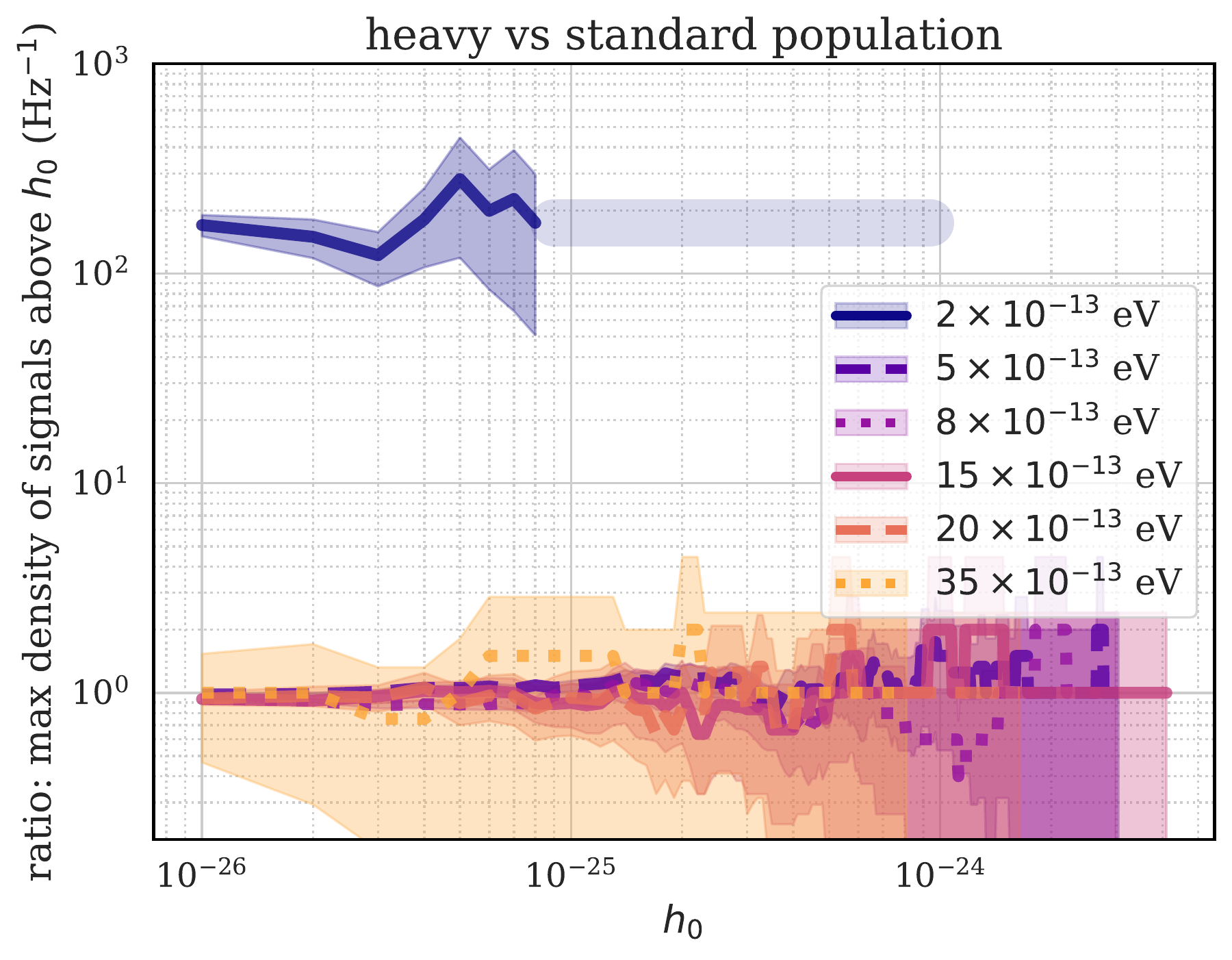}
  \caption{For the six boson masses plotted here, the ratio of the maximum
    signal density above a given $h_0$ for the heavy versus standard black
    hole populations is similar to the ratio of the mean signal density.
    The maximum signal density for the lightest boson ($\mu_\mathrm{b} =
    2\times10^{-13}$~eV) is over a hundred times larger for the heavier
  black hole population.}
  \label{fig:ratio_densities_max_heavyVsStandard}
\end{figure}

We show the ratio of the heavy to the standard population in the number of signals (Fig.~\ref{fig:ratio_numSignals_heavyVsStandard}), mean density of signals (Fig.~\ref{fig:ratio_densities_heavyVsStandard}), and maximum density of signals (Fig.~\ref{fig:ratio_densities_max_heavyVsStandard}).  The ratios are generally consistent with one. The exceptions are the number and density of signals for the lightest boson mass of $2\times 10^{-13}$~eV, which is a factor of several hundred larger for the heavy population, and for  $5\times 10^{-13}$~eV, a factor of approximately two larger. In addition, the mean density of signals is reduced by $\sim20\%$ for the intermediate boson mass of $8\times 10^{-13}$~eV, due to the larger range of frequencies covered by the heavier black hole population for a fixed black hole number.

\begin{figure*}[h]
  \includegraphics[width=0.8\textwidth]{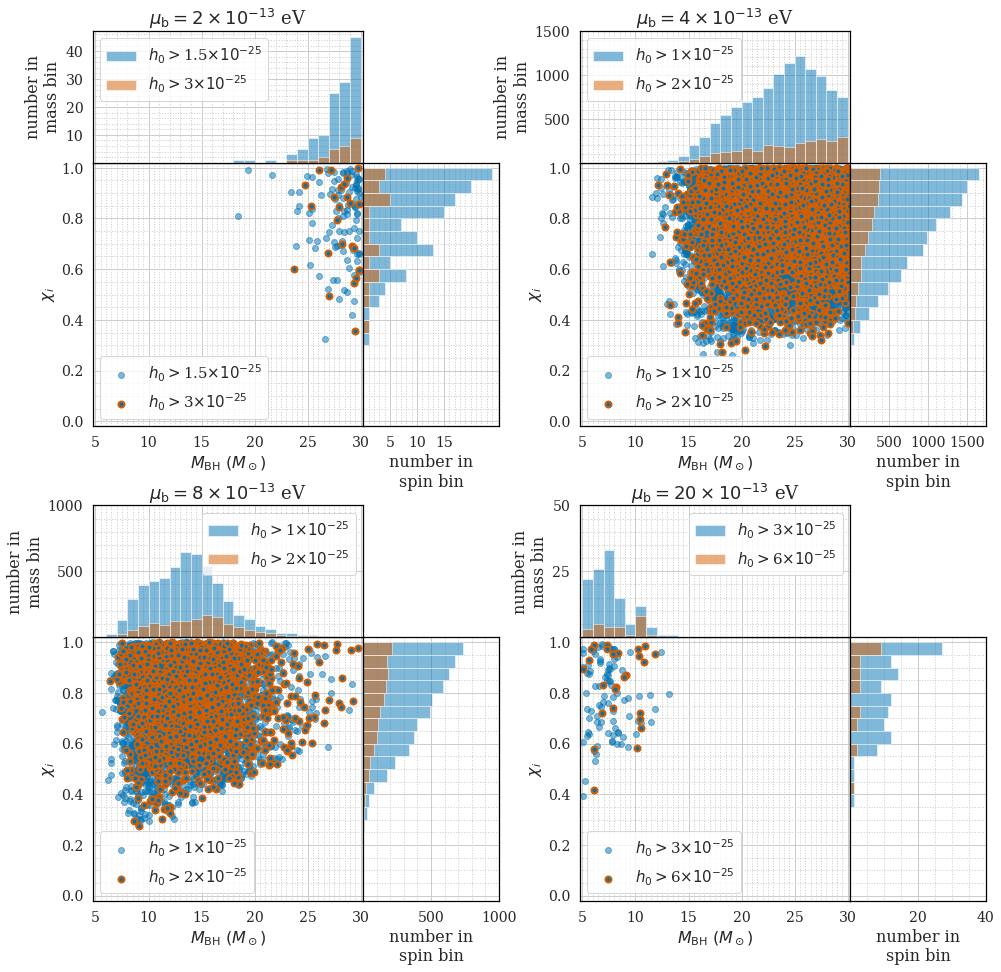}
  \caption{Mass and spin properties of the heavy black hole population.
    For the two heavier boson masses ($\mu_\mathrm{b}= 8\times10^{-13}$~eV
    and $2\times10^{-12}$~eV, the black holes that produce the potentially detectable ensemble
    signals     have similar properties as the black holes that produce the
    signals in Figure~\ref{fig:ensembleExamples} (the standard population),
    due to the fact that systems with large values of $\alpha$ have
    dramatically shortened lifetimes. 
    For the two lighter boson masses ($\mu_\mathrm{b} = 2\times10^{-13}$~eV
    and $4\times10^{-13}$~eV), the addition of black holes with $M_\mathrm{BH} > 20\mathrm{M}_\odot$
  greatly increases the number of detectable signals.}
  \label{fig:ensembleBHs_massSpin_heavy}
\end{figure*}

\begin{figure*}[h]
  \includegraphics[width=0.8\textwidth]{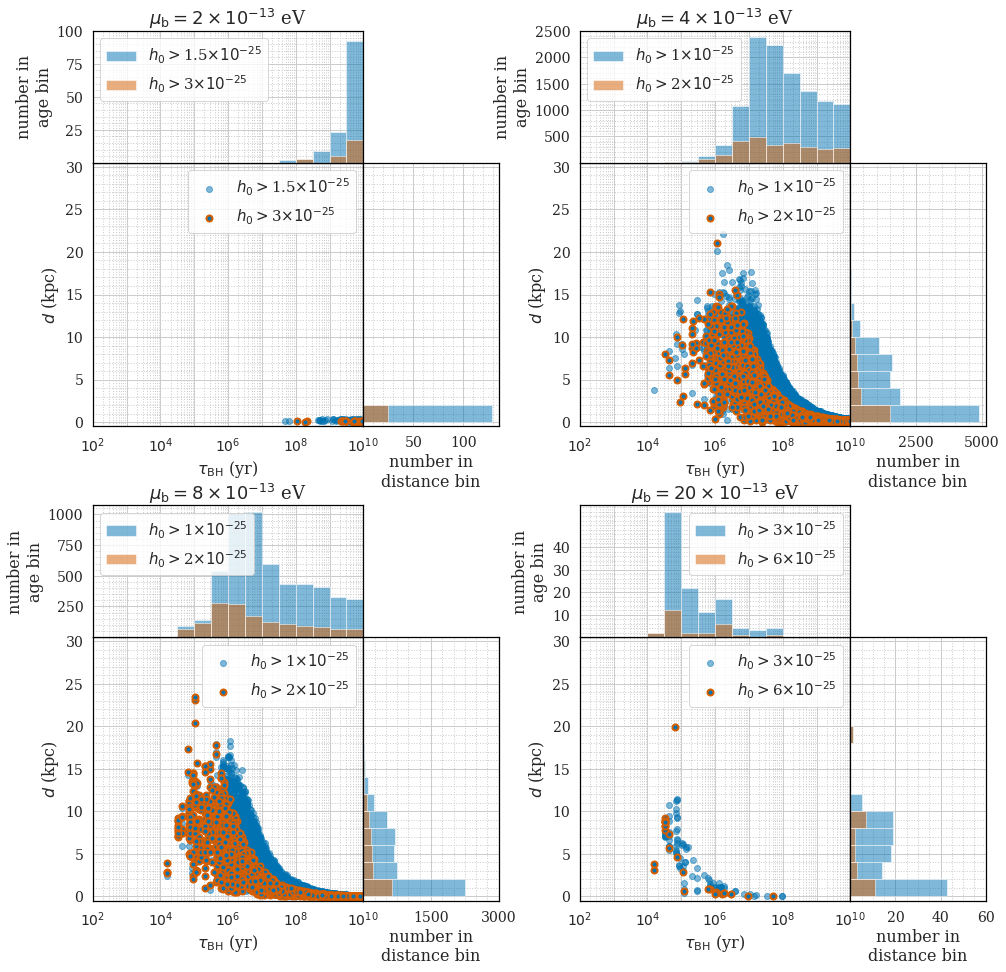}
  \caption{
    Ages and distances of the heavy black hole population. As in Fig.~\ref{fig:ensembleBHs_massSpin_heavy}, the addition of black holes with $M_\mathrm{BH} > 20 \mathrm{M}_\odot$ increases the number of detectable systems for the lighter bosons ($\mu_\mathrm{b} = 2\times10^{-13}$ and $4\times10^{-13}$~eV)    but has little to no effect on the heavier bosons.
    }
  \label{fig:ensembleBHs_ageDistance_heavy}
\end{figure*}

We also show the mass and spin distributions (Fig.~\ref{fig:ensembleBHs_massSpin_heavy}) as well as the age and distance distributions (Fig.~\ref{fig:ensembleBHs_ageDistance_heavy}) for the heavy black hole population. In the heavy population, black holes from further away are able to source loud signals, as compared to the standard population.

\section{Ensemble signal: signals above the detection threshold}
\label{appendix:signalsAboveUL}

\begin{figure*}
\includegraphics[width=\columnwidth]{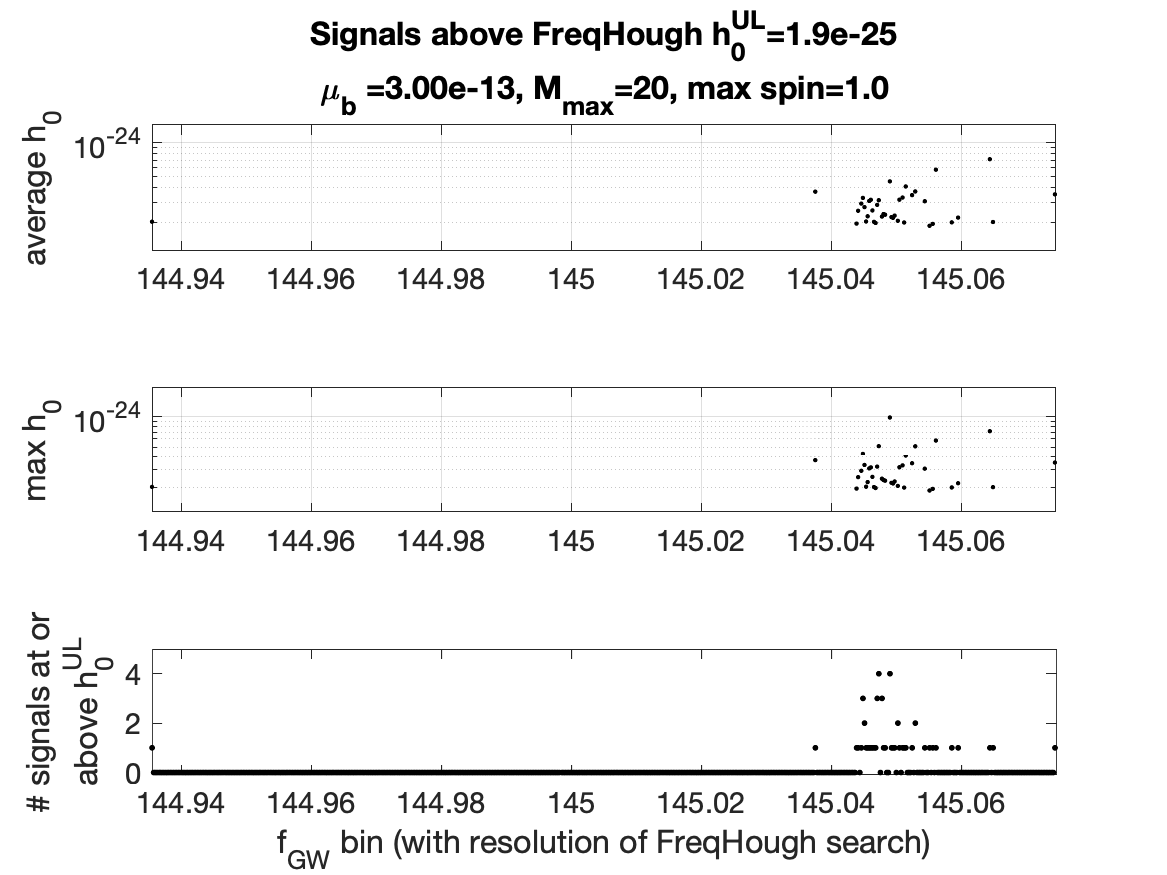}
\includegraphics[width=\columnwidth]{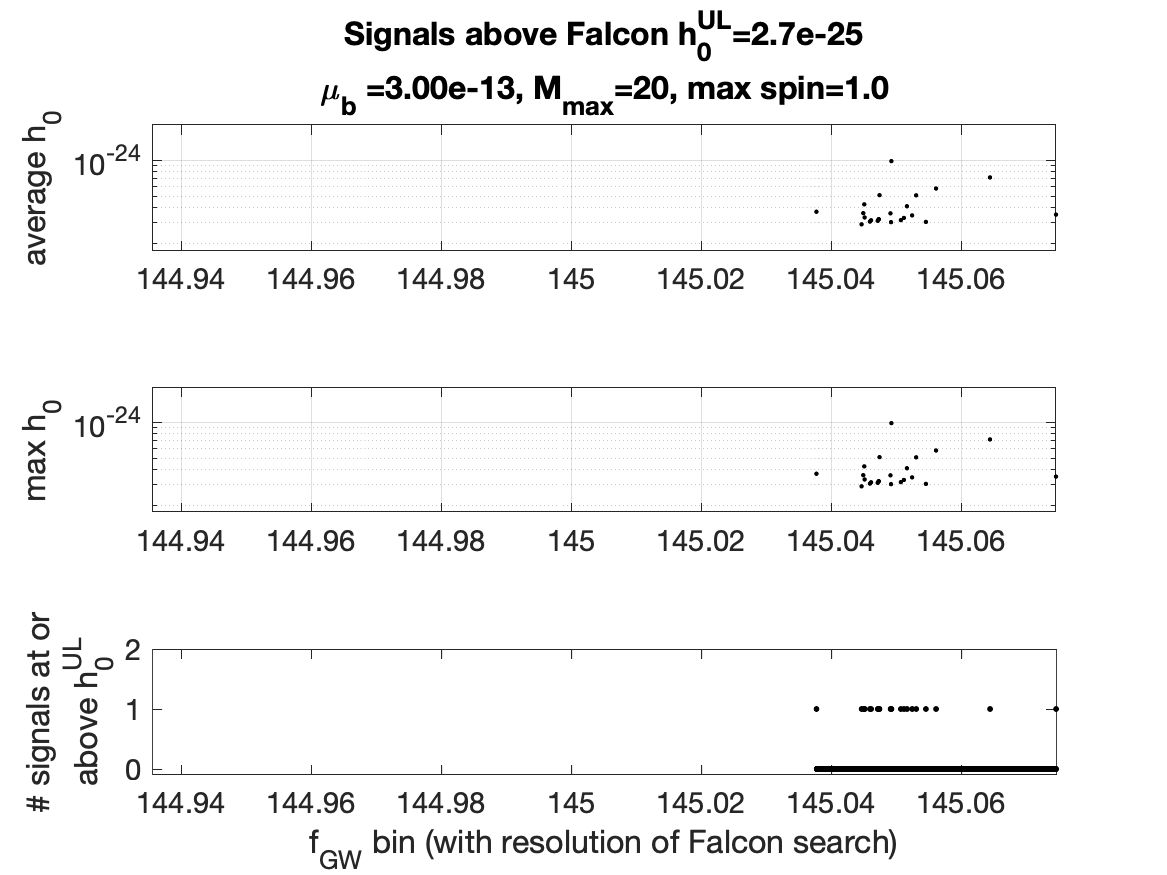}
\includegraphics[width=\columnwidth]{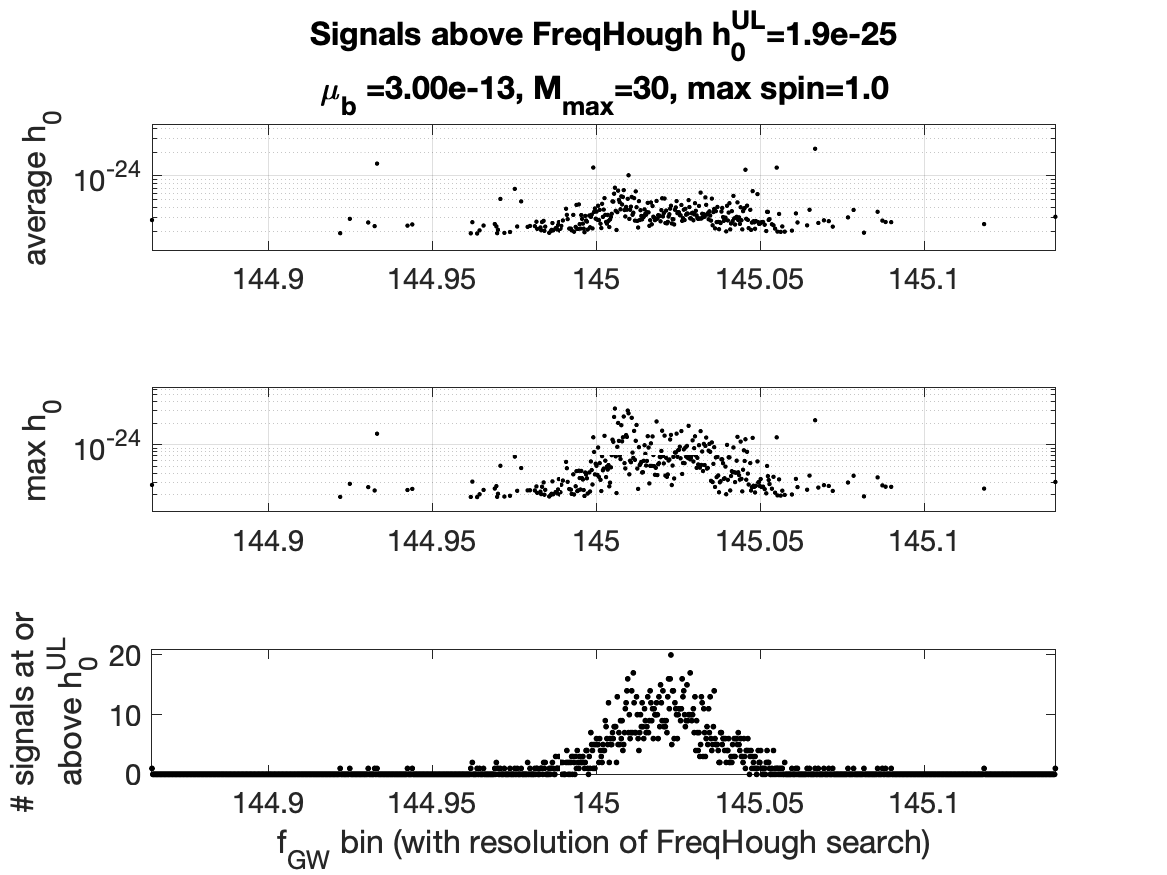}
\includegraphics[width=\columnwidth]{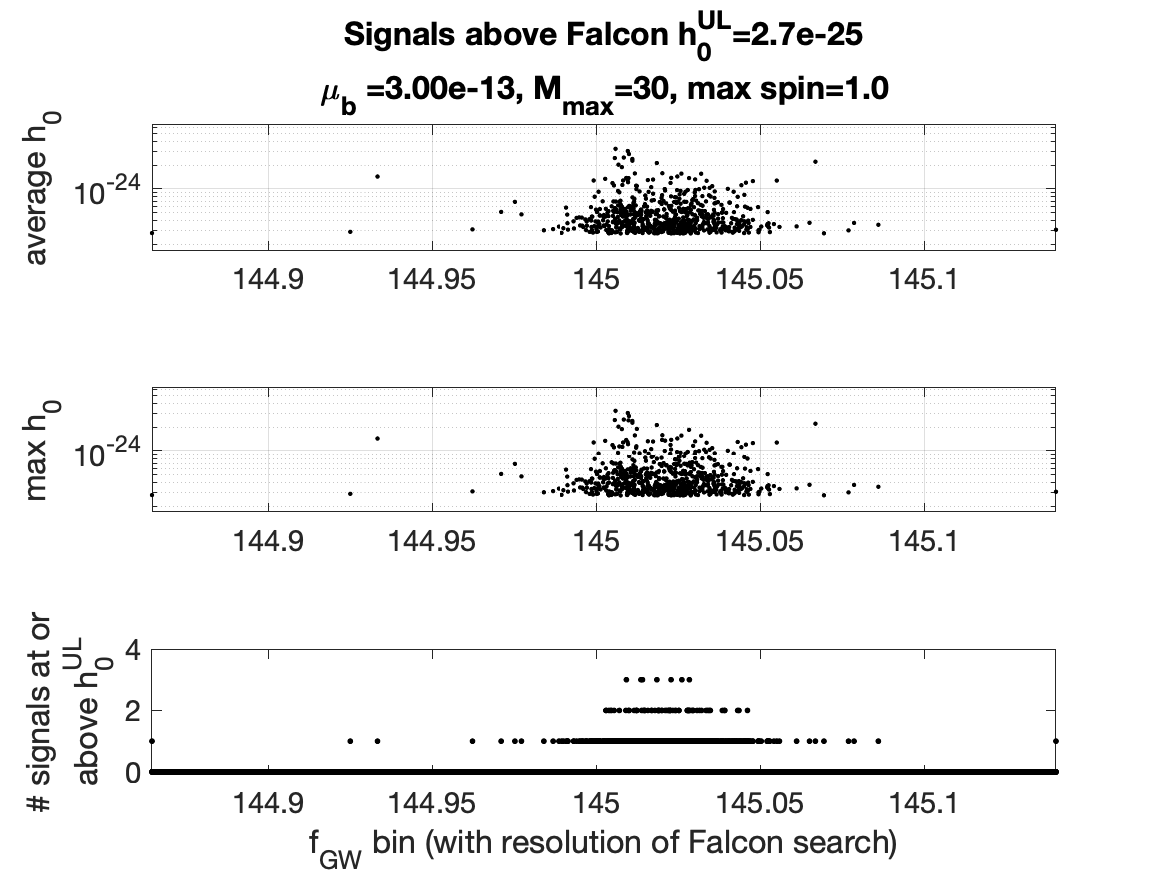}
\caption{
The x-axes show the signal frequency binned with the resolution of the two searches: Freq Hough, $2.4 \times 10^{-4}$ Hz  \cite{RomeBosonClouds}  ({\it{left}}) and Falcon  \cite{FalconO1,Dergachev:2019oyu}, $1.7 \times 10^{-5}$ Hz ({\it{right}}). From the bottom panel going up, the y-axis shows the number of detectable signals in each bin, their maximum intrinsic amplitude and their average amplitude.\label{fig:aboveULmub3}
}

\end{figure*}

\begin{figure*}
\includegraphics[width=\columnwidth]{FreqHough-mub4p00e-13Mmax20spin1p0.png}
\includegraphics[width=\columnwidth]{Falcon-mub4p00e-13Mmax20spin1p0.png}
\includegraphics[width=\columnwidth]{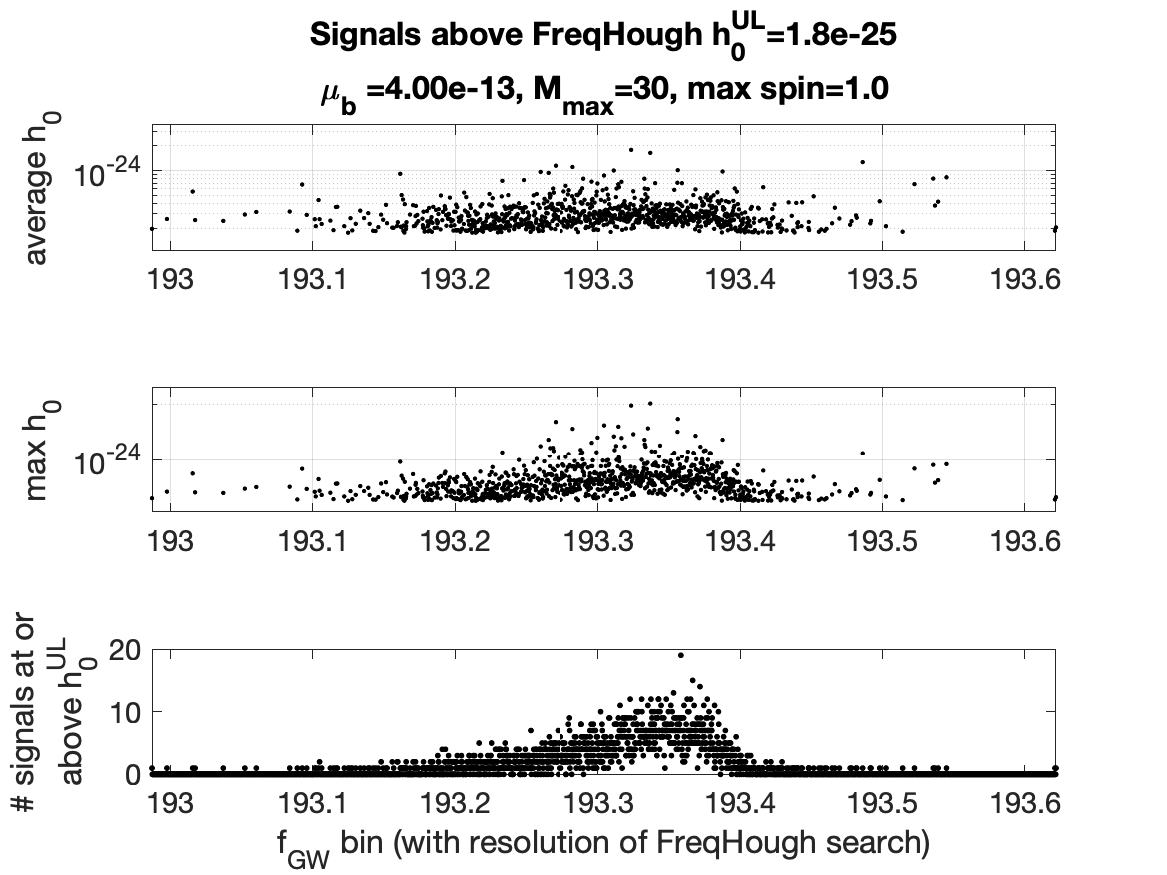}
\includegraphics[width=\columnwidth]{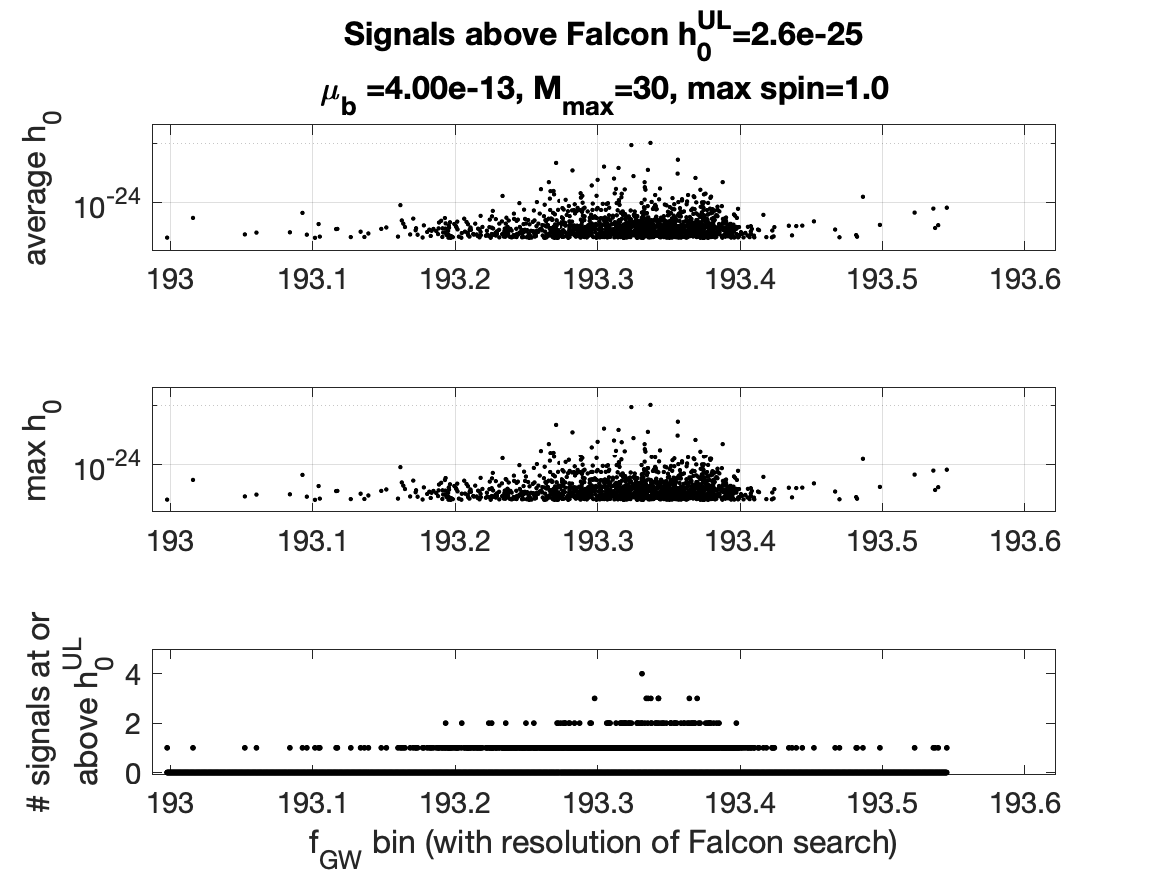}
  \caption{The x-axes show the signal frequency binned with the resolution of the two searches: Freq Hough, $2.4 \times 10^{-4}$ Hz  \cite{RomeBosonClouds}  ({\it{left}}) and Falcon  \cite{FalconO1,Dergachev:2019oyu}, $1.7 \times 10^{-5}$ Hz ({\it{right}}). From the bottom panel going up, the y-axis shows the number of detectable signals in each bin, their maximum intrinsic amplitude and their average amplitude. \label{fig:aboveULmub4}}
\end{figure*}

\begin{figure*}
\includegraphics[width=\columnwidth]{FreqHough-mub4p00e-13Mmax20spin0p5.png}
\includegraphics[width=\columnwidth]{Falcon-mub4p00e-13Mmax20spin0p5.png}
\includegraphics[width=\columnwidth]{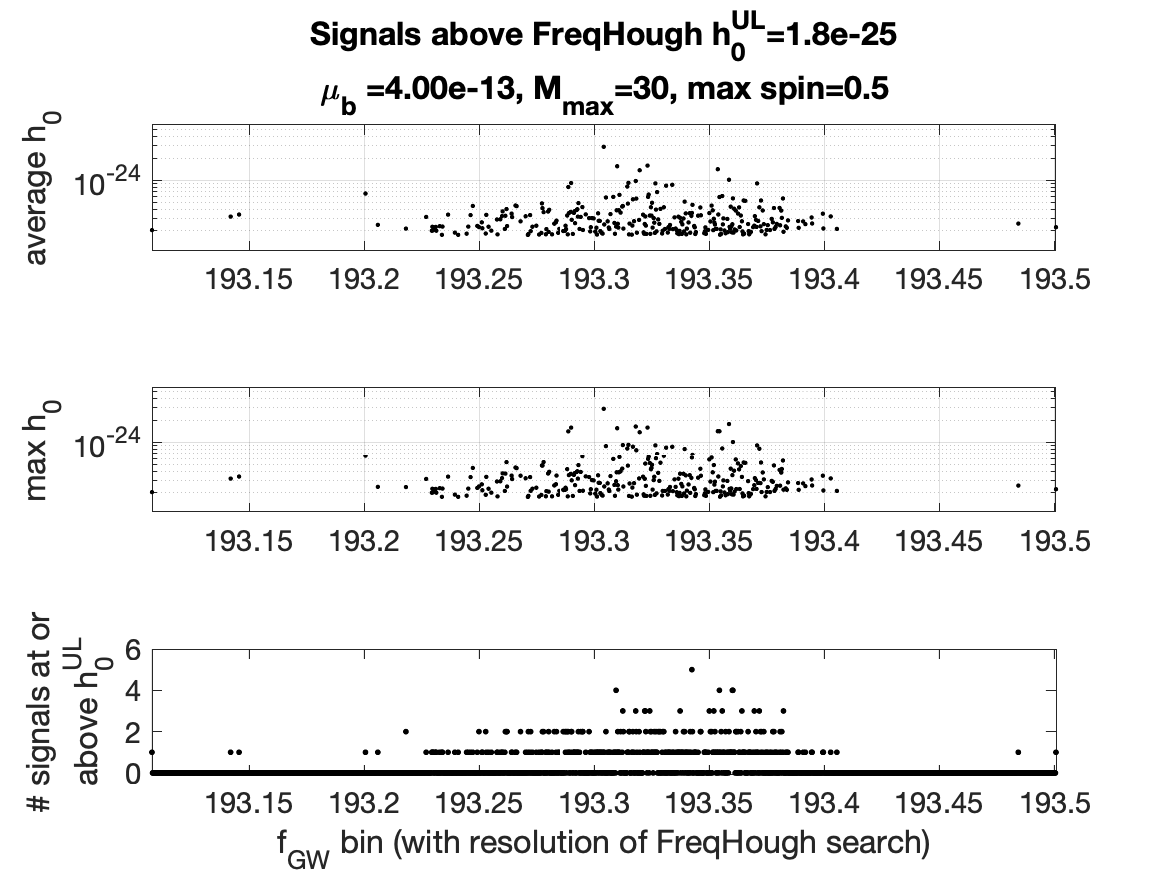}
\includegraphics[width=\columnwidth]{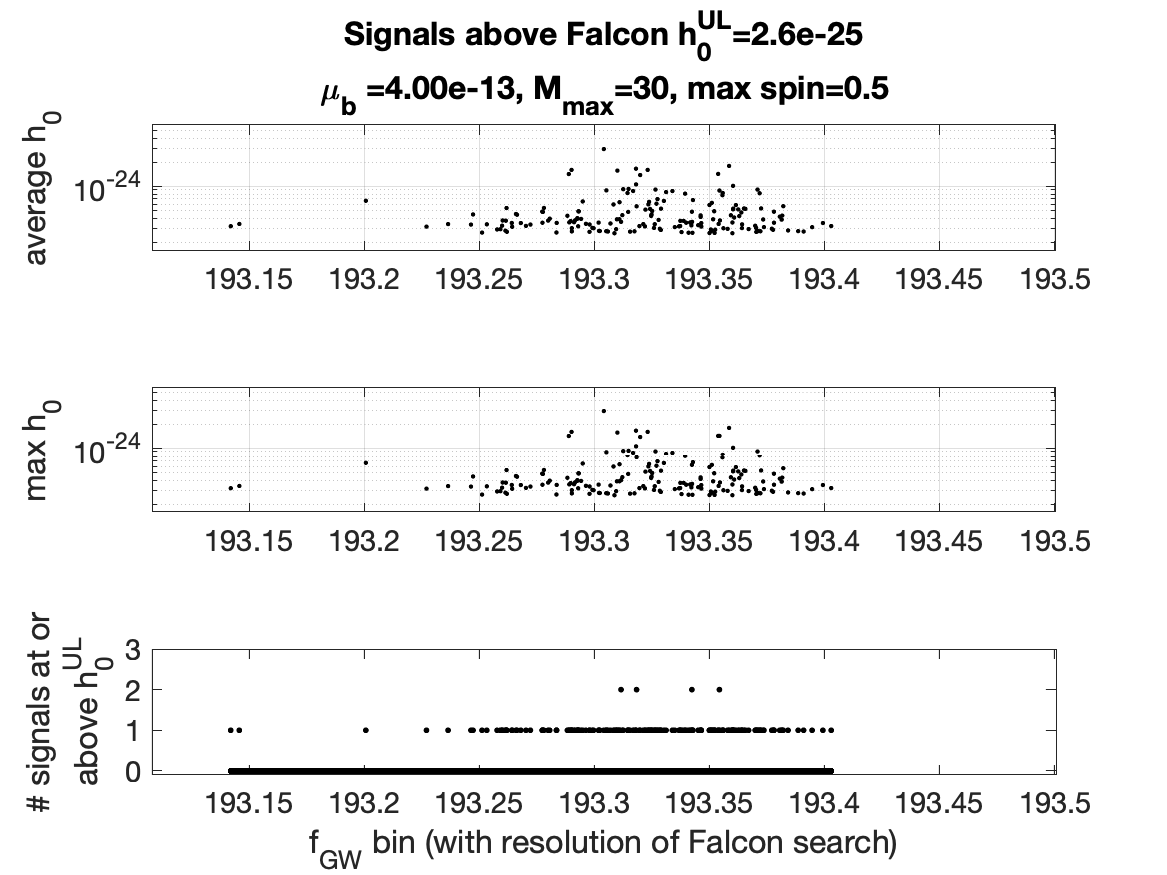}
  \caption{The x-axes plots show the signal frequency binned with the resolution of the two searches: Freq Hough, $2.4 \times 10^{-4}$ Hz  \cite{RomeBosonClouds}  ({\it{left}}) and Falcon  \cite{FalconO1,Dergachev:2019oyu}, $1.7 \times 10^{-5}$ Hz ({\it{right}}). From the bottom panel going up, the y-axis shows the number of detectable signals in each bin, their maximum intrinsic amplitude and their average amplitude. \label{fig:aboveULmub4M20}}
\end{figure*}

\begin{figure*}
\includegraphics[width=\columnwidth]{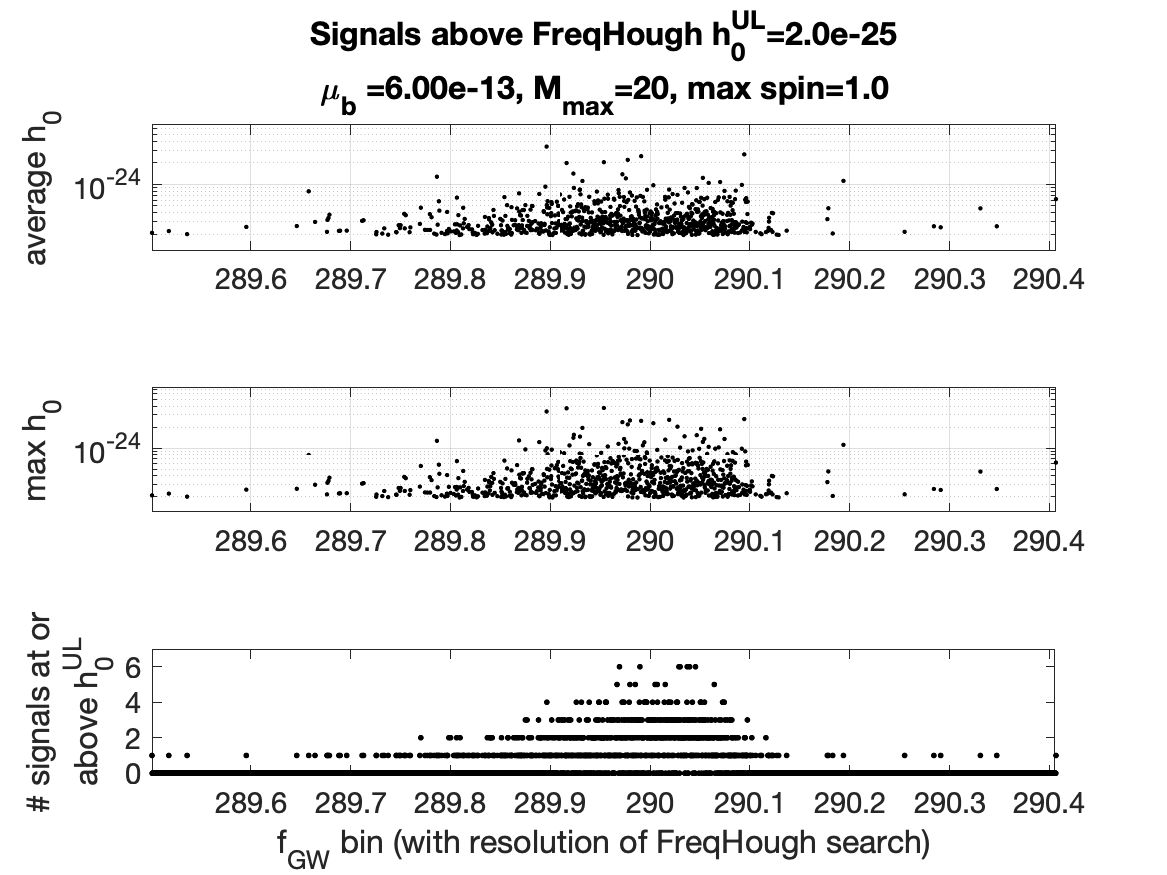}
\includegraphics[width=\columnwidth]{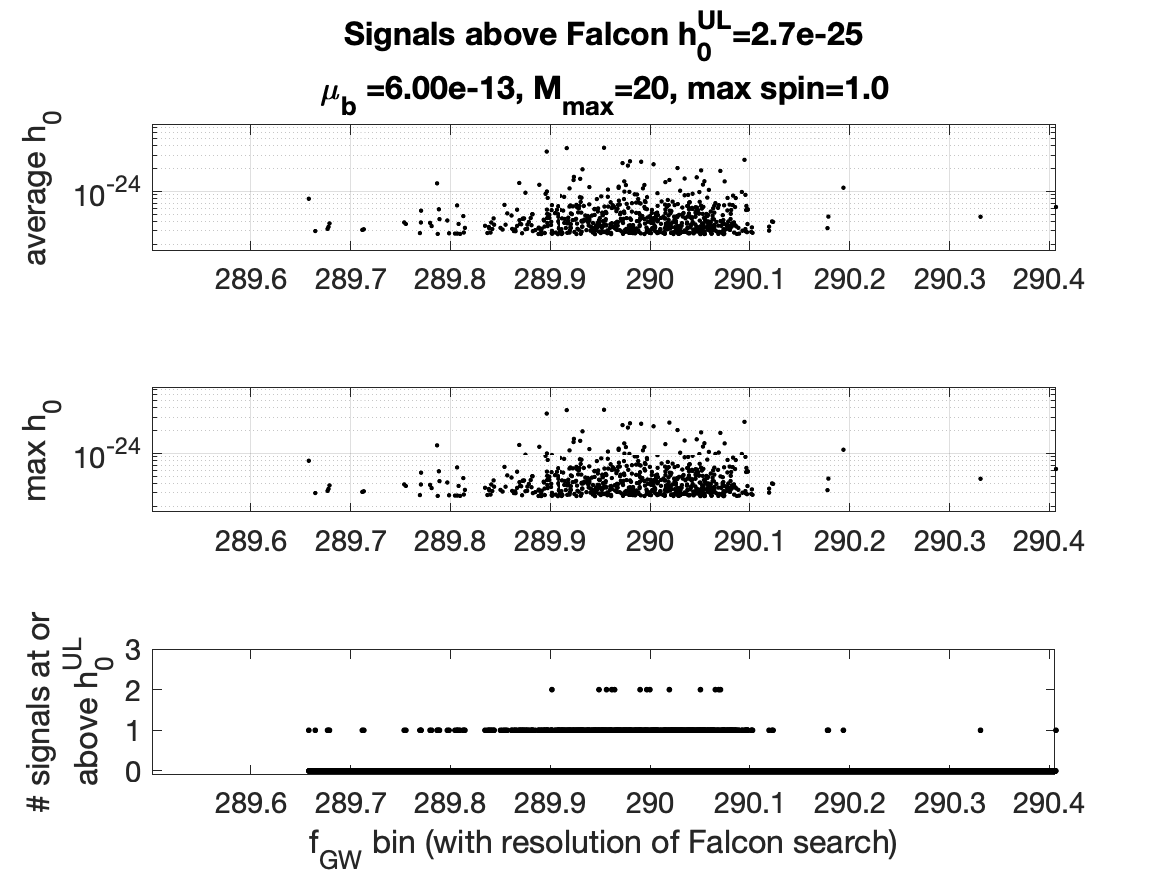}
\includegraphics[width=\columnwidth]{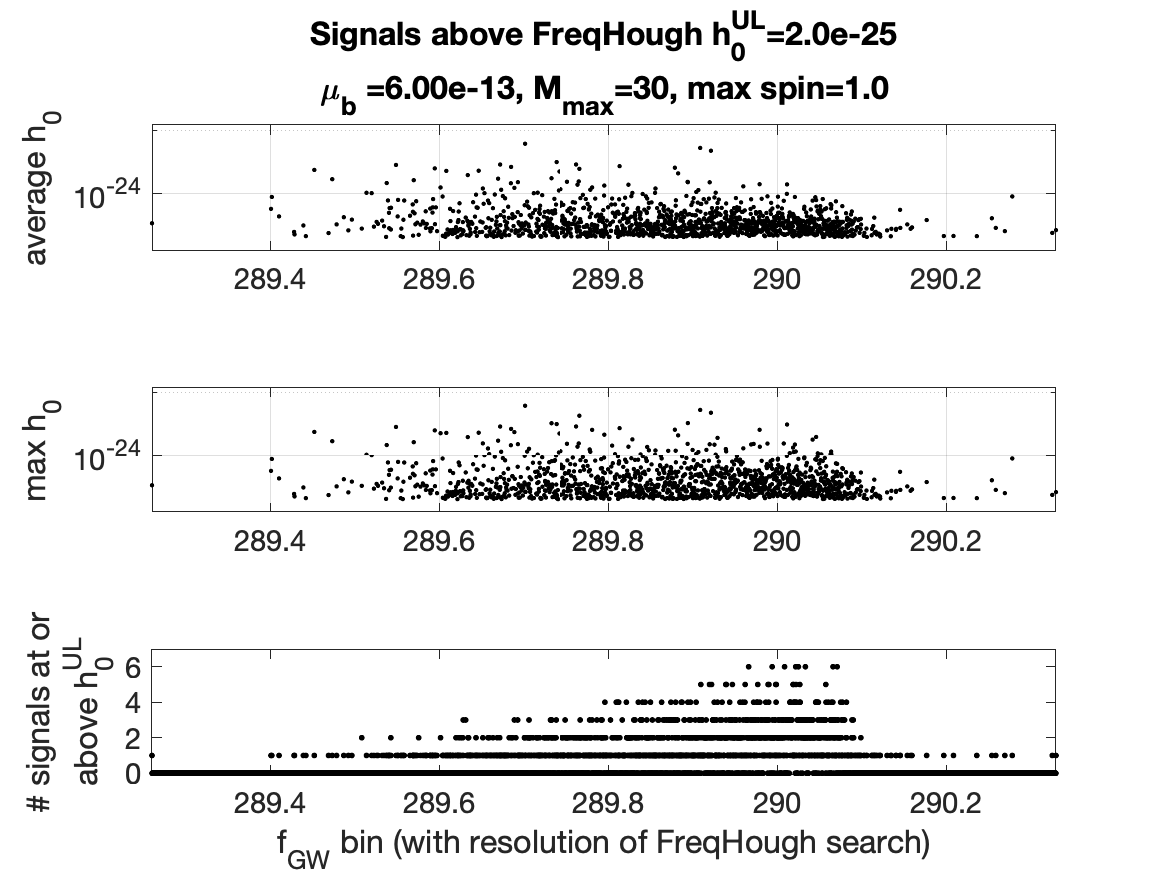}
\includegraphics[width=\columnwidth]{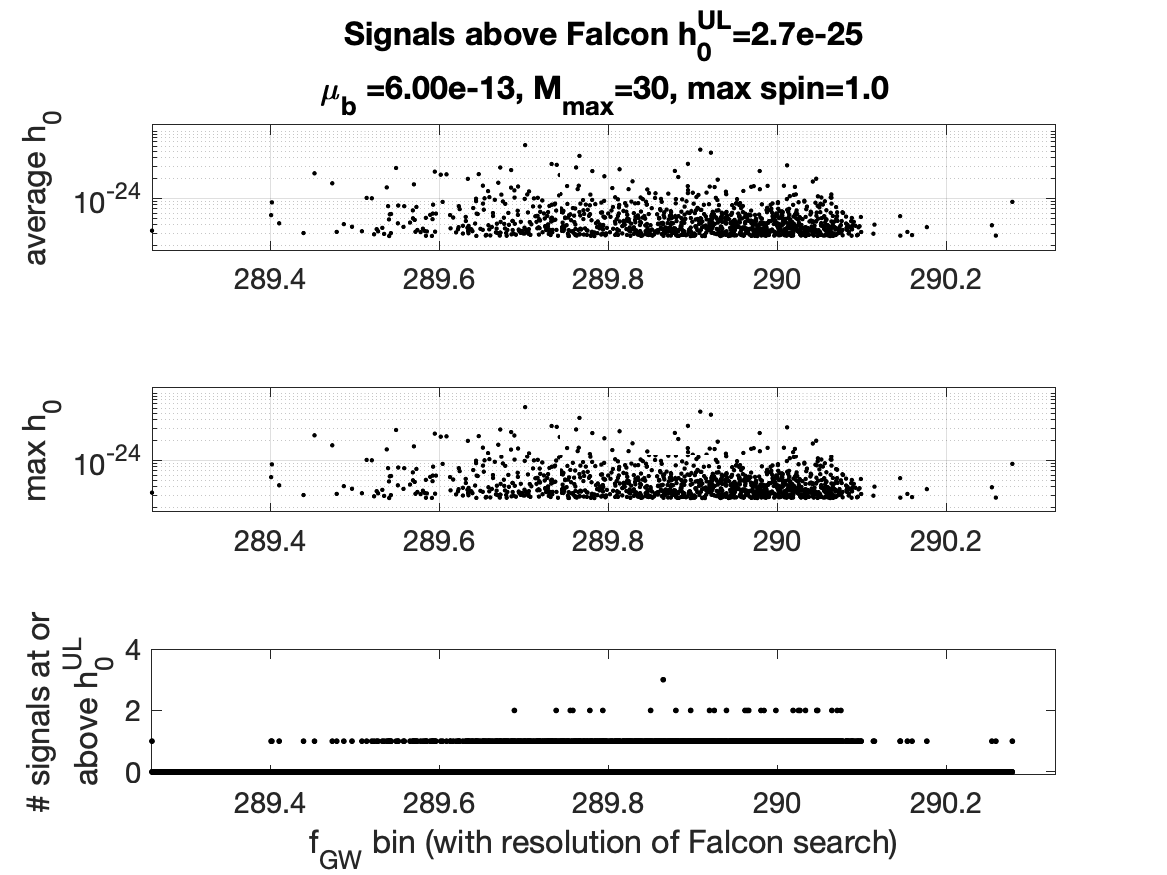}
 \caption{The x-axes show the signal frequency  binned with the resolution of the two searches: Freq Hough, $2.4 \times 10^{-4}$ Hz  \cite{RomeBosonClouds}  ({\it{left}}) and Falcon  \cite{FalconO1,Dergachev:2019oyu}, $1.7 \times 10^{-5}$ Hz ({\it{right}}). From the bottom panel going up, the y-axis shows the number of detectable signals in each bin, their maximum intrinsic amplitude and their average amplitude. \label{fig:aboveULmub6}}
\end{figure*}

\begin{figure*}
\includegraphics[width=\columnwidth]{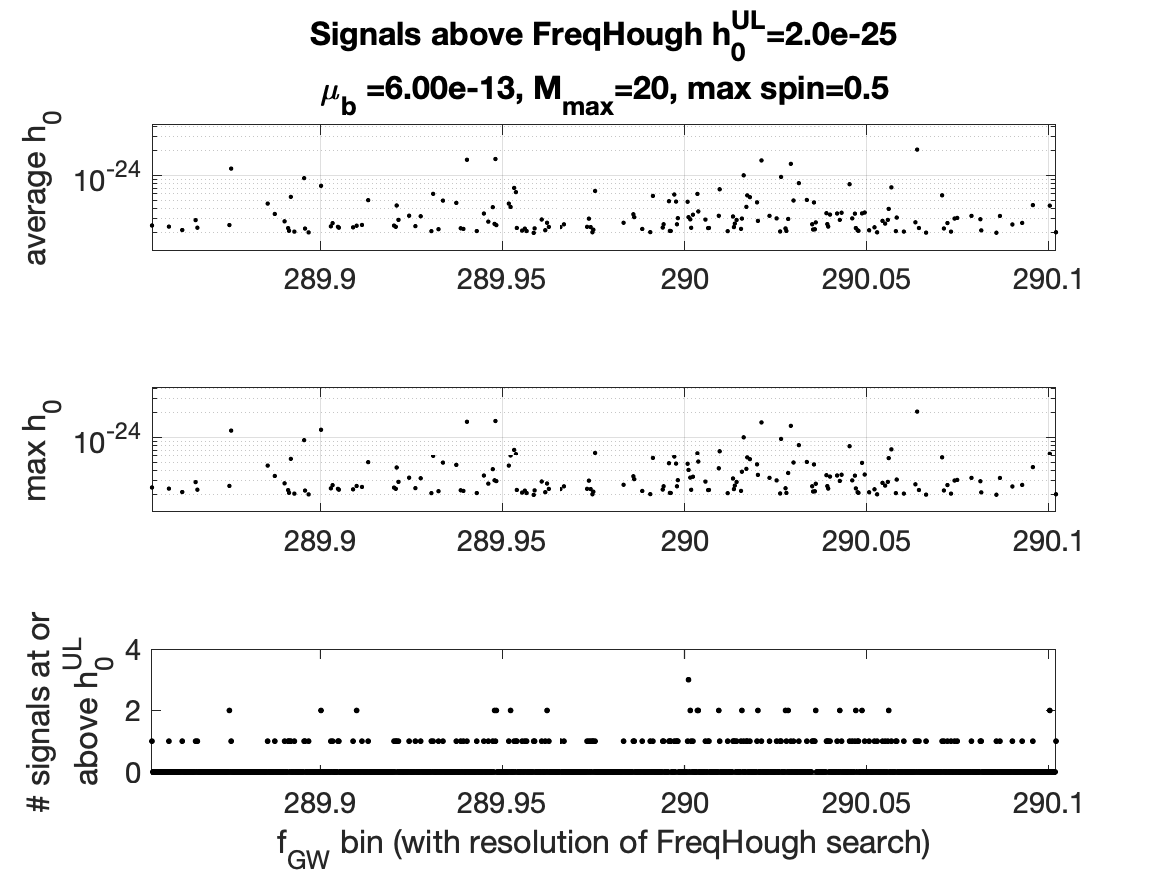}
\includegraphics[width=\columnwidth]{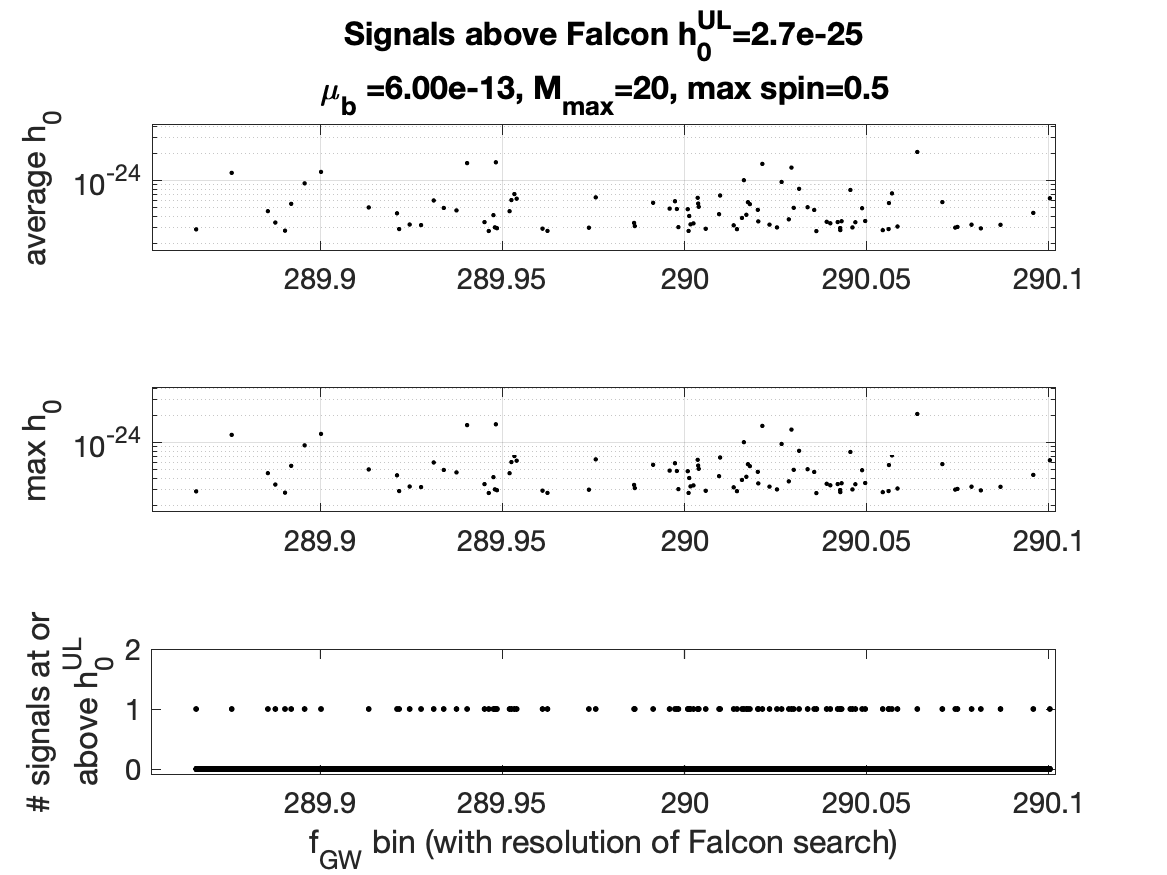}
\includegraphics[width=\columnwidth]{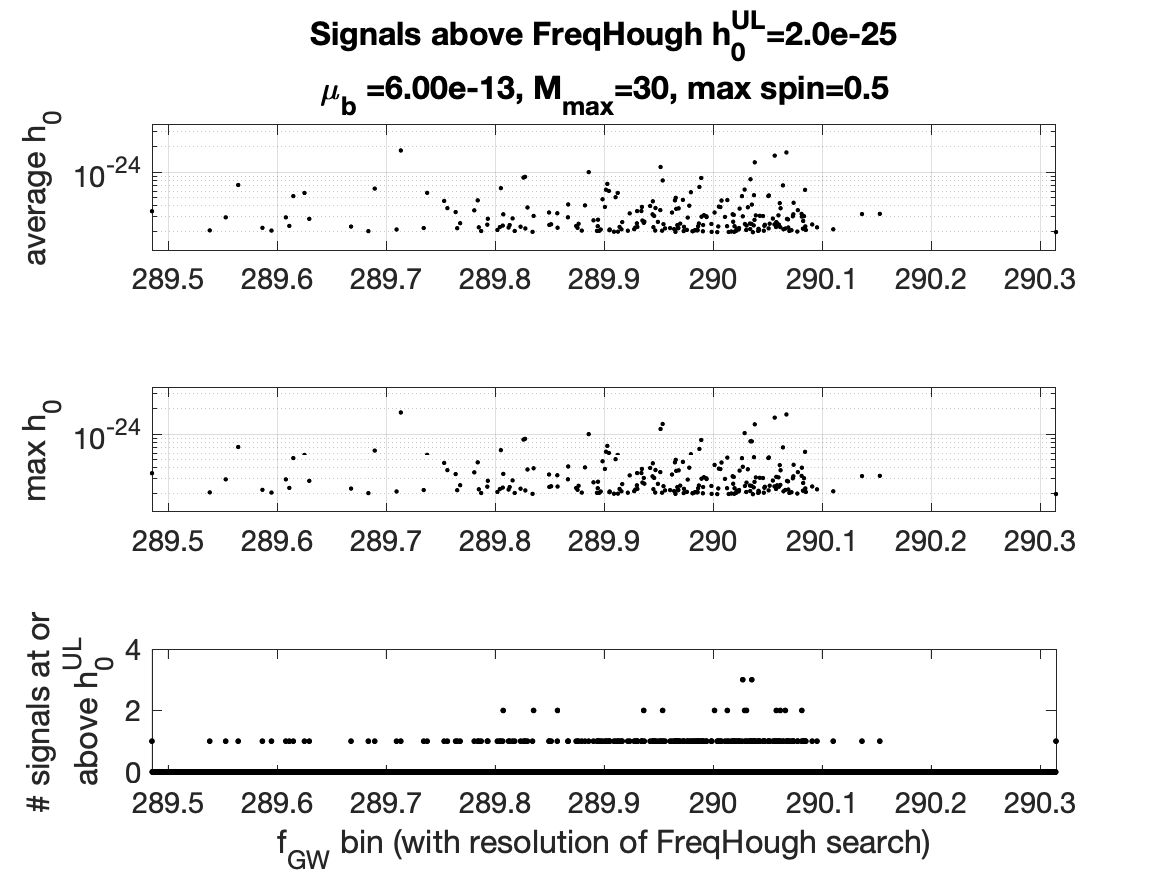}
\includegraphics[width=\columnwidth]{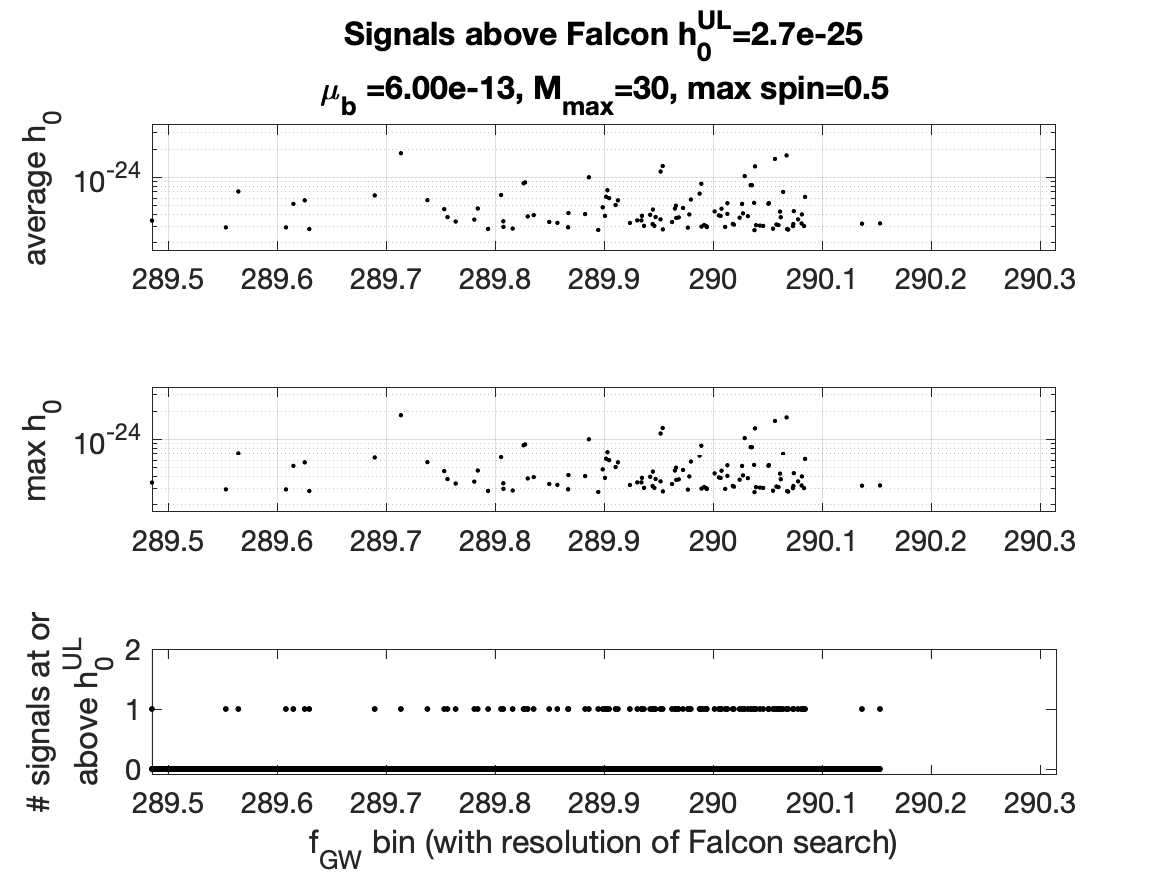}
  \caption{The x-axes show the signal frequency binned with the resolution of the two searches: Freq Hough, $2.4 \times 10^{-4}$ Hz  \cite{RomeBosonClouds}  ({\it{left}}) and Falcon  \cite{FalconO1,Dergachev:2019oyu}, $1.7 \times 10^{-5}$ Hz ({\it{right}}). From the bottom panel going up, the y-axis shows the number of detectable signals in each bin, their maximum intrinsic amplitude and their average amplitude. \label{fig:aboveULmub6M20}}
\end{figure*}

\begin{figure*}
\includegraphics[width=\columnwidth]{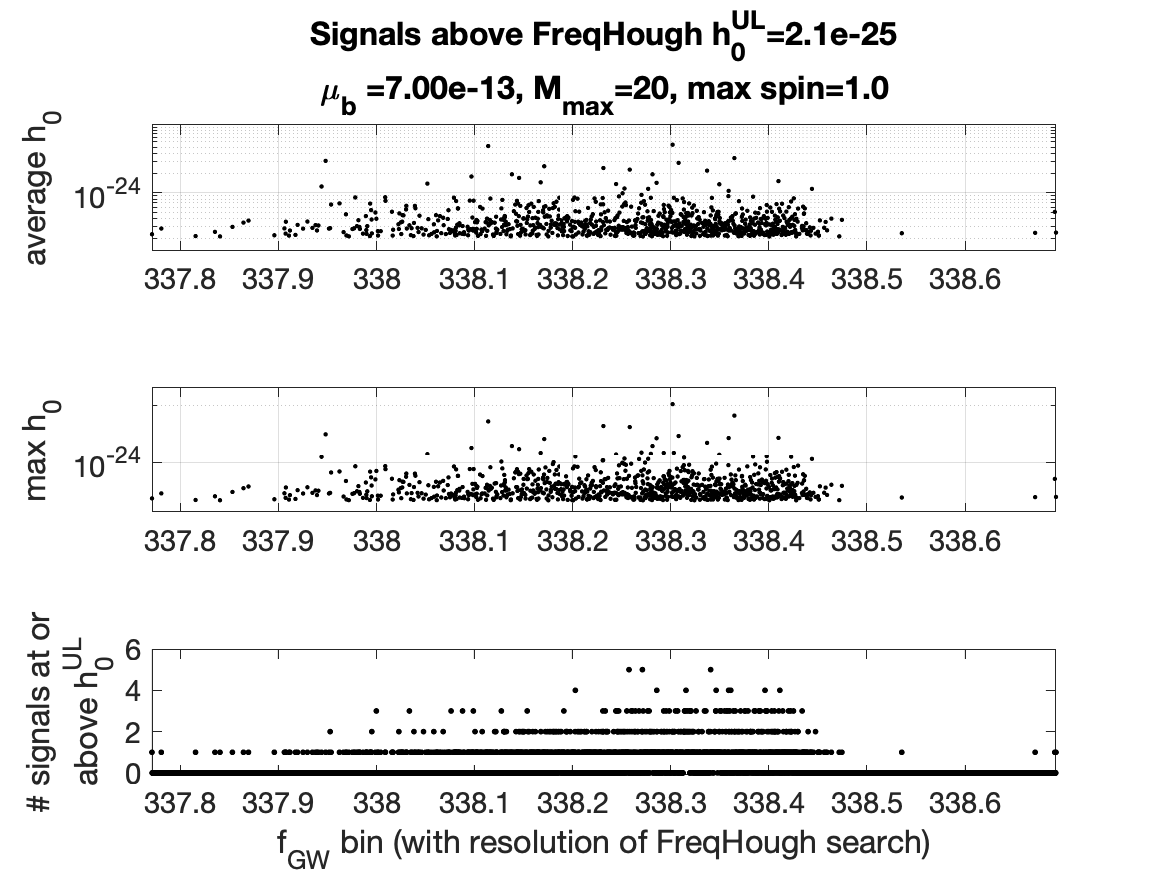}
\includegraphics[width=\columnwidth]{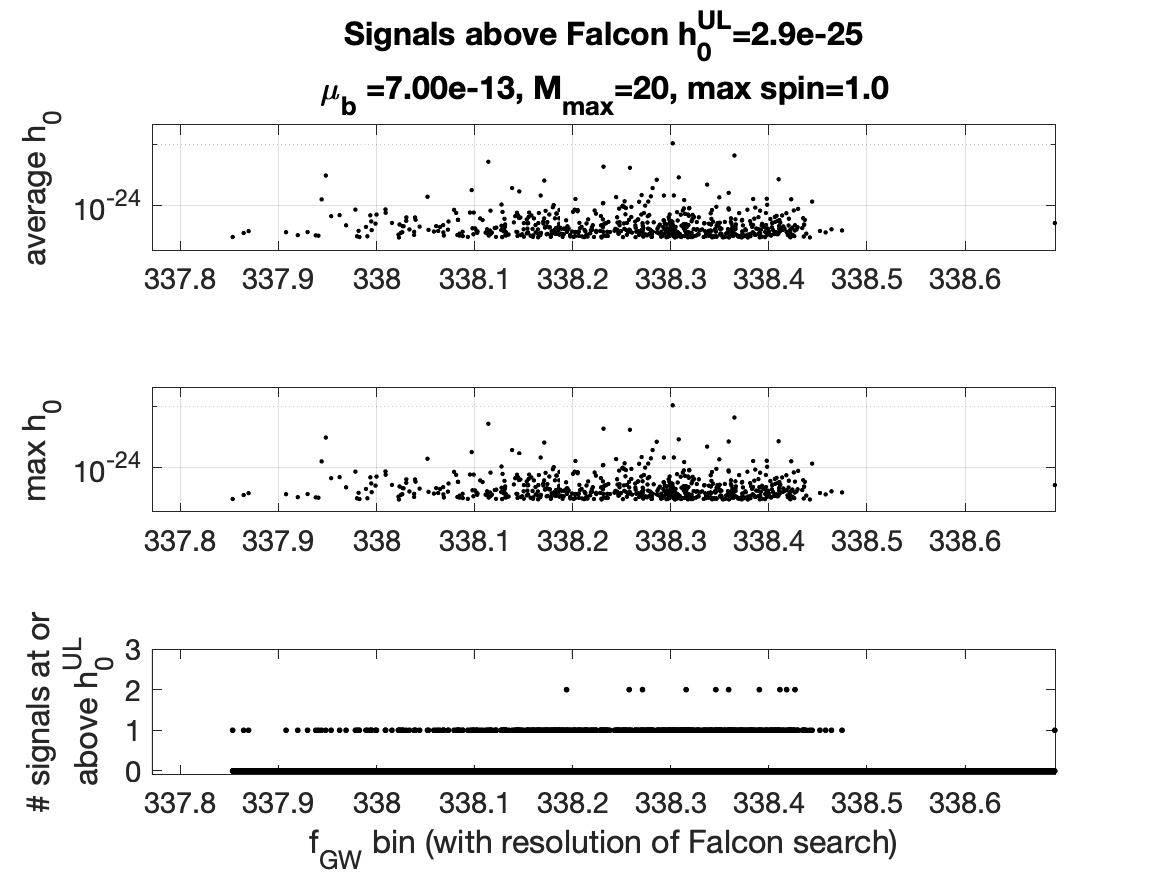}
\includegraphics[width=\columnwidth]{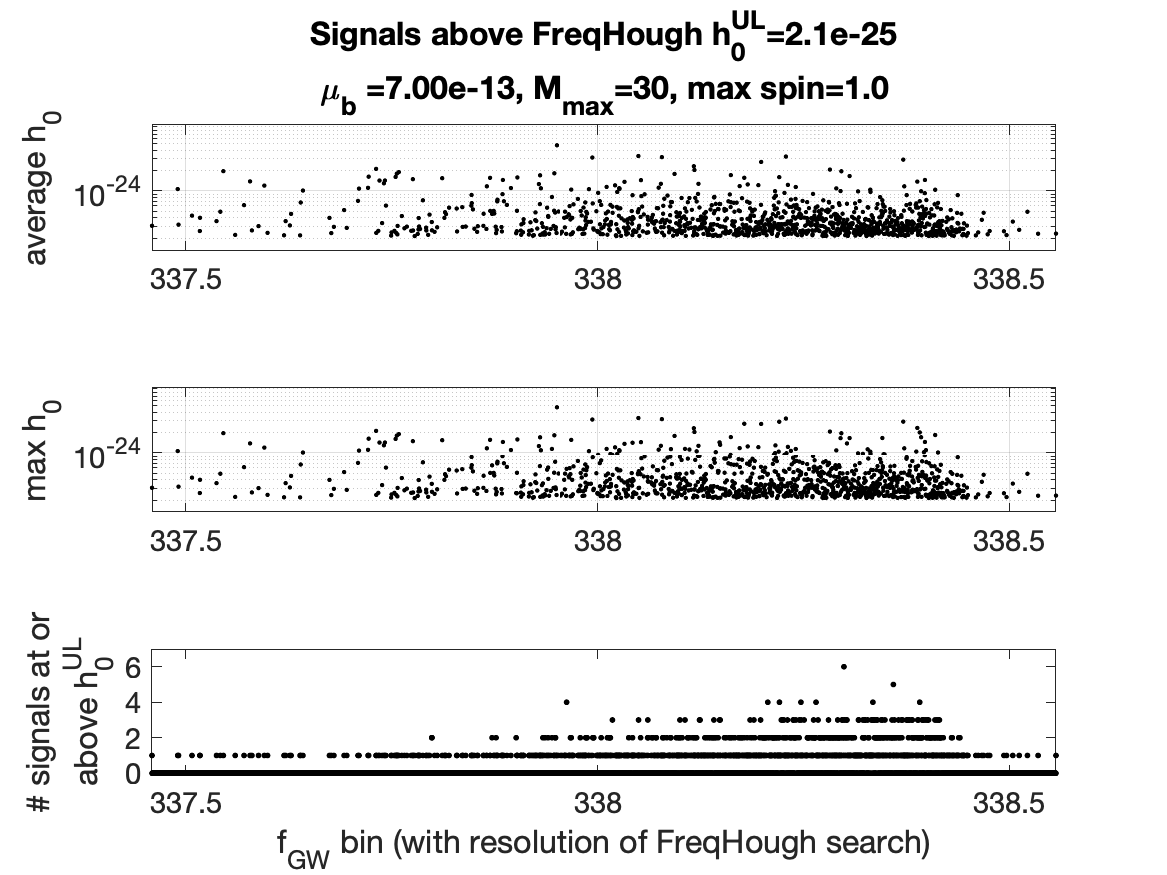}
\includegraphics[width=\columnwidth]{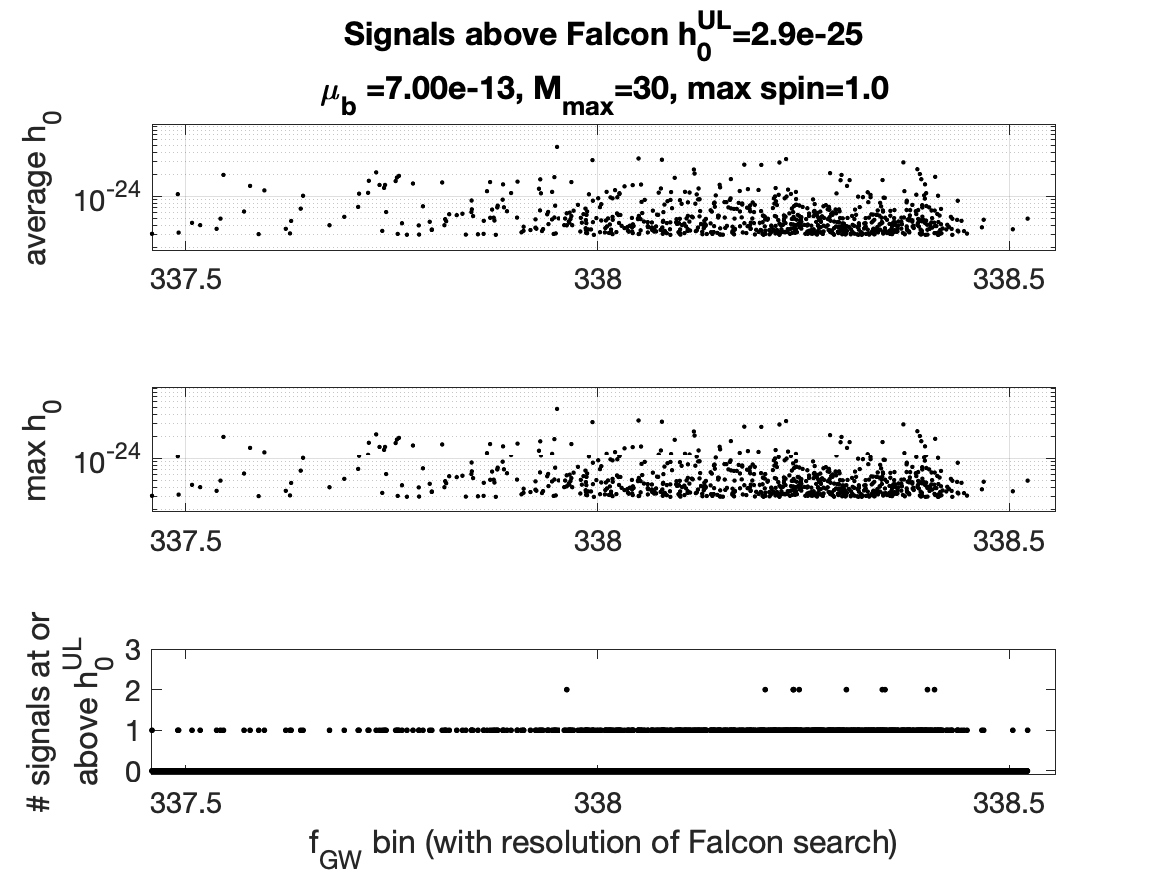}
\caption{The x-axes show the signal frequency binned with the resolution of the two searches: Freq Hough, $2.4 \times 10^{-4}$ Hz  \cite{RomeBosonClouds}  ({\it{left}}) and Falcon  \cite{FalconO1,Dergachev:2019oyu}, $1.7 \times 10^{-5}$ Hz ({\it{right}}).  From the bottom panel going up, the y-axis shows the number of detectable signals in each bin, their maximum intrinsic amplitude and their average amplitude.  \label{fig:aboveULmub7}}
\end{figure*}

\begin{figure*}
\includegraphics[width=\columnwidth]{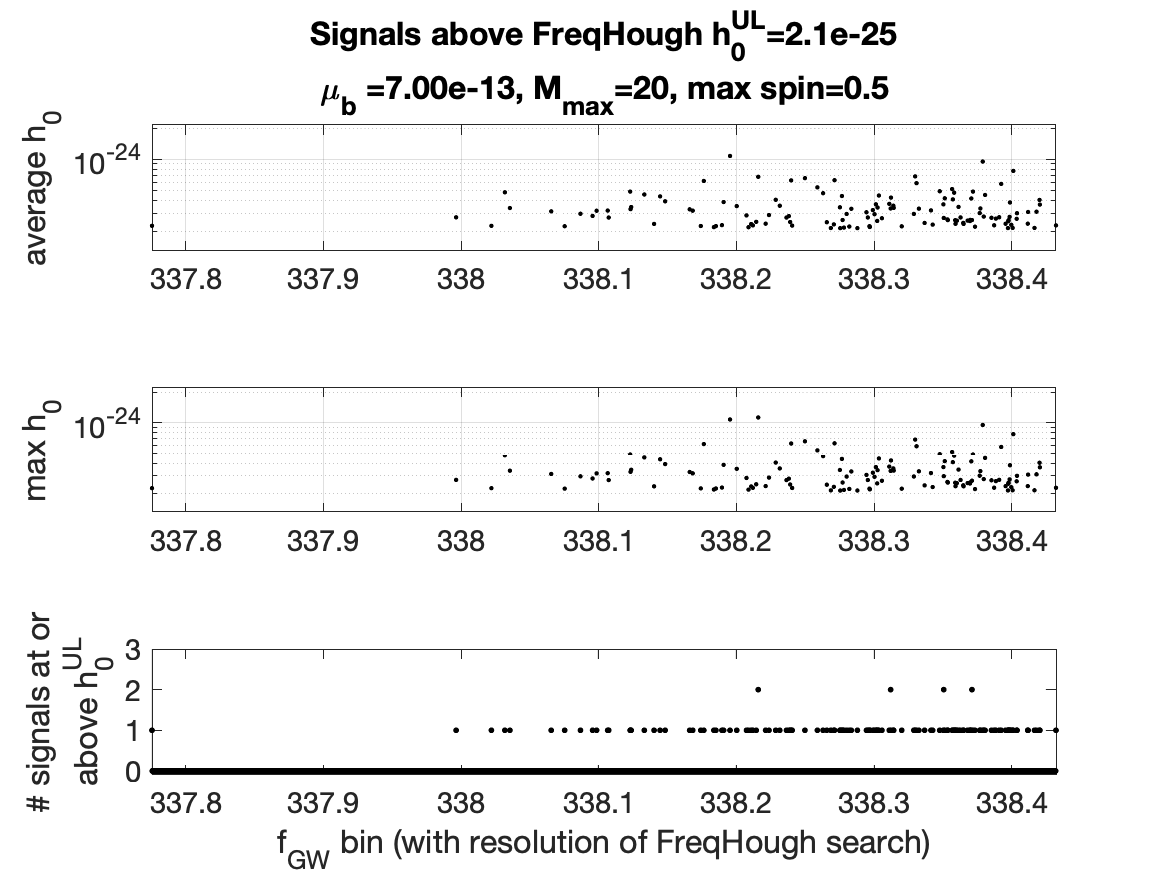}
\includegraphics[width=\columnwidth]{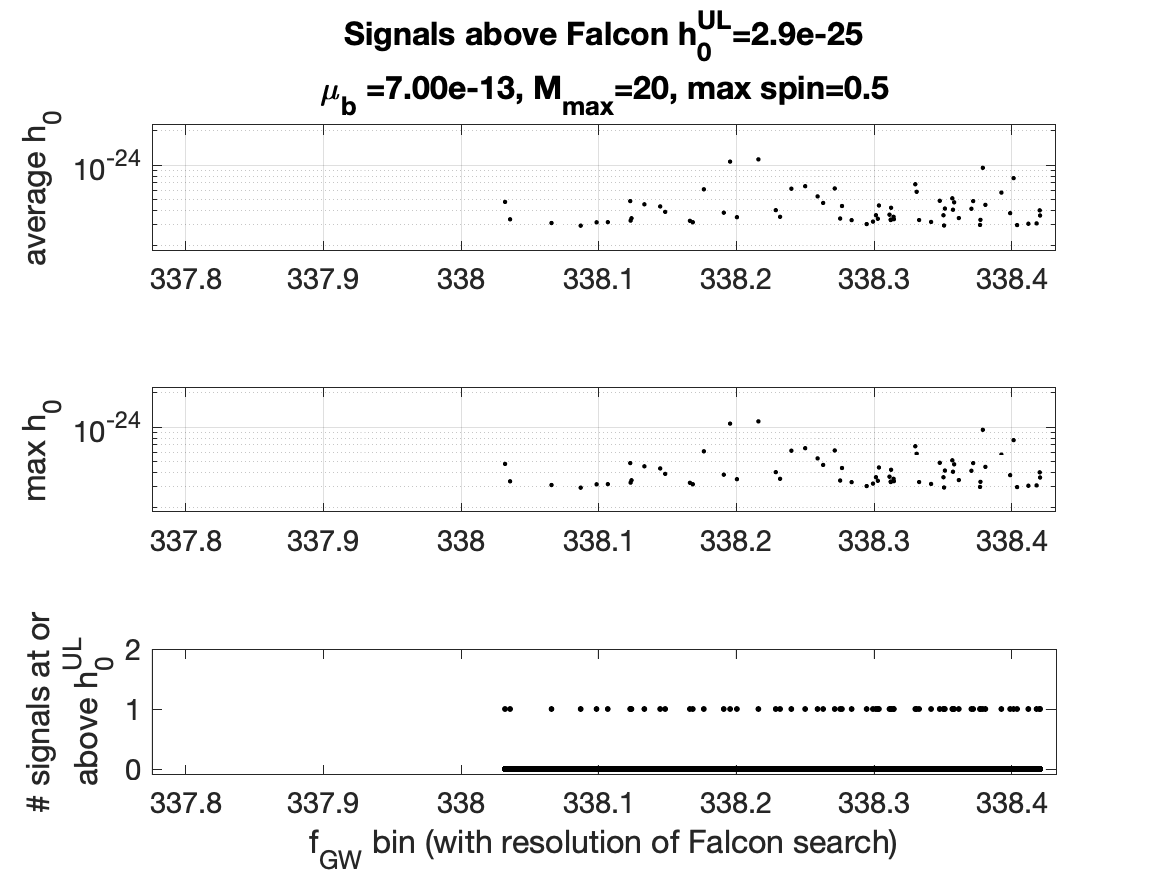}
\includegraphics[width=\columnwidth]{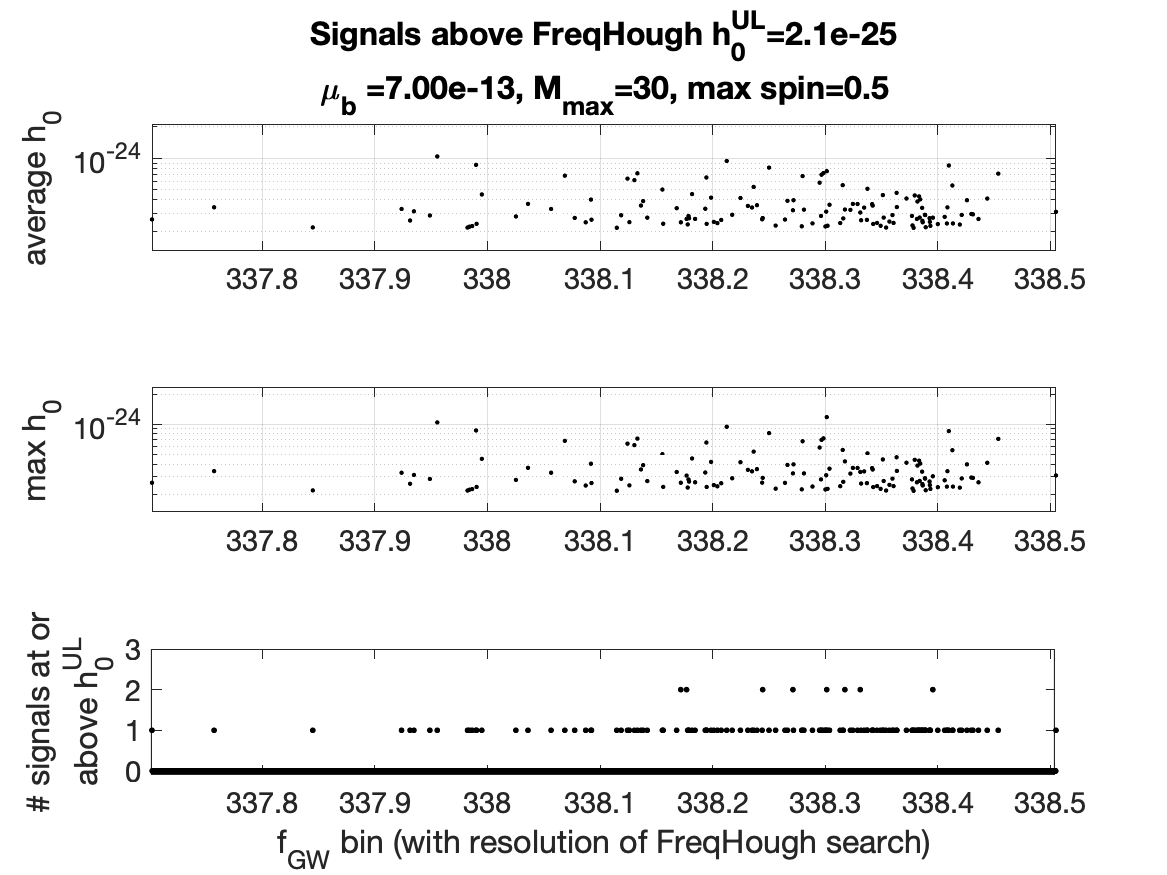}
\includegraphics[width=\columnwidth]{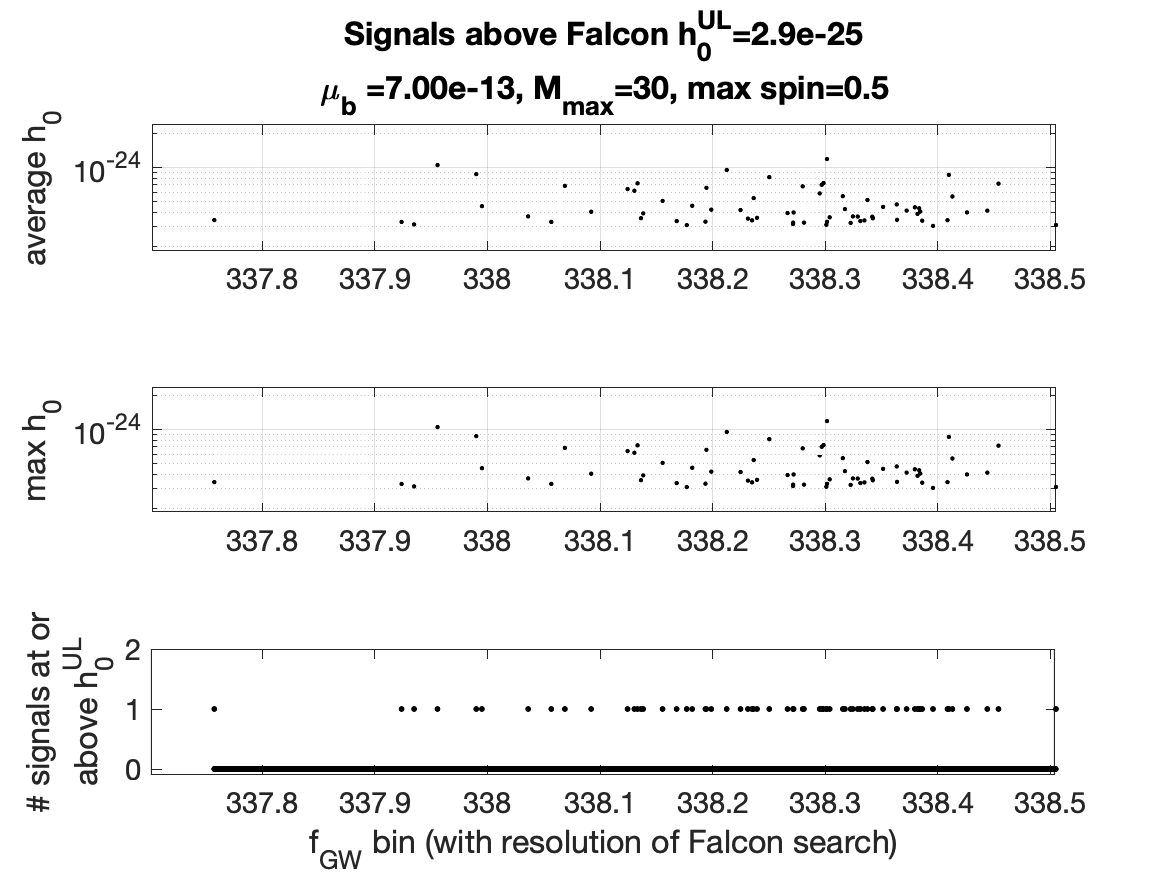}
\caption{The x-axes   show the signal frequency binned with the resolution of the two searches: Freq Hough, $2.4 \times 10^{-4}$ Hz  \cite{RomeBosonClouds}  ({\it{left}}) and Falcon  \cite{FalconO1,Dergachev:2019oyu}, $1.7 \times 10^{-5}$ Hz ({\it{right}}).  From the bottom panel going up, the y-axis shows the number of detectable signals in each bin, their maximum intrinsic amplitude and their average amplitude. \label{fig:aboveULmub7M20}}
\end{figure*}

\begin{figure*}
\includegraphics[width=\columnwidth]{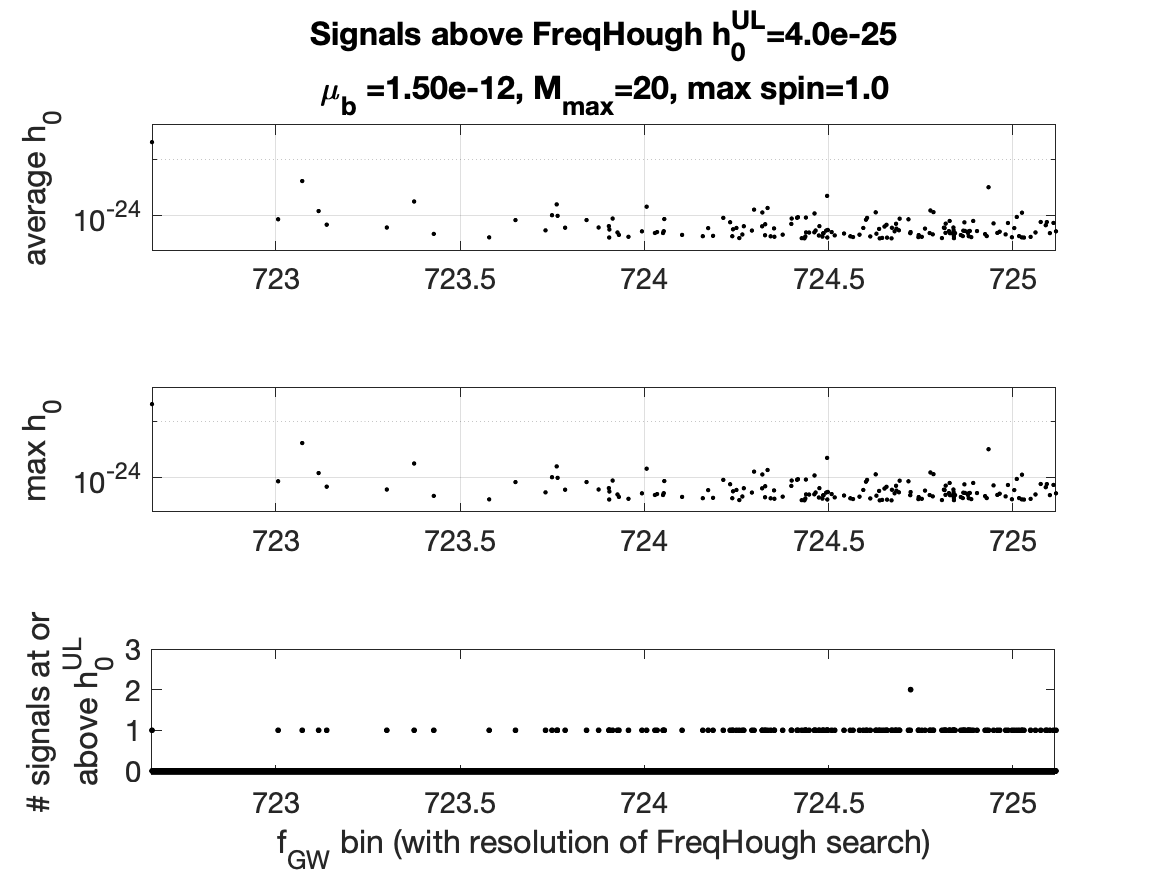}
\includegraphics[width=\columnwidth]{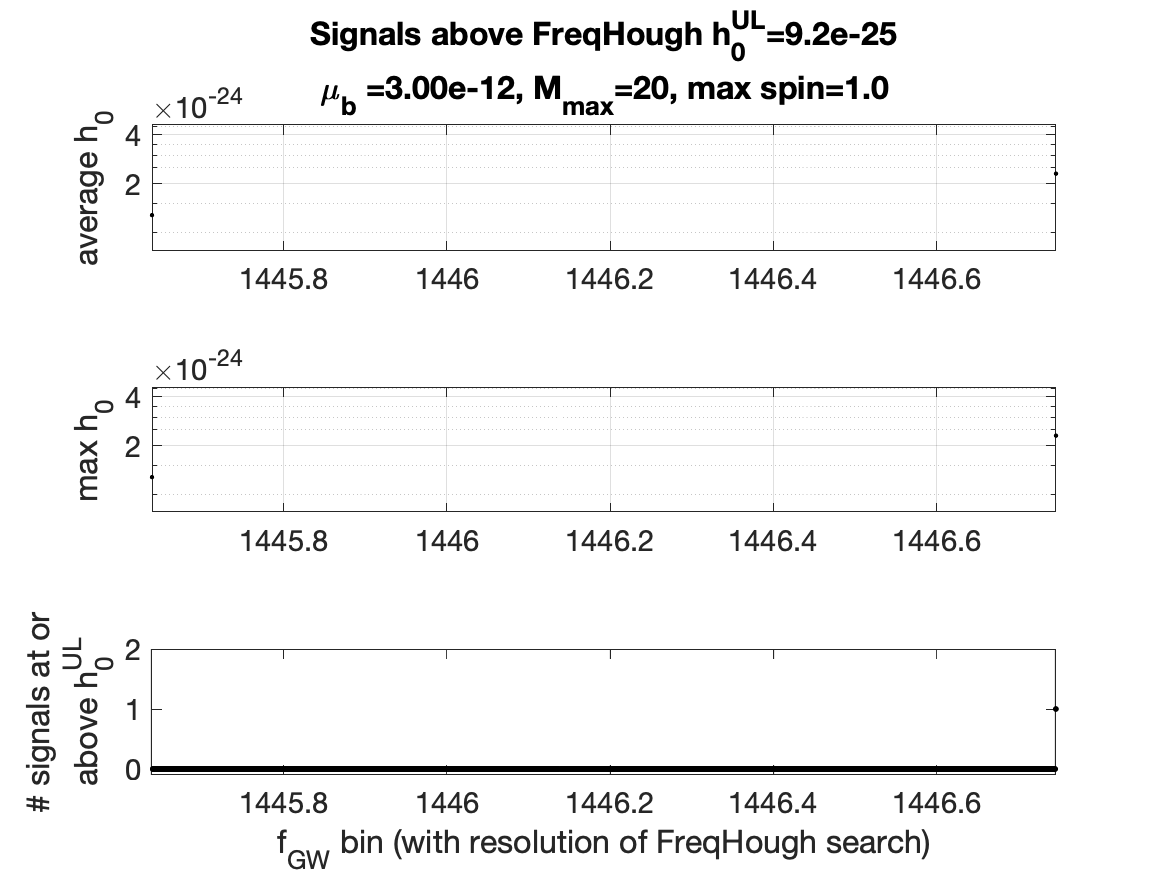}
\includegraphics[width=\columnwidth]{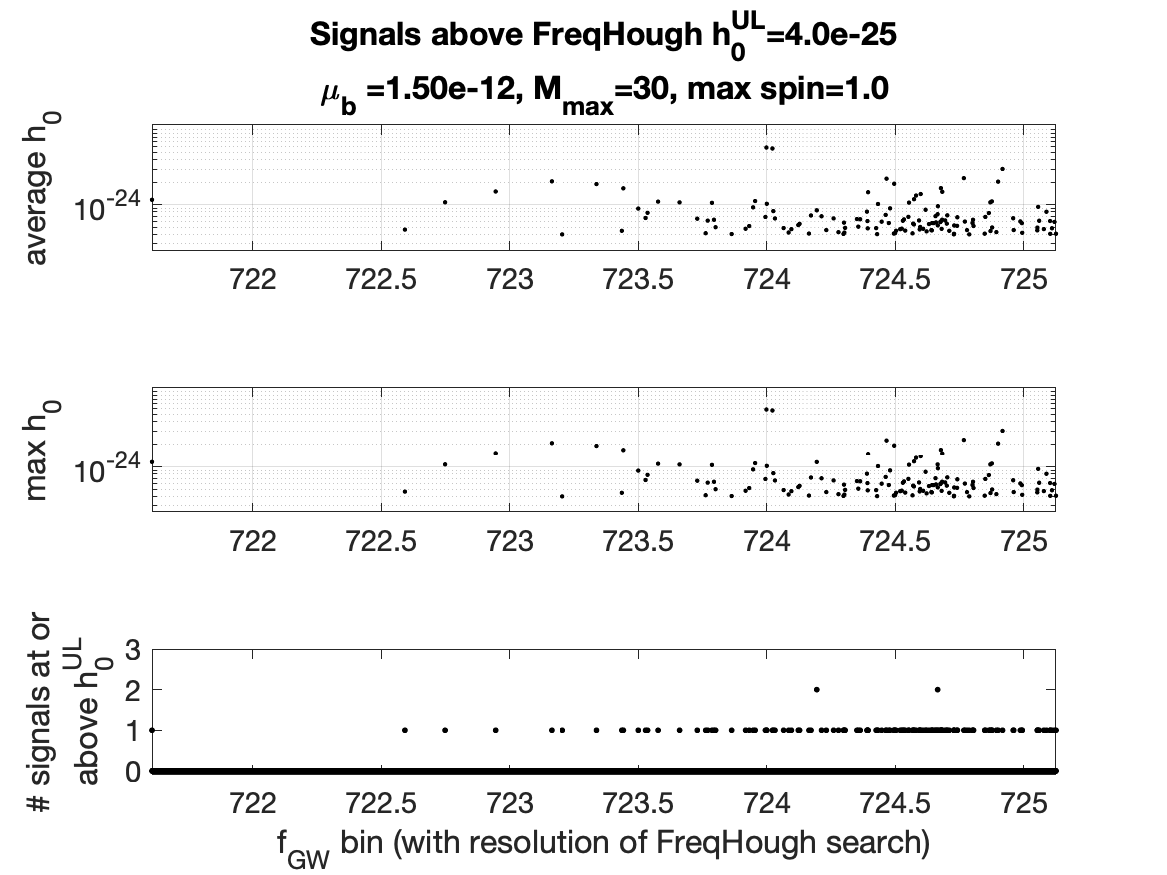}
\includegraphics[width=\columnwidth]{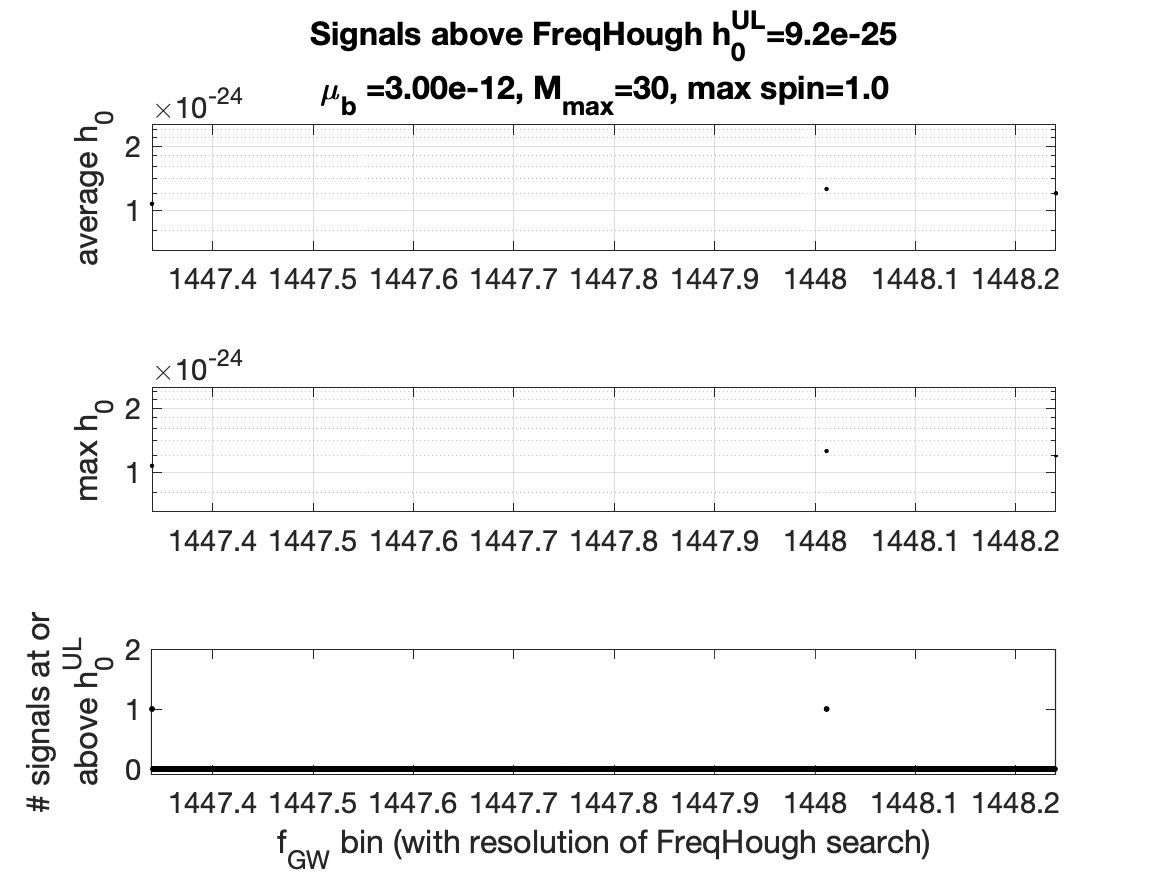}
  \caption{The x-axes   show the signal frequency binned with the resolution of the two searches: Freq Hough, $2.4 \times 10^{-4}$ Hz  \cite{RomeBosonClouds}  ({\it{left}}) and Falcon  \cite{FalconO1,Dergachev:2019oyu}, $1.7 \times 10^{-5}$ Hz ({\it{right}}). From the bottom panel going up, the y-axis shows the number of detectable signals in each bin, their maximum intrinsic amplitude and their average amplitude.  \label{fig:aboveULmub15}}
\end{figure*}

As discussed in Section~\ref{sec:density}, large signal numbers and densities can affect the search efficiency. Figures~\ref{fig:aboveULmub3} --\ref{fig:aboveULmub15} show  the of number of signals, as well as the maximum and average strain of signals, with amplitude $h_0$ above the upper limit values set by two recent continuous wave all-sky searches,   Freq Hough \cite{RomeBosonClouds} and Falcon  \cite{FalconO1,Dergachev:2019oyu,Dergachev:2020fli} for boson masses between $3\times 10^{-13}$~eV and $1.5\times 10^{-12}$~eV. We consider both standard and heavy black hole mass distributions, and both standard and moderate spin distributions. 

\section{Contribution to the noise amplitude spectral density from the ensemble signal}
\label{app:ASDPlots}
\begin{figure*}
\includegraphics[width=0.6\columnwidth]{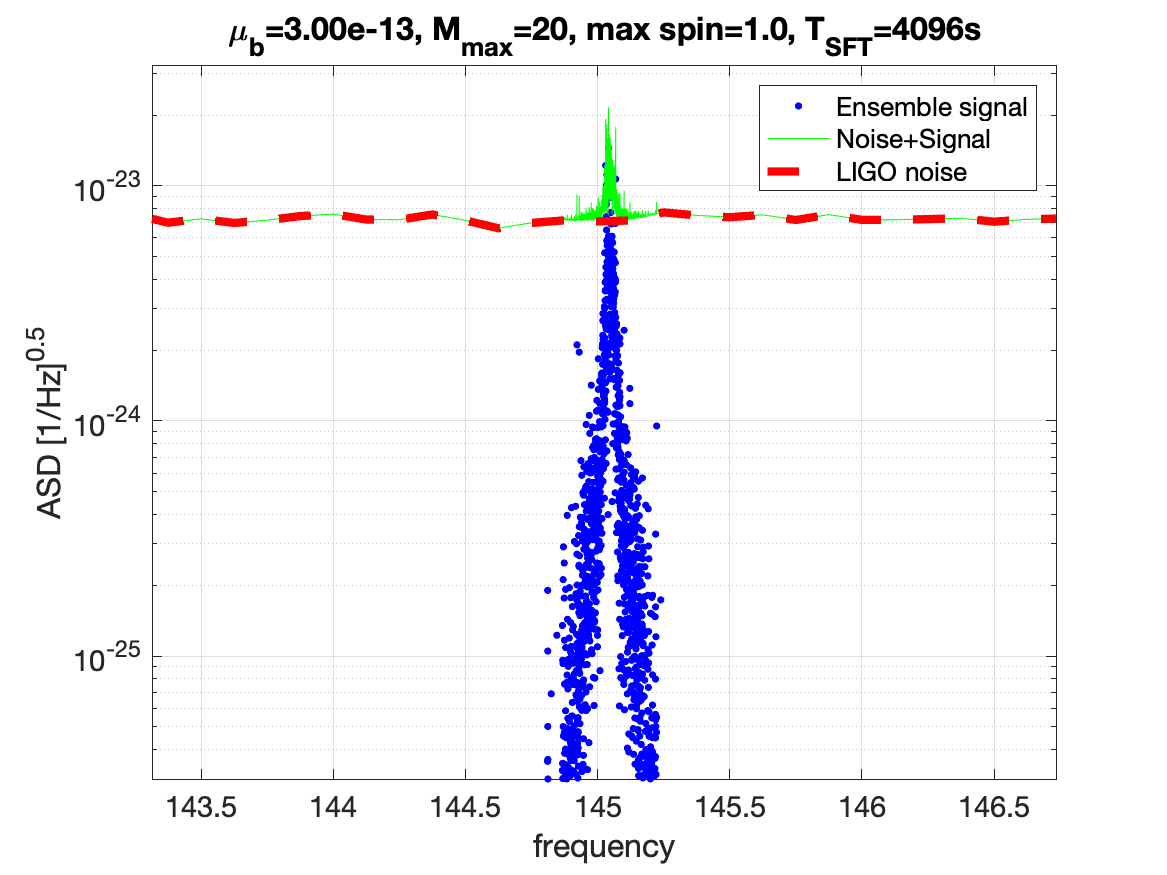}
\includegraphics[width=0.6\columnwidth]{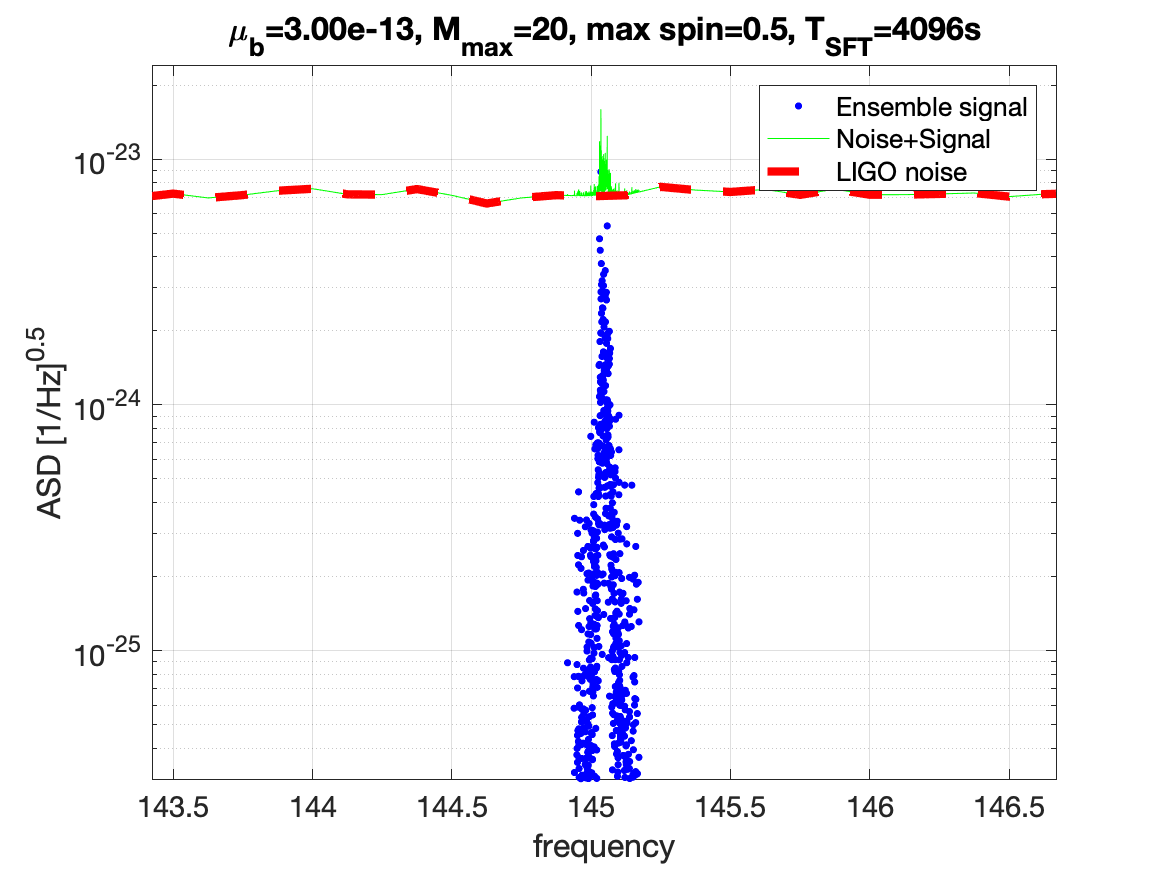}
\includegraphics[width=0.6\columnwidth]{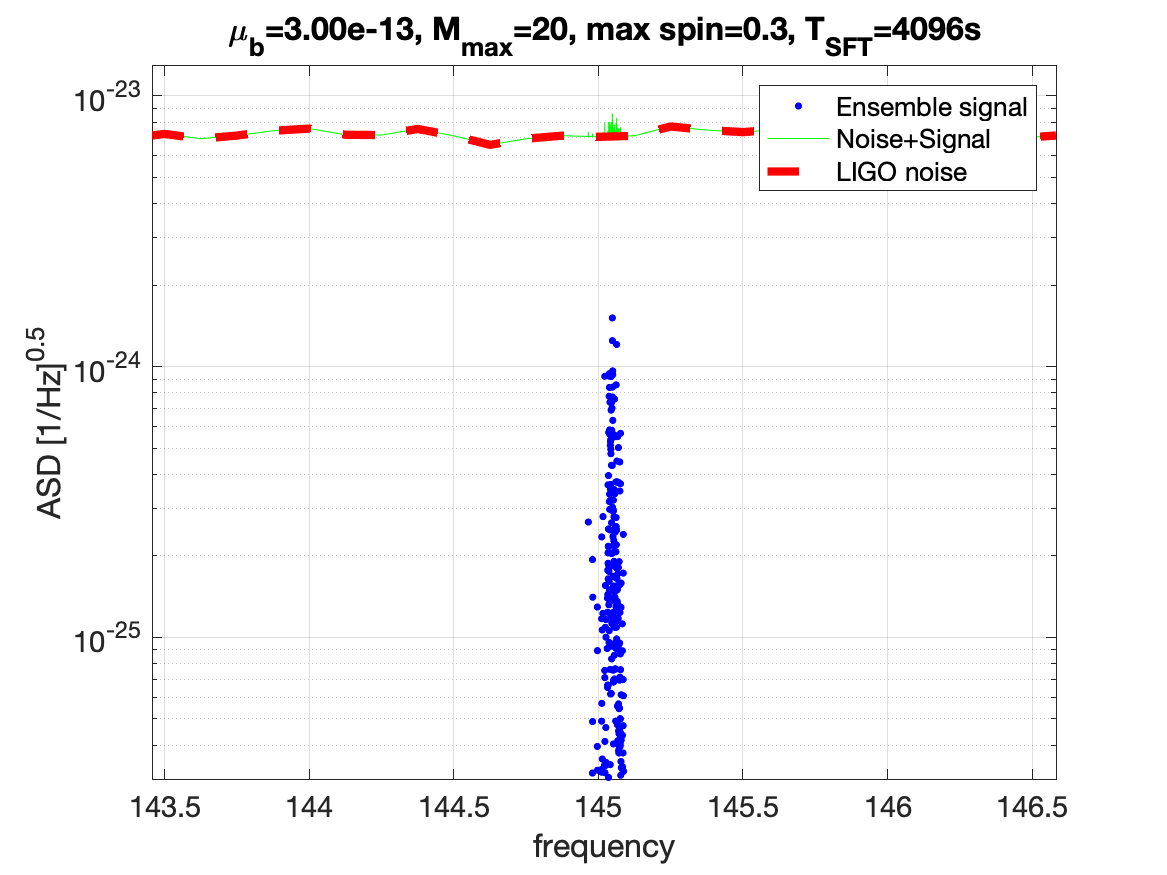}
\includegraphics[width=0.6\columnwidth]{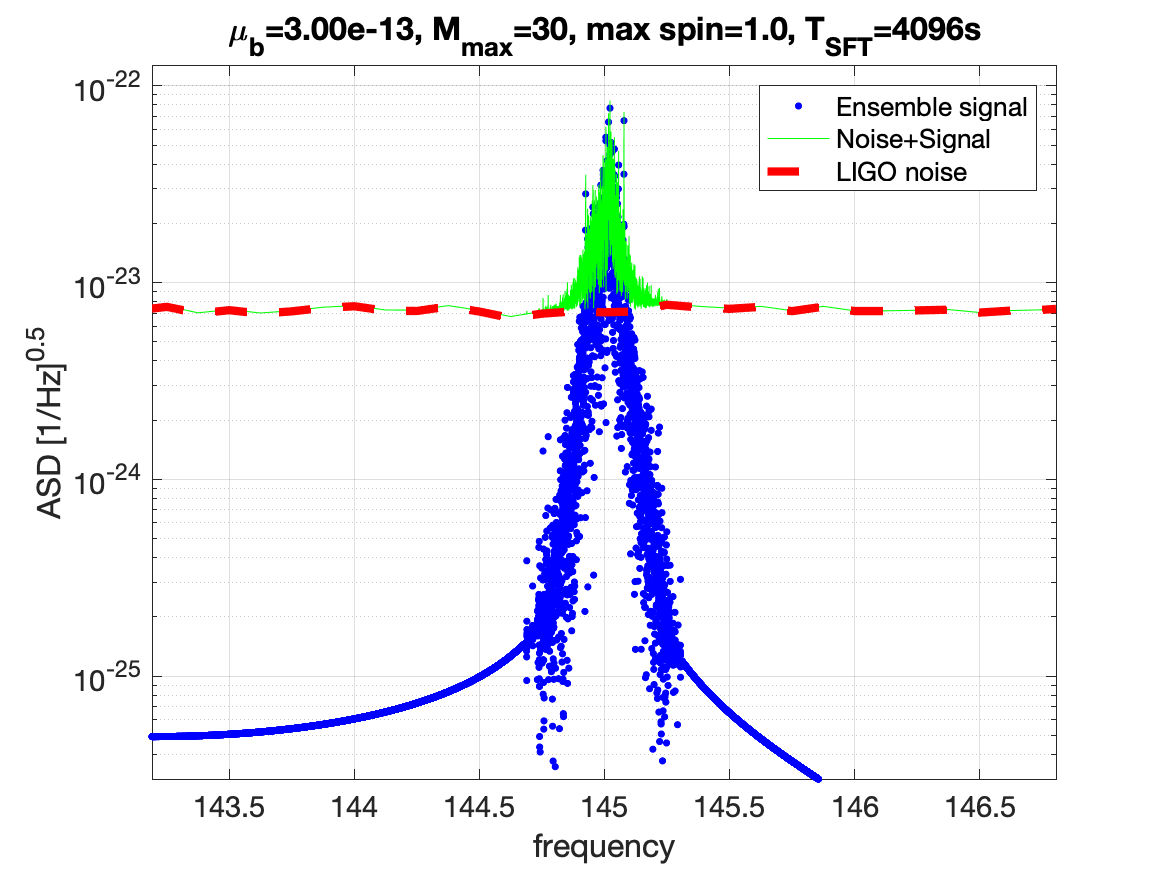}
\includegraphics[width=0.6\columnwidth]{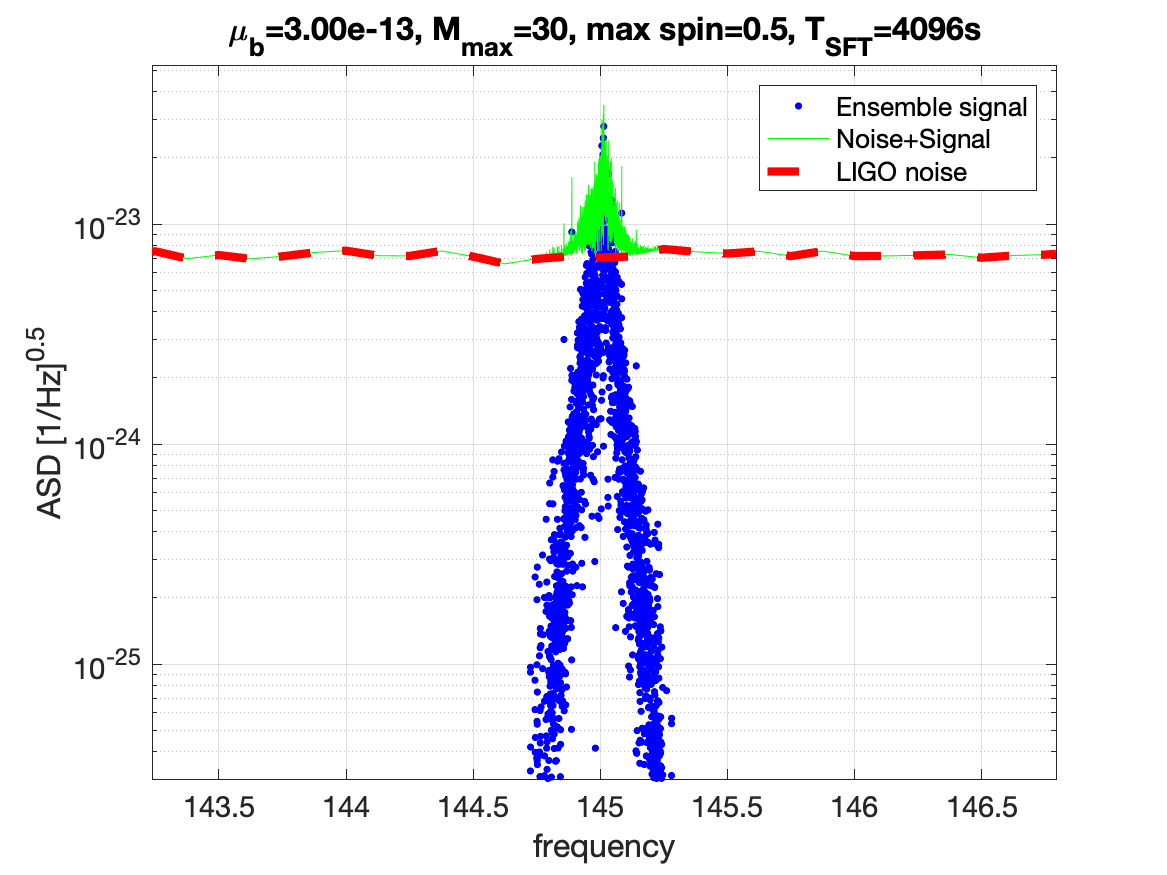}
\caption{The amplitude spectral density of the LIGO O2 data alone (dashed red line) and of an ensemble signal (top green points) assuming $\mu_b=3\times 10^{-13}$ eV and a galactic black hole population with maximum mass 20 M$_\odot$ (top plots) and max initial spin of 0.3, 0.5 and 1; and maximum mass 30 M$_\odot$ (bottom plots) and max initial spin of 0.5 and 1. The time-baseline assumed is 4096 s as used by the Freq.Hough search in this frequency range \cite{RomeBosonClouds}. \label{fig:ASD3}}
\end{figure*}

\begin{figure*}
\includegraphics[width=0.6\columnwidth]{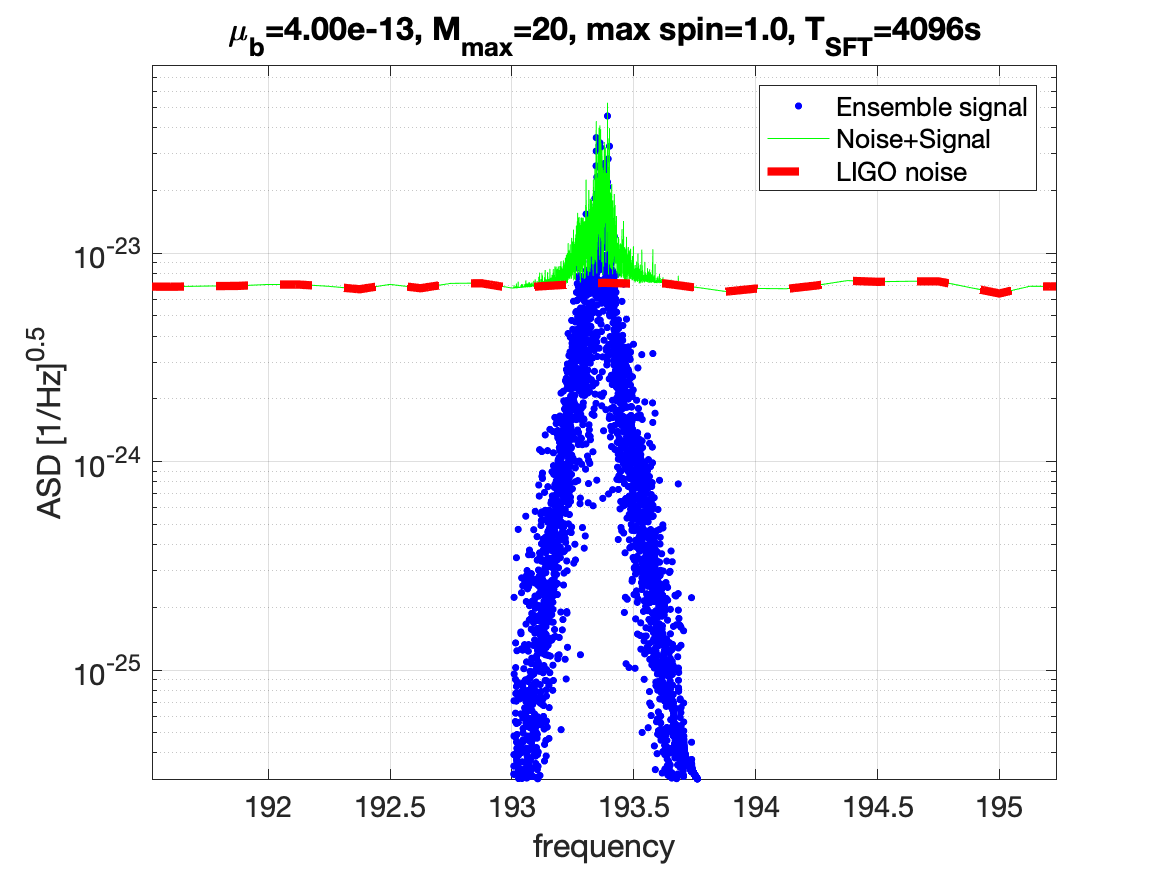}
\includegraphics[width=0.6\columnwidth]{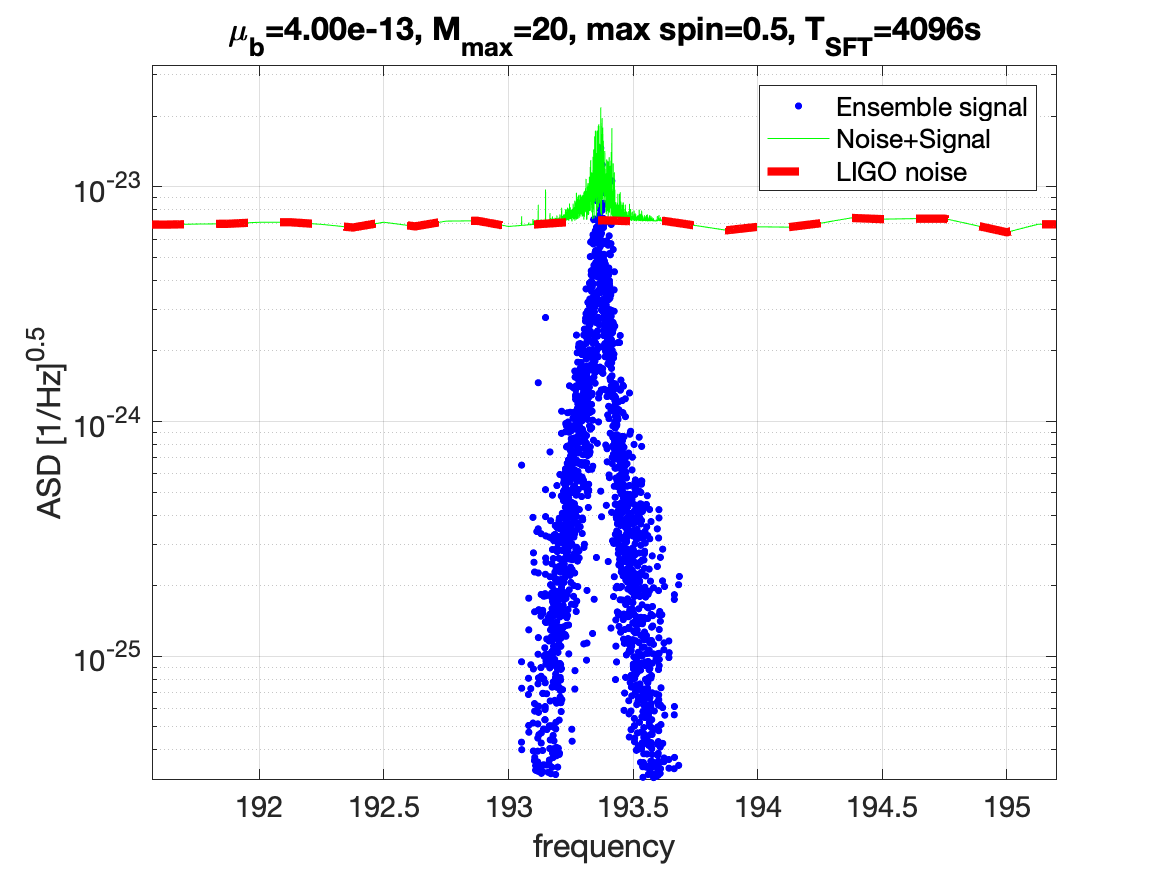}
\includegraphics[width=0.6\columnwidth]{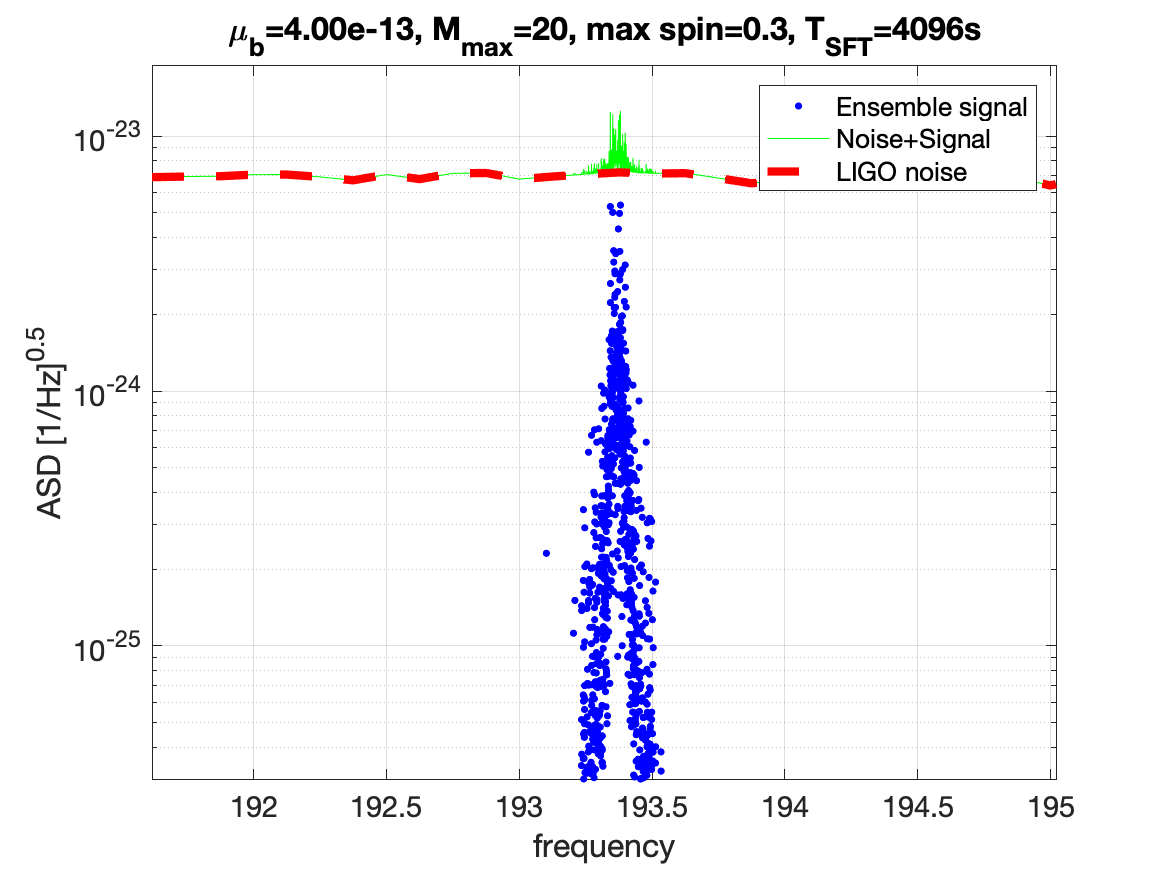}
\includegraphics[width=0.6\columnwidth]{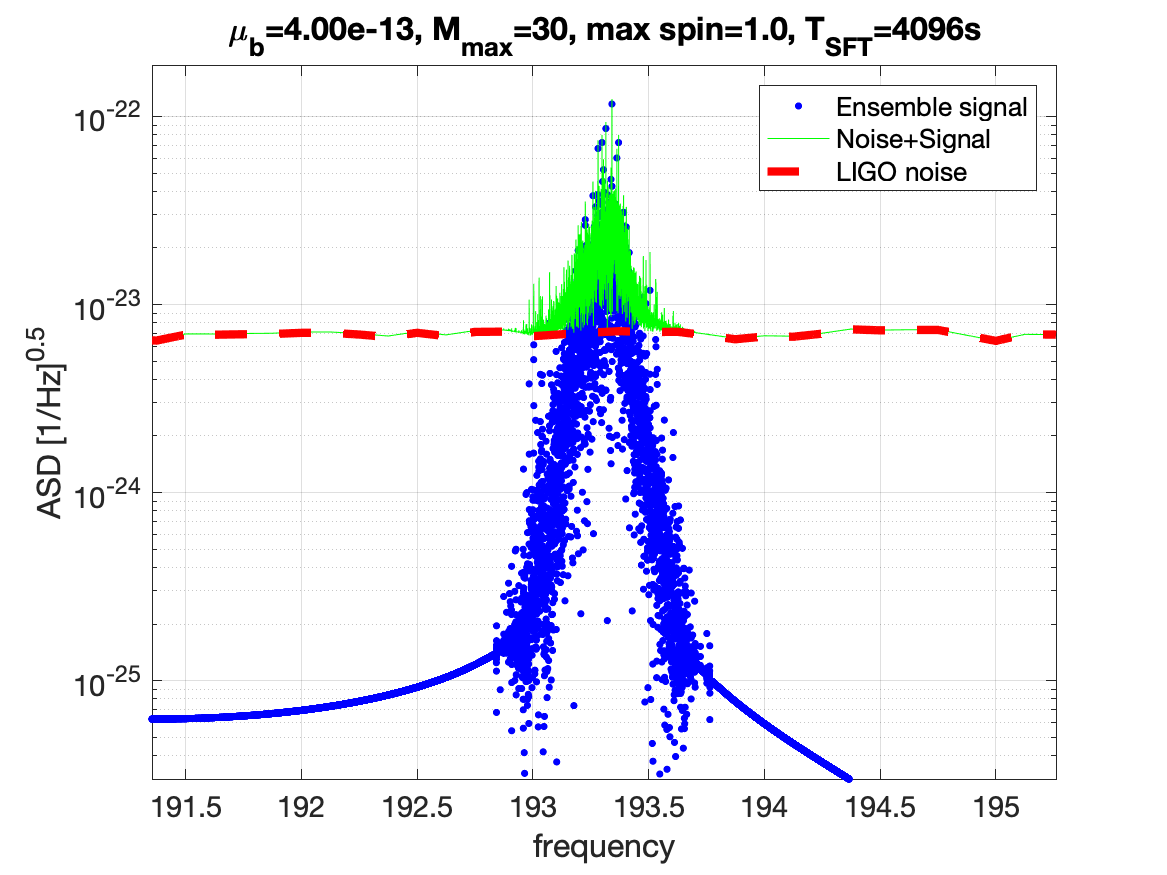}
\includegraphics[width=0.6\columnwidth]{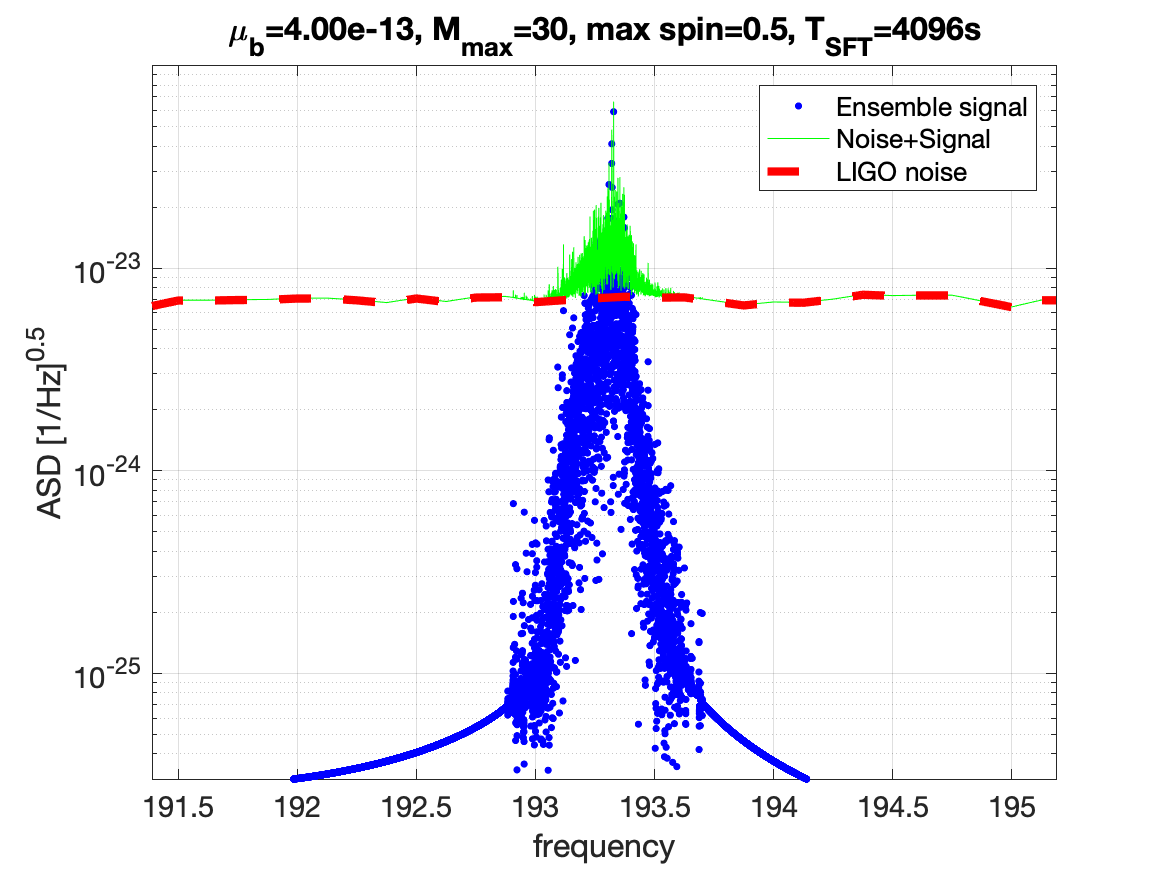}
\label{fig:EnsembleSignalASDlowcentralmub}
\caption{The amplitude spectral density of the LIGO O2 data alone (dashed red line) and of an ensemble signal (top green points) assuming $\mu_b=4\times 10^{-13}$ eV and a galactic black hole population with maximum mass 20 M$_\odot$ (top plots) and max initial spin of 0.3, 0.5 and 1; and maximum mass 30 M$_\odot$ (bottom plots) and max initial spin of 0.5 and 1. The time-baseline assumed is 4096 s as used by the Freq.Hough search in this frequency range \cite{RomeBosonClouds}. \label{fig:ASD4}}
\end{figure*}

\begin{figure*}
\includegraphics[width=0.6\columnwidth]{EnsembleSignalASD-mub7p00e-13Mmax20Tsft4096maxSpin1p0.png}
\includegraphics[width=0.6\columnwidth]{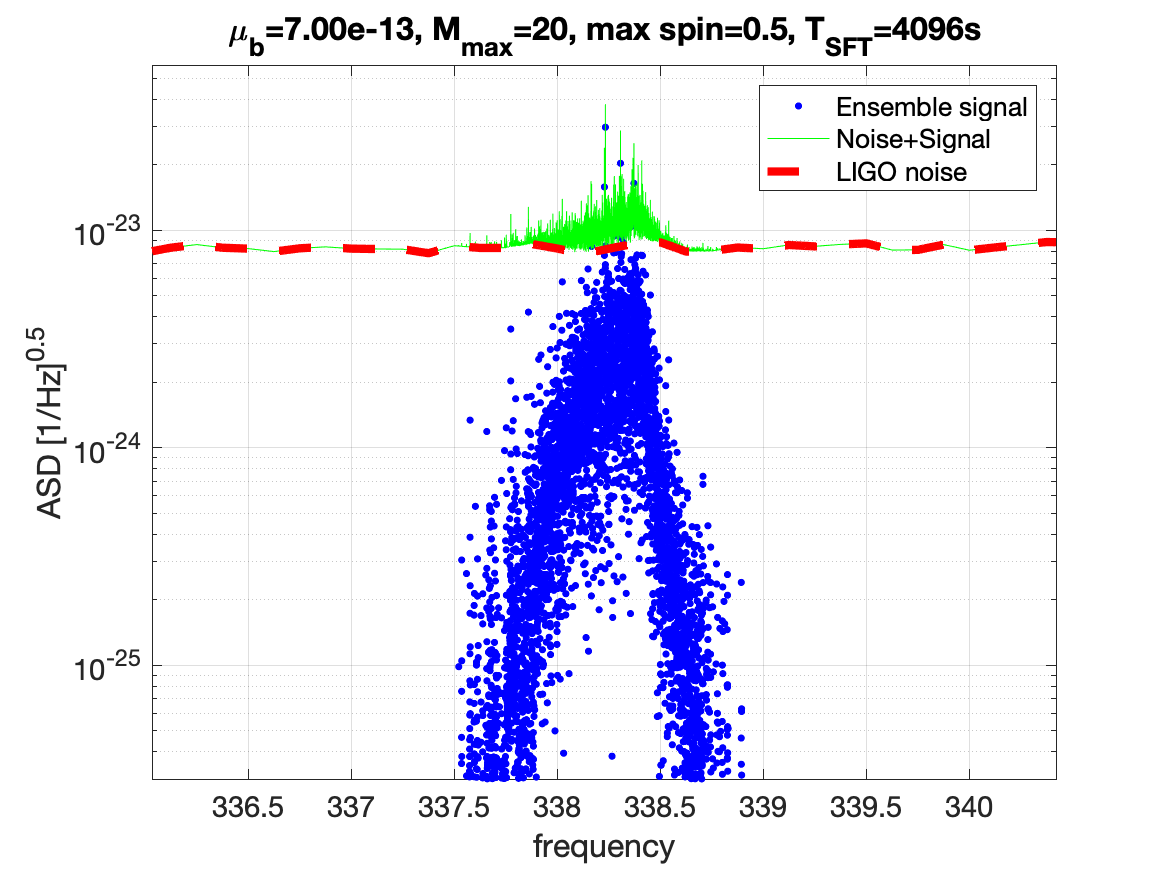}
\includegraphics[width=0.6\columnwidth]{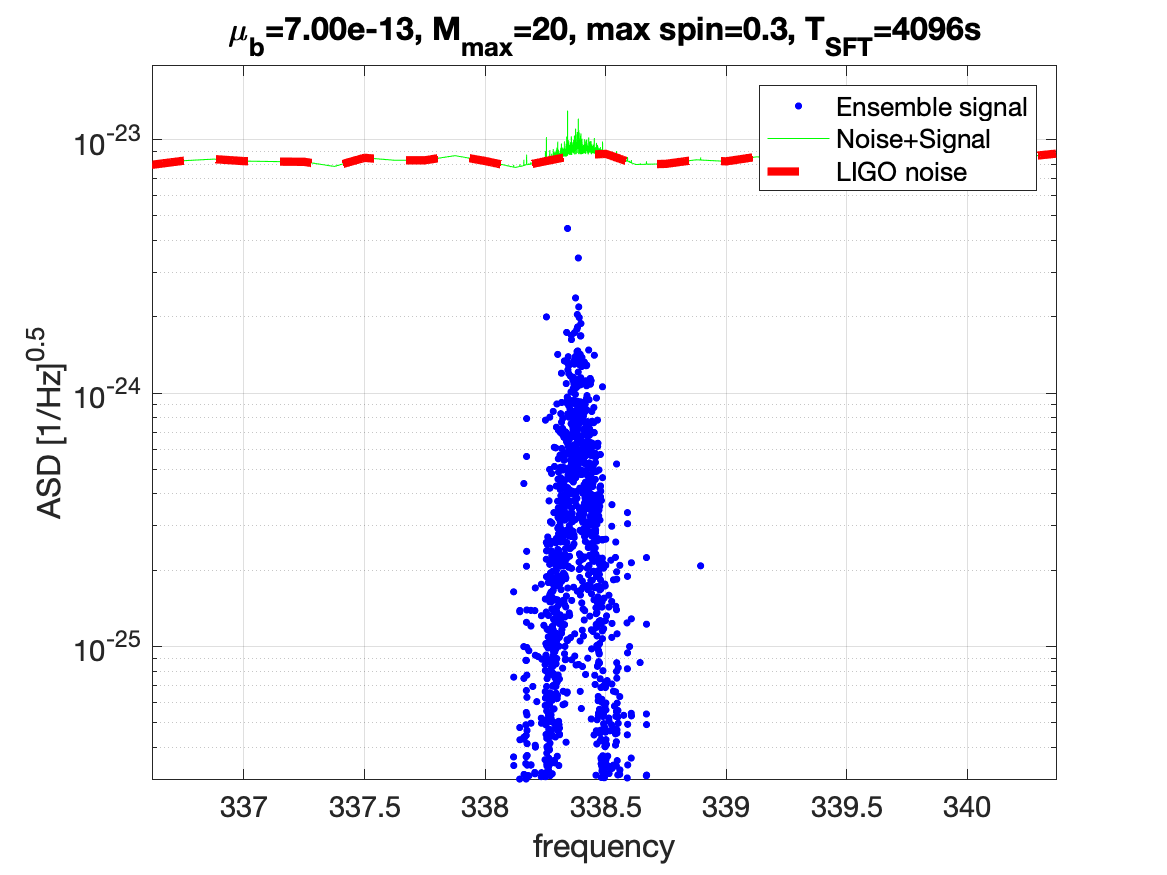}
\includegraphics[width=0.6\columnwidth]{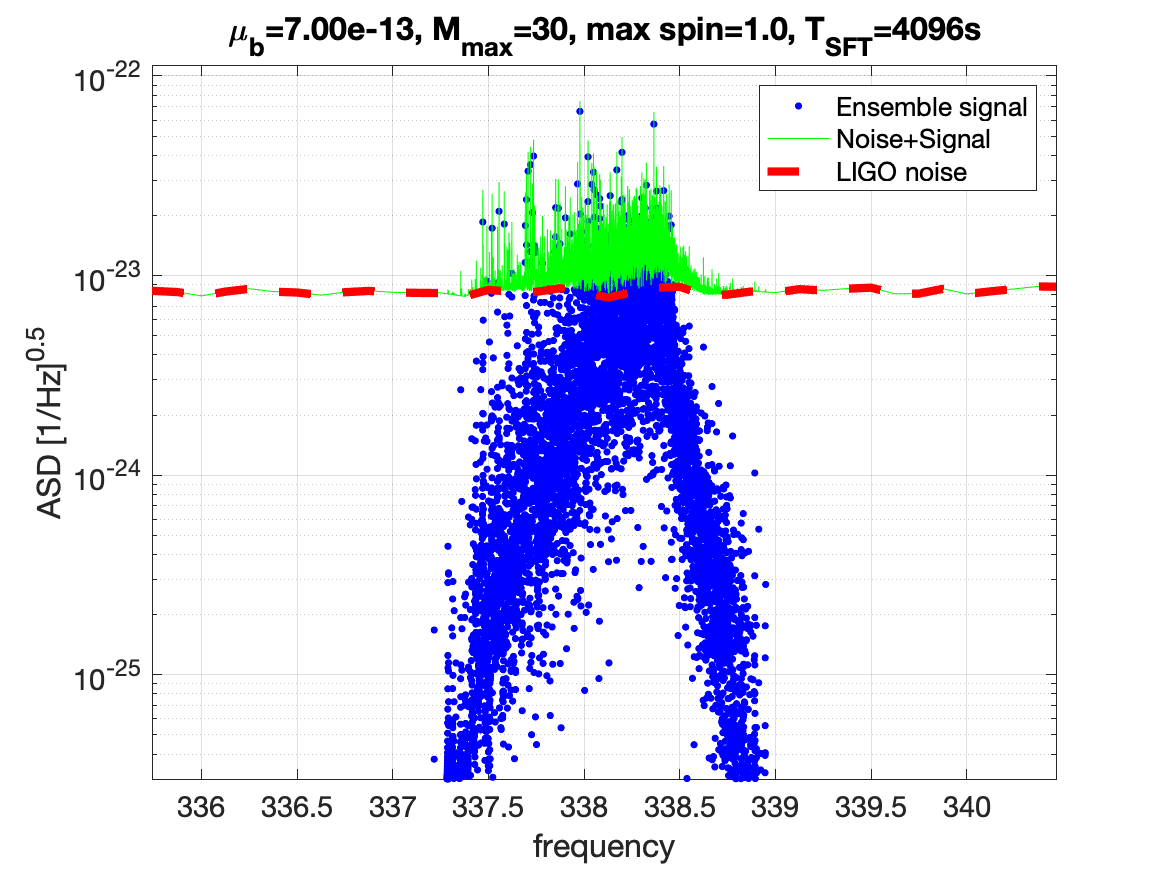}
\includegraphics[width=0.6\columnwidth]{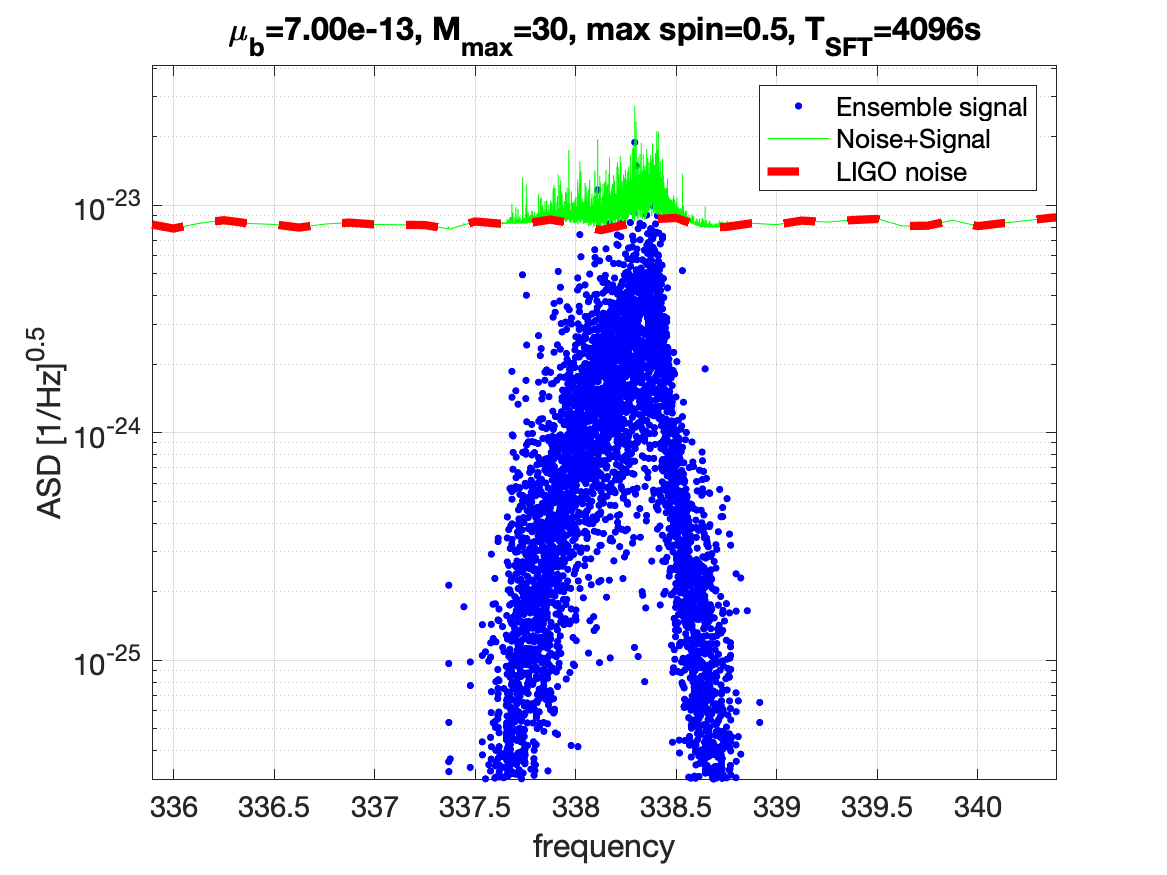}
\caption{The amplitude spectral density of the LIGO O2 data alone (dashed red line) and of an ensemble signal (top green points) assuming $\mu_b=7\times 10^{-13}$ eV and a galactic black hole population with maximum mass 20 M$_\odot$ (top plots) and max initial spin of 0.3, 0.5 and 1; and maximum mass 30 M$_\odot$ (bottom plots) and max initial spin of 0.5 and 1. The time-baseline assumed is 4096 s as used by the Freq.Hough search in this frequency range \cite{RomeBosonClouds}. \label{fig:ASD7}}
\end{figure*}

\begin{figure*}
\includegraphics[width=0.6\columnwidth]{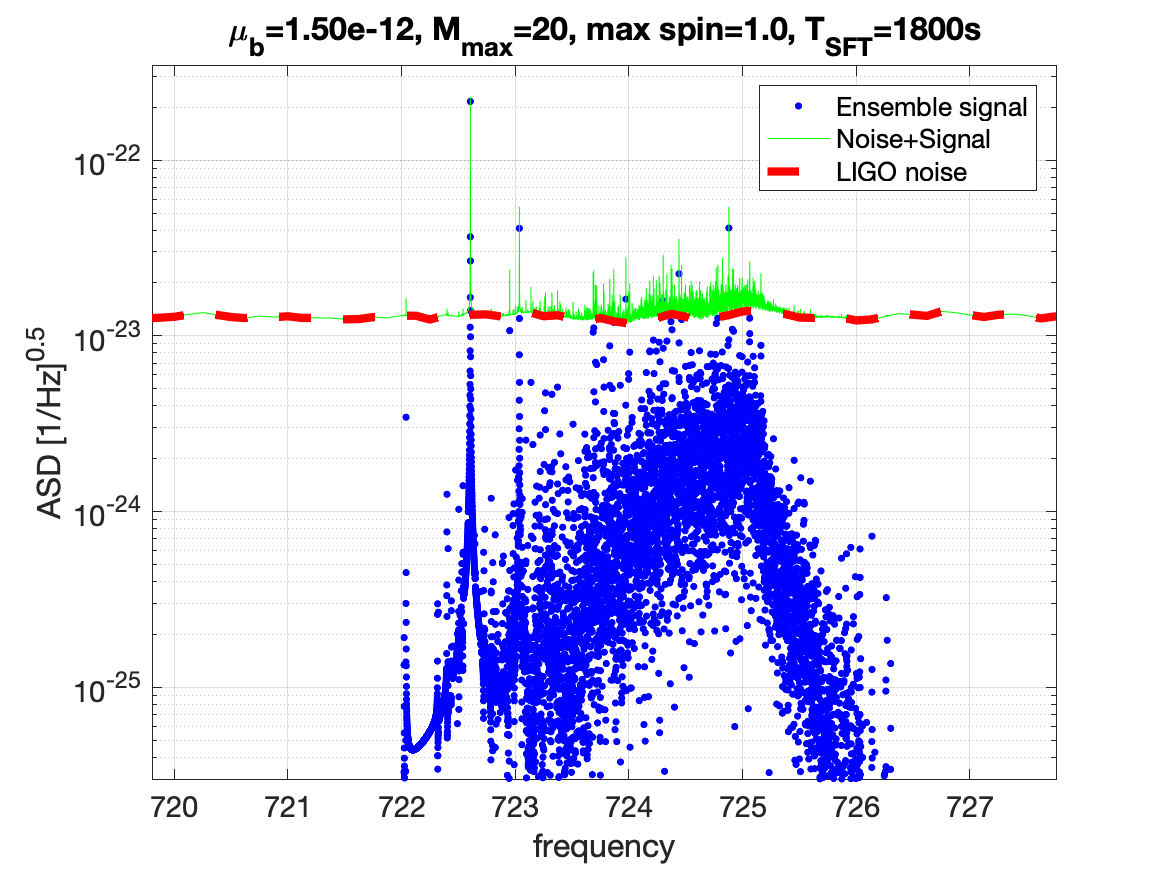}
\includegraphics[width=0.6\columnwidth]{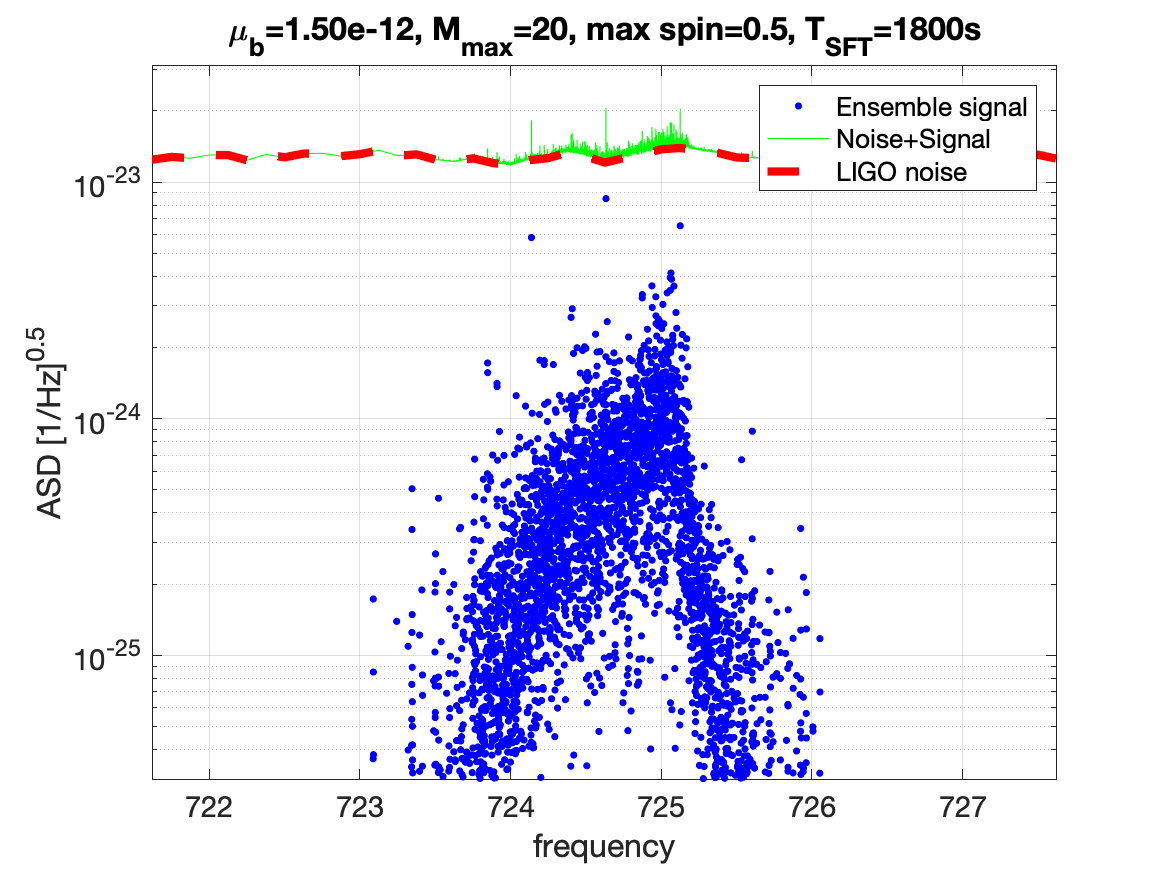}
\includegraphics[width=0.6\columnwidth]{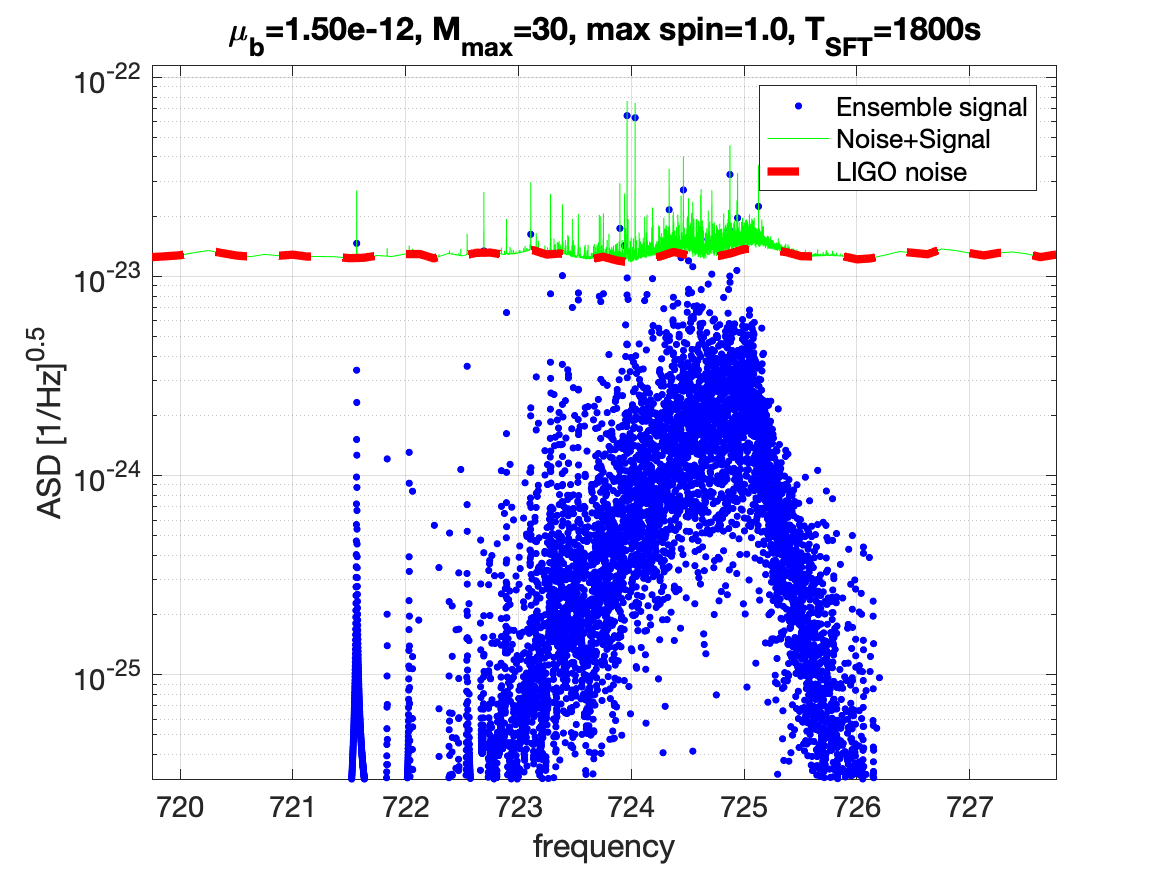}
\includegraphics[width=0.6\columnwidth]{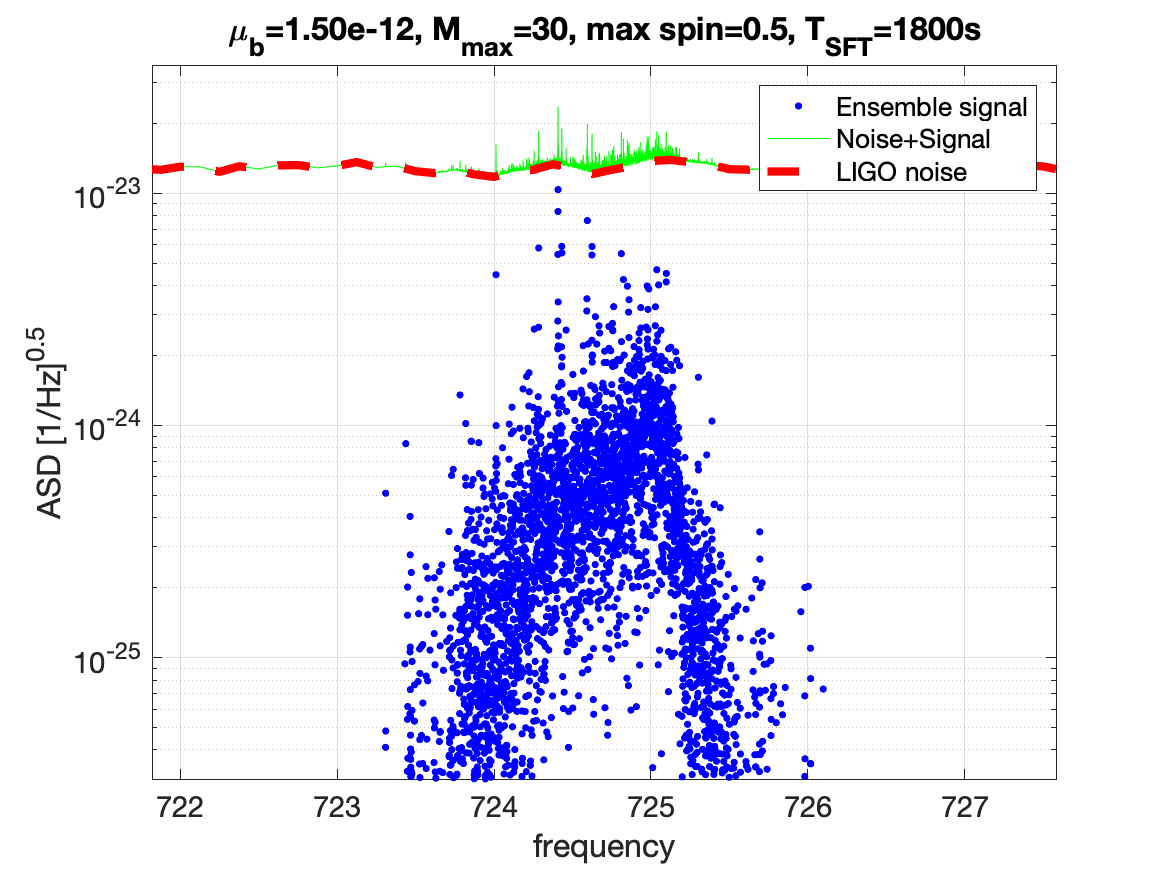}
\caption{The amplitude spectral density of the LIGO O2 data alone (dashed red line) and of an ensemble signal (top green points) assuming $\mu_b=1.5\times 10^{-12}$ eV and a galactic black hole population with maximum mass 20 M$_\odot$ or 30 M$_\odot$  and max initial spin of 0.5 and 1. The time-baseline assumed is 1800 s close to what is used by the Freq.Hough search in this frequency range \cite{RomeBosonClouds}. The distinct excess at low frequencies arises from an exceptionally young and nearby black hole. \label{fig:ASD15}}
\end{figure*}

\begin{figure*}
\includegraphics[width=0.6\columnwidth]{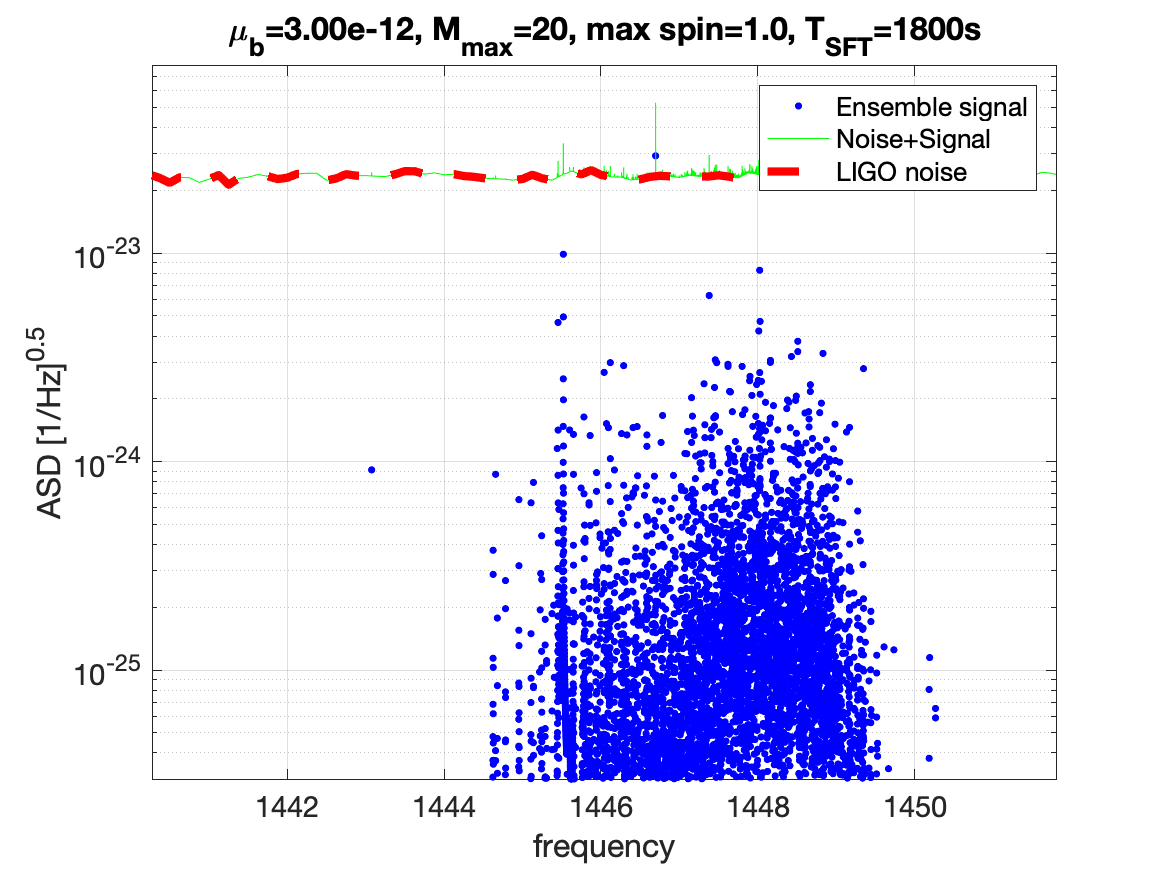}
\includegraphics[width=0.6\columnwidth]{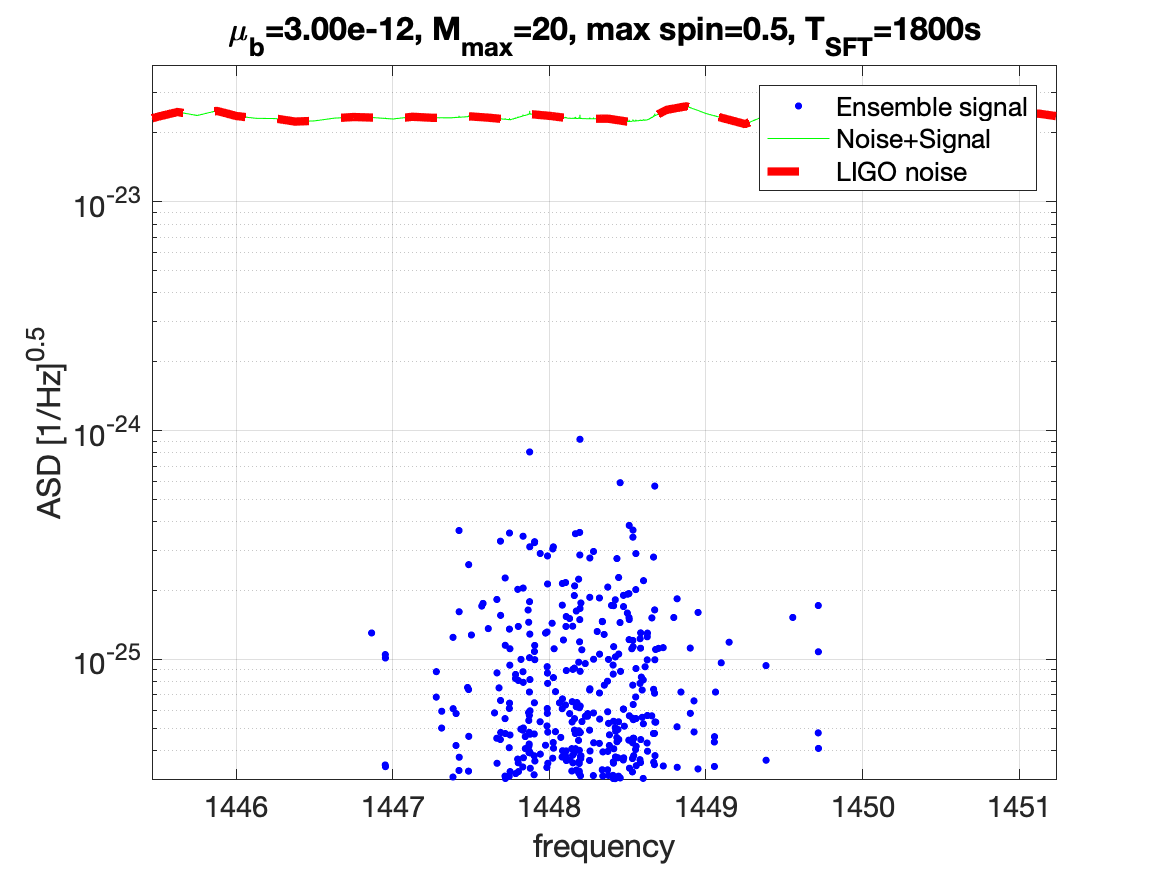}
\includegraphics[width=0.6\columnwidth]{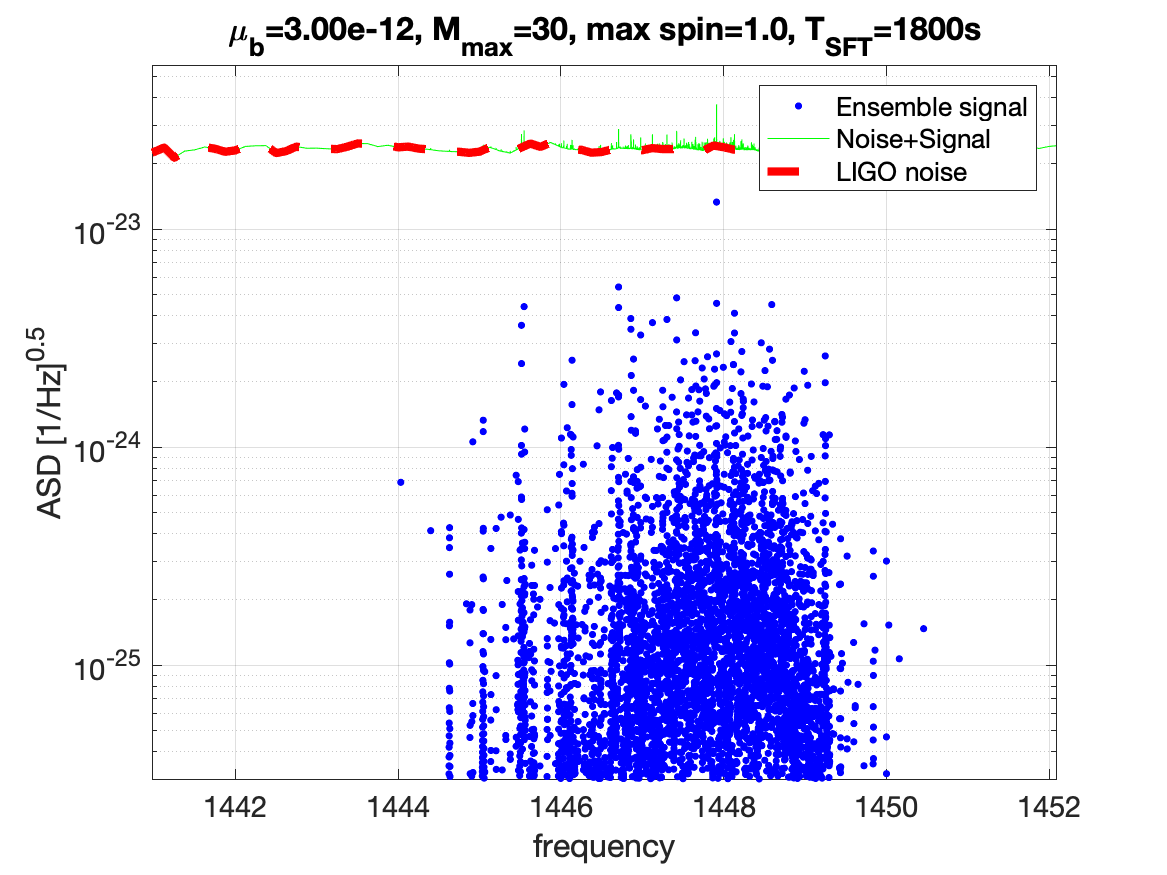}
\includegraphics[width=0.6\columnwidth]{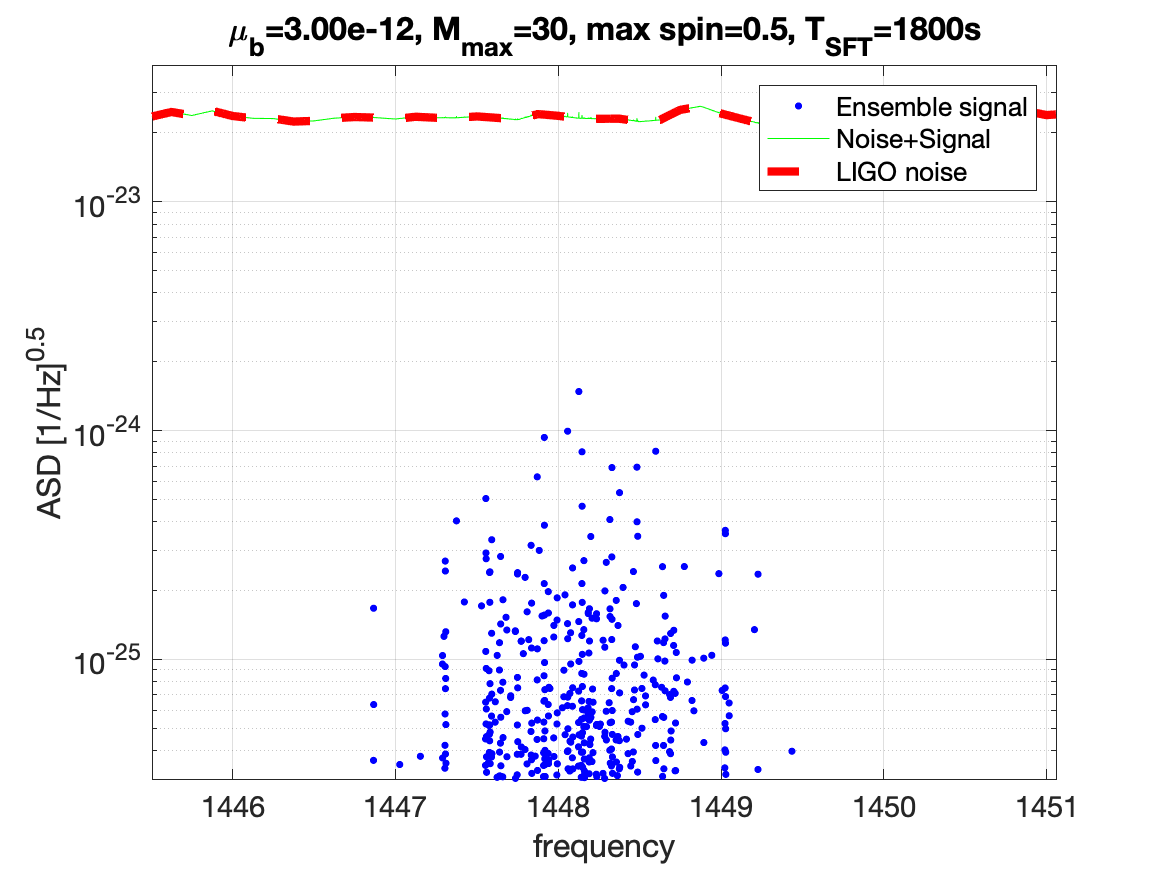}
\caption{The amplitude spectral density of the LIGO O2 data alone (dashed red line) and of an ensemble signal (top green points) assuming $\mu_b=3 \times 10^{-12}$ eV and a galactic black hole population with maximum mass 20 M$_\odot$ or 30 M$_\odot$  and max initial spin of 0.5 and 1. The time-baseline assumed is 1800 s close to what is used by the Freq.Hough search in this frequency range \cite{RomeBosonClouds}. \label{fig:ASD30}} 
\end{figure*}

As discussed in Section~\ref{sec:density}, the amplitude spectral density can be significantly altered from search expectations by the ensemble signal. Figures~\ref{fig:ASD3}-- \ref{fig:ASD30} show the expected amplitude spectral density of the ensemble signal from a Fourier transform of varying time-baseline ($T_{SFT}$), under different assumptions on the in particular the maximum black hole mass and maximum initial black hole spin. The time-baseline $T_{SFT}$ at the range of frequencies corresponding to the range of boson masses is chosen to reflect existing search strategies. The values of the boson mass, maximum black hole mass and maximum black hole spin, as well as  $T_{SFT}$  are indicated in the figure titles.
\\



\end{document}